\documentclass[a4paper,11pt]{article}

\pdfoutput=1

\usepackage[usenames,dvipsnames]{color}
\usepackage{jcappub}
\usepackage{epsfig}
\usepackage{amssymb}
\usepackage{amsmath}
\usepackage{graphicx}
\usepackage{multirow}
\usepackage{verbatim}
\usepackage{xspace}
\usepackage{placeins}
\usepackage{tabularx}

\definecolor{darkblue}{rgb}{0,0,0.5}
\definecolor{darkgreen}{rgb}{0.1,0,0.3}
\definecolor{darkred}{rgb}{0.6,0,0}

\defcitealias{GS98}{GS98}
\defcitealias{AGSS}{AGSS09}
\defcitealias{AGS05}{AGS05}

\newcommand{\nc}{\newcommand}
\newcommand{\vect}[1]{\boldsymbol{#1}}
\nc{\ba}{\begin{eqnarray}}
\nc{\ea}{\end{eqnarray}}

\newcommand\s{\sigma}

\nc{\ga}{\gamma}
\nc{\om}{\omega}

\nc{\x}{{\bf x }}
\nc{\mx}{m_\chi}
\nc{\mnuc}{m_{\rm nuc}}
\nc{\kk}{{\bf k }}
\nc{\f}{{\bf f }}
\nc{\e}{{\bf e }}
\nc{\T}{ \theta (s_i (t)- \s) }
\nc{\TT}{ \theta (s_i (t_{ r \, i } )- \s) }
\nc{\br}{   (s_i (t)- \s)  }
\nc{\gta}{\gamma \rightarrow a}
\nc{\Dag}{\Delta_{a \gamma}}
\nc{\Dosc}{\Delta_{osc}}
\nc{\Dpl}{\Delta_{pl}}
\nc{\Da}{\Delta_a}
\nc{\gag}{g_{a \gamma}}
\nc{\wpl}{\omega_{pl}}
\nc{\hr}{$h_{res}$}
\nc{\ud}{\,\mathrm{d}}
\nc{\dr}{\delta h_{res}}
\nc{\igev}{GeV$^{-1}$}
\nc{\ssi}{\sigma_{\mathrm{SI}}}
\nc{\ssd}{\sigma_{\mathrm{SD}}}
\nc{\tq}{\tilde \q}
\nc{\qmin}{q_{\mathrm{min}}}
\nc{\qmax}{q_{\mathrm{max}}}
\nc{\dmin}{\delta_{\mathrm{min}}}
\nc{\dmax}{\delta_{\mathrm{max}}}
\nc{\DS}{\textsf{DarkStec}\xspace}
\nc{\GS}{\textsf{GARSTEC}\xspace}
\nc{\dstars}{\textsf{DarkStars}\xspace}
\nc{\vrel}{v_{\rm rel}}
\nc{\qref}{q_{\rm ref}}
\nc{\sv}{\langle \sigma v \rangle_{\rm ann}}
\nc{\er}{E_\mathrm{R}}
\begin{document}

%%%%%%%%%%%%%%%%%%%%%%%%%%%%%%%%%%%%%%%%%%%%%%%%%%%%%%%%%%%%%%%%%%%%%%%
\title{Generalised form factor dark matter in the Sun}

\author[1]{Aaron C. Vincent,}
 \emailAdd{aaron.vincent@durham.ac.uk}
 \affiliation[1]{Institute for Particle Physics Phenomenology (IPPP), Department of Physics, Durham University,   Durham DH1 3LE, UK}

\author[2]{Aldo Serenelli}
 \emailAdd{aldos@ice.csic.es}
 \affiliation[2]{Institut de Ci\`encies de l'Espai (ICE-CSIC/IEEC), Campus UAB, Carrer de Can Magrans s/n, 08193 Cerdanyola del VallŽs, Spain}

\author[3]{and Pat Scott} 
 \emailAdd{p.scott@imperial.ac.uk}
 \affiliation[3]{Department of Physics, Imperial College London, Blackett Laboratory, Prince Consort Road, London SW7 2AZ, UK}

\abstract{We study the effects of energy transport in the Sun by asymmetric dark matter with momentum and velocity-dependent interactions, with an eye to solving the decade-old Solar Abundance Problem.  We study effective theories where the dark matter-nucleon scattering cross-section goes as $\vrel^{2n}$ and $q^{2n}$ with $n = -1, 0, 1 $ or $2$, where $\vrel$ is the dark matter-nucleon relative velocity and $q$ is the momentum exchanged in the collision.  Such cross-sections can arise generically as leading terms from the most basic nonstandard DM-quark operators. We employ a high-precision solar simulation code to study the impact on solar neutrino rates, the sound speed profile, convective zone depth, surface helium abundance and small frequency separations. We find that the majority of models that improve agreement with the observed sound speed profile and depth of the convection zone also reduce neutrino fluxes beyond the level that can be reasonably accommodated by measurement and theory errors. However, a few specific points in parameter space yield a significant overall improvement.  A 3--5 GeV DM particle with $\sigma_{SI} \propto q^2$ is particularly appealing, yielding more than a $6\sigma$ improvement with respect to standard solar models, while being allowed by direct detection and collider limits. We provide full analytical capture expressions for $q$- and $\vrel$-dependent scattering, as well as complete likelihood tables for all models.}

\maketitle
%%%%%%%%%%%%%%%%%%%%%%%%%%%%%%%%%%%%%%%%%%%%%%%%%%%%%%%%%%%%%%%%%%%%%%%

\section{Introduction}
A non-relativistic relic dark matter particle, with a mass of a few GeV or more, is the leading candidate to explain astrophysical and cosmological phenomena ranging from cluster kinematics and galactic rotation curves, to gravitational lensing and the heights and positions of the acoustic peaks in the cosmic microwave background. Dark matter (DM) may have been produced in a similar way to Standard Model (SM) particles, either via chemical freeze-out (as in the weakly-interacting massive particle -- WIMP -- scenario) or via an initial asymmetry, analogous to baryogenesis (as in the asymmetric DM -- ADM -- scenario).  If either of these scenarios is correct, it is possible that DM interacts weakly with SM particles. Such interactions would be seen most easily as a small elastic scattering cross-section between DM and quarks.

The search for DM-quark interactions has been the focus of terrestrial direct detection experiments such as DAMA \cite{Bernabei08}, CoGeNT \cite{CoGeNTAnnMod11}, CRESST-II \cite{CRESST11} and CDMS II \cite{Agnese:2013rvf}, who have all reported excess events above their expected backgrounds.  On the other hand, XENON10 \cite{Angle:2011th}, XENON100 \cite{XENON2013}, COUPP \cite{COUPP12}, SIMPLE \cite{SIMPLE11}, LUX \cite{LUX13} and SuperCDMS \cite{Agnese:2014aze} have all established strong limits on the DM-nucleon cross-section, seemingly in contradiction to the excesses observed by other experiments. These underground detectors are typically most sensitive to DM particles with masses of order $\mx = $ 50 to 100 GeV, which lead to the largest recoil energies on the heavy nuclei used as targets.  For the same reason, they are also best suited to probing fast-moving DM particles, resulting in a threshold velocity of a few tens of\,km\,s$^{-1}$ for an incoming DM particle to create a nuclear recoil event. 

Collisions between DM and nuclei can also lead to capture and accumulation of DM in the solar core.  For this to occur, collisions between DM and nuclei in the Sun must result in sufficient energy transfer for the DM velocity to be brought below the local escape velocity. The kinematic region probed by the Sun is quite different to the one probed by direct detection: optimal energy transfer, leading to optimal capture rates, occurs for DM particle masses closely matching the solar composition, i.e.\ a few GeV. Because DM gains speed as it falls into the solar potential well, the low-velocity tail of the local DM distribution is the dominant contributor to solar capture; the opposite is true for direct detection experiments. Direct detection and solar physics are therefore highly complementary laboratories for the study of DM-quark scattering.
 
\subsection{Dark matter in the Sun}

The effect of DM on stars has been the subject of investigation for some time (see \cite{Scott09,Turck12,Zurek13} for reviews).  There are two main scenarios where DM might create observable effects on the Sun: 1) the capture and annihilation of WIMP-like particles; and 2) the accumulation of an ``asymmetric'' species. The first case is characterised mainly by a search for high-energy neutrinos with $E_\nu \sim m_\chi$ \cite{Krauss86, Gaisser86, Griest87, Gandhi:1993ce, Bottino:1994xp, Bergstrom98b, Barger02, Desai04, Desai08, IceCube09, IceCube09_KK, IC40DM, SuperK11, IC22Methods, Silverwood12, IC79, Guo2013, Catena15b} and MeV-scale decay products \cite{Rott13, Bernal:2012qh}, using detectors such as IceCube and SuperKamiokande. Annihilation of DM in a star can also provide an energy source through the release of other SM particles \cite{SalatiSilk89, BouquetSalati89a, Moskalenko07, Bertone07, Spolyar08, Fairbairn08, Scott08a, Iocco08a, Iocco08b, Scott09, Casanellas09, Ripamonti10, Zackrisson10a, Zackrisson10b, Scott11}, leading to changes in the core structure.  The capture rate of DM in the Sun is however far too low for the energy released this way to have any significant effect on solar structure \cite{Scott09}.

Our focus in this paper is therefore on the case where DM either cannot annihilate, or does so far more slowly than it is captured.  In this case DM is effectively asymmetric (i.e.\ ADM, \cite{Petraki13, Zurek14}), and large quantities of DM may have built up inside the Sun over its lifetime.  The weak interactions of DM with quarks give the DM particles relatively large inter-scattering distances, making them potentially significant conductors of energy.  Just like the capture process, energy transfer is most efficient for lighter DM masses ($1-10$\,GeV), as momentum transfer is maximised when the masses of the colliding particles are equal, and the Sun is mostly H ($A=1$) and He ($A=4$). It is interesting to note that this is roughly the mass range expected in the most common models of ADM \cite{Kaplan:2009ag}, in order to explain the 1:5 cosmological relic baryon-to-DM density ratio. In contrast, earth-based direct detection experiments lose sensitivity in this range, as they make use of high-mass elements such as germanium ($A \sim 72$) and xenon ($A \sim 131$), for which high incoming DM velocities are necessary to create a measurable recoil.  

There is a window of elastic DM-nucleon scattering cross-sections \cite{Lopes02a,Bottino02,Cumberbatch:2010hh,Taoso10,Lopes:2012} at \mbox{$\sigma \sim 10^{-36}$} cm$^2$ for spin-independent (SI) couplings ($10^{-34}$\,cm$^2$ in the spin-dependent -- SD -- case) for which $\sigma$ is large enough to allow sizeable capture of DM, but small enough that the mean interscattering distance is still large.  A large inter-scattering distance allows heat to be redistributed away from the solar core by DM-nucleon scattering.  This has the effect of reducing the temperature of the solar core $T_{\rm c}$, and increasing the central density and pressure.  Changes of state variables near the solar centre affect the production of neutrinos from fusion processes, especially the flux of $^8$B neutrinos, which goes as $T_{\rm c}^\beta$ with $\beta \sim$ 20--25.  The temperature and density changes in the core reduce the local sound speed, and force an overall mass redistribution that impacts the solar structure at other radii.  This leads to modifications of the sound speed profile over the entire depth profile of the Sun, and shifts the height of the base of the convection zone.  Both the sound speed profile and the depth of the convection zone have been independently measured using helioseismology.

It has been shown that high-precision solar evolution models including capture, transport and (minimal) annihilation of DM can be built to satisfy the observed solar age, radius, and luminosity \cite{Taoso10,Cumberbatch:2010hh,Lopes:2012}.  At the same time, it appears possible for the inclusion of DM in such models to affect the (less well constrained) $^8$B flux in an observable way, and even improve agreement with the observed sound speed profile and the depth of the convection zone.  Neither of these latter two observables are well reproduced by standard solar models computed with the latest surface compositions \cite{Bahcall:2004yr, Basu:2004zg, Bahcall06, Yang07, Basu08, Serenelli:2009yc}.  This issue, known as the ``solar composition'' or ``solar abundance'' problem, has been brought about by the 20--25\% reduction in the measured solar metallicity in recent years \cite{APForbidO,CtoO,AspIV,AspVI,AGS05,ScottVII,Melendez08,Scott09Ni,AGSS,AGSS_NaCa,AGSS_FePeak,AGSS_heavy}, and is one of our motivations for this paper.

Although several studies have indicated that ADM can alleviate some of this tension, the cross-sections required to do so are typically far higher than allowed by limits from direct detection.  Here we investigate whether broader consideration of the kinematic structure of the DM-nucleon vertex might provide a way around this.  In the process, we provide first rigorous limits from solar physics on such DM models, which we refer to as `generalised form factor dark matter'.  In a separate paper \cite{Vincent2014} we discussed a specific realisation of momentum-dependent dark matter that leads to a 6$\sigma$ improvement over the Standard Solar Model (SSM). We revisit this model in Sec.~\ref{sec:bestcases}.

\subsection{Generalised form factor dark matter}
 
The kinematic differences between direct detection and the Sun become even more marked if the DM-nucleon interaction is not assumed to be independent of the DM-nucleon relative velocity, or the momentum transferred in the collision.  There is indeed no guarantee that the standard SI and SD operators correctly represent the DM-quark interaction. In particle physics, the interaction cross-section generally depends on the centre of mass energy and the transferred momentum, parameterised using the Lorentz-invariant Mandelstam variables $s$, $t$ and $u$. In the non-relativistic limit, these become the centre of mass momentum, proportional to the relative velocity $\vrel$, and the transferred momentum $q = \Delta p$. As these are small quantities, the constant term usually dominates in a series expansion of the cross-section. However, many models with non-trivial dependencies on $\vrel$ and $q$ exist, typically motivated by theoretical arguments or attempts to reconcile experimental results. 

To make quantitative predictions, a specific form of $\sigma(\vrel,q)$ must be chosen.  Because we wish to remain as general as possible, we choose to focus on couplings of the form $\sigma \propto \vrel^{2n}$ and $\sigma \propto q^{2n}$, with $n = \{-1,1,2\}$. The $\vrel^2$ and $\vrel^4$ forms are respectively called $p$-wave and $d$-wave interactions, and correspond to the cases where the initial state particles possess 1 and 2 units of relative angular momentum, respectively.  These are always present, but normally only dominate when all lower-order terms in the scattering matrix element -- including the constant ($s$-wave) term -- are suppressed due to cancellations. Cross-sections depending on $q$ can arise, for example, from a non-zero particle radius (the analogue of a nuclear form factor), from parity-violating couplings like $\bar \chi \gamma_5 \chi \bar Q Q$, $\bar \chi \gamma_\mu \gamma_5 \chi \bar Q \gamma^\mu Q$ and $\bar \chi \sigma_{\mu\nu}\gamma_5 \chi \bar Q \sigma^{\mu\nu} Q$, or from a small anapole or dipole interaction between the dark and visible sectors \cite{Pospelov00,Sigurdson04,Chang:2009yt,Feldstein:2009tr,Feldstein10,Chang:2010en,Fitzpatrick10,Barger11,Frandsen:2013cna,DelNobile:2013cva,DelNobile:2014eta,KKS,Catena14}.  We refer to the class of models where DM-nucleon scattering cross-sections depend on some combination of $q$ and/or $\vrel$ as `generalised form factor DM' because it generalises the effects of form factors to arbitrary powers of $q$ and $\vrel$.

Concretely, we focus on the couplings\vspace{1mm}\\
\begin{subequations}
\label{qdepvdep}
\begin{minipage}{0.4\linewidth}
  \begin{equation}
    \label{qdep}
    \sigma = \sigma_0 \left(\frac{q}{q_0}\right)^{2n}
  \end{equation}
\end{minipage}
\begin{minipage}{0.18\linewidth}
  \flushright and
\end{minipage}
\begin{minipage}{0.4\linewidth}
  \begin{equation}
    \label{vdep}
    \sigma = \sigma_0 \left(\frac{\vrel}{v_0}\right)^{2n}.
  \end{equation}
\end{minipage}
\end{subequations}\vspace{0.5mm}\\
These can lead to either spin-dependent (SD) or spin-independent (SI) interactions, depending on the axial structure of the DM-nucleon interaction vertex. The normalisation $\sigma_0$ must be defined with respect to some reference velocity $v_0$ or momentum $q_0$. We will choose $v_0 = 220$\,km\,s$^{-1}$, the typical halo DM velocity, and $q_0 = 40$ MeV, corresponding to a nuclear recoil energy of around 10 keV in an underground direct detection experiment.

The DM-\textit{nucleus} cross-section is related to the above DM-nucleon cross-sections via:
\begin{equation}
\sigma_{N,i} = \frac{\mnuc^2 (\mx + m_p)^2}{m_p^2(\mx + \mnuc)^2}\left[\sigma_{\rm SI} A_i^2 + \sigma_{\rm SD}\frac{4(J_i +1)}{3J_i}|\langle S_{p,i}\rangle +\langle S_{n,i}\rangle |^2 \right],
\label{DMnucleus}
\end{equation}
where $A_i$ and $J_i$ are respectively the atomic number and total angular momentum of nuclear species $i$;  $\langle S_{p,i}\rangle$ and $\langle S_{n,i}\rangle$ are the spin expectation values of its proton and neutron systems. Given the $A^2$ dependence of Eq.\ \ref{DMnucleus} we will find that, in spite of the Sun's small metallicity, spin-independent DM can have a significantly larger effect than a spin-dependent DM candidate, which couples mostly to hydrogen.\footnote{See also caveats to this treatment in Sec. \ref{sec:FFsection}. }  Beyond a few studies, e.g. \cite{Ellis09}, this fact has not been emphasised very much in the literature. The full effect of a momentum-dependent cross-section will furthermore depend crucially on the composition of heavier elements. 

The full impact of momentum and velocity-dependent DM on solar observables has not been studied before. The authors of Ref.~\cite{Guo2013} computed the effect of such couplings on the capture rate of spin-dependent DM, and computed neutrino fluxes from an annihilating species. However, they did not include a treatment of heavier elements, nor of the crucial energy transport by DM once captured.  In Refs.~\cite{Lopes:2014,Lopes14}, the authors computed capture and transport rates for ADM models with long-range interactions.  To account for the impacts of the non-trivial scaling of these cross-sections with momentum and velocity, they employed effective cross-section scaling factors to decouple the cross-sections entering the capture and transport calculations, and avoid modifying the standard velocity-and-momentum-independent treatment of capture and energy transport.  As we show later, it happens that this rescaling can indeed be done without any loss of generality for capture in the velocity-dependent case, but it is not possible in the momentum-dependent case.  It is also not possible to account for the effects on energy transport of either a velocity or momentum dependence in this manner; rather, a full recalculation of the transport coefficients must be performed \cite{VincentScott2013}.

The structure of this paper is as follows: in Section \ref{sec:cap}, we review the capture equations for DM in the Sun, and present the necessary modifications for velocity and momentum-dependent scattering of DM with nucleons.  In Section \ref{sec:transport} we review the theory of conductive heat transport by DM developed in Refs.\ \cite{GouldRaffelt90a,VincentScott2013}, along with its application to solar modelling.  Section \ref{sec:code} describes the \DS computer code that we have developed for simulating the effects of generalised form factor DM on the Sun.  We present results in Section \ref{sec:results}, and discuss their implications with regards to the Solar Abundance Problem, current experimental limits and previous work on the topic in Section \ref{sec:discussion}. We summarise in Section \ref{sec:conclusion}.

\section{Capture of dark matter by the Sun}
\label{sec:cap}

\subsection{Standard (velocity and momentum independent) treatment}

The population of DM particles in the Sun $N_\chi(t)$ is follows the differential equation
\begin{equation}
\frac{d N_\chi(t)}{dt} = C_\odot(t) - A(t) - E(t),
\label{DMpopEqn}
\end{equation}
where $C_\odot(t)$ is the capture rate, $A(t)$ is rate at which annihilations occur and $E(t)$ represents evaporation.  Unless DM is strongly self-interacting (not the case we consider here, but discussed in Ref. \cite{Albuquerque2013}), $C_\odot(t)$ does not depend on the DM population already captured by the Sun.  $A(t)$ is the rate of annihilation, and is proportional to  the square of the DM population. Here we consider the case where DM is fully asymmetric, so $A(t)=0$, although we comment briefly in Section \ref{sec:annihilation} on the implications of allowing DM to  self-annihilate.

Evaporation occurs when a DM particle gains enough energy from a scattering  event to overcome the Sun's gravitational potential and escape, so $E(t)$ is linear in the DM population (being simply the product of the single-particle evaporation probability and the number of candidates for evaporation).  Evaporation requires a significant gain in momentum, as the typical velocity of a thermalised DM particle is $\sim$$100$\,km\,s$^{-1}$, whereas the escape velocity in the solar core approaches 1400\,km\,s$^{-1}$. This means that evaporation is only significant if DM is similar in mass to the nucleus with which it scatters in the Sun. In practice this means that DM about the mass of helium (4\,GeV) and lighter is typically most prone to evaporation, but other masses closely matched to significant elements in the Sun (C, N, O, Fe $\implies m_\chi \sim 12$, 14, 16, 56\,GeV) can also in principle be affected \cite{Gould87a}. The evaporation rate depends on the interaction cross-section, mean free path and thermal regime (LTE vs Knudsen); extending the standard analyses to generalised form factor DM is therefore non-trivial.  In practice the rate of evaporation is extremely low in almost all cases where the nuclear scattering cross-section is allowed by direct detection; for very specific analyses it should be taken into account, but for the purposes of this paper we assume that the evaporation rate is zero.  We intend to return to this point in detail in a future paper.

We now turn to the capture rate as it is implemented in our simulations. As we are using the \dstars code \cite{Scott09b}, we closely follow Refs. \cite{Scott09,Gould87b}. We take the local distribution function $f(u)$ of DM to be Maxwell-Boltzmann, with dispersion $u_0 = 270$\,km\,s$^{-1}$. In the frame of the Sun, moving at $u_\odot=220$\,km\,s$^{-1}$ relative to the Galactic rest frame,
\begin{equation}
f_\odot(u) = \left(\frac32\right)^{3/2}\frac{4\rho_\chi u^2}{\pi^{1/2}m_\chi u_0^3}\exp\left(-\frac{3 (u_\odot^2+u^2)}{2u_0^2} \right) \frac{\sinh (3 u u_\odot/u_0^2)}{3 u u_\odot /u_0^2}.
\end{equation}
As it falls into the gravitational potential well of the Sun, a DM particle acquires a velocity $w = \sqrt{u^2 + v_{\rm esc}^2(r,t)}$.  It becomes gravitationally captured by the Sun if it loses enough kinetic energy in a scattering event for $w$ to fall below the local escape velocity $v_{\rm esc}(r,t)$. For this to occur, the fractional energy lost by the DM particle $\Delta=2\er/m_\chi w^2$, corresponding to nuclear recoil energy $\er$, must be in the interval
\begin{equation}
 \frac{u^2}{w^2} \leq \frac{2\er}{\mx w^2} \leq \frac{\mu}{\mu_+^2},
 \label{scatint}
\end{equation}
where $\mu \equiv \mx/\mnuc$ and $\mu_\pm \equiv (\mu \pm 1)/2$. The local capture rate of particles with velocity $w$ is the sum over nuclear species at radius $r$, of the rate of scattering in the interval of Eq.\ \ref{scatint}, i.e.
\begin{equation}
 \Omega(w) = \frac{2}{m_\chi w} \sum_i \sigma_{N,i} n_i(r,t) \frac{\mu_{i,+}^2}{\mu_i}\Theta\left(\frac{\mu_i v_{\rm esc}^2}{\mu_{i,-}^2} - u^2 \right) \int_{\mx u^2/2}^{\mx w^2 \mu_i/2\mu_{i,+}^2} |F_i(\er)|^2 \ud \er,
 \label{eq:omega}
\end{equation} 
where $|F_i(\er)|^2$ is the nuclear form factor. For hydrogen this is a constant, whereas for heavier nuclei we use the usual Helm form factor
\begin{equation}
|F_i(\er)|^2 = \exp{\left(-\frac{\er}{E_i}\right)}.
\label{heavyformfactor}
\end{equation}
Here $E_i$ is a constant quantity for each nuclear species $i$, given by
\begin{equation}
E_i = \frac{5.8407 \times 10^{-2}}{m_{N,i} (0.91 m_{N,i}^{1/3} + 0.3)^2} \mathrm{GeV}.
\end{equation}
Integrating over the phase space, the total capture rate of DM in the Sun is then
\begin{equation}
C_\odot(t) = 4\pi \int_0^{R_\odot} r^2 \int_0^\infty \frac{f_\odot(u)}{u} w \Omega(w) \ud u\ud r.
\label{caprate}
\end{equation}
For a constant cross-section and a Maxwell-Boltzmann velocity distribution, this can be solved analytically \cite{Scott09,Gould87b}. However, in the following we generalise the above equations to allow momentum and velocity-dependent $\sigma_{\rm SI}$ and $\sigma_{\rm SD}$.\footnote{We will use $\sigma$ as a shorthand for either $\sigma_{\rm SI}$ or $\sigma_{\rm SD}$, as the spin and kinematic dependence can be factorised. For an explicit treatment, see \cite{Catena:2015uha}.} In this case, numerical integration becomes necessary.   

\subsection{Velocity and momentum-dependent treatment}
\begin{figure}
\begin{tabular}{c c}
\includegraphics[width=0.5\textwidth]{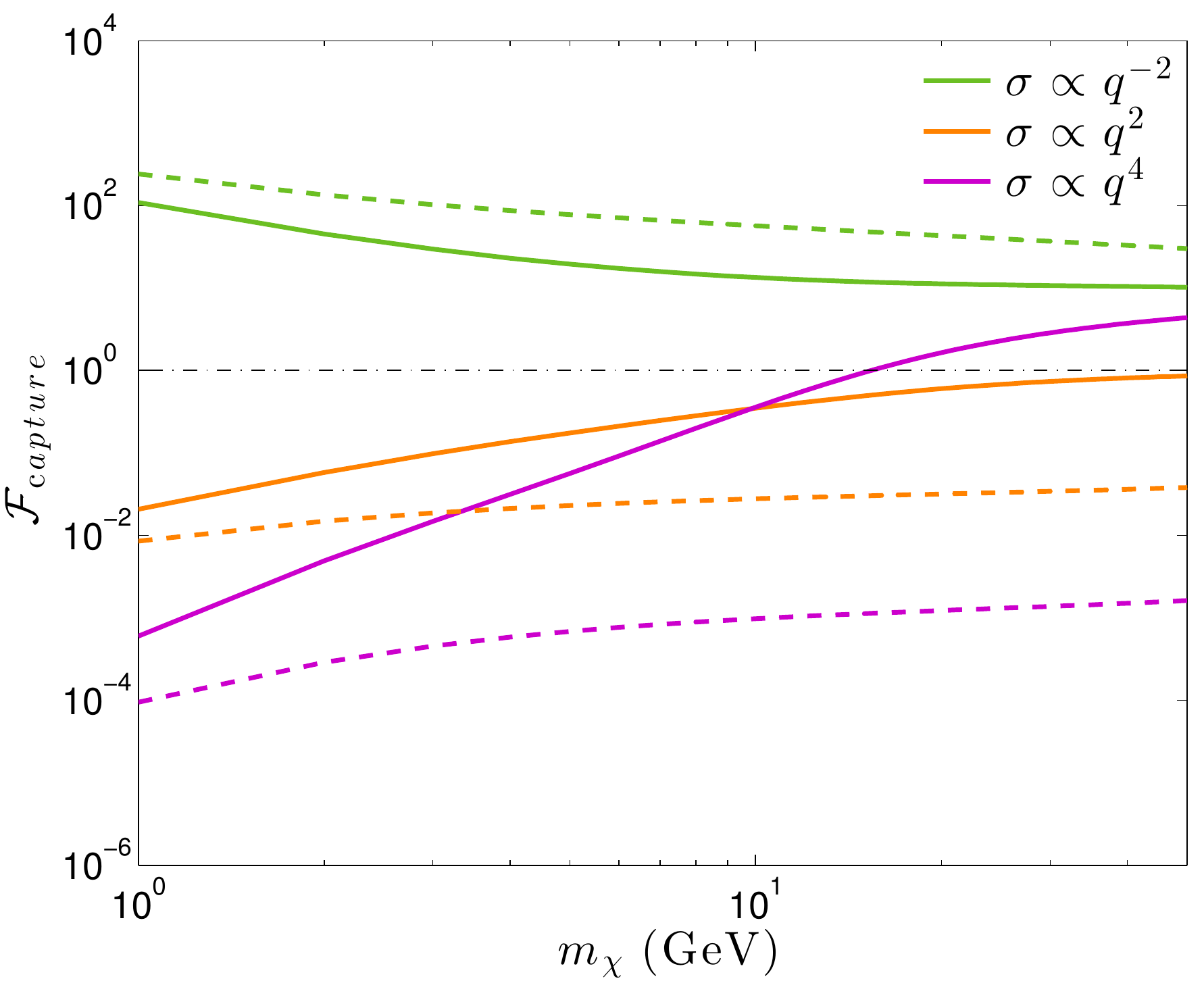} & \includegraphics[width=0.5\textwidth]{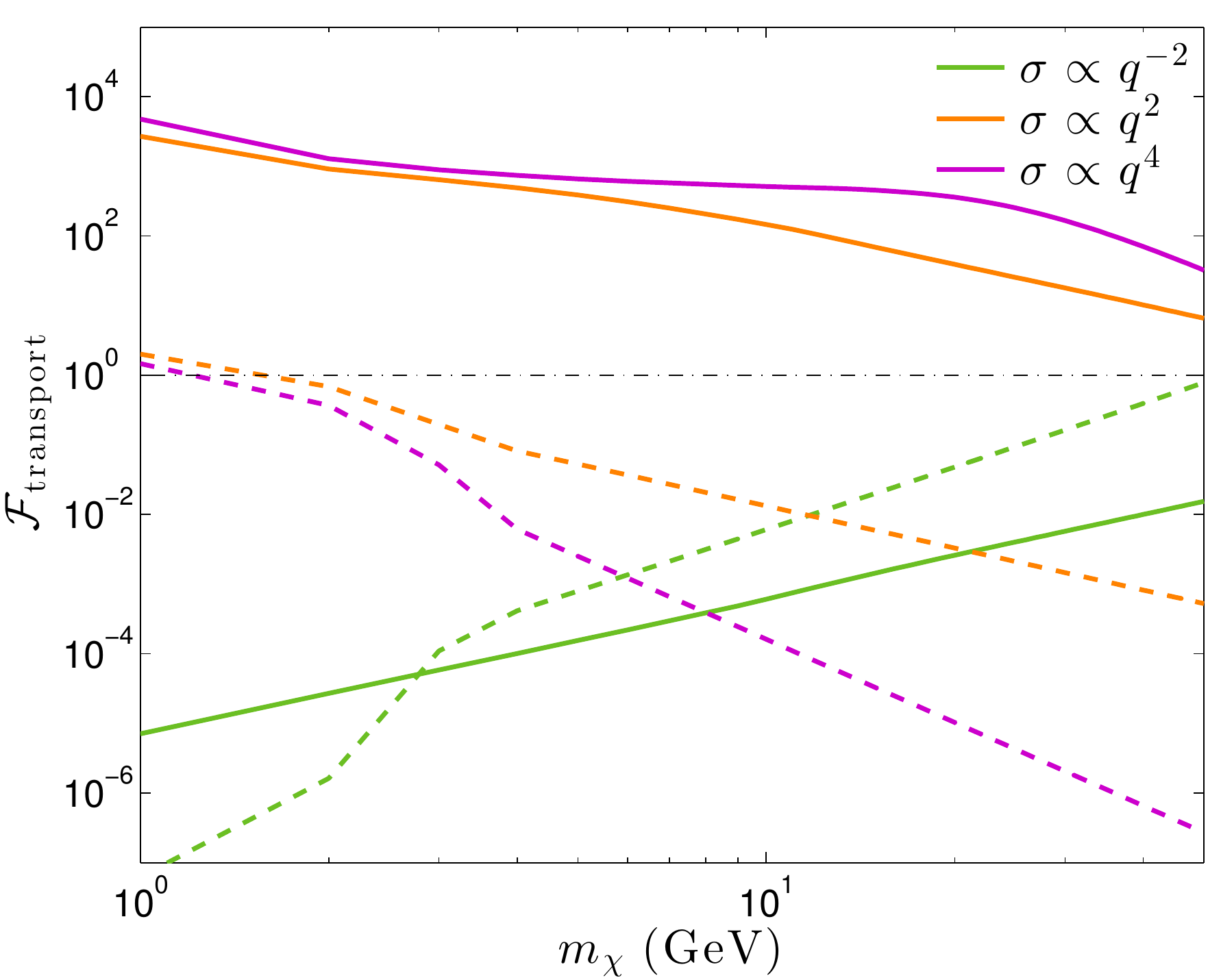} 
\end{tabular}
\caption{Left: effective  enhancement/suppression of DM capture in a standard model of the Sun (AGSS09ph; \cite{Serenelli:2009yc}), due to a momentum-dependent cross-section, as computed with Eq.\ \ref{effcap}. The black line at $\mathcal F = 1$ corresponds to $\sigma = const.$ Solid lines are spin-independent (SI) and dashed lines are spin-dependent (SD), coupling only to hydrogen. Right: effective enhancement/suppression of energy transport as defined in Eq.\ \ref{trFdef} , in the LTE regime, with $\sigma_0 = 10^{-35}$ cm$^2$. We note that away from LTE, the behaviours reverse, and a $q^{-2}$ cross-section actually yields an enhancement with respect to the constant case. This can be seen in Fig.~\ref{fig:knudsen}. The full effect of a $q$-dependent cross-section is then the combined effect of the left and right-hand panels.}
\label{capfig}
\end{figure}

We begin with a momentum-dependent cross-section of the form Eq.\ \ref{qdep}, with $\sigma \propto q^{2n}$. To include such a cross-section in Eq.\ \ref{eq:omega}, the constant cross-section must be replaced with Eq.\ \ref{qdep} and the dependence on the nuclear recoil energy moved inside the form-factor integral.  This explicitly illustrates the equivalence of a momentum-dependent cross-section to a change in form factor, and we refer to the corresponding integral and associated multiplicative factors as the `generalised form factor integral' ($GFFI$).  For hydrogen, $|F(\er)|^2 = 1$, so the change required is
\begin{equation}
  \label{ffHintermediate} 
  GFFI_{n=0} = \int |F(\er)|^2 \ud \er \ \rightarrow\ GFFI_{n\ne0,\mathrm{H}} = \left(\frac{p}{q_0}\right)^{2n} \frac{m_\chi w^2}{2\mu^n}\int_{u^2/w^2}^{\mu/\mu_+^2} \Delta^n \ud \Delta,
\end{equation} 
where we have expressed the form factor integral in terms of $\Delta$ instead of $\er$ for compactness.  We do this by noting that the transferred momentum is $q = \sqrt{2m_{\rm nuc}\er}$, so the fractional energy change can be written $\Delta = \mu q^2 / p^2$, where $p = m_\chi w$ is DM particle's incoming momentum.  Performing the integral in Eq.\ \ref{ffHintermediate} yields
\begin{equation}
GFFI_{n\ne0,\mathrm{H}} = \left(\frac{p}{q_0}\right)^{2n}\frac{m_\chi w^2}{2\mu^n}
\begin{cases}
 \frac{1}{1+n}\left[\left(\frac{\mu}{\mu_+^2}\right)^{n+1} - \left(\frac{u^2}{w^2}\right)^{n+1}\right],         &(n \neq -1) \\
 \ln \left(\frac{\mu }{\mu_+^2}\frac{w^2}{u^2}\right),        &(n = -1) \\ 
\end{cases}
\label{ffHresult}
\end{equation}
For heavier elements, the integrand includes the Helm form factor (Eq.\ \ref{heavyformfactor}), so
\begin{equation}
GFFI_{n\ne0,i\ne\mathrm{H}} = \left(\frac{p}{q_0}\right)^{2n}\frac{E_i}{(B \mu)^n}\left[ \Gamma\left(1+n,B\frac{u^2}{w^2}\right) - \Gamma\left(1+n,B\frac{\mu}{\mu_+^2}\right) \right],
\label{ffZresult}
\end{equation}
where $B\equiv \frac12 \mx w^2 / E_i$, and $\Gamma(m,x)$ is the (upper) incomplete gamma function. To gain a more intuitive understanding of Eqs. \ref{ffHresult} and \ref{ffZresult} we show in Fig.\ \ref{capfig} the overall enhancement or suppression to the capture rate with respect to the constant case,
\begin{equation}
\mathcal{F}_{\rm capture} = \frac{\sum_i f_i A_i^2 GFFI_{n\ne0,i}}{  \sum_i f_i A_i^2 GFFI_{n=0,i}},
\label{effcap}
\end{equation}
for three values of $n\neq 0$.  In this example we just take $w = w(r = 0)$ and use the present-day AGSS09ph Standard Solar Model\footnote{Publicly available at \url{http://www.ice.csic.es/personal/aldos/Solar_Models.html}.} \cite{AGSS,Serenelli:2009yc}. Here $f_i$  represents the fractional composition of each species with atomic number $A_i$.  For the spin-dependent cross-sections, the sum over the species $i$ only includes hydrogen.  Of course the actual capture rate must be accurately integrated over the entire star, and we  compute it precisely with Eq.\ \ref{caprate} in our simulation.   In each case, the overall behaviour roughly follows the $(p/q_0)^{2n}$ dependence, as heavier DM will have gained more momentum as it falls into the solar gravitational well.

In the velocity-dependent case (Eq.\ \ref{vdep}), where $\sigma \propto \vrel^{2n}$, the modification to the capture rate is much simpler. The partial capture rate $\Omega(w) \propto \sigma$ is simply transformed by the replacement
\begin{equation}
\label{vresult}
\sigma \rightarrow \sigma_0 \left(\frac{w}{v_0}\right)^{2n}.
\end{equation}
The integral Eq.\ \ref{caprate} can then be evaluated to obtain the modified capture rate.  We later show with our solar simulation code that the enhancement due to velocity-dependence agrees very well with what is obtained using the standard ($n=0$) treatment and an average of the cross-section throughout the volume of the star
\begin{equation}
\sigma(\vrel) \simeq \langle \sigma(w_0) \rangle = \frac{1}{M_\odot}4 \pi \int_0^{R_\odot} \sigma(w_0) \rho_\odot(r)r^2  \ud r,
\end{equation}
where $w_0 \equiv \sqrt{v_0^2 + v_{\rm esc}^2(r)}$ is the typical velocity acquired by an in-falling WIMP as it reaches radius $r$. This indicates that a cross-section proportional to $v^2$ or $v^4$ -- typically thought of as a suppression --  actually yields a considerable enhancement.  We note that $\sigma(\vrel)$ has been erroneously approximated to $\sigma(\langle w_0 \rangle)$ rather than $\langle \sigma(w_0) \rangle$ in the literature \cite{Lopes14}, although $\left[v_{\rm esc}(r=0)/v_0\right]^{2n}$ does at least provide the correct enhancement to the capture rate to within a factor of a few.

\subsubsection{Form Factors}
\label{sec:FFsection}
We finally comment on the use of the Helm form factor \ref{heavyformfactor} in our capture equations. Very recent work \cite{Catena:2015uha} has shown that a more accurate computation of the nuclear response functions to DM-nucleus scattering can have a substantial impact on the capture rate. These response functions amount to corrections to the Helm form factor by terms proportional to powers of $y = E_R/2E_i$. These will dominate at large momentum transfers. We have checked that in most cases considered here -- and in all cases where the effect of DM is large enough to modify the solar observables -- this quantity is much smaller than one. This is because the low DM masses that yield the largest conduction effects are mainly sensitive to the leading, constant term in the NR expression. However, we caution the reader that accurate modification of the capture rate for DM with larger masses than those considered here, should make use of the full nuclear response functions detailed in Ref. \cite{Catena:2015uha}.

We have also used the standard approach for spin-dependent capture in the Sun, wherein it is assumed that hydrogen is the dominant contributor to the capture rate. It was shown in \cite{Catena:2015uha} that, while this is strictly true for velocity-dependent cross sections (non-relativistic operator $ \mathcal{O}_7$) and holds to a few percent accuracy for a constant cross section ($ \mathcal{O}_4$), momentum-dependent interactions can lead to a different behaviour. In the $q^4$ case ($\mathcal{O}_6$), nitrogen dominates capture by an order of magnitude or more for all DM masses. The net effect is that the spin-dependent capture rates computed here for $q^4$ have been underestimated with respect to the case when the full nuclear response functions are included.

\subsection{The geometric limit}

In all cases, the total effective cross-section of the Sun to collisions with DM particles cannot exceed the geometric ``cutoff''
\begin{equation}
\sigma_\mathrm{max} = \pi R_\odot^2(t), 
\label{eq:sigmacutoff}
\end{equation}
which corresponds to the case where the Sun is optically thick to DM.  This places a fundamental limit on the capture rate
\begin{eqnarray}
  C_\mathrm{max}(t) &=& \pi R_\odot^2(t) \int_0^\infty \frac{f_\odot(u)}{u}w^2(u,R_\odot) \ud u \label{eq:capcutoff} \\
  &=& \frac{1}{3}\pi\frac{\rho_\chi}{m_\chi}R_\odot^2(t)\left(e^{-\frac{3}{2}\frac{u_\odot^2}{u_0^2}}\sqrt\frac{6}{\pi}u_0 + \frac{6G_{\rm N}M_\odot + R_\odot(u_0^2 + 3 u_\odot^2)}{R_\odot u_\odot}\mathrm{Erf}{\left[\sqrt{\frac{3}{2}}\frac{u_\odot}{u_0}\right]} \right). \nonumber
\end{eqnarray}
The actual capture rate to be used must therefore be the lesser of Eqs.\ \ref{eq:capcutoff} and \ref{caprate}.  Assuming a steady radius and a local DM density of $\rho_\chi \simeq 0.38$ GeV cm$^{-3}$, the maximum amount of DM that can accumulate in the Sun by its current age is therefore 
\begin{equation}
\label{nmax}
N_{\chi, \mathrm{max}} \simeq  1.5 \times 10^{47} \left(\frac{\rm GeV}{\mx}\right) \simeq 1.3 \times 10^{-11} \left(\frac{\rm GeV}{\mx}\right) N_{\rm b},
\end{equation}
where $N_{\rm b} \simeq M_\odot/m_p$ is the total number of baryons in the Sun.

\section{Conductive energy transport by dark matter}
\label{sec:transport}
If enough DM is captured by the Sun, its large typical inter-scattering distance $l_\chi$ means that it is a more efficient carrier of heat over long distances than ordinary baryonic material, so it can act as an additional mechanism for heat transport alongside photons. The formalism we use to compute the effect of microscopic energy transport by DM conduction was developed by Gould and Raffelt \cite{GouldRaffelt90a}. By solving a perturbative expansion of the Boltzmann collision equation (BCE) in the Sun's gravitational potential, they showed that the thermal conduction by a weakly-interacting species can be expressed in terms of two quantities: a dimensionless molecular diffusivity $\alpha(\mu)$, and thermal conductivity $\kappa(\mu)$, where $\mu \equiv \mx/\mnuc$ is again the ratio between the DM and nucleon masses. If more than one nuclear species is present, $\alpha$ and $\kappa$ get replaced with effective values, which are weighted by the number densities of each species in the plasma mixture at each height in the Sun. 

In the local thermal equilibrium (LTE) regime, where $l_\chi (r)$ is much smaller than both the inverse of the local temperature 
gradient $|\nabla \ln T(r)|$ and the DM scale height $r_\chi$, the values of $\alpha$ and $\kappa$ are found by solving the first order expansion of the BCE.  This is done in terms of the quantity $\varepsilon \equiv l_\chi(r) |\nabla \ln T(r)|$, via the formal inversion of the Boltzmann collision operator $C(\vect{u},\vect{r})$, where $C(\vect{u},\vect{r})F(\vect{u},\vect{r})$ represents the net change in the DM phase space distribution $F(\vect{u},\vect{r})$ due to collisions with nuclei. The collision operator is defined via phase space integrals over collision rates, meaning that the dependence of the collisional cross-section $\sigma$ on $\vrel$ and $q$ must be explicitly included in the calculation. In Ref. \cite{GouldRaffelt90a} Gould and Raffelt computed and tabulated $\alpha(\mu)$ and $\kappa(\mu)$ for a constant scattering cross-section ($n=0$). In Ref. \cite{VincentScott2013}, we extended the formalism to include velocity and momentum-dependent cross-sections ($n\ne0$), showing that the resulting changes in $\alpha$, $\kappa$ and $l_\chi$ can have potentially large effects on heat transfer in the Sun. 

The equilibrium distribution of DM particles in the gravitational potential $\phi(r)$ of the sun is given by \cite{GouldRaffelt90a,Scott09,VincentScott2013}
\begin{equation}
\label{LTEdens}
    n_{\chi,{\rm LTE}}(r) = n_{\chi,\mathrm{LTE}}(0)\left[\frac{T(r)}{T(0)}\right]^{3/2} \exp\left[-\int^r_0 \ud r'\,\frac{k_{\rm B}\alpha(r')\frac{\ud T(r')}{\ud r'} + 
      m_\chi\frac{\ud \phi(r')}{\ud r'}}{k_{\rm B}T(r')}\right],
\end{equation}
where $r = 0$ represents the centre of the Sun. The conductive luminosity is:
\begin{equation}
\label{LTEtransport}
    L_{\chi,{\rm LTE}}(r)= 4\pi r^2 \zeta^{2n}(r) \kappa(r)n_{\chi,{\rm LTE}}(r)l_\chi(r) \left[\frac{k_\mathrm{B}T(r)}{m_\chi}\right]^{1/2}k_\mathrm{B}\frac{\ud T(r)}{\ud r}.
\end{equation}
where the factor $\zeta^{2n}$ accounts for a velocity-dependent $\left[\zeta = v_0/v_T(r)\right]$ or momentum-dependent $\left[\zeta = q_0/ \mx v_T(r)\right]$ cross-section, and $v_T(r) \equiv \sqrt{2 k_{\rm B} T(r)/\mx}$ is related to the typical thermal velocity \cite{GouldRaffelt90a}; $v_0$ and $q_0$ are respectively the reference velocity and momentum defined in Eq.\ \ref{qdepvdep}. The rate of energy transported per unit mass of stellar material is:
\begin{equation}
\label{epsLTE}
    \epsilon_{\chi,{\rm LTE}}(r) = \frac{1}{4\pi r^2 \rho(r)}\frac{\ud L_{\chi,{\rm LTE}}(r)}{\ud r}.
\end{equation}
This quantity is usually expressed in units of ergs g$^{-1}$ s$^{-1}$.

\begin{figure}[tb]
\begin{tabular}{c c}
\includegraphics[width=0.5\textwidth]{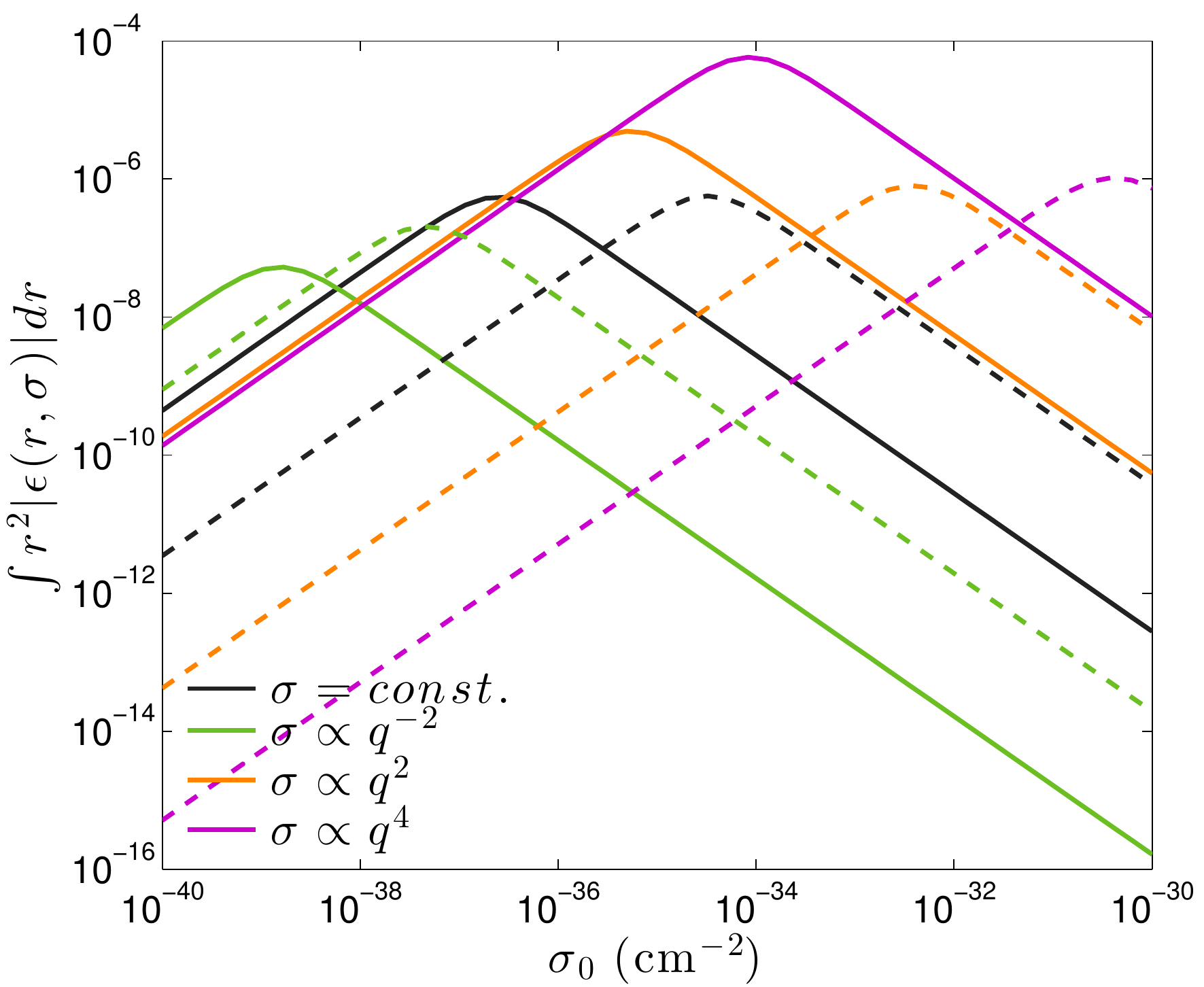} & \includegraphics[width=0.5\textwidth]{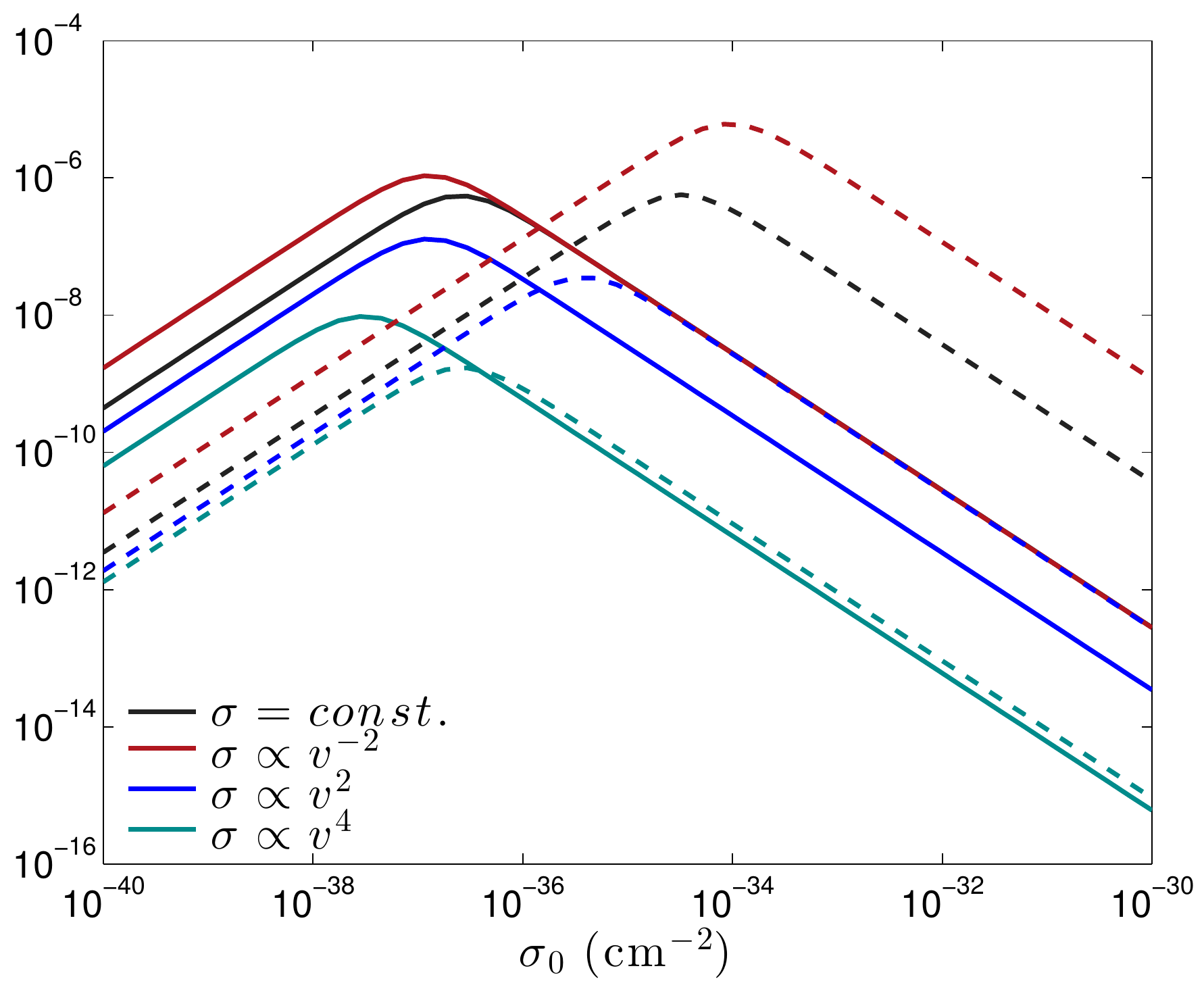} 
\end{tabular}
\caption{Illustration of the transition from the local thermal equilibrium (LTE) to the Knudsen (non-local, isothermal) regime of energy transport by DM scattering, for momentum-dependent (\textit{left}) and velocity-dependent scattering (\textit{right}).   Leftwards of the peak in each curve corresponds to the Knudsen regime, whereas rightwards is the LTE regime.  The total energy transport is plotted for a fixed DM mass ($\mx = 10$\,GeV) and number ratio of DM to baryons ($n_\chi/n_{\rm b} = 10^{-15}$). Solid lines are spin-independent couplings, whereas dashed lines represent the spin-dependent case where the DM scatters only on hydrogen. This has the effect of increasing the mean free path, leading to a transition to the Knudsen regime at a much higher value of $\sigma_0$.}
\label{fig:knudsen}
\end{figure}

If the condition $l_\chi \ll r_\chi$ is violated, LTE is no longer valid and the system exists in the Knudsen regime of non-local transport, where the DM distribution is essentially isothermal. It was shown in Refs.\ \cite{GouldRaffelt90a,GouldRaffelt90b} by Monte Carlo simulation that Eqs.\ \ref{LTEdens} and \ref{LTEtransport} can be corrected to properly account for a large Knudsen number $K \equiv l_\chi/r_\chi$.  Here we have formally defined
\begin{equation}
r_\chi = \left(\frac{3 k_{\rm B} T_c}{2\pi G \rho_c \mx}\right)^{1/2}.
\label{scaleheight}
\end{equation}
Following \cite{Bottino02,Scott09}, we define:
\begin{eqnarray}
\mathfrak{h}(r) &=& \left(\frac{r - r_\chi}{r_\chi}\right)^3 + 1, \label{hr}\\
\mathfrak f(K) &=& \frac{1}{1+ \left(\frac{K}{K_0}\right)^{1/\tau}}, \label{fK}
\end{eqnarray}
where $K_0 = 0.4$ and $\tau = 0.5$ are empirical values taken from the numerical results of \cite{GouldRaffelt90b}. The DM distribution is then a combination of the isothermal and LTE distributions:
\begin{equation}
\label{nchi}
n_\chi(r) = \mathfrak f(K) n_{\chi,{\rm LTE}} + \left[1 - \mathfrak f(K) \right] n_{\chi,{\rm iso}},
\end{equation}
where
\begin{equation}
\label{isodens}
n_{\chi,{\rm iso}}(r,t) = N(t)\frac{e^{-\frac{r^2}{r_\chi^2}}}{\pi^{3/2}r_\chi^3}.
\end{equation}
Finally, the Knudsen-corrected luminosity is:
\begin{equation}
\label{ltransport}
L_{\chi,\rm total}(r,t) = \mathfrak f(K) \mathfrak h(r,t)L_{\chi,{\rm LTE}}(r,t).
\end{equation}
This should then be used in place of $L_{\chi,\rm LTE}$ in Eq.\ \ref{epsLTE} to compute the energy injected or removed at each radius by DM-nucleon collisions. In Fig.\ \ref{capfig} we illustrate the effective enhancement or suppression of transport in a toy solar model due to a momentum-dependent cross-section, plotting
\begin{equation}
\mathcal F_{\rm transport} \equiv \frac{\int |\epsilon(r,n\ne0)|r^2dr }{\int |\epsilon(r,n=0)|r^2dr},
\label{trFdef}
\end{equation}
for the three cases of momentum-dependent cross-section, with $\sigma_0 = 10^{-35}$ cm$^2$. Once again, this is illustrated using a present-day SSM with the AGSS09ph abundances \cite{AGSS,Serenelli:2009yc}. Note that this cross-section leads mainly to transport near the LTE regime.  As $\sigma_0$ is decreased further towards the Knudsen regime, the behaviour reverses, as the enhancement provided by the longer inter-scattering distance is overcome by the Knudsen suppression. We illustrate the Knudsen behaviour in Fig.~\ref{fig:knudsen} by plotting the total energy transport for different types of generalised form factor DM as a function of the cross-section $\sigma_0$, for a constant DM-to-baryon ratio $n_\chi/n_{\rm b} = 10^{-15}$. This shows how the peak energy transport varies in each model. We note that this peak occurs for a much smaller cross-section in the spin-independent case, due to the reduced mean free path caused by scattering with helium and metals rather than just hydrogen.

The full effect of ADM on energy transport is then a combination of the effects illustrated in the left and right panels of Fig.\ \ref{capfig}, keeping in mind the degree of non-locality indicated by Fig.\ \ref{fig:knudsen}.

\section{The \DS solar dark matter code}
\label{sec:code}
In order to accurately model the effects of generalised form factor DM on solar observables, we implemented full velocity- and momentum-dependent DM capture and energy transport in the high-precision solar evolution code \GS \cite{weiss:2008, Serenelli11}.  We took DM routines from the public dark stellar evolution code \dstars \cite{Scott09b}, producing a hybrid code \DS.

\GS is the descendant of the legendary Kippenhahn code. Numerical aspects and physics inputs are described in detail in Ref.~\cite{weiss:2008} and the modified version of \GS used for this work is the same described in Ref.\ \cite{Serenelli11}.  Here we just give a summary of the most relevant physical inputs. It includes the nuclear energy generation routine \textsf{exportenergy.f}\footnote{Publicly available at \url{http://www.sns.ias.edu/~jnb}.}, updated with the astrophysical factors recommended in the Solar Fusion II~\cite{adelberger:2011} compilation. It makes use of the Opacity Project radiative opacities~\cite{badnell:2005}, complemented at low temperatures with those from Ref.~\cite{ferguson:2005}. The equation of state is the 2005 update of OPAL \cite{rogers:2002}. Microscopic diffusion of elements, including gravitational settling, thermal and concentration mixing, is treated according to Ref. \cite{thoul:1994}. 

The calibration of a solar model generally implies adjusting a number of free parameters in the model to match an equal number of observables. In the present case, the observables are the present-day ($\tau_\odot=4.57$~Gyr) solar luminosity $L_\odot$, solar radius $R_\odot$ and the metal-to-hydrogen mass fraction $(Z/X)_\odot$. The latter is a critical quantity, as it determines the composition, namely the metallicity, of the calibrated model. In this work, we adopt the photospheric solar abundances from Ref.~\cite{AGSS}, for which $(Z/X)_\odot=0.0180$. The free parameters in the model are the mixing length parameter $\alpha_{\rm MLT}$ and initial helium and metal mass fractions $Y_{\rm ini}$ and $Z_{\rm ini}$ respectively. The latter two suffice to determine the initial abundances of all elements in the model because the relative metal abundances are taken from Ref.~\cite{AGSS} with $Z_{\rm ini}$ acting as the normalisation factor, and $X_{\rm ini} + Y_{\rm ini} + Z_{\rm ini}=1$ by definition ($X_{\rm ini}$ being the initial hydrogen abundance).

In practice, the solar model calibration starts with a homogeneously-mixed 1~M$_\odot$ pre-main sequence model that is evolved assuming no mass loss until it reaches $\tau_\odot$, the current solar system age. At that age, model predictions are compared with the observables, and a Newton-Raphson scheme is implemented to iteratively find the solution. This is generally achieved to 1 part in 10$^5$ within two to three iterations for standard solar models (i.e.\ with no DM). More iterations are necessary as the effects of DM become more important. In the most extreme cases, no physical solutions are found (e.g.\ resulting in negative $Y_{\rm ini}$). Note that each iteration requires four evolutionary calculations to evaluate the partial derivatives needed for the Newton-Raphson scheme. Each evolutionary calculation requires 700--800 timesteps. At each timestep, the solar structure is discretized in about 2000 shells. These requirements for the integration of solar models, both in spatial and time resolution, guarantee a numerical precision better than 1\% in all model predictions \cite{Bahcall06}.

\dstars \cite{Scott09b} is a \textsf{Fortran95} package that implements capture, annihilation and energy transport by regular ($n=0$) WIMP dark matter in a general stellar evolution code, as described in Refs.\ \cite{Scott09,Scott08a,Fairbairn08,Scott08b}.  The capture routines were originally adapted from \textsf{DarkSUSY} \cite{DarkSUSY}.  The underlying evolutionary code \cite{Paxton04} is a \textsf{Fortran90} rewrite of the the venerable \textsf{Fortran77} Cambridge \textsf{STARS} package \cite{Eggleton71, Eggleton72, Pols95}.  These codes use the relaxation method to solve the coupled 1D ordinary differential equations of stellar structure over an adaptive grid.  \dstars is the state of the art in dark stellar evolution for the $n=0$ case, as it features the full capture calculation (Eq.\ \ref{caprate}) for SI and SD scattering on the 22 most important elements, various options for the DM velocity distribution (including user-defined distributions), and proper Gould-Raffelt treatment of conductive energy transport (Eq.\ \ref{LTEtransport}), including self-consistent density profiles and the Knudsen-dependent interpolation between the LTE and isothermal (non-local) regimes (Eqs.\ \ref{nchi}, \ref{ltransport}).  

For \DS, we adapted the \dstars capture routines to implement capture of generalised form factor DM (Eq.\ \ref{ffHresult}--\ref{ffZresult}, \ref{vresult}) rather than just the $n=0$ case.  We used the conductive transport routines from \dstars essentially unaltered, except that we included the additional factor of $\zeta^{2n}$ in Eq.\ \ref{LTEtransport} and utilised the $\alpha$ and $\kappa$ tables that we computed earlier for $n\ne0$ \cite{VincentScott2013}.

At each regular \GS timestep, \DS computes the total DM capture rate by solving Eq. \ \ref{caprate}, assuming a local halo DM density of 0.38 GeV/cm$^3$.  This input rate is then used to update the DM population in the Sun following Eq.\ \ref{DMpopEqn}.  \DS uses the numerical version of the capture routines from \dstars to evaluate the modified capture equation Eq.\ \ref{caprate}.  It computes the thermal diffusivity and conductivity coefficients $\alpha(r,t)$ and $\kappa(r,t)$ at each height in the star by interpolating in the tables of Ref.\ \cite{VincentScott2013}, as these quantities depend on the specific mixture of plasma species with which the DM particles interact. The code then computes the DM density $n_\chi(r,t)$ using $\alpha(r,t)$ and $\kappa(r,t)$, which it uses to determine energy transport. \DS then interpolates the resulting values of $\epsilon_\chi(r)$ (Eq.\ \ref{epsLTE}) to the grid used by \GS, where they are treated as an additional energy source at each height in the star. 

We considered seven ADM models with SI interactions with nucleons, and seven with SD interactions: the constant cross-section case, $\sigma \propto q^{2n}$ and $\sigma \propto \vrel^{2n}$ with $n = \{-1,1,2\}$. For each coupling, we simulated one solar model for each point on a grid of $\mx$ and $\sigma_0$. We computed a subset of models using a stringent convergence criterion of one part change in $10^{5}$ for the the solar luminosity, radius and surface metallicity ($Z/X$), with 4\,kyr and 10\,Myr minimum and maximum time steps. Because it was extremely computationally expensive to do every simulation including the full treatment of capture and transport at this accuracy, we carried out all other simulations with a more relaxed convergence criterion of a part in $10^{3}$, with minimum and maximum time steps of 40\,kyr and 40\,Myr. We then corrected these lower-accuracy results to consistently reproduce the observables and chi-squared values of the higher-accuracy models, using the systematic differences we saw between models computed both ways. In total, our calculations took $\sim 1.5$ CPU years.

\subsection{Annihilation}
\label{sec:annihilation}

DM models with momentum- and velocity-dependent nuclear scattering need not necessarily be entirely asymmetric.  To investigate the implications of energy injection from annihilation in such models, we also implemented annihilation in \DS, following \cite{Scott09} and \dstars.  In general however, allowing annihilation simply weakens the limits that we obtain from solar physics on generalised form factor DM, and does not improve the overall fit to solar data in any significant way. We therefore do not show these results, nor discuss annihilation beyond this subsection.  It is worth noting that although we assume zero annihilation cross-section for all the results we show, some small (sub-thermal) annihilation cross-section is certainly still allowed in each model, and would make no impact on the solar observables.

\section{Limits on generalised form factor dark matter from solar physics}
\label{sec:results}
In this section we present a systematic overview of the results obtained from our simulations. In each subsection, we review the effects of velocity and momentum-dependent asymmetric dark matter on a specific observable. These are: the boron-8 and beryllium-7 neutrino fluxes, the depth of the convection zone $r_{\rm CZ} $, the sound speed profile $c_{\rm s} (r)$, the small frequency separations and the surface helium abundance $Y_{\rm S}$. Agreement between the predicted and observed values of the neutrino fluxes is typically unchanged or worsened by thermal transport by DM.  The same is true for $Y_{\rm S}$.  In contrast, the predictions of $r_{\rm CZ} $, the sound speed profile and the small frequency separations are often closer to the observed values when conductive energy transport by DM is included. In Sec.\ \ref{sec:combined} we construct a combined likelihood, which encompasses the overall improvement or degradation in the fit to solar data for each cross-section form and combination of $\sigma_0$ and $m_\chi$.  We also present a few interesting benchmark cases in which the agreement is significantly improved.

For every form of the cross-section, we ran simulations over a grid of DM masses and cross-sections: $\mx =$ 5, 10, 15, 20 and 25 GeV, and each decade in $\sigma_0$ between $10^{-40}$ and $10^{-30}$ cm$^2$. The colour scales in the figures of this section are interpolations from this grid. For some specific cases, where low-mass points gave a good overall improvement over the Standard Solar Model, we extended the mass axis down to 1 GeV. We once again caution that although evaporation is expected to have an effect near these small masses, a full kinematic analysis would be necessary to determine its exact $m_\chi$ and $\sigma_0$ dependence for each model; the importance of evaporation for these models remains essentially unexplored.  Given the importance of kinematic matching with individual nuclei, we caution against placing too much store in quick estimates of this effect.

\begin{figure}[tb]
\begin{tabular}{c c}
\hspace{-.5cm} \includegraphics[width=0.5\textwidth]{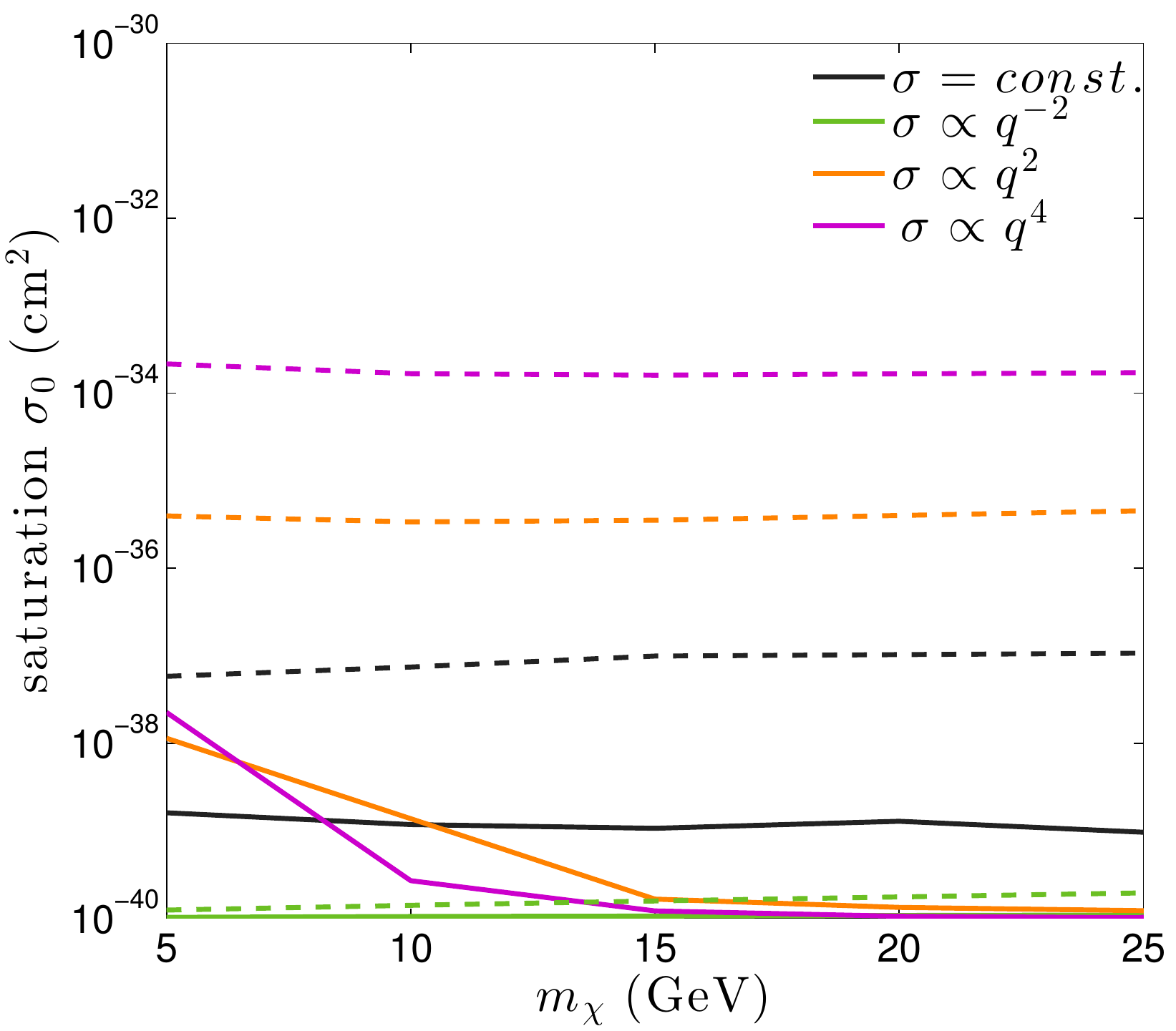}&\includegraphics[width=0.5\textwidth]{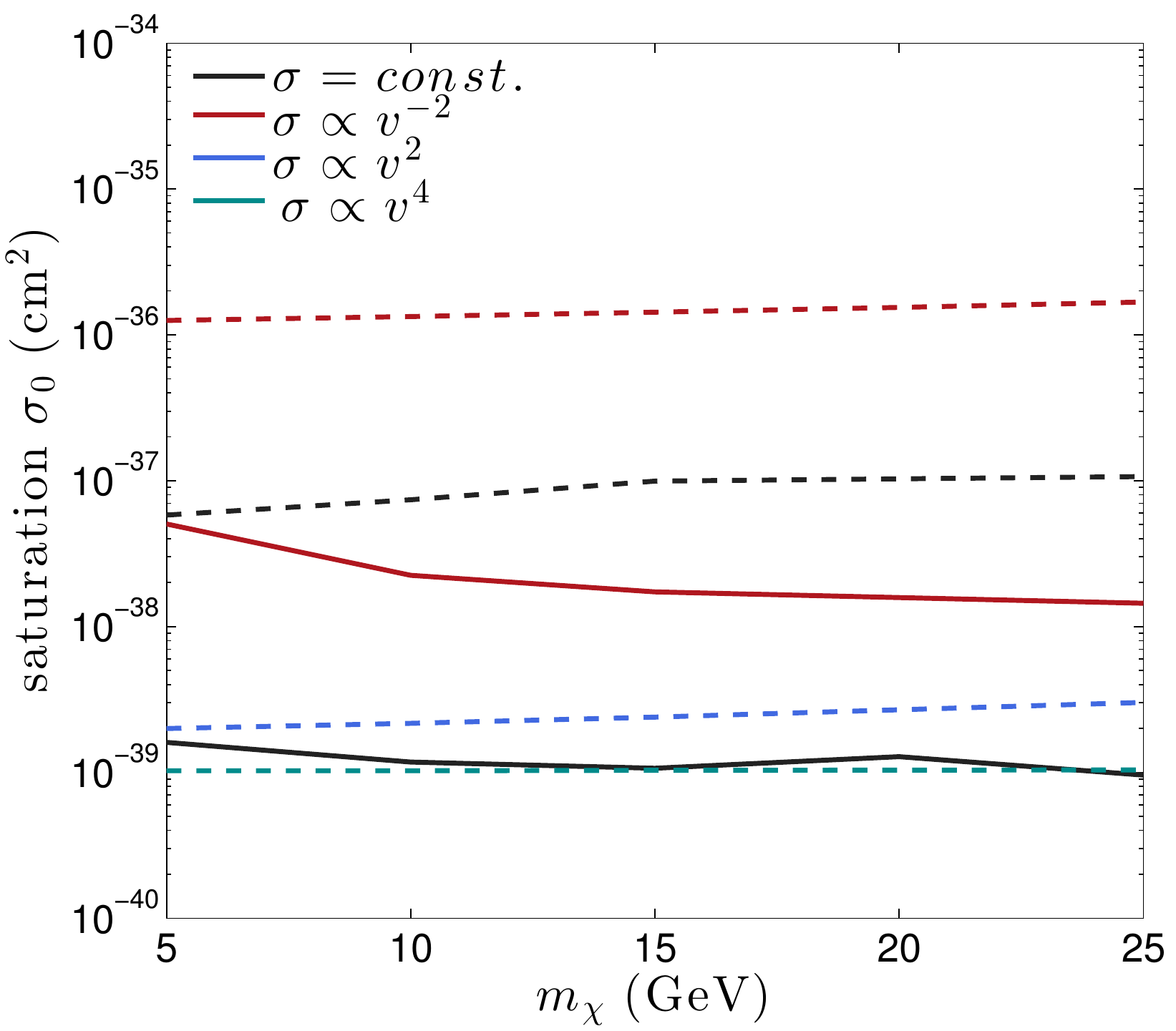}
\end{tabular}
\caption{Cross-section normalisation $\sigma_0$ at which dark matter capture saturates the geometric limit of Eq.\ \ref{eq:capcutoff} for momentum-dependent (left) and velocity-dependent (right) cross-sections. Spin-independent couplings are shown with solid lines, and spin-dependent with dashed lines. The SI $q^{-2}$, $v^2$ and $v^4$ cases are not shown, as they saturate the capture rate below $\sigma_0  = 10^{-40}$ cm$^2$ (the smallest cross-section that we simulated). We note that by using the standard assumption that spin-dependent DM is only captured by hydrogen, we have underestimated the capture rate for $q^4$; the suppression to capture could therefore be as much as an order of magnitude smaller in this specific case. }
\label{fig:cappedDM}
\end{figure}

We show the impacts of generalised form factor DM on capture rates and their saturation in Fig.\ \ref{fig:cappedDM}, and on observables in Figs.\ \ref{SIboronfluxes} to \ref{SDchisq}.

Simulations that led to a reduction in $\phi^\nu_{\rm B}$ of approximately $\sim$$60\%$ or more did not converge, because the effects on the overall solar evolution are too large for the numerical solver to handle. In many cases, the excess energy transport actually led to a density inversion in the core. In some of these models, a solution probably exists, and could be found with a better solver.  In others, the Sun probably cannot exist as a stable body.  We simply mask the the non-converged region in blue in all our plots.

\subsection{Saturation of capture}
\label{sec:saturation}

A velocity or momentum dependence in the nuclear scattering cross-section has a striking impact on the rate of DM capture by the Sun. The minimal $\sigma_0$ required to render the Sun opaque to dark matter -- and thus to saturate the capture rate Eq.\ \ref{eq:capcutoff} -- can be lowered by as much as two orders of magnitude for a $\vrel^4$ cross-section, and approximately one order of magnitude in the $q^{-2}$ and $\vrel^2$ cases. On the other hand, negative powers of $\vrel$ and positive powers of a spin-dependent $q$-coupling yield a large suppression in the capture rate, seen as a much larger required cross-section to achieve saturation. We illustrate the required value of $\sigma_0$ in Fig.\ \ref{fig:cappedDM}, based on the output of our simulations. Interestingly, a spin-independent $q^2$ or $q^4$ cross-section can yield an enhancement or suppression of the capture rate depending on the DM mass; this is a consequence of the behaviour illustrated in Fig.\ \ref{capfig}, and simply reflects the fact that $q$-dependent scattering is most efficient for large momentum transfers, which is much easier when the DM mass is closely matched with the masses of heavier elements. 

\subsection{Solar neutrino fluxes}
\label{sec:neutrinos}

In general, energy transport by DM removes energy from the solar core and reduces its temperature, causing a reduction in neutrino production rates.  Given its strong temperature-dependence, the $^8$B neutrino production rate is the first place to look for changes in solar observables. 

\begin{figure}[p]
\begin{tabular}{c@{\hspace{0.04\textwidth}}c}
\multicolumn{2}{c}{\includegraphics[height = 0.32\textwidth]{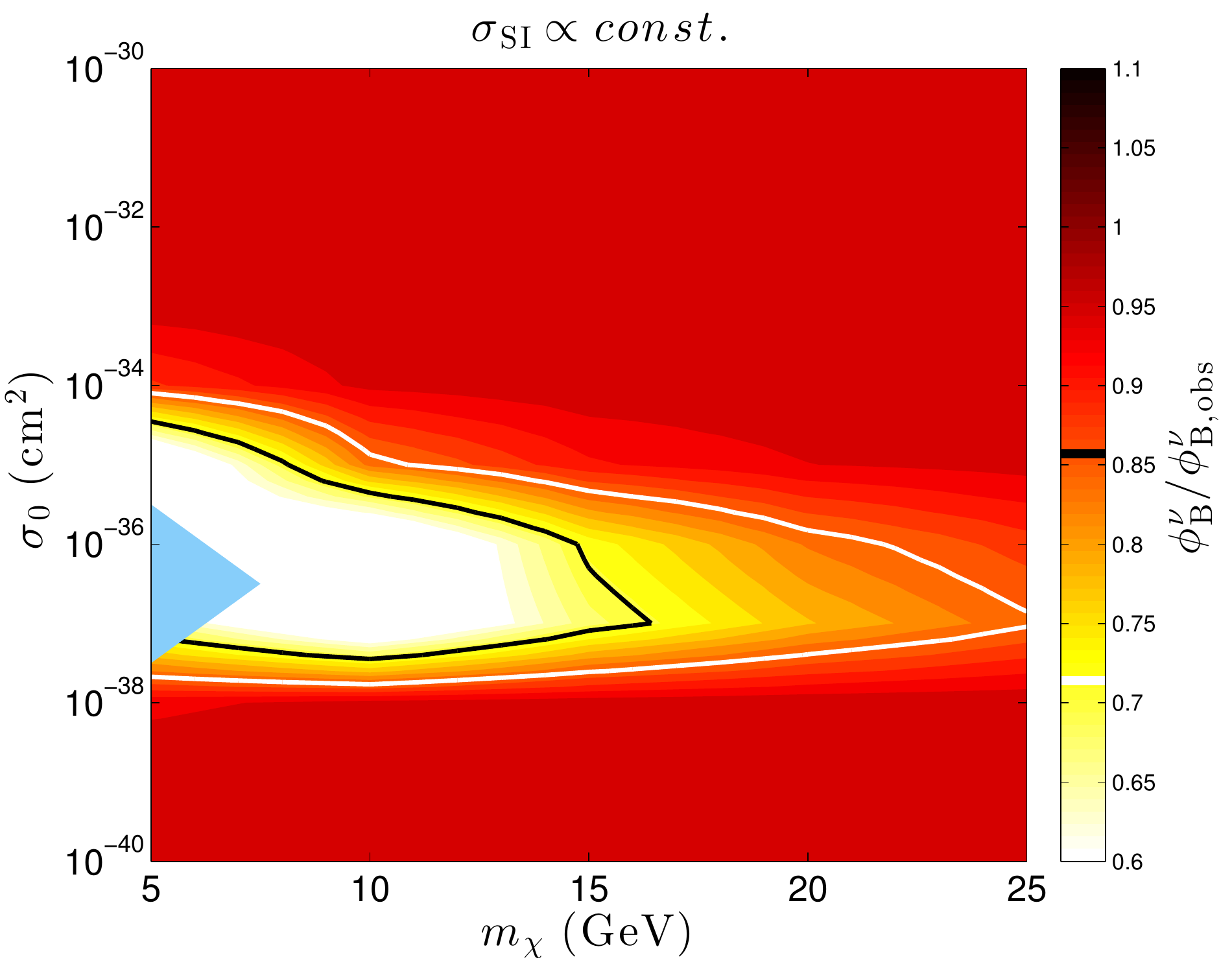}} \\
\includegraphics[height = 0.32\textwidth]{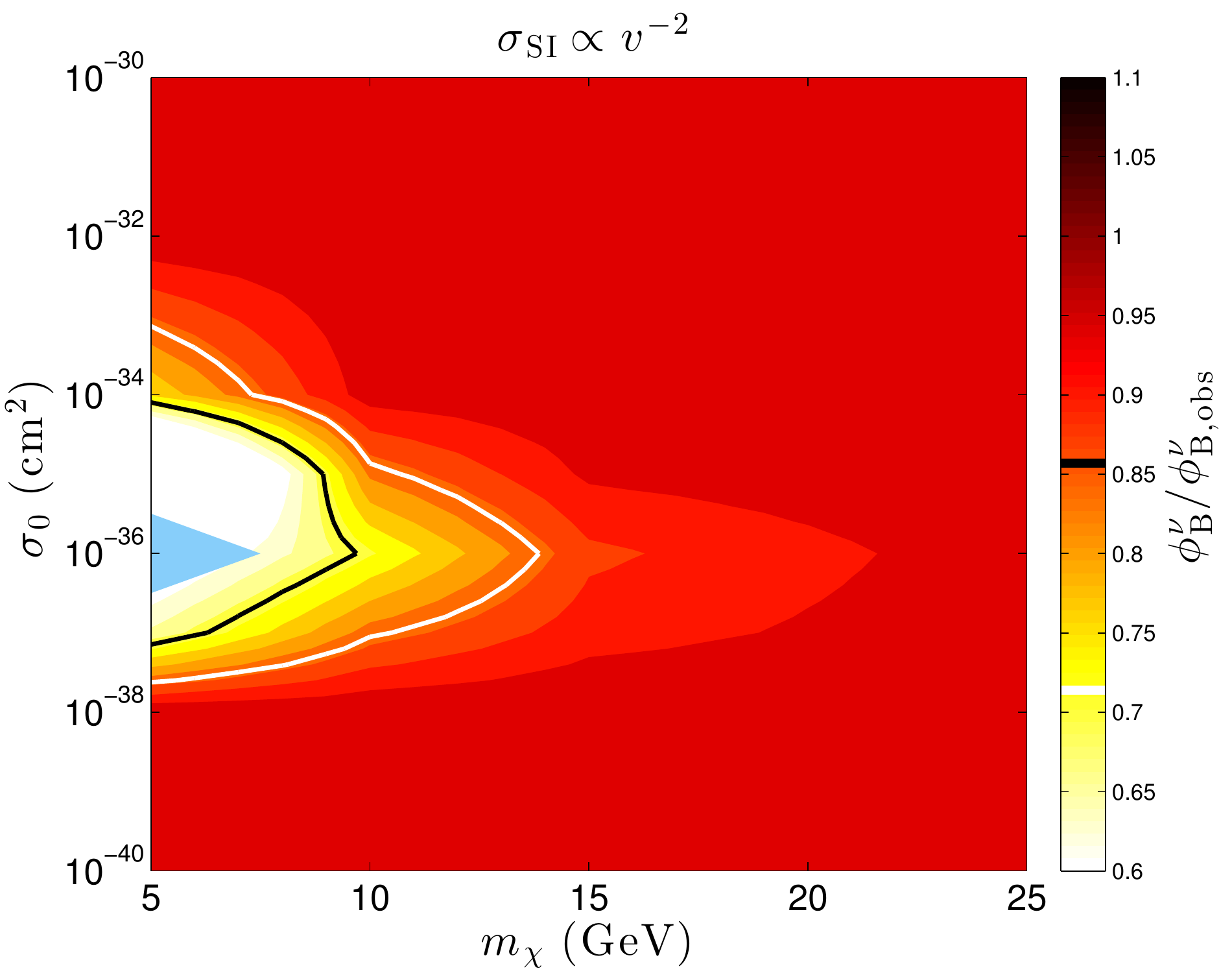} & \includegraphics[height = 0.32\textwidth]{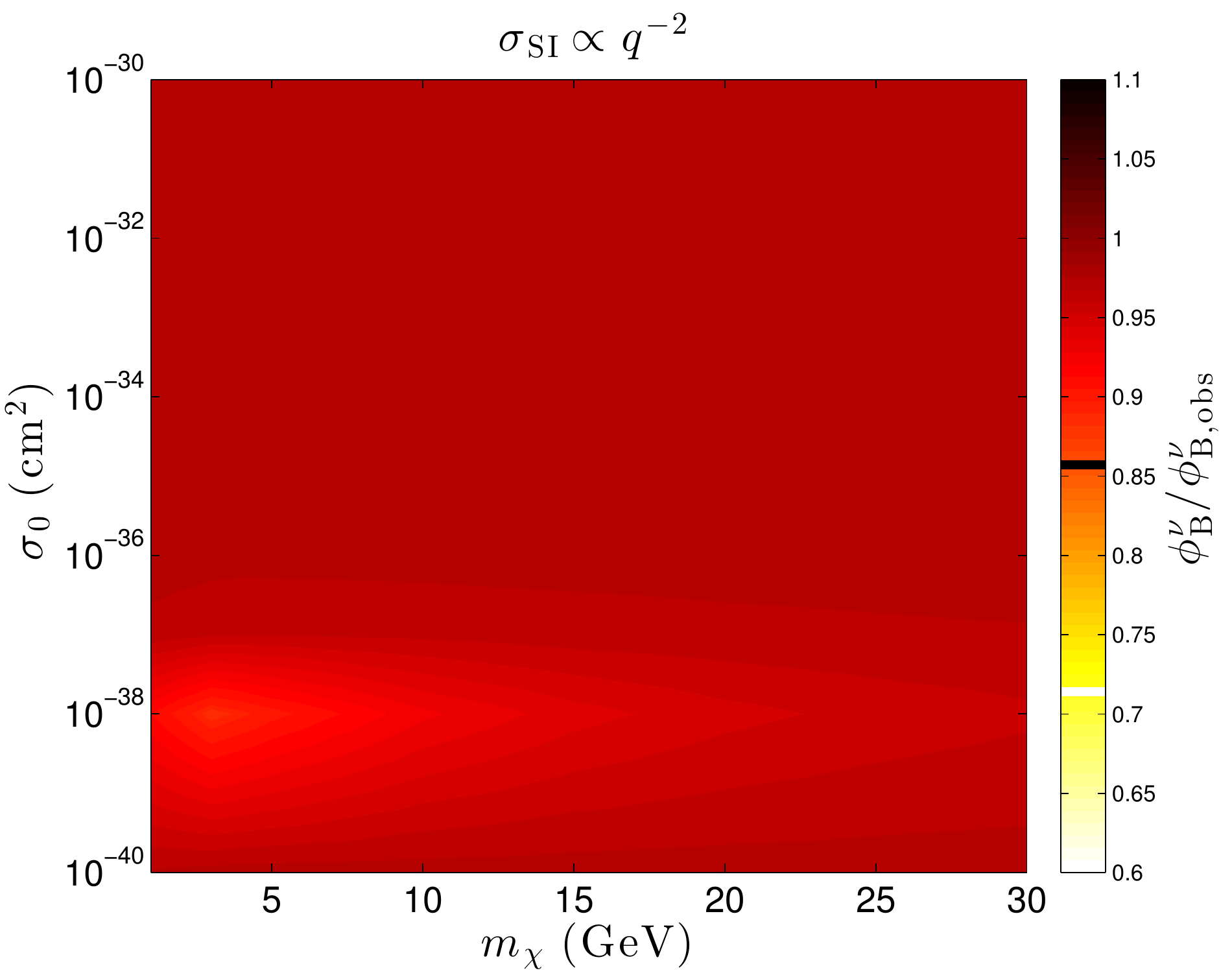} \\
\includegraphics[height = 0.32\textwidth]{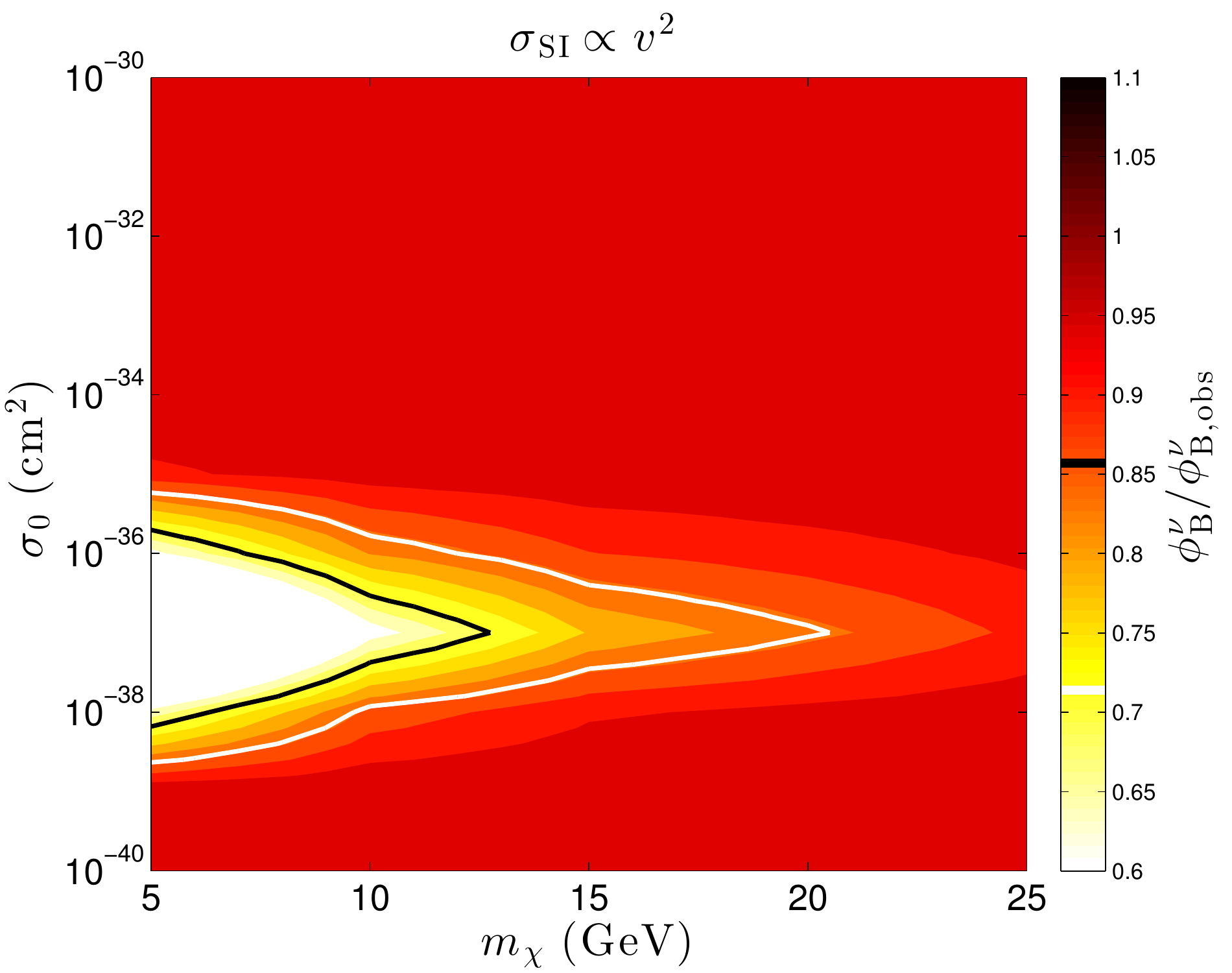} & \includegraphics[height = 0.32\textwidth]{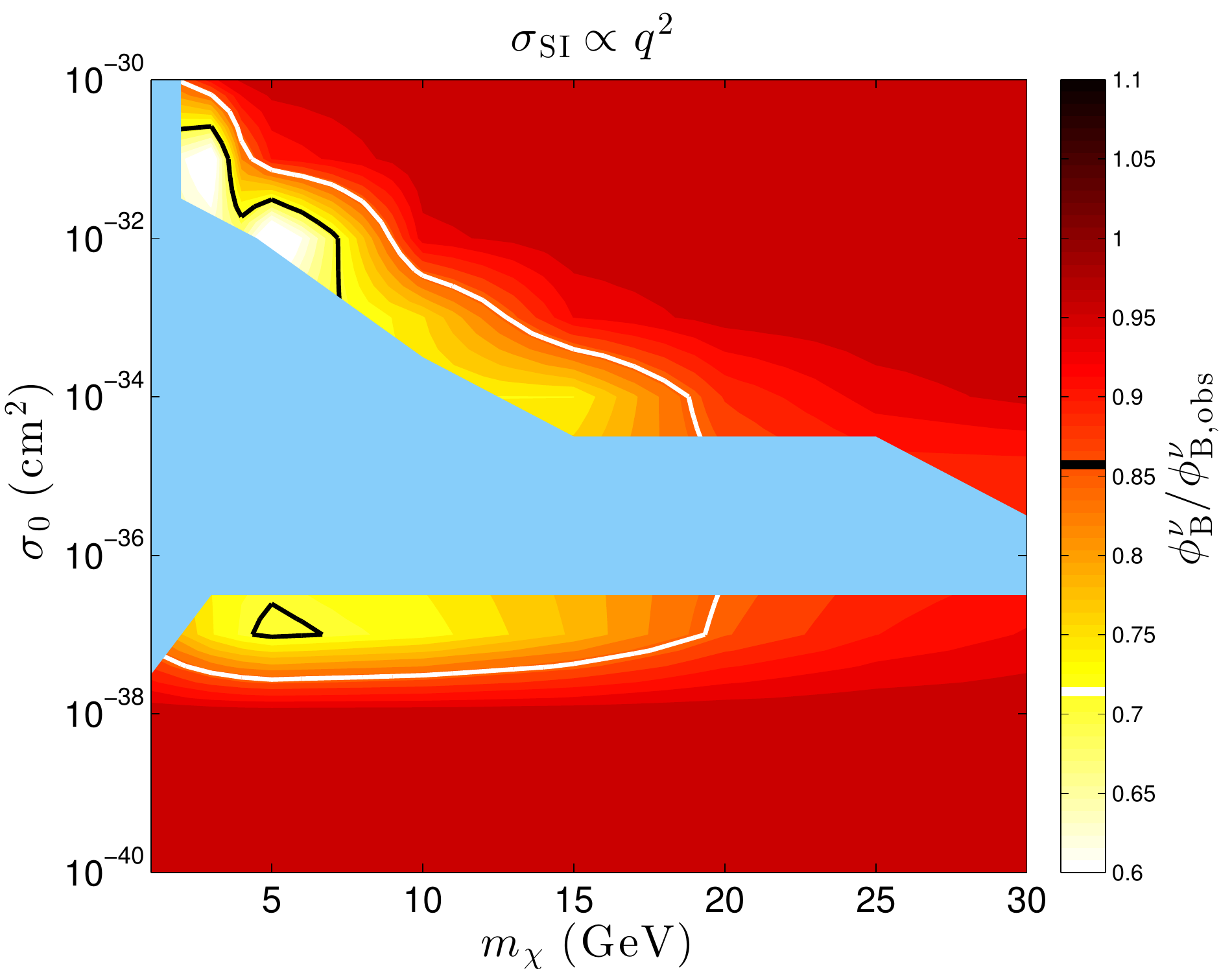} \\
\includegraphics[height = 0.32\textwidth]{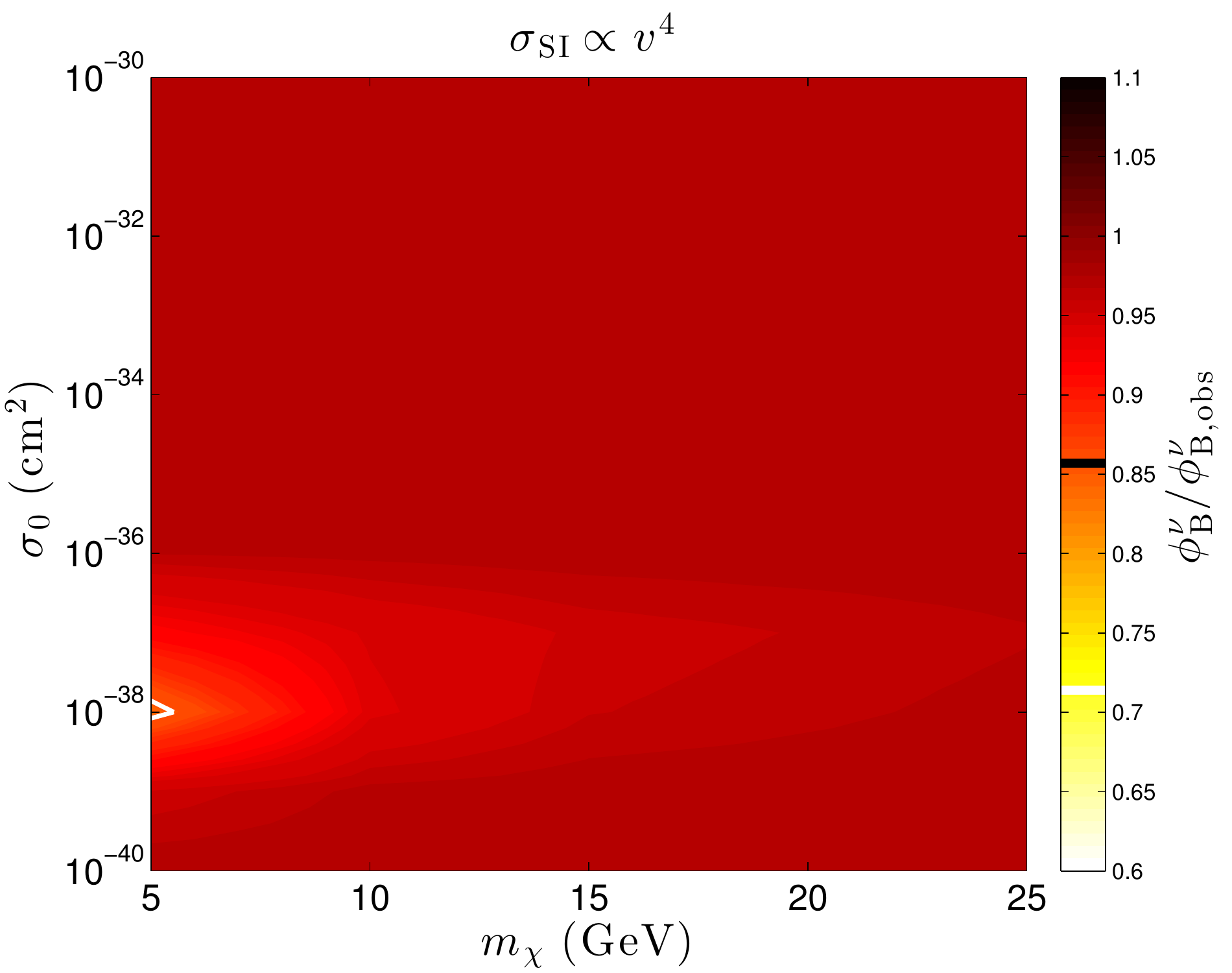} & \includegraphics[height = 0.32\textwidth]{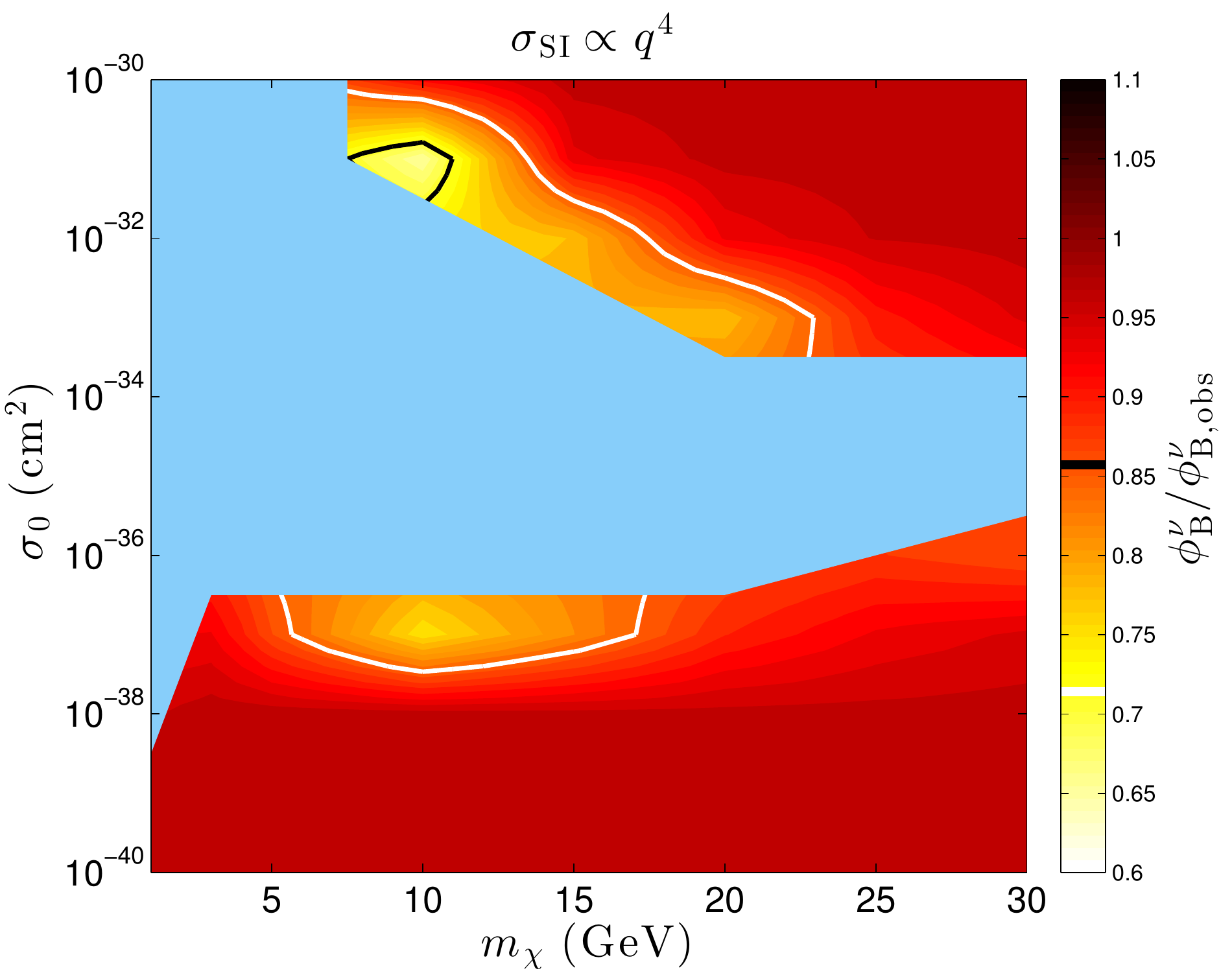} \\
\end{tabular}
\caption{The ratio of the predicted $^8$B neutrino flux to the  measured value $\phi^\nu_{\rm B,obs} = 5 \times 10^6$\,cm$^{-2}$s$^{-1}$, for each type of spin-independent dark matter coupling defined in Eq.\ \ref{qdepvdep}.  In every case the white and black lines show the isocontours where the flux is respectively 1 and 2$\sigma$ lower than the observed values, based on observational (3\%) and modelling (14\%) errors, added in quadrature. The cross-sections are normalized such that $\sigma = \sigma_0 (v/v_0)^{2n}$ or $\sigma = \sigma_0 (q/q_0)^{2n}$, with $v_0 = 220$\,km\,s$^{-1}$ and $q_0 = 40$\,MeV. Simulations carried out in the masked regions did not converge, due to the significant heat conduction by the DM particles, leading in extreme cases to density inversions in the core. }
\label{SIboronfluxes}
\end{figure}

\begin{figure}[p]
\begin{tabular}{c@{\hspace{0.04\textwidth}}c}
\multicolumn{2}{c}{\includegraphics[height = 0.32\textwidth]{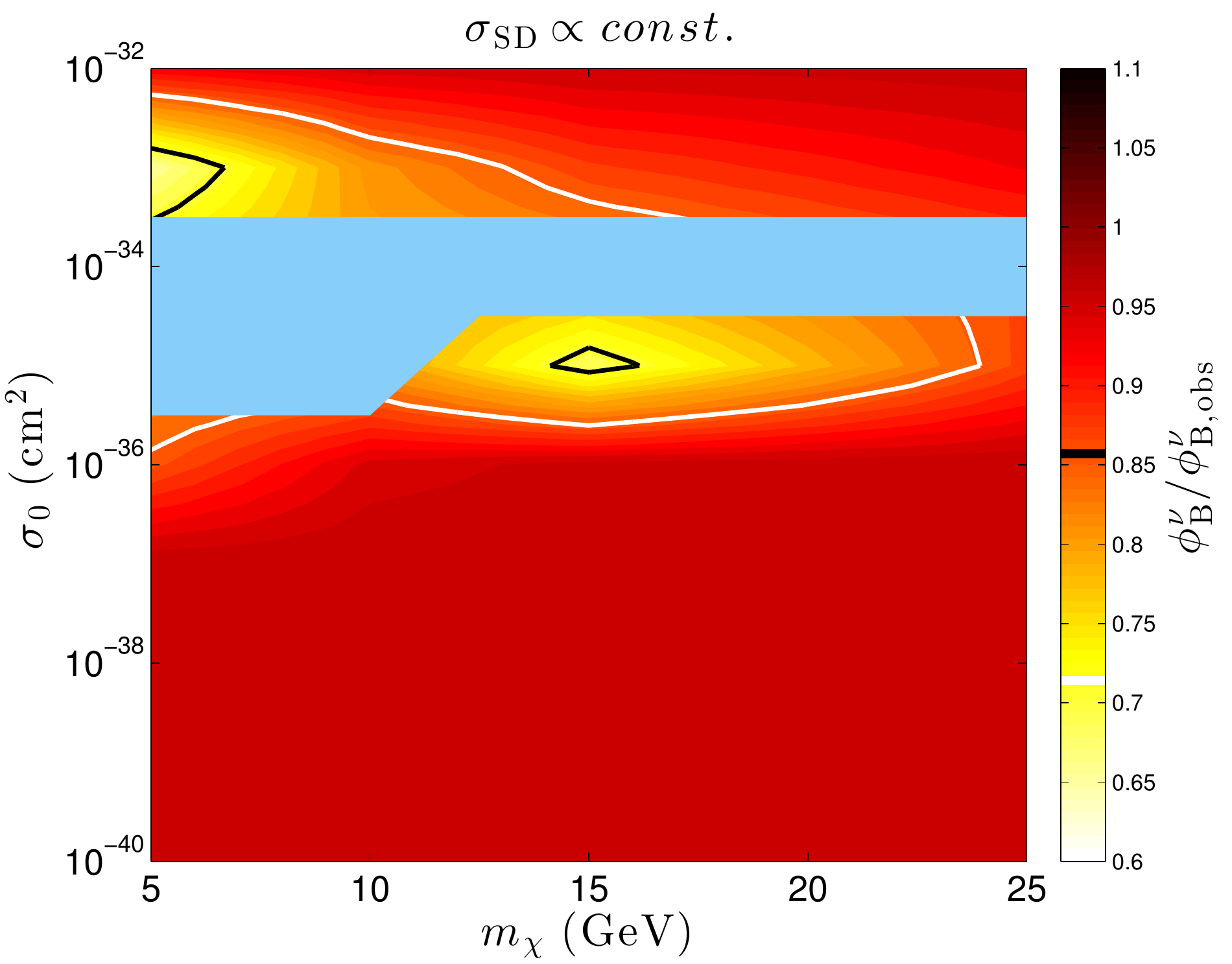}} \\
\includegraphics[height = 0.32\textwidth]{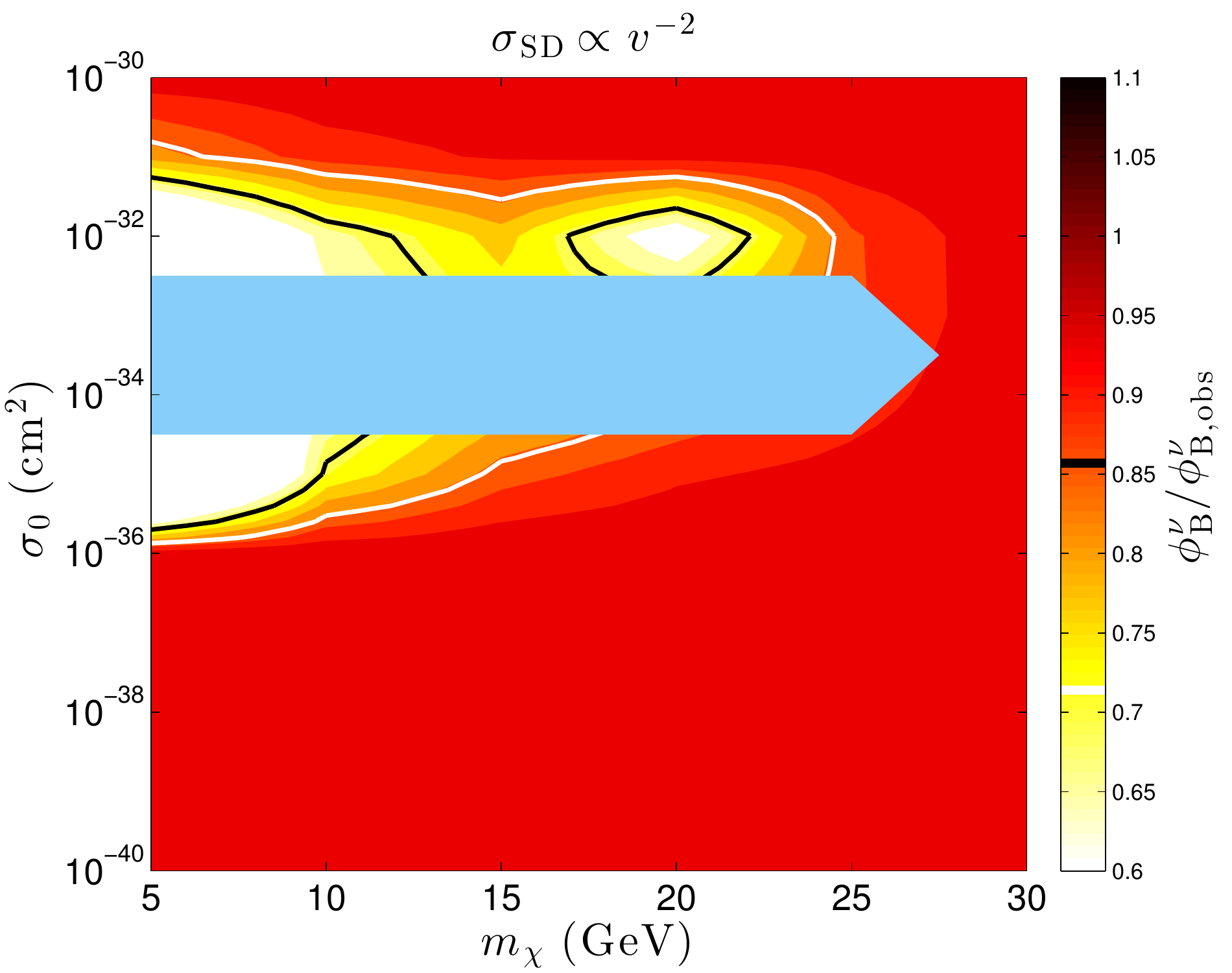} & \includegraphics[height = 0.32\textwidth]{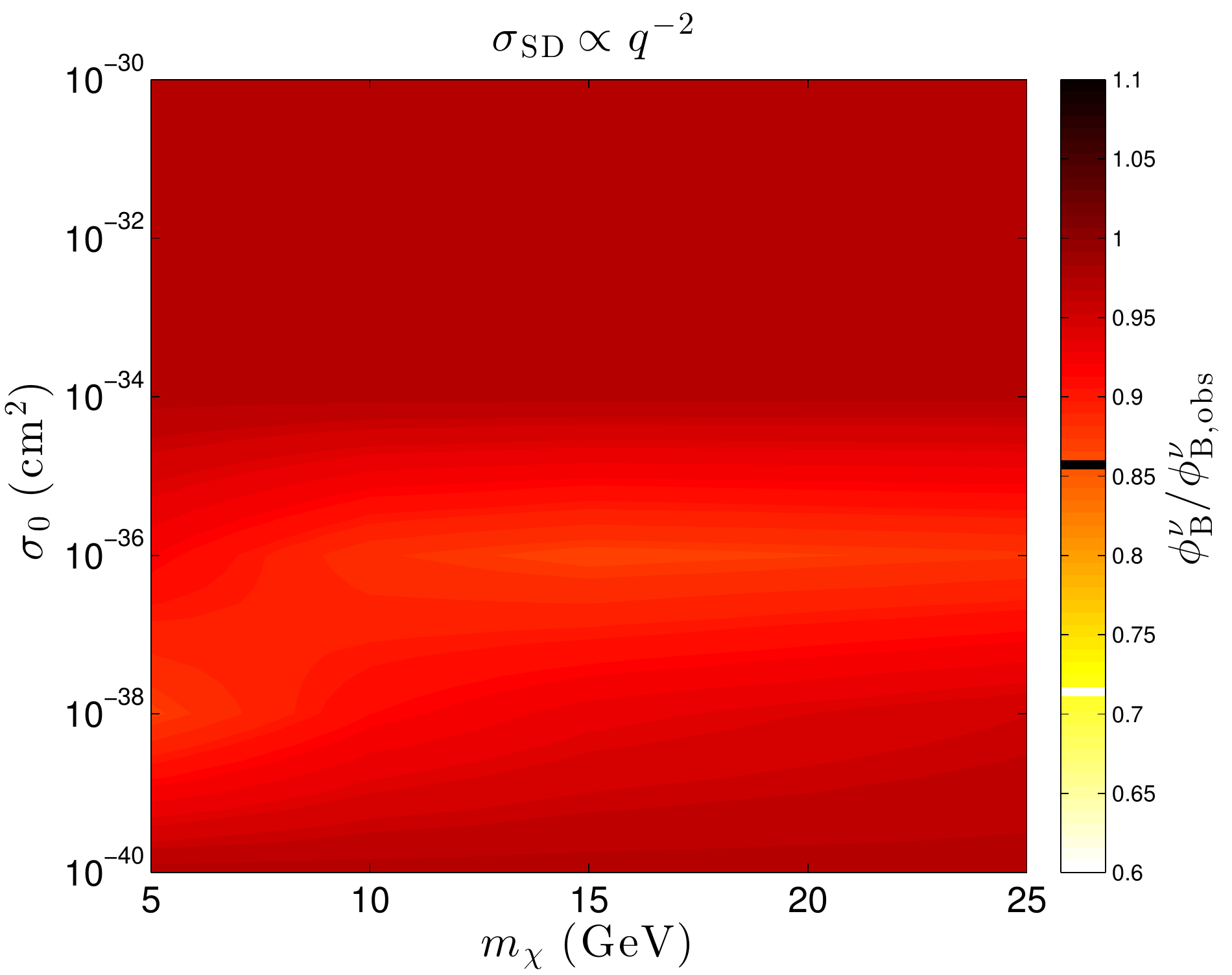} \\
\includegraphics[height = 0.32\textwidth]{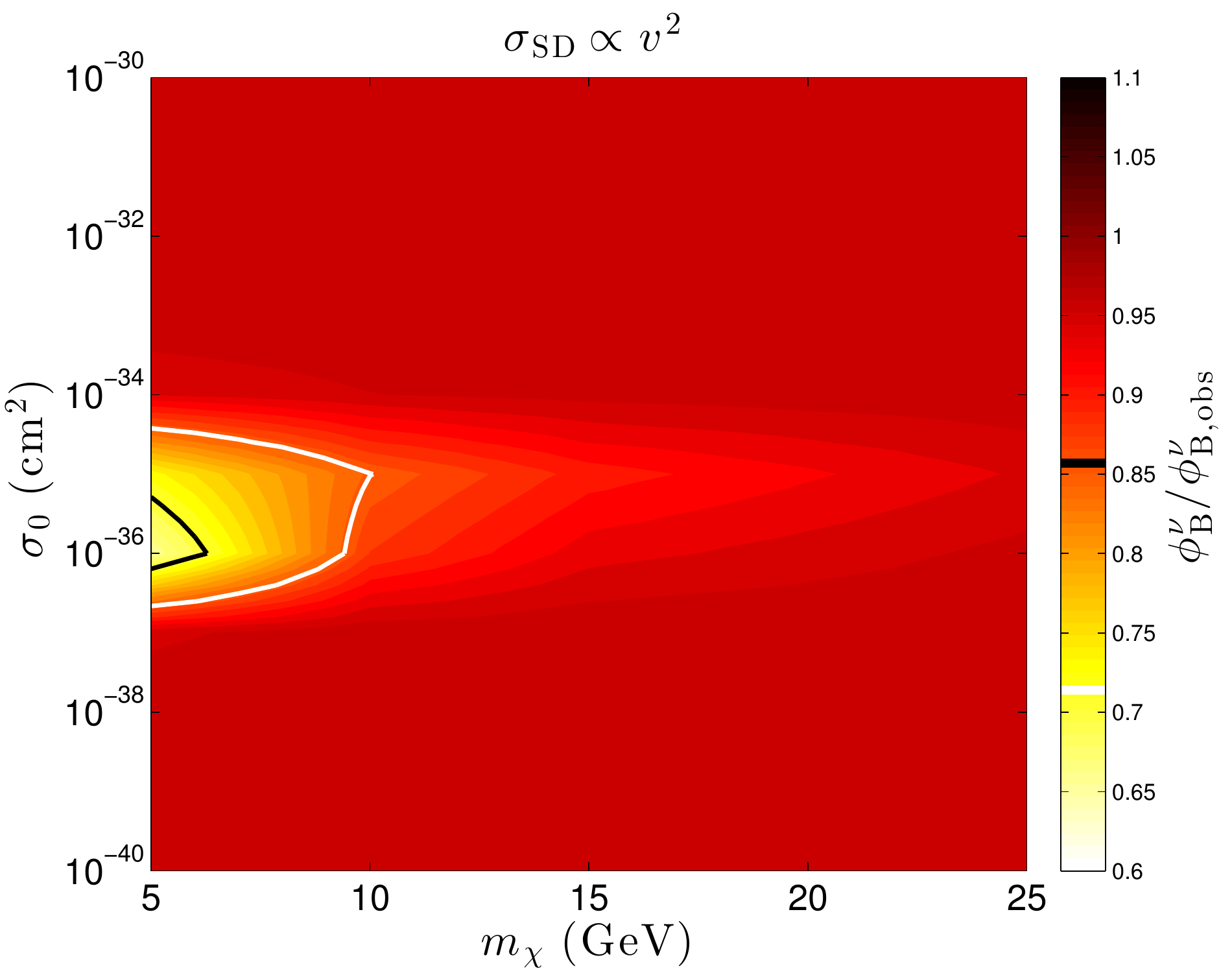} & \includegraphics[height = 0.32\textwidth]{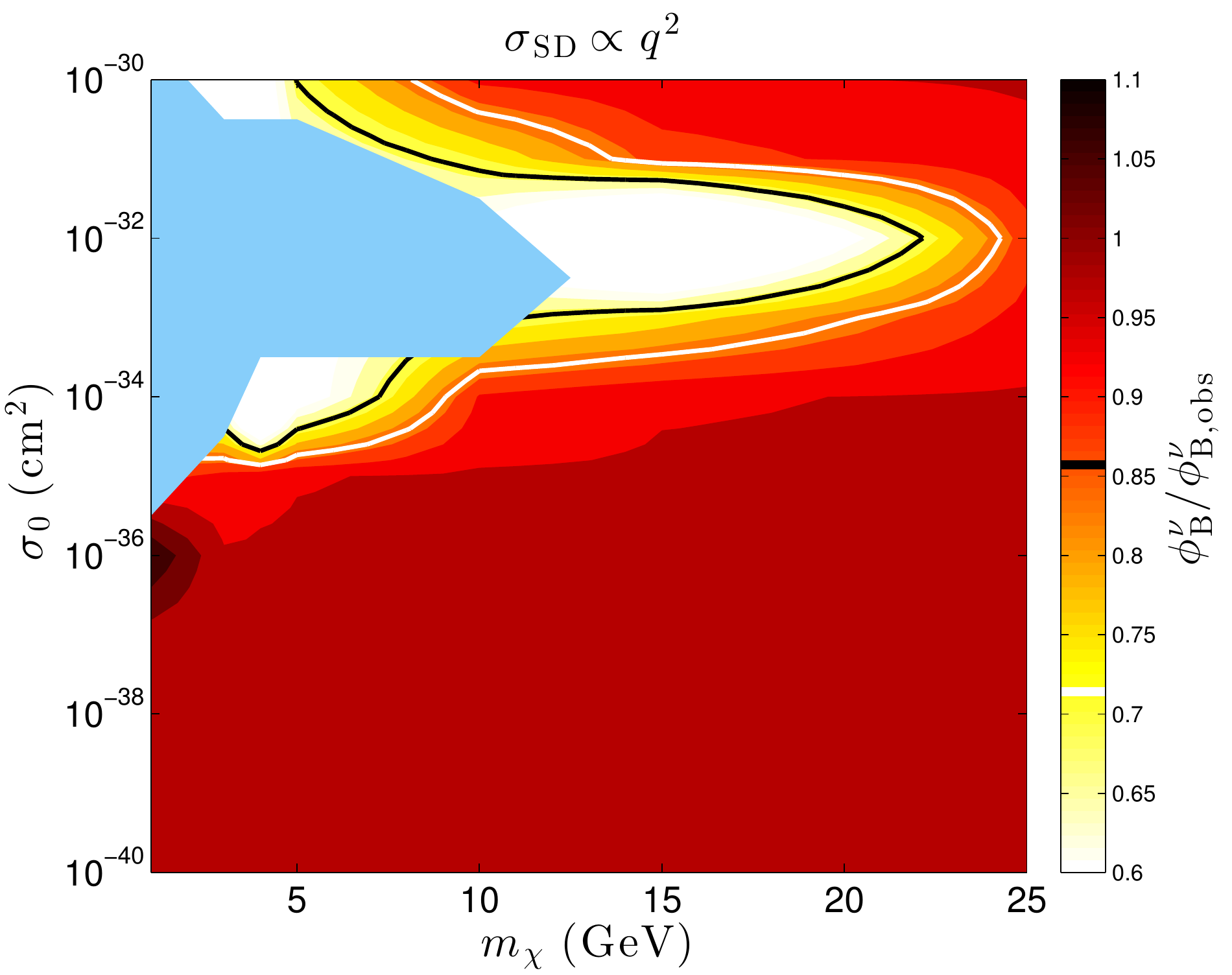} \\
\includegraphics[height = 0.32\textwidth]{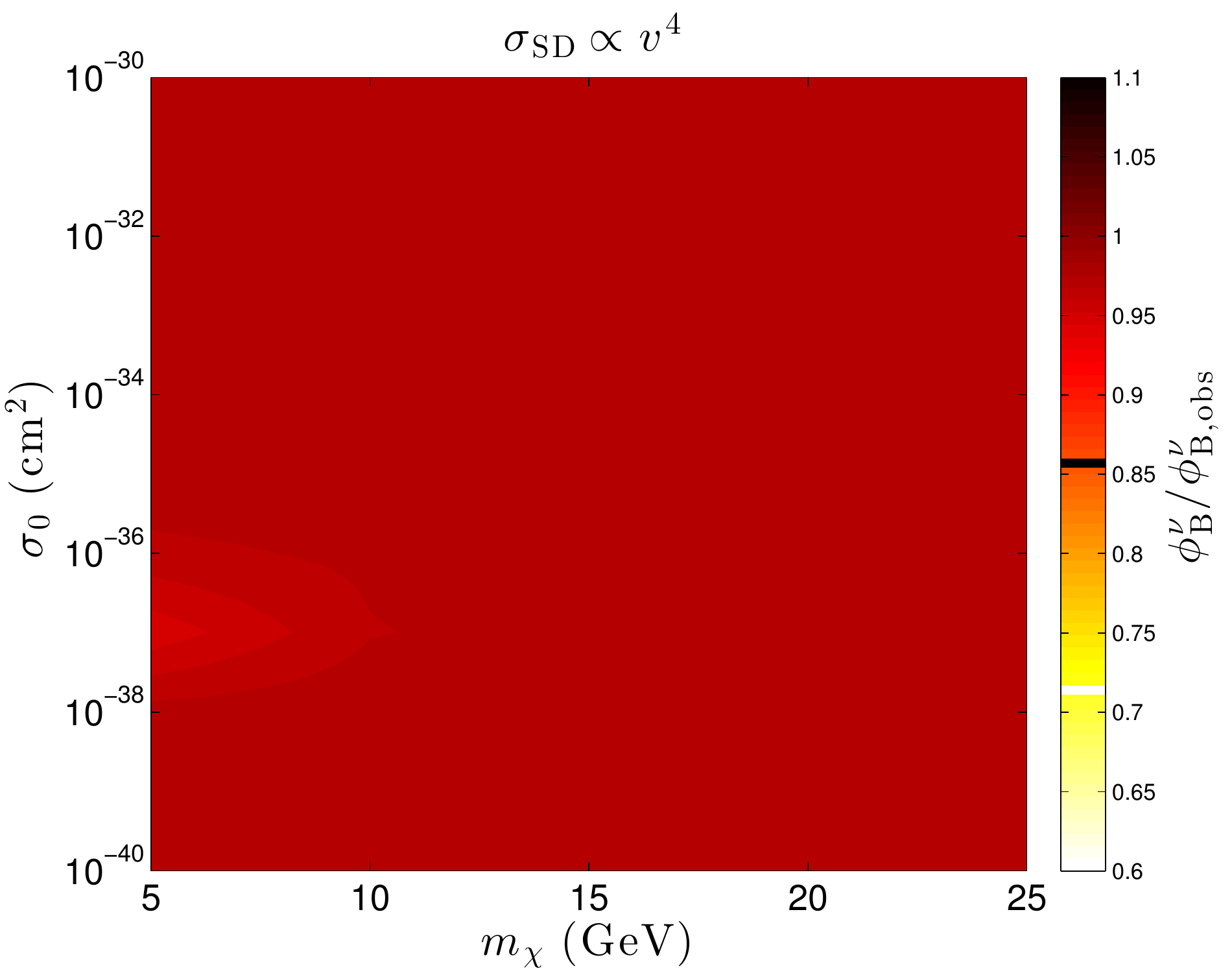} & \includegraphics[height = 0.32\textwidth]{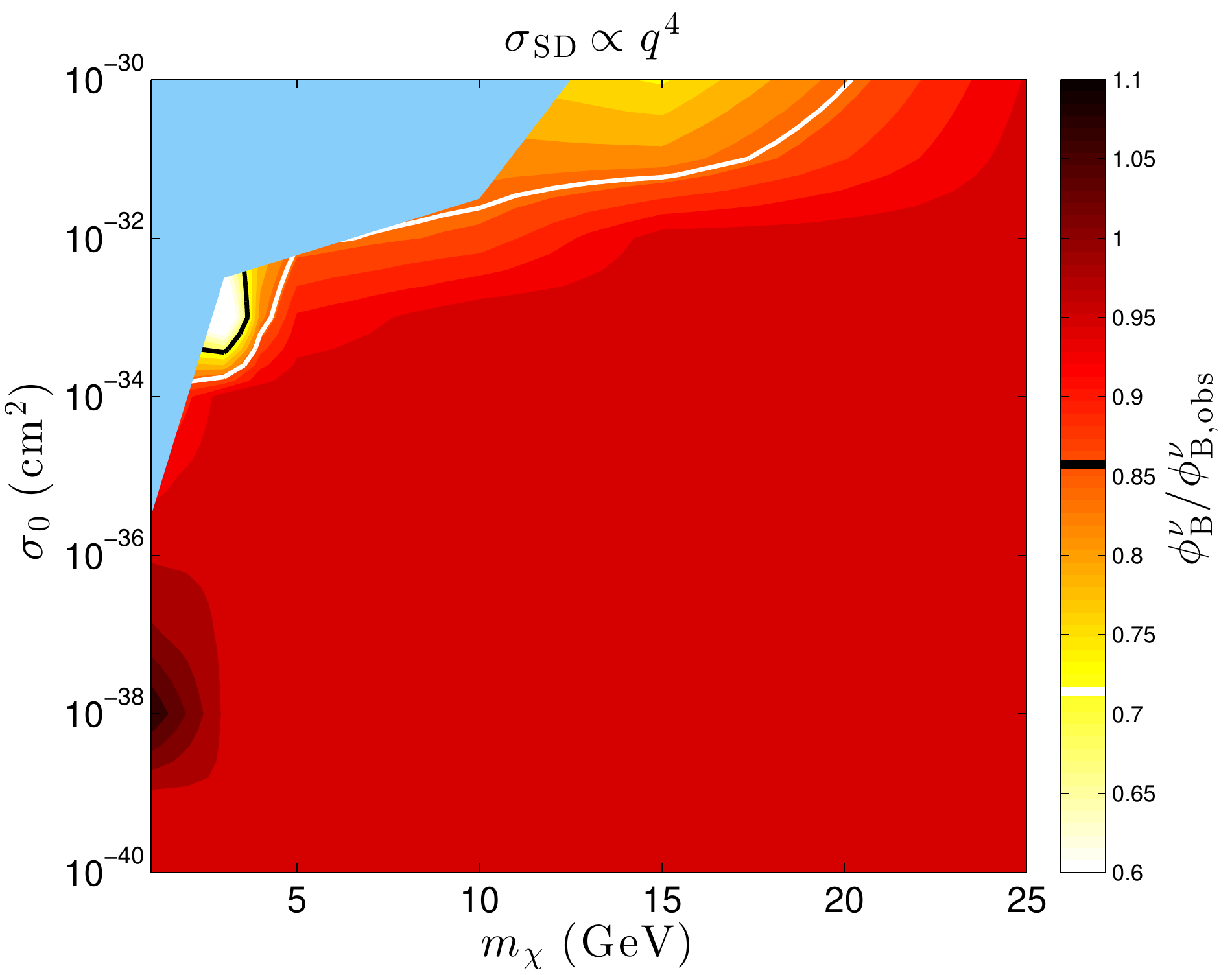} \\
\end{tabular}
\caption{As per Fig.\ \ref{SIboronfluxes}, but for spin-dependent couplings. We note that by neglecting capture on elements other than hydrogen in the SD case, we have underestimated the capture rate (and net effect) for the $q^4$ case. See discussion in Sec. \ref{sec:FFsection}}
\label{SDboronfluxes}
\end{figure}

In Figs.\ \ref{SIboronfluxes} and  \ref{SDboronfluxes}, we show the effect on the $^8$B neutrino flux for spin-independent and spin-dependent dark matter with velocity and momentum-dependent couplings. Fluxes are normalized to the observed value, $\phi^\nu_{\rm B,obs} = 5.0 \times 10^{6}$ cm$^{-2}$s$^{-1}$; lighter colouring represents a reduced flux. Although the measurement error on the $^8$B neutrino flux is only 3\% \cite{superk:2011}, the overall uncertainty from modelling is around 14\%. We add the absolute uncertainties in quadrature to obtain the 1$\sigma$ region. White lines therefore represent the $1\sigma$ isocontour where the neutrino flux falls below $\sim$ 85\% of the measured value; black lines are the 2$\sigma$ (71\%) contours. 

\begin{figure}[p]
\begin{tabular}{c@{\hspace{0.04\textwidth}}c}
\multicolumn{2}{c}{\includegraphics[height = 0.32\textwidth]{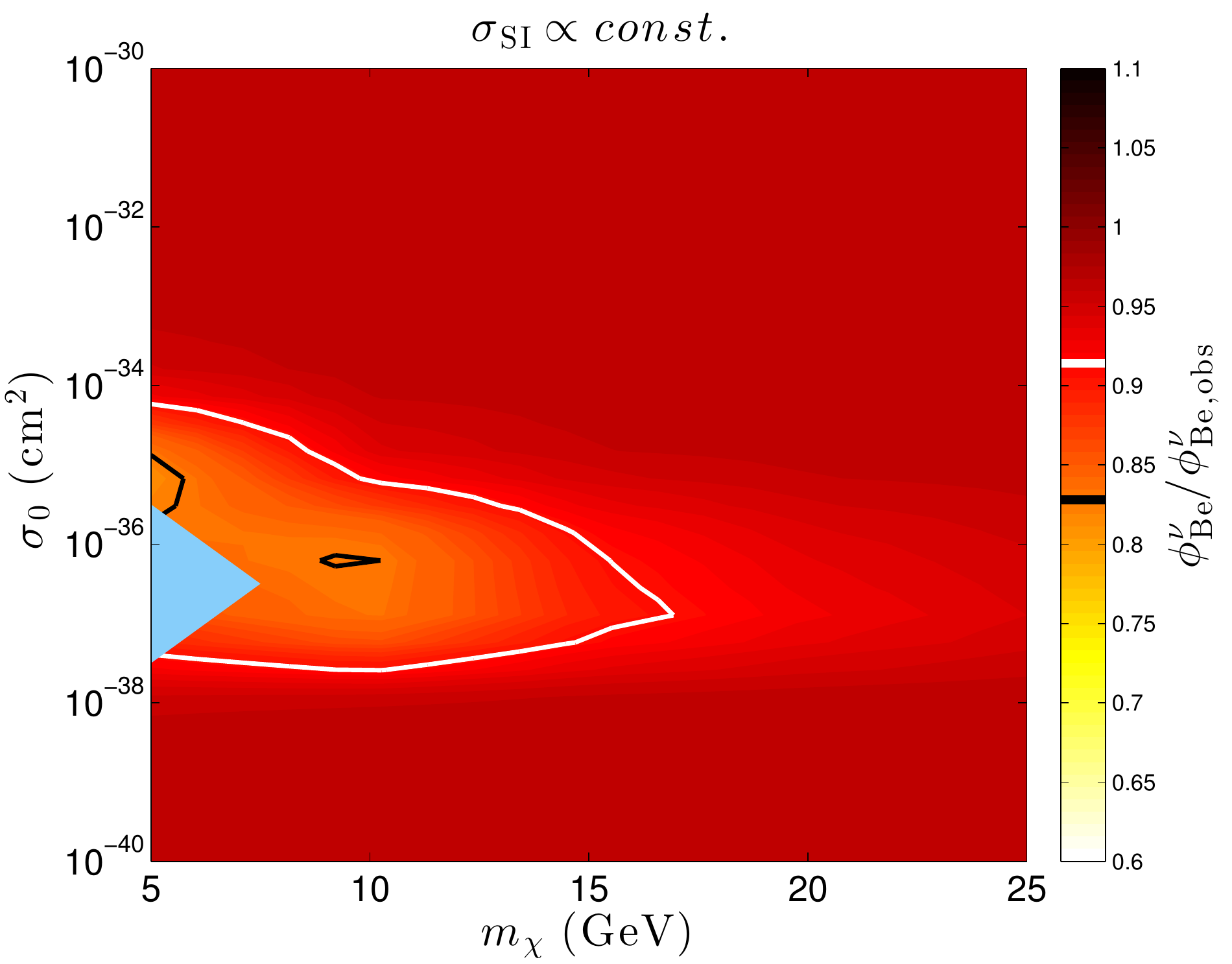}} \\
\includegraphics[height = 0.32\textwidth]{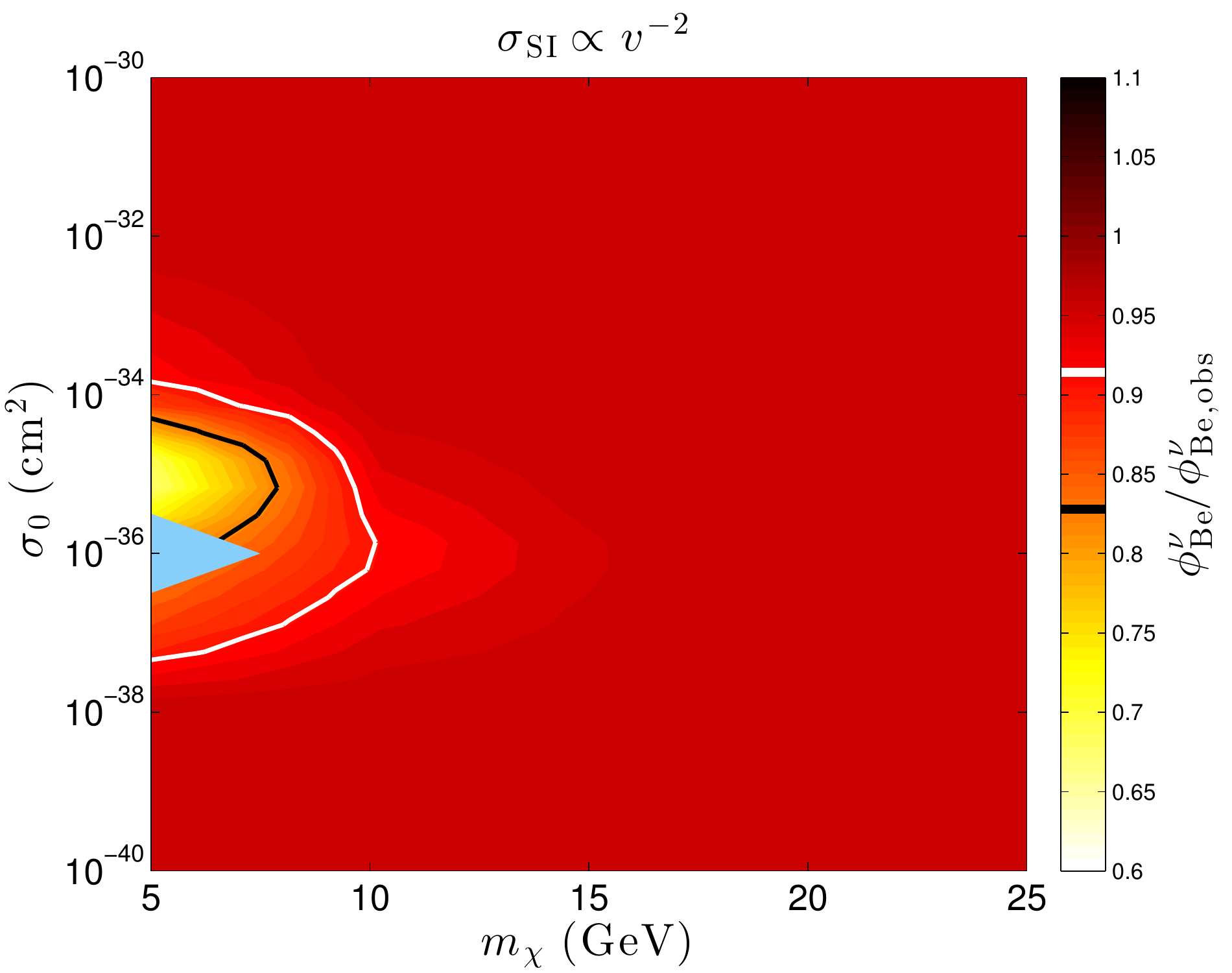} & \includegraphics[height = 0.32\textwidth]{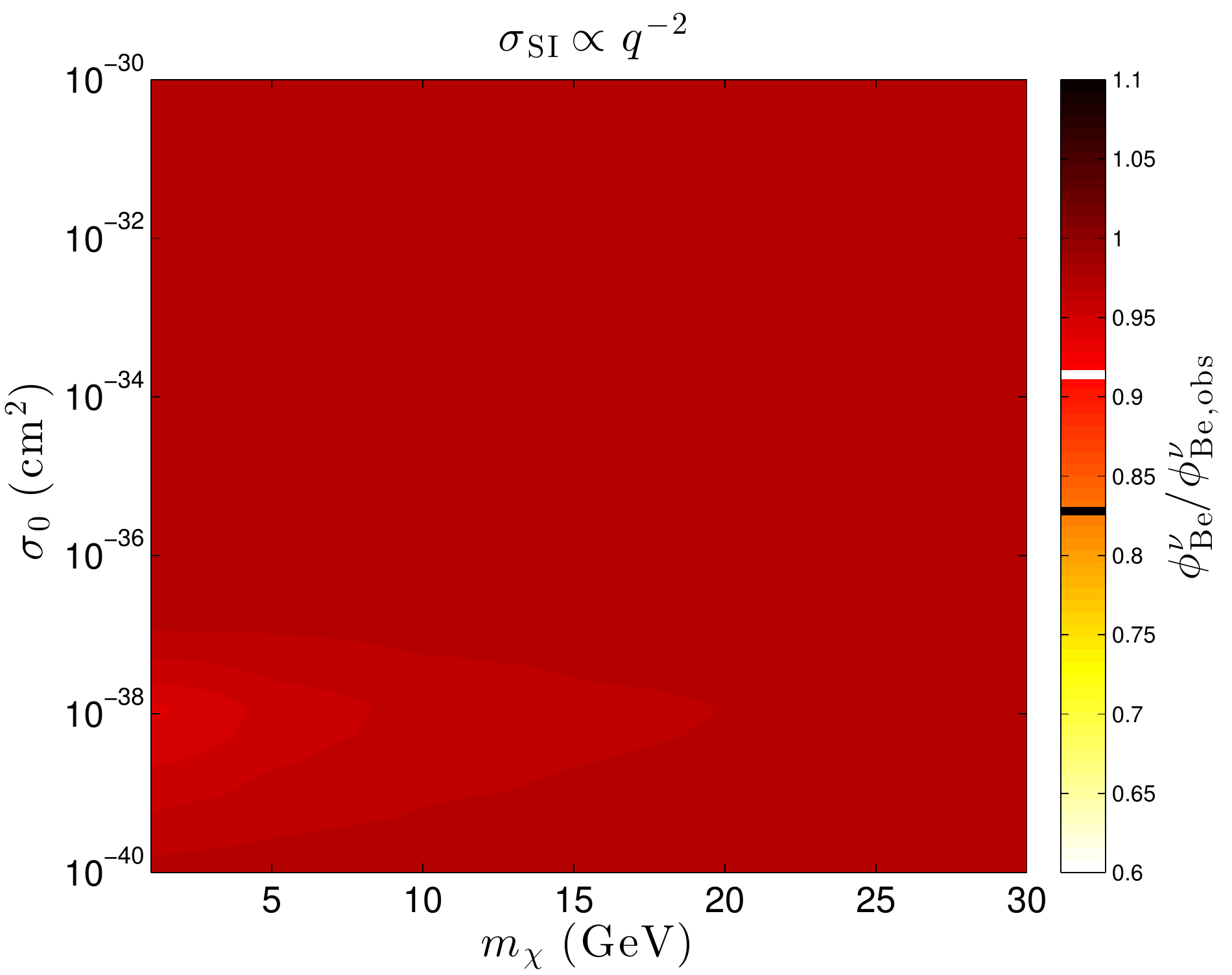} \\
\includegraphics[height = 0.32\textwidth]{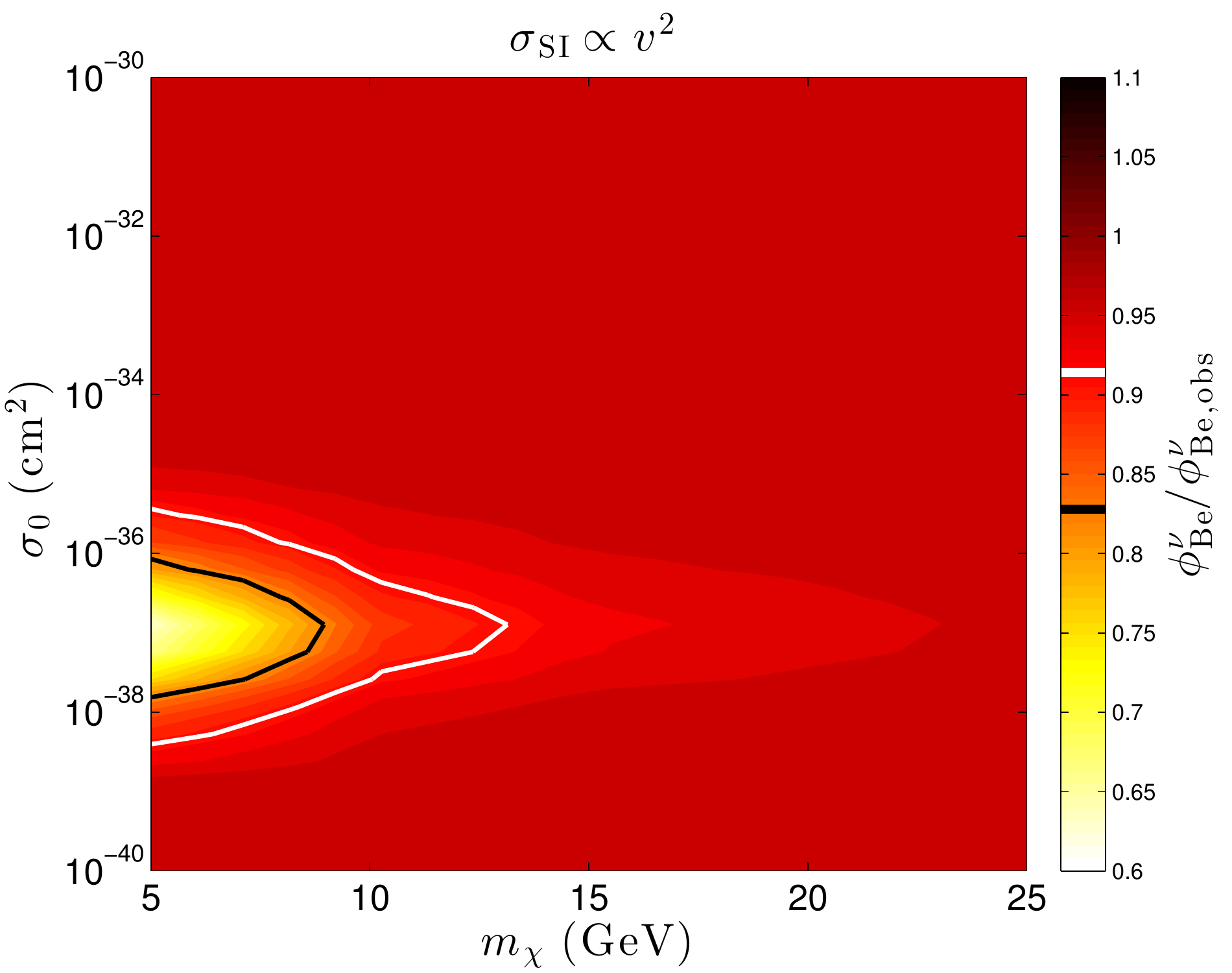} & \includegraphics[height = 0.32\textwidth]{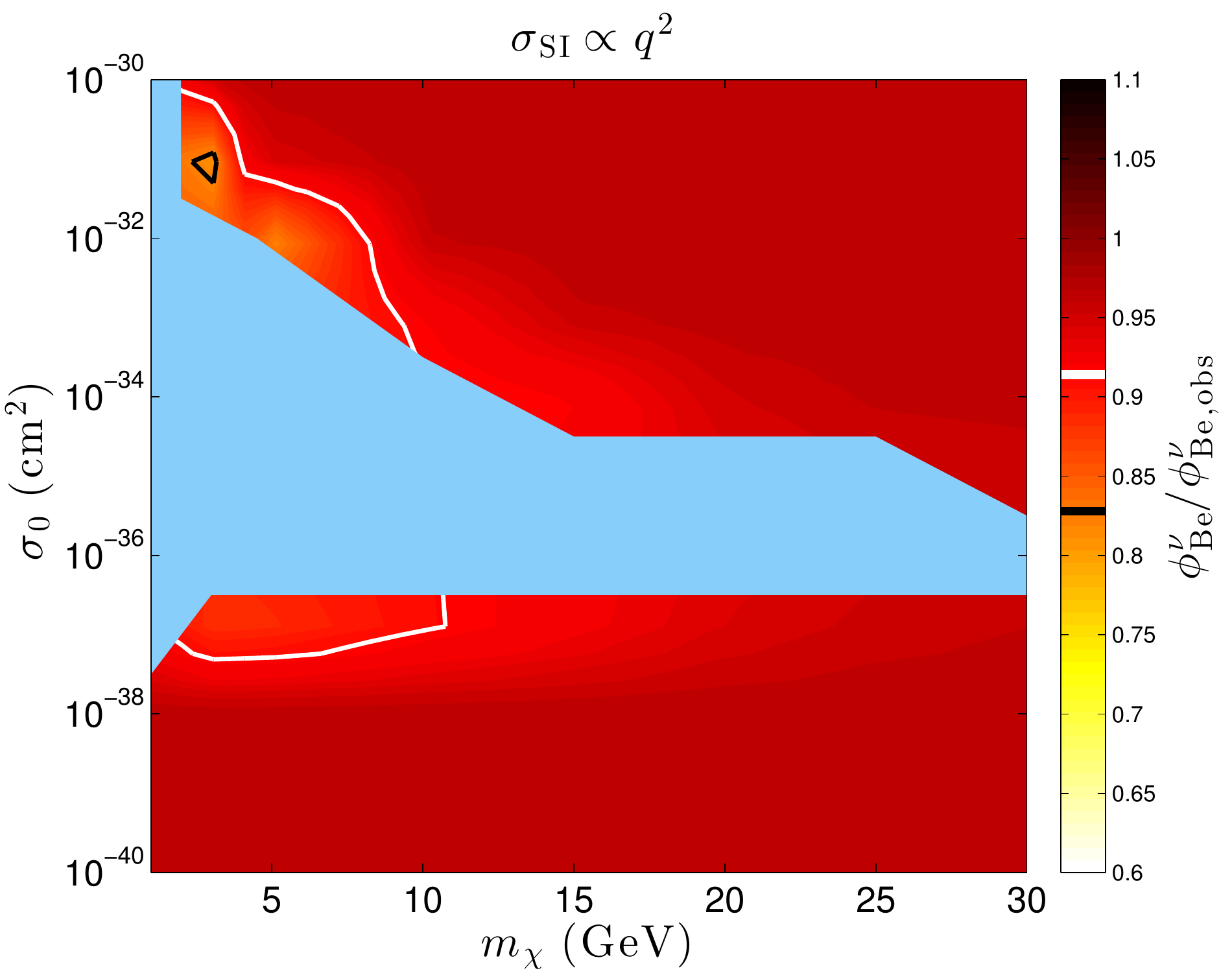} \\
\includegraphics[height = 0.32\textwidth]{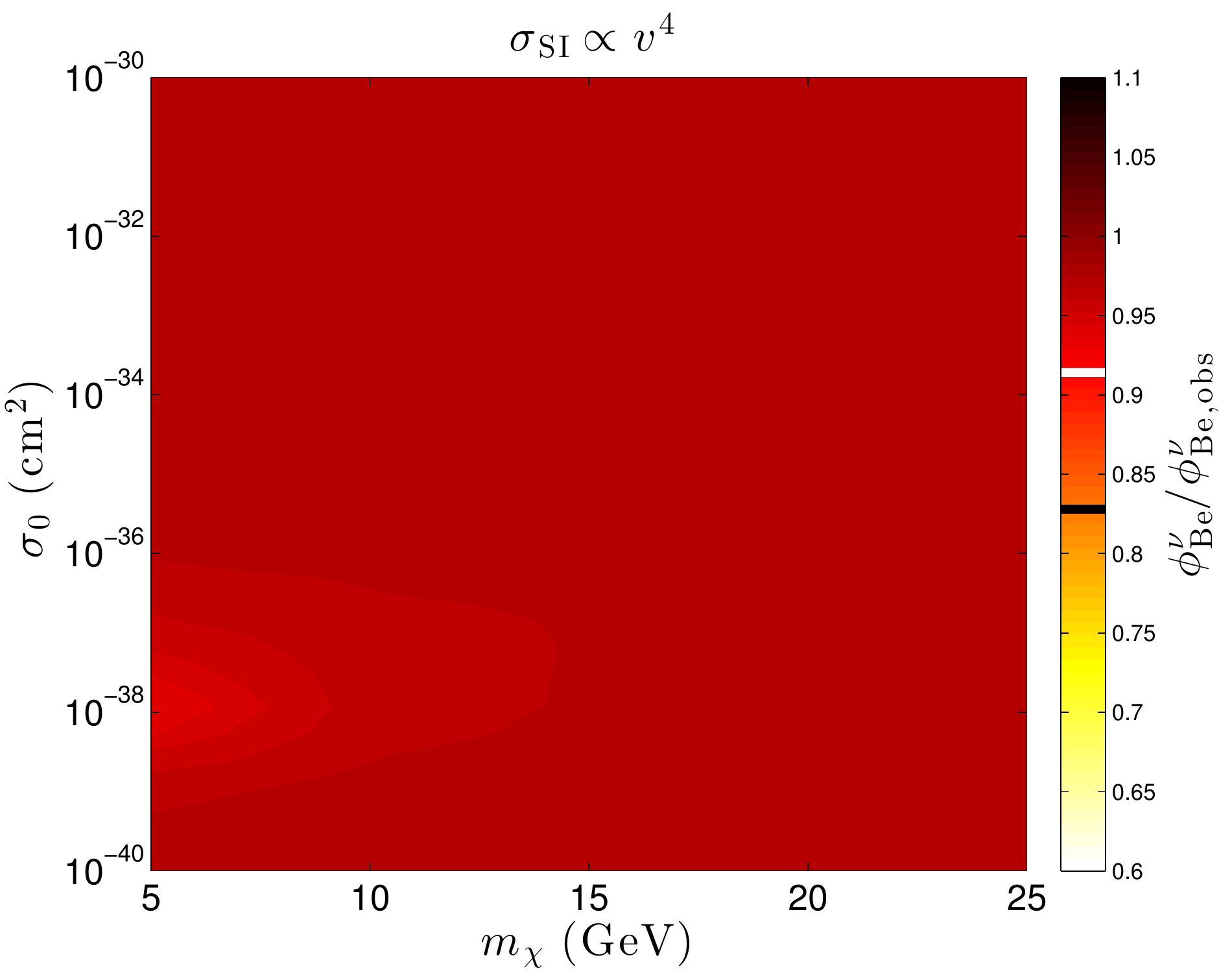} & \includegraphics[height = 0.32\textwidth]{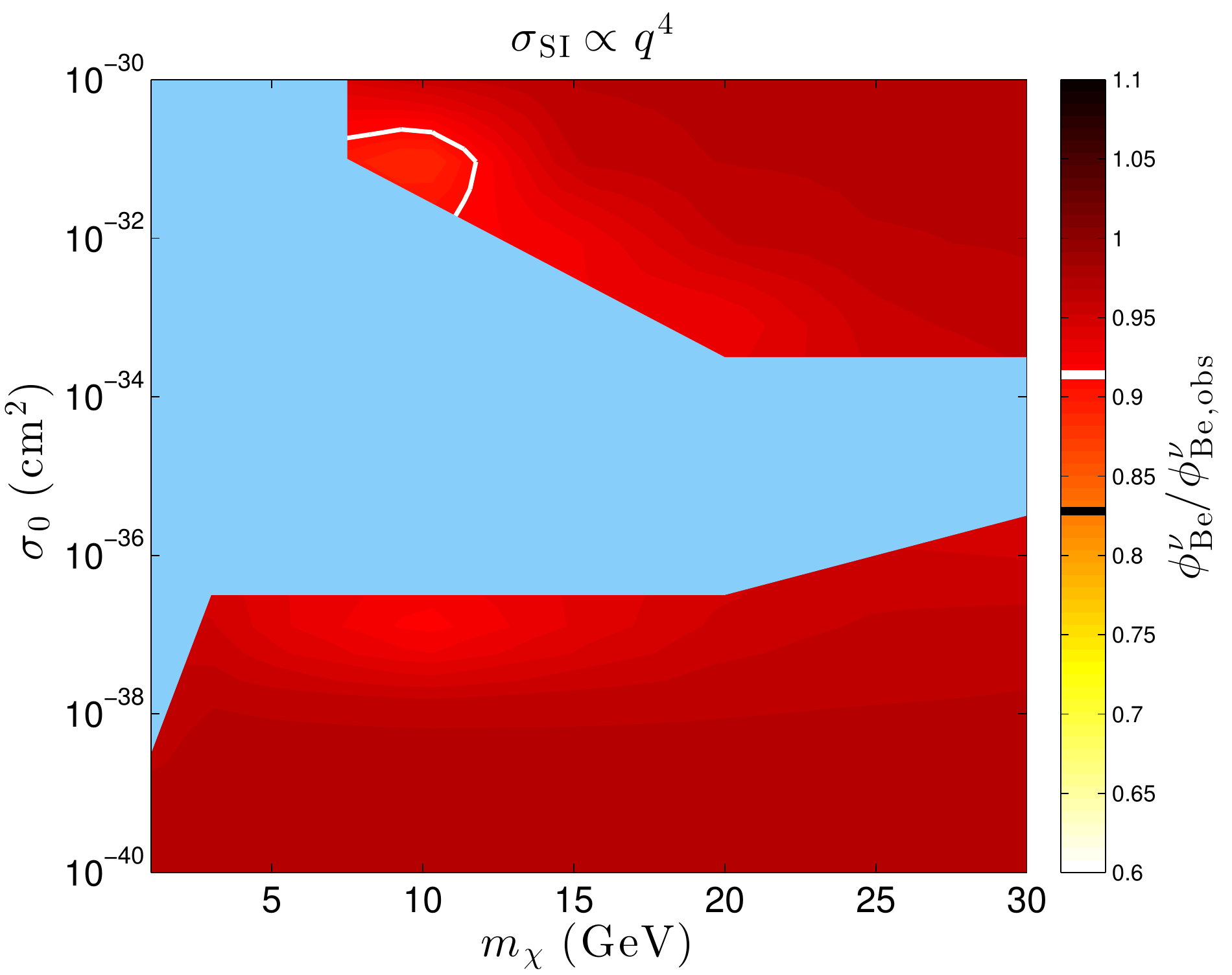} \\
\end{tabular}
\caption{The ratio of the predicted $^7$Be neutrino flux to the  measured value $\phi^\nu_{\rm Be,obs} = 4.82 \times 10^9$\,cm$^{-2}$s$^{-1}$, for each type of spin-independent dark matter coupling defined in Eq.\ \ref{qdepvdep}.  In every case the white and black lines show the isocontours where the flux is respectively 1 and 2$\sigma$ lower than the observed values, based on observational (5\%) and modelling (7\%) errors, added in quadrature. The cross-sections are normalized such that $\sigma = \sigma_0 (v/v_0)^{2n}$ or $\sigma = \sigma_0 (q/q_0)^{2n}$, with $v_0 = 220$\,km\,s$^{-1}$ and $q_0 = 40$\,MeV.}
\label{SIBerylliumfluxes}
\end{figure}

\begin{figure}[p]
\begin{tabular}{c@{\hspace{0.04\textwidth}}c}
\multicolumn{2}{c}{\includegraphics[height = 0.32\textwidth]{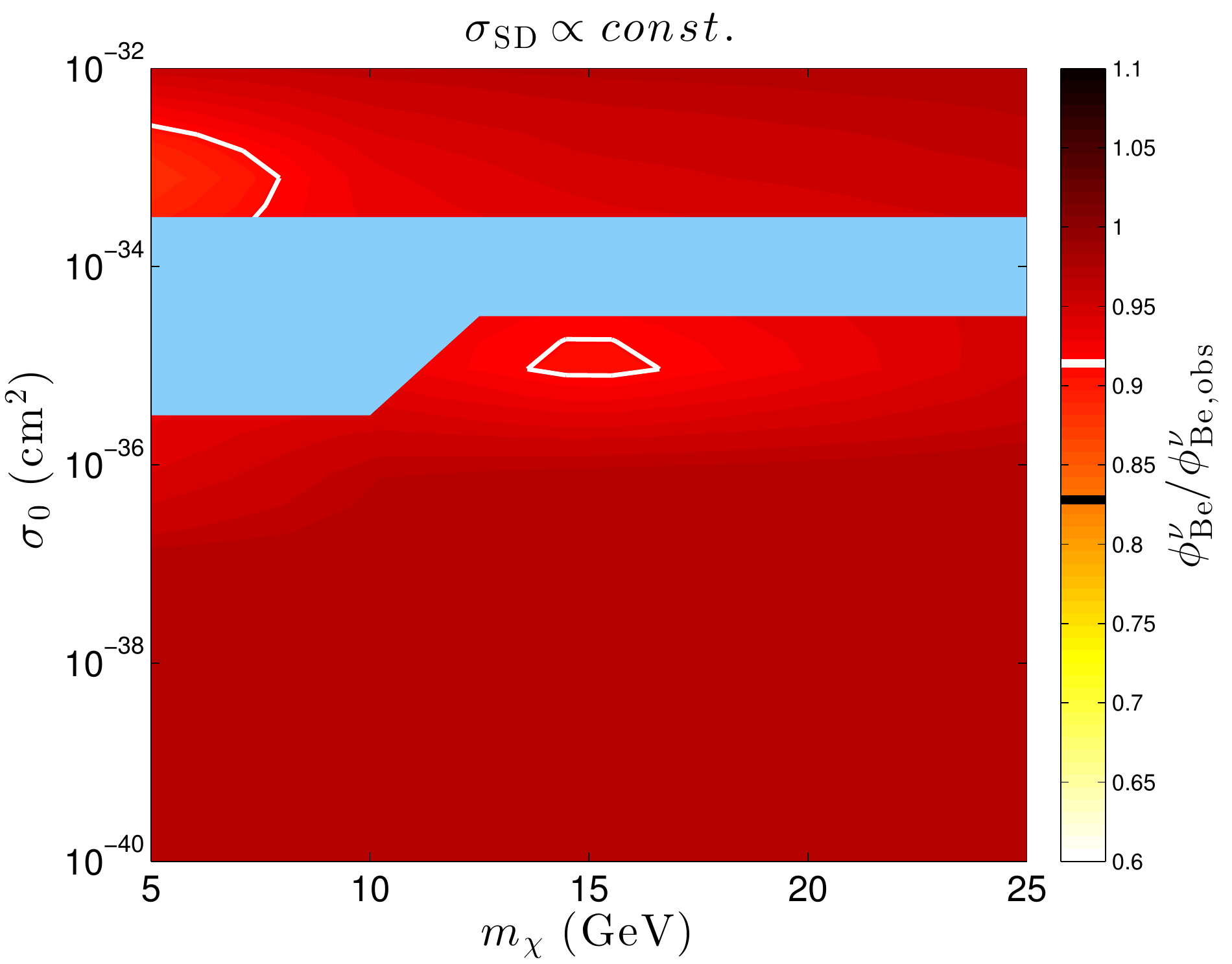}} \\
\includegraphics[height = 0.32\textwidth]{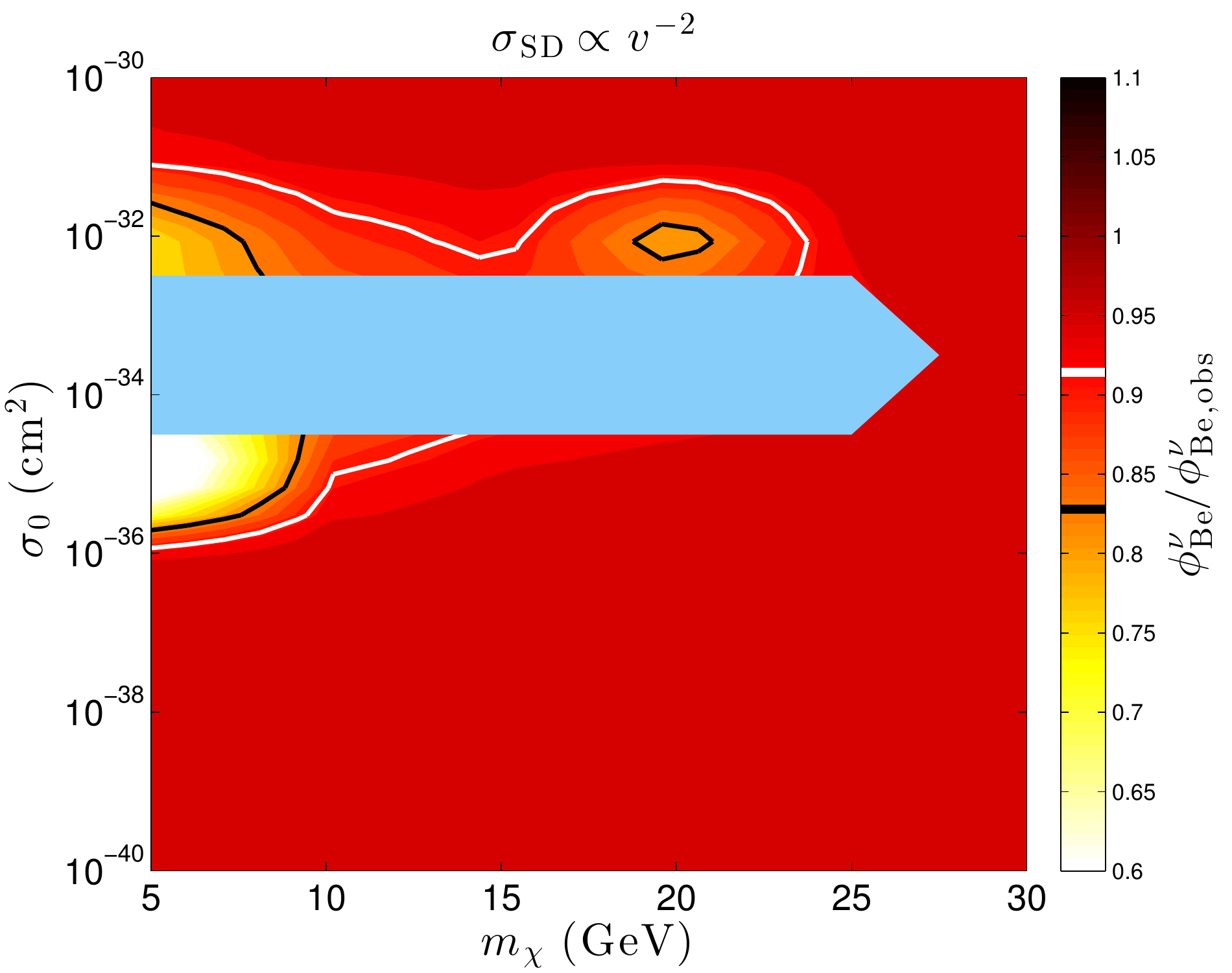} & \includegraphics[height = 0.32\textwidth]{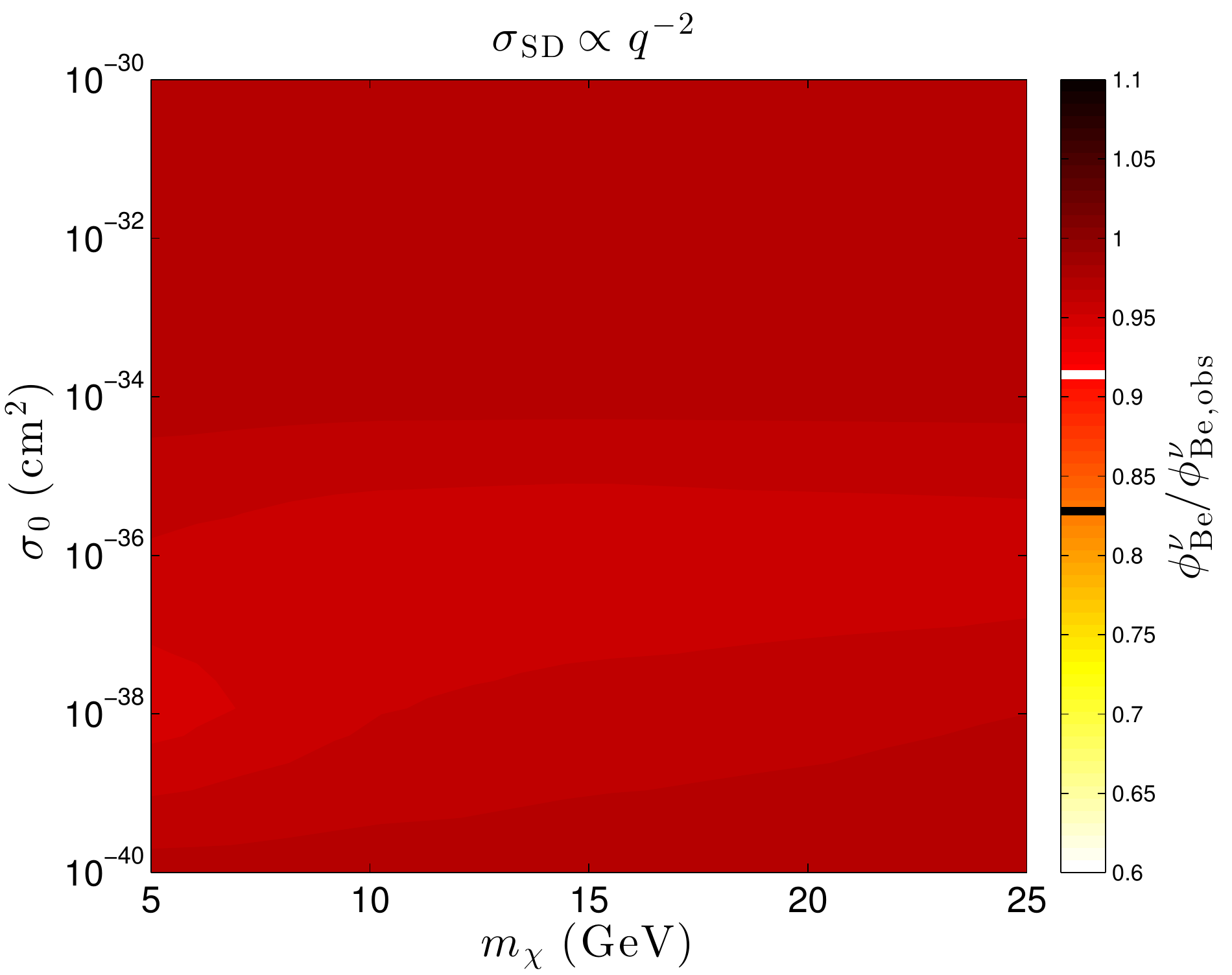} \\
\includegraphics[height = 0.32\textwidth]{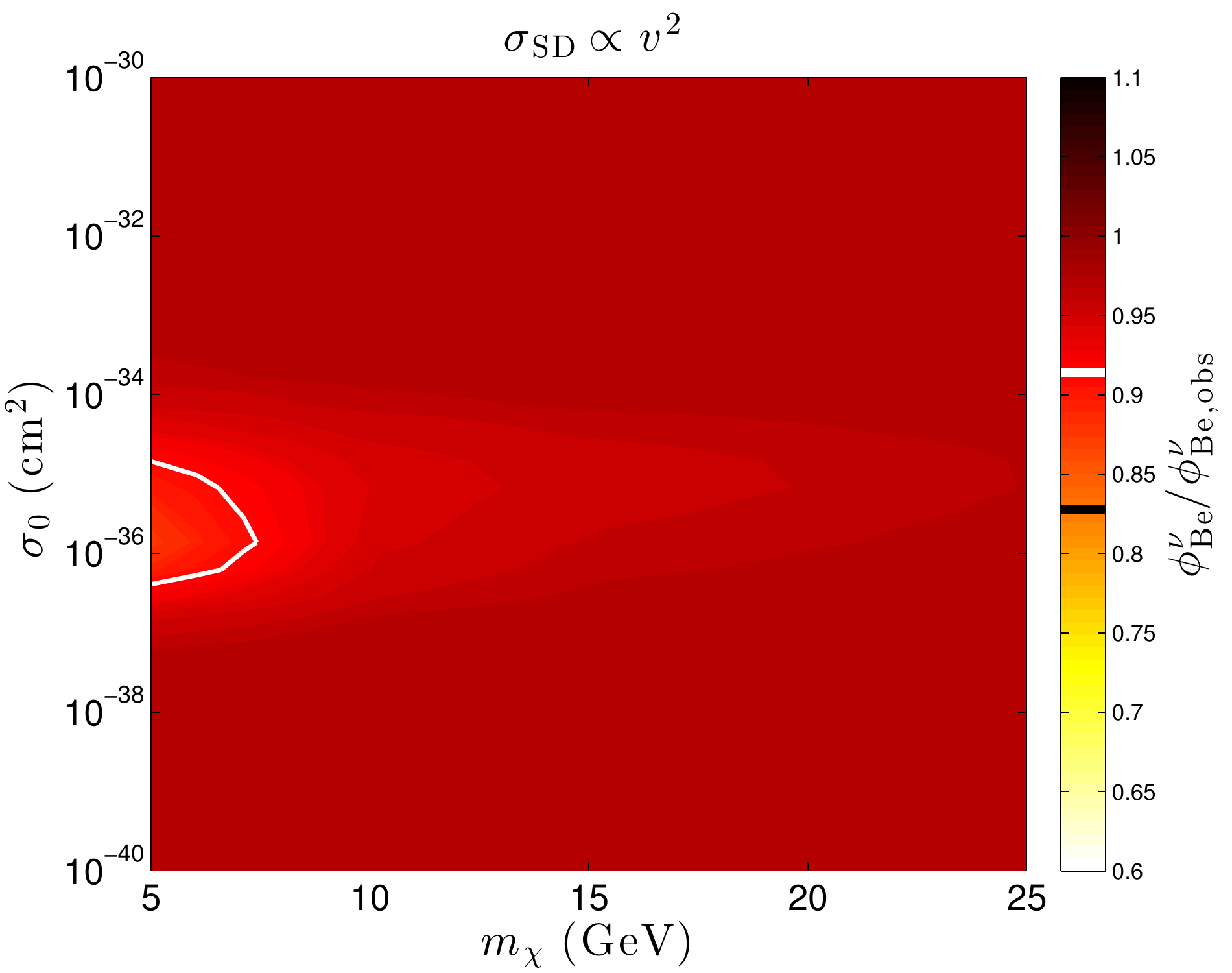} & \includegraphics[height = 0.32\textwidth]{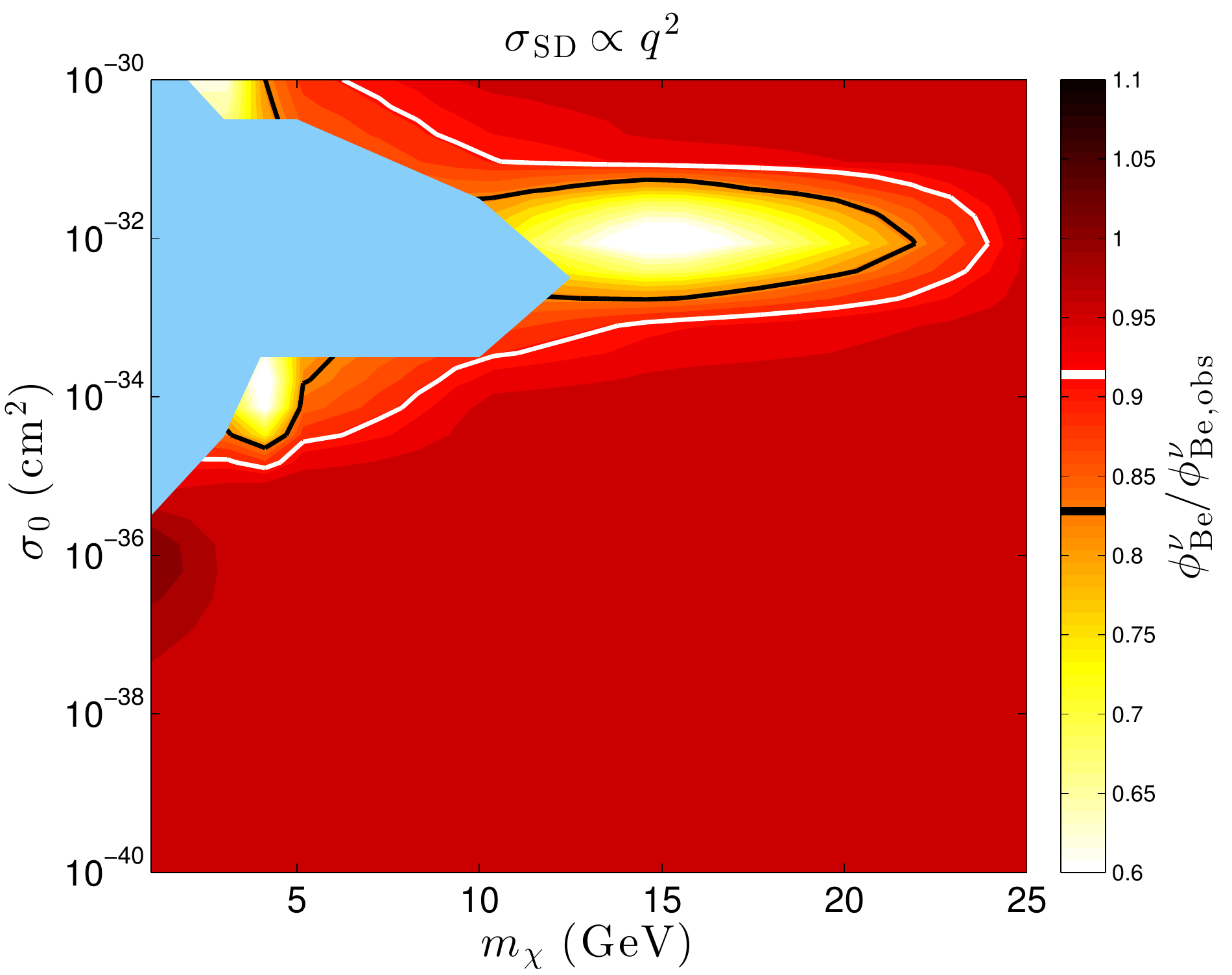} \\
\includegraphics[height = 0.32\textwidth]{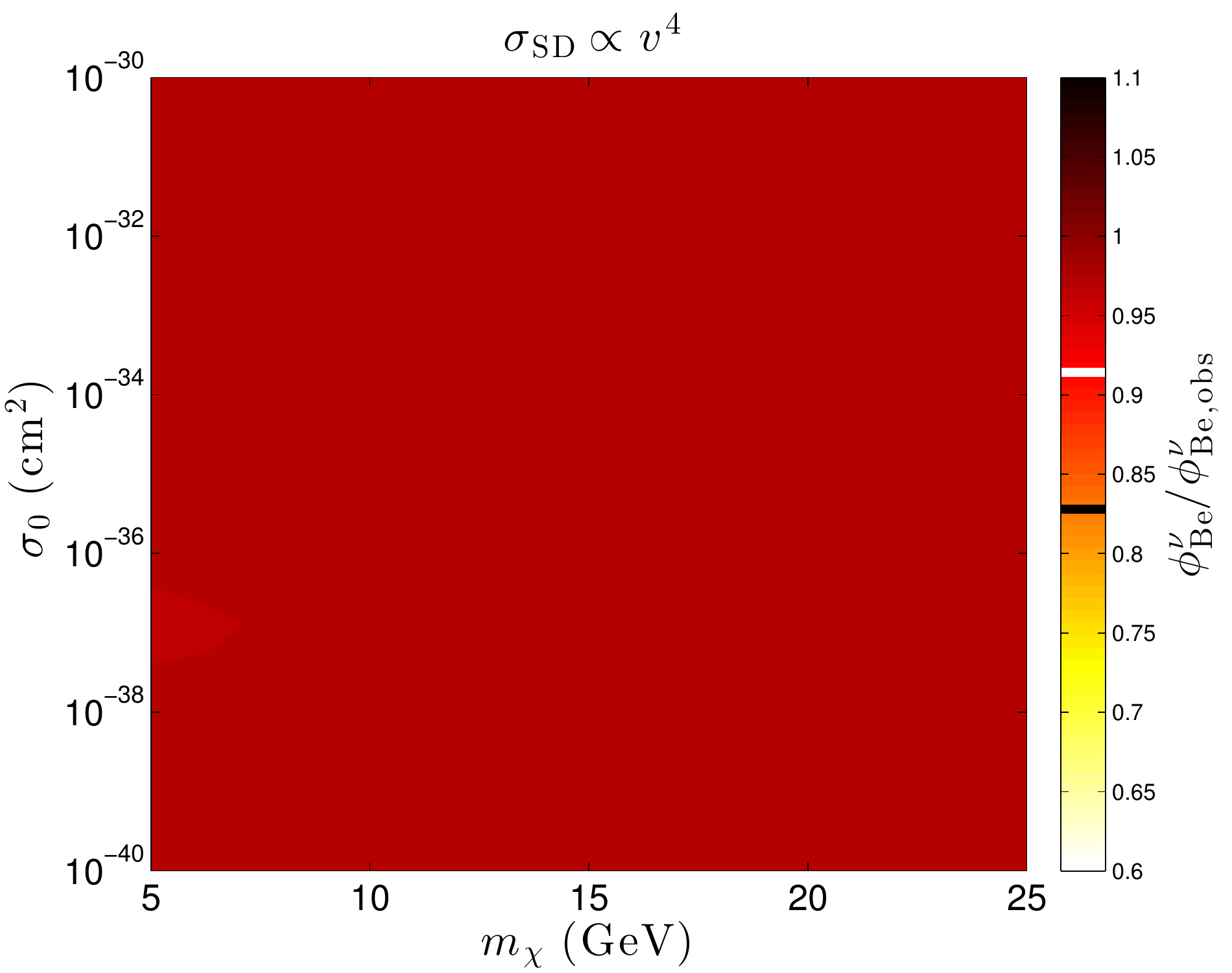} & \includegraphics[height = 0.32\textwidth]{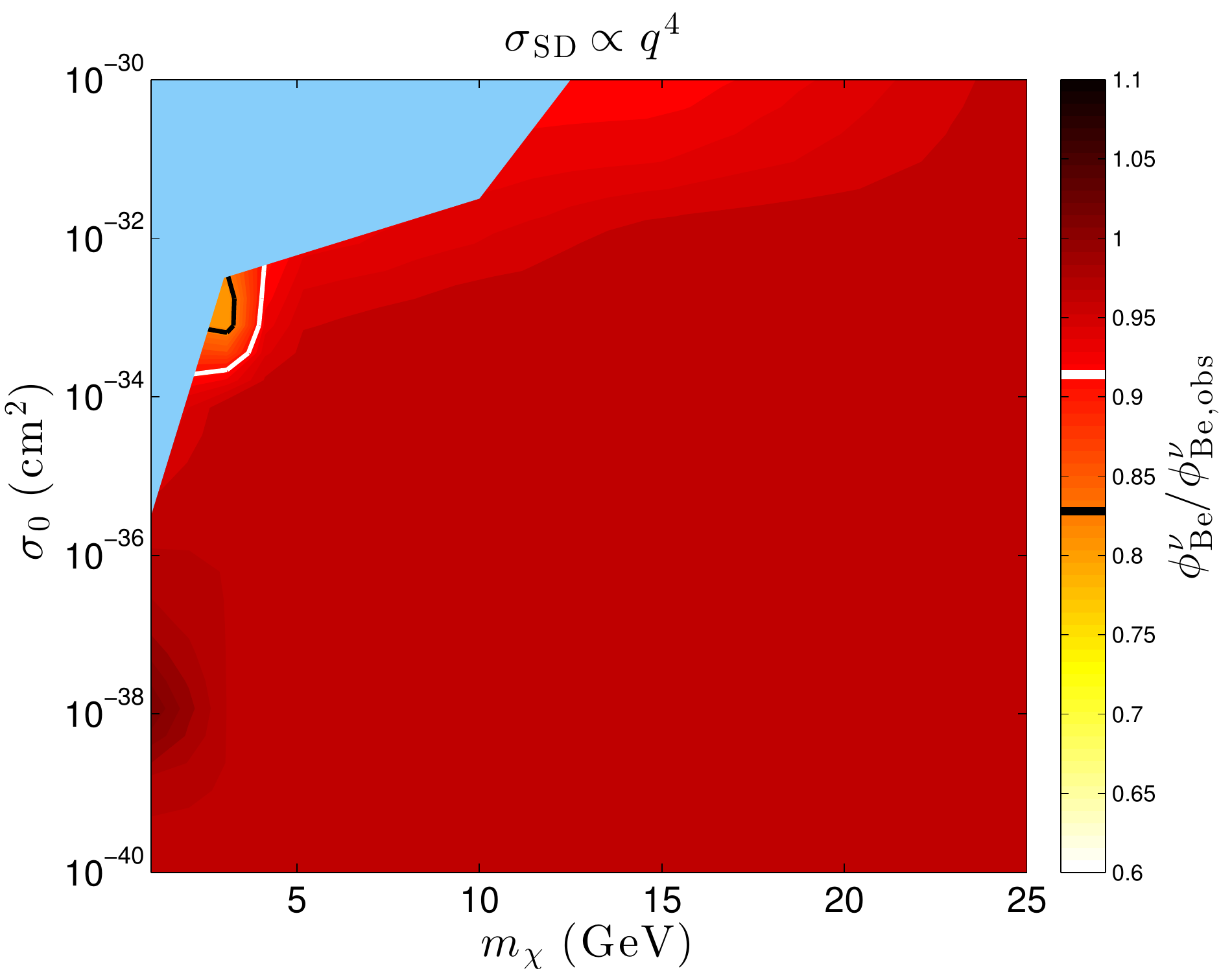} \\
\end{tabular}
\caption{As per Fig.\ \ref{SIBerylliumfluxes}, but for spin-dependent couplings.}
\label{SDBerylliumfluxes}
\end{figure}

Although the flux of $^7$Be neutrinos is not as temperature-sensitive as that of $^8$B neutrinos, they can also be used as a weaker, independent, probe of the solar core temperature.  In Figs.\ \ref{SIBerylliumfluxes} and \ref{SDBerylliumfluxes} we show the ratio of the predicted $^7$Be neutrino fluxes to the observed value $\phi^\nu_{\rm Be,obs} = 4.82 \times 10^{9}$ cm$^{-2}$s$^{-1}$.  Here again we plot $1\sigma$ and $2\sigma$ contours using the theoretical and observational uncertainties, which are respectively  7\% and 5\% for $^7$Be neutrinos.

\FloatBarrier

As expected, the effect of ADM on neutrino fluxes is larger for low DM masses, mainly due to the increased efficiency of momentum transfer between DM and H/He as the DM takes on a similar mass to these two dominant nuclei. Ignoring evaporation, transport becomes even more efficient as mass is decreased to 1 GeV for SD dark matter; in the SI case, the largest effect is seen at $\mx \simeq$ 3 to 4 GeV, around the average nucleus mass in the Sun.

The dependence on the cross-section is also as expected: below a minimum value of $\sigma_0$, not enough DM can be captured to significantly affect the $^8$B production rate. For constant cross-sections, this is around $10^{-38}$ cm$^2$ in the SI case, and $10^{-36}$ cm$^2$ for SD dark matter. This highlights the importance of heavier elements, to which only the SI DM couples in our simulations: the large fraction of helium, combined with the $A^2$ dependence of the DM-nucleus cross-section (Eq.\ \ref{DMnucleus}) means that the behaviour of DM inside a star is highly sensitive to elements heavier than hydrogen. 

The upper edges of the regions shown in Figs.\ \ref{SIboronfluxes}--\ref{SDBerylliumfluxes} highlight the transport behaviour: if the cross-section falls below a certain value, it allows the DM to more efficiently penetrate the dense plasma, carrying energy from hot to cooler regions. The location of the maximum reflects a combination of being near the transport efficiency peak (where the transition from the LTE to the Knudsen regime occurs) while simultaneously maintaining a large capture rate through a large enough cross-section.

The impacts of velocity and momentum-dependent scattering of dark matter can be understood in terms of this balance between capture and transport.  Positive powers of $\vrel$ lead to an enhancement in the capture rate, moving the window in which DM transport has a significant effect on neutrino fluxes to lower cross-sections. However, the important suppression in the overall transport rate, as seen in Fig.\ 5 of \cite{VincentScott2013}, yields an overall suppression in the effect relative to the constant case. The opposite behaviour is seen for $\sigma \propto \vrel^{-2}$, although the enhanced transport is not sufficient to compensate for the $\sim$$10^{-2}$ suppression in capture. 

Again, the effect of a momentum-dependent cross-section is more subtle. Although the boosted capture rate for a $q^{-2}$ cross-section moves the region of interest to lower cross-sections, closer to what is allowed by underground experiments, the overall suppression in transport means that there is very little overall effect on solar observables. Positive powers of $q$ fare much better, with an enhancement both in capture and transport.

Comparison of Figs.\ \ref{SIboronfluxes} and \ref{SDboronfluxes} highlights the importance of heavier elements in such processes. In most cases, they lead to a significant enhancement of the capture and transport rate; however, we note in comparing the $q^{-2}$ plots that they can also inhibit energy transport, by reducing the mean free path for conduction. 

\subsection{Depth of the convection zone}
\label{sec:rcz}

The boundary between the convective and radiative zones $r_{\mathrm{CZ}}$ is the location at which the radiative and adiabatic temperature gradients are equal. Above this location, the temperature gradient becomes super-adiabatic, hydrostatic equilibrium breaks down and convection sets in. The amount of energy transported by DM is negligible at heights $r \sim r_{\rm CZ}$. However, the changes in the density and temperature gradient near the core lead to a small but significant increase in the radiative temperature gradient at much larger radii, causing it to exceed the adiabatic gradient at a slightly lower depth and shift the lower boundary of the convection zone downwards.

\begin{figure}[!p]
\begin{tabular}{c@{\hspace{0.04\textwidth}}c}
\multicolumn{2}{c}{\includegraphics[height = 0.32\textwidth]{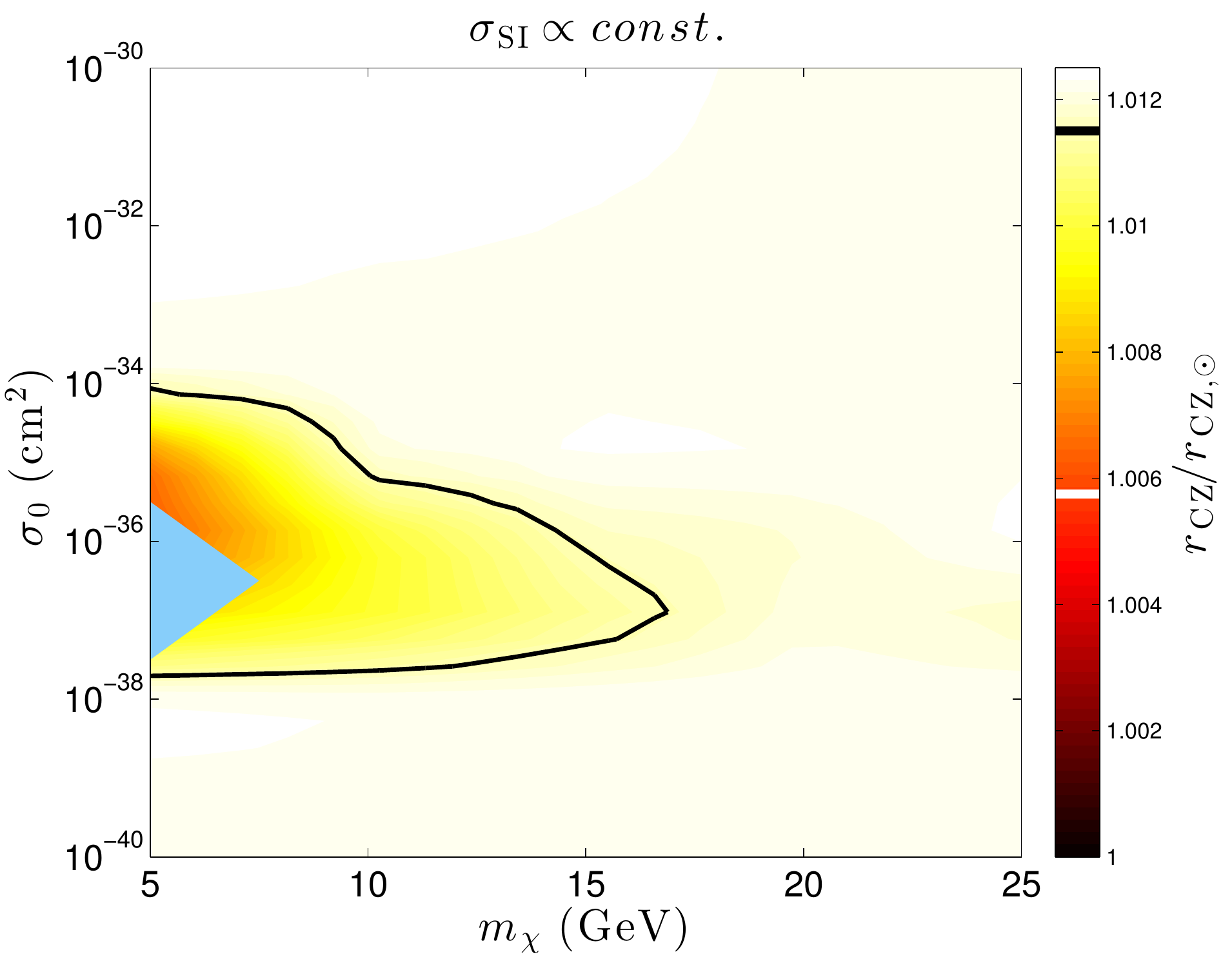}} \\
\includegraphics[height = 0.32\textwidth]{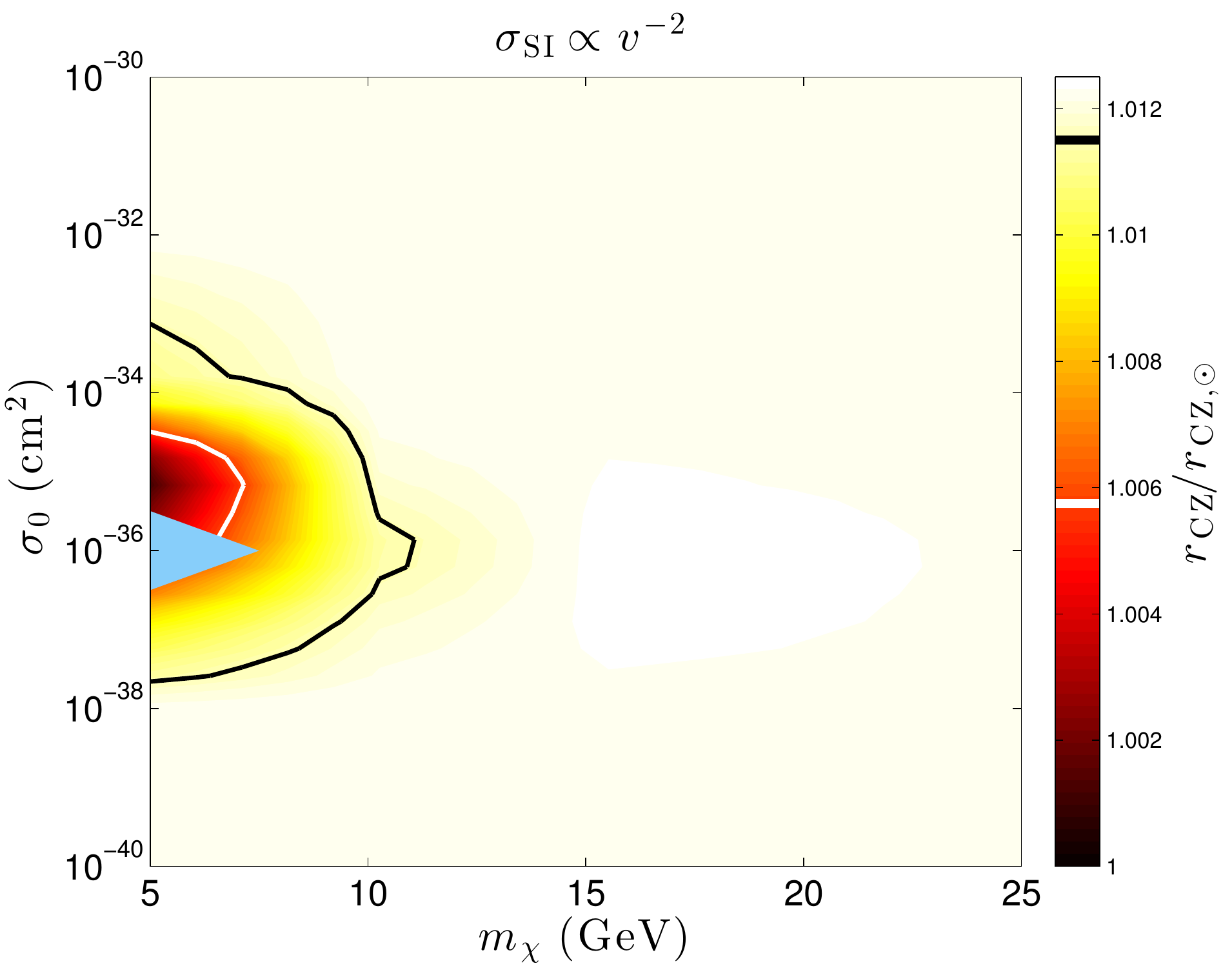} & \includegraphics[height = 0.32\textwidth]{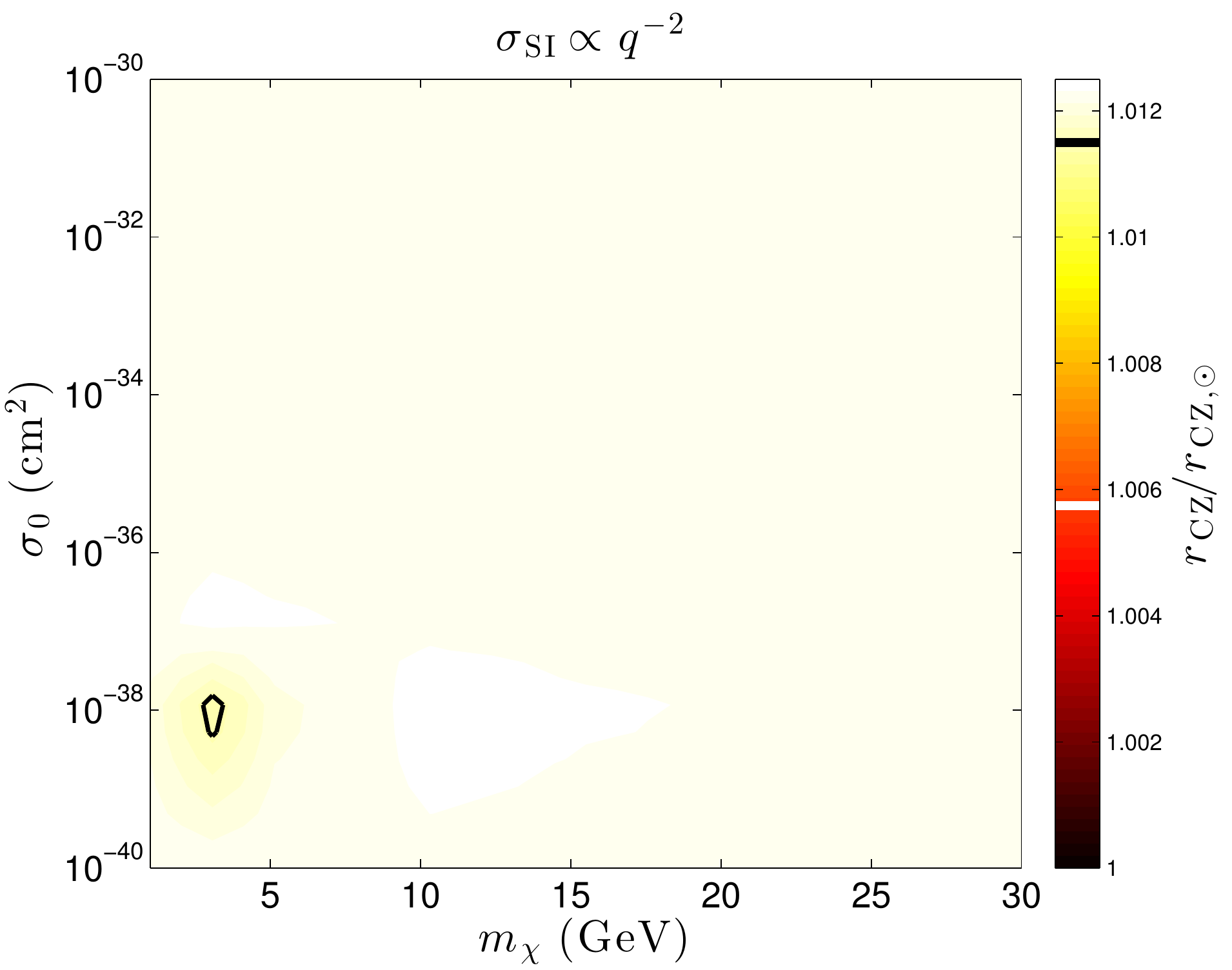} \\
\includegraphics[height = 0.32\textwidth]{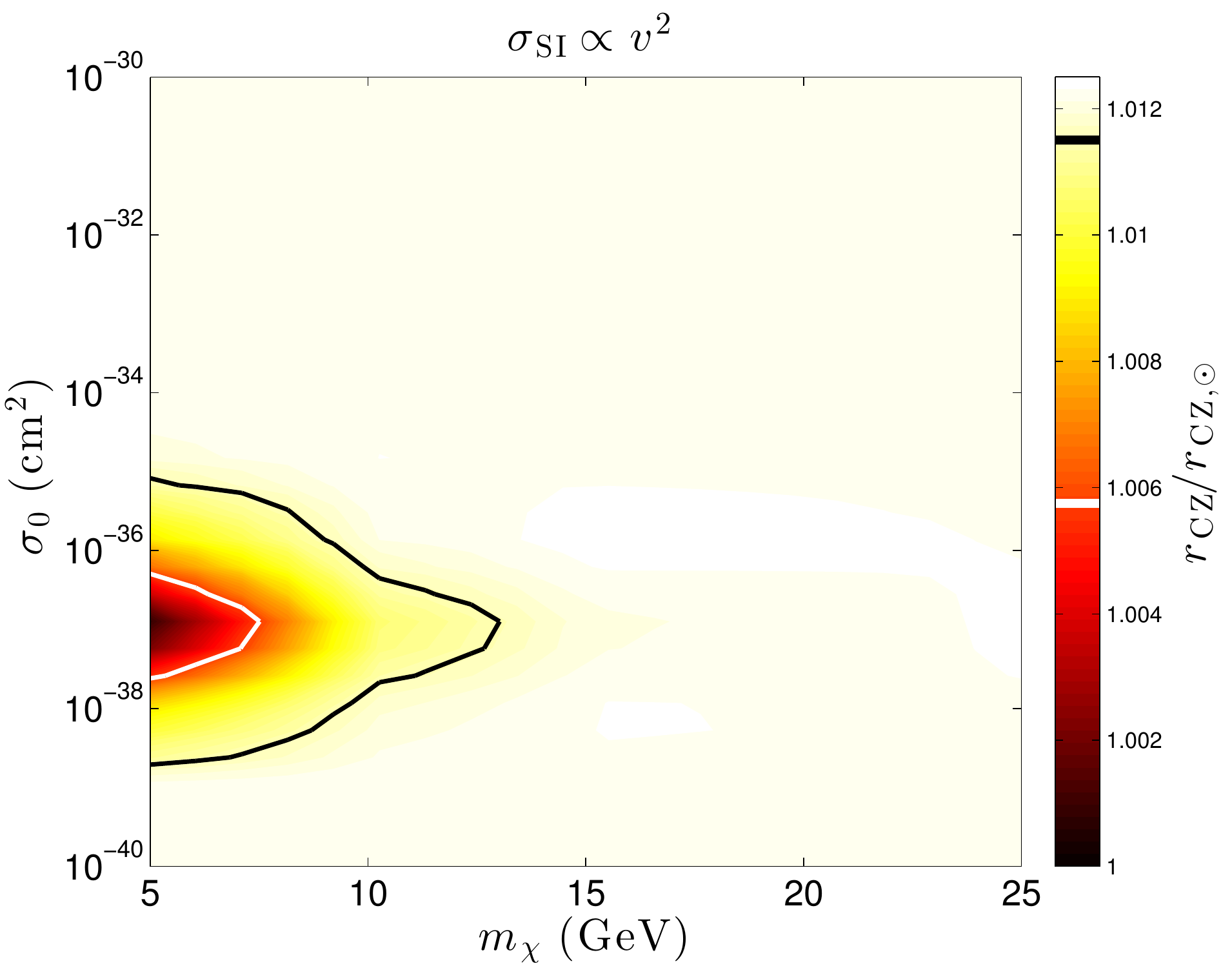} & \includegraphics[height = 0.32\textwidth]{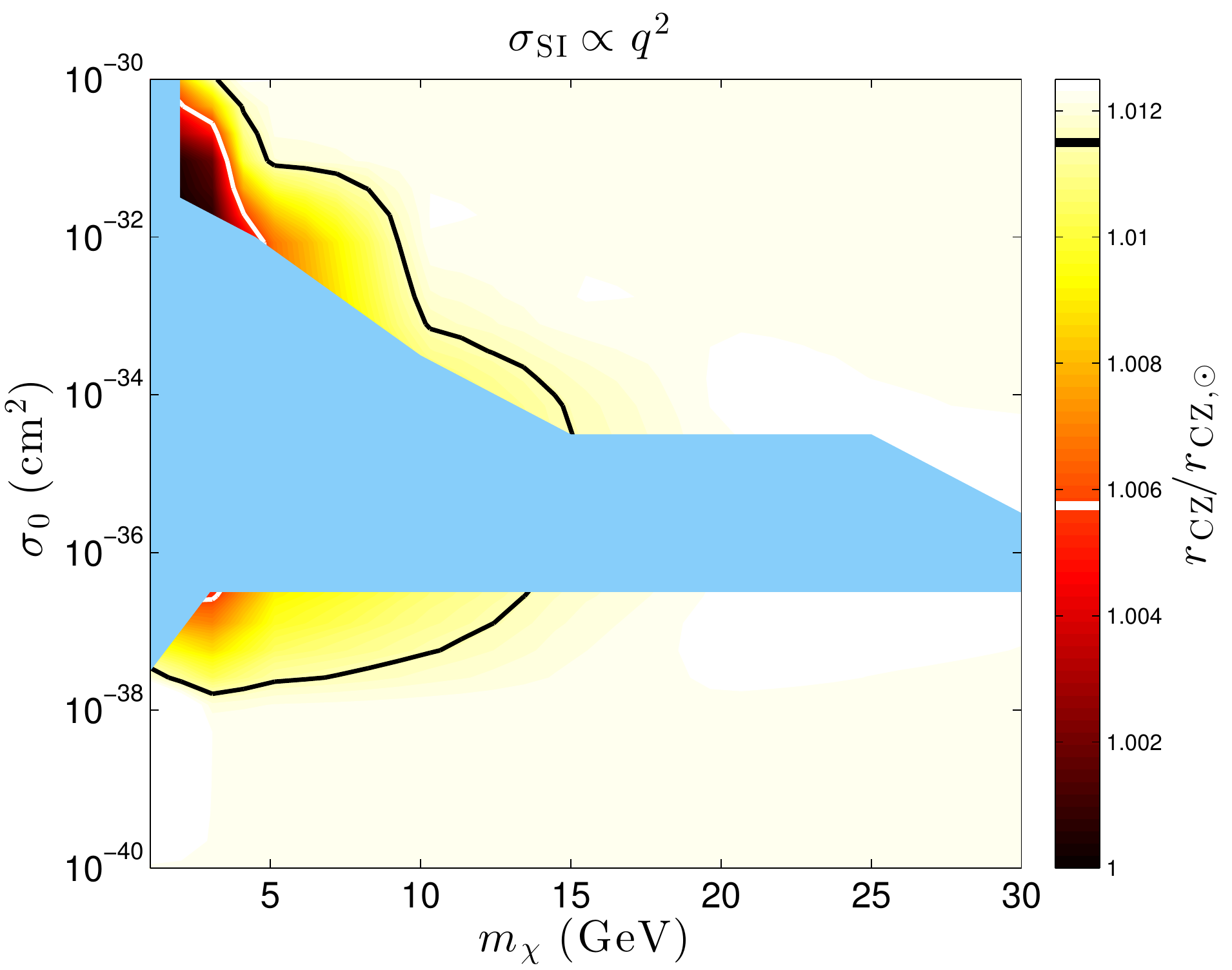} \\
\includegraphics[height = 0.32\textwidth]{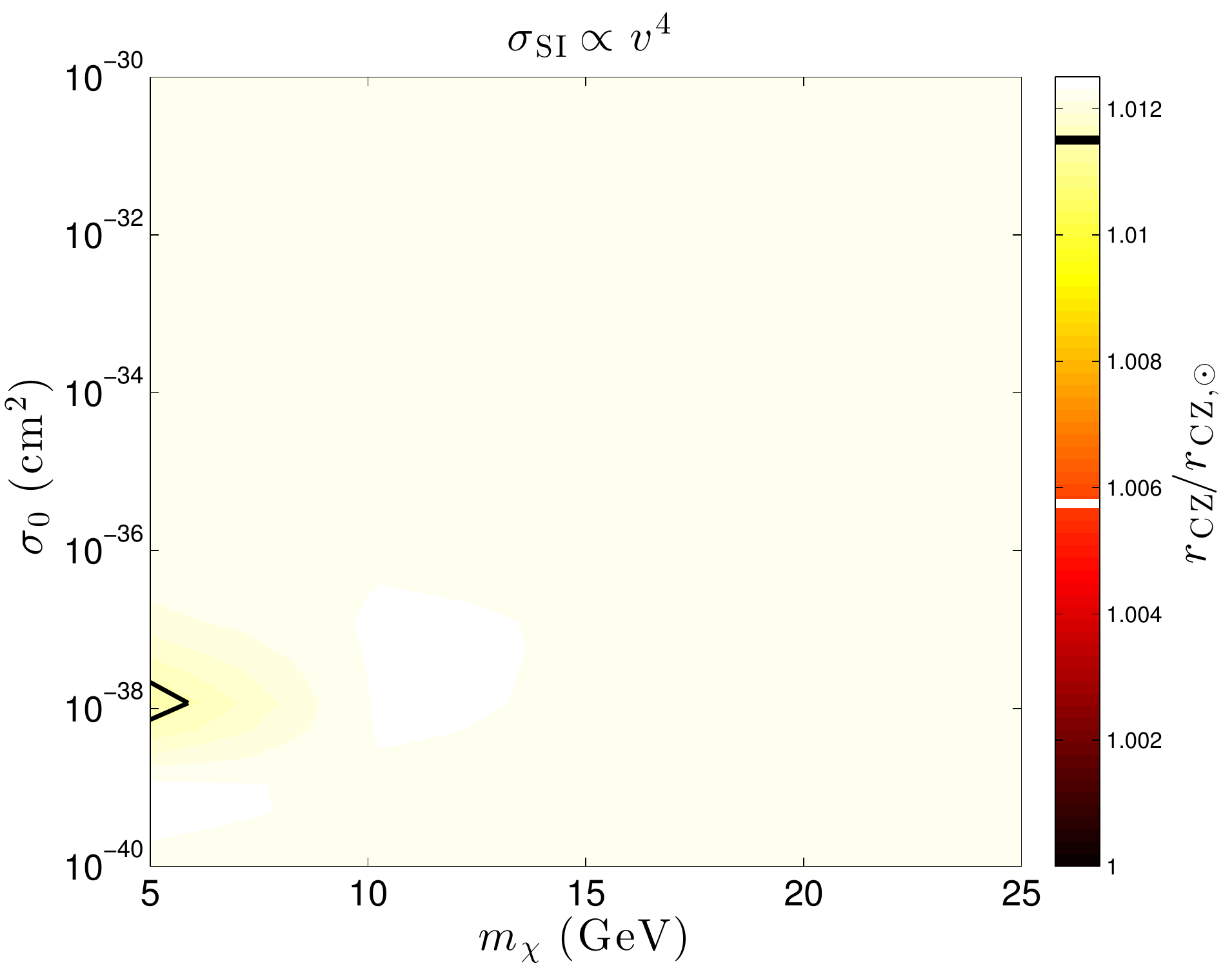} & \includegraphics[height = 0.32\textwidth]{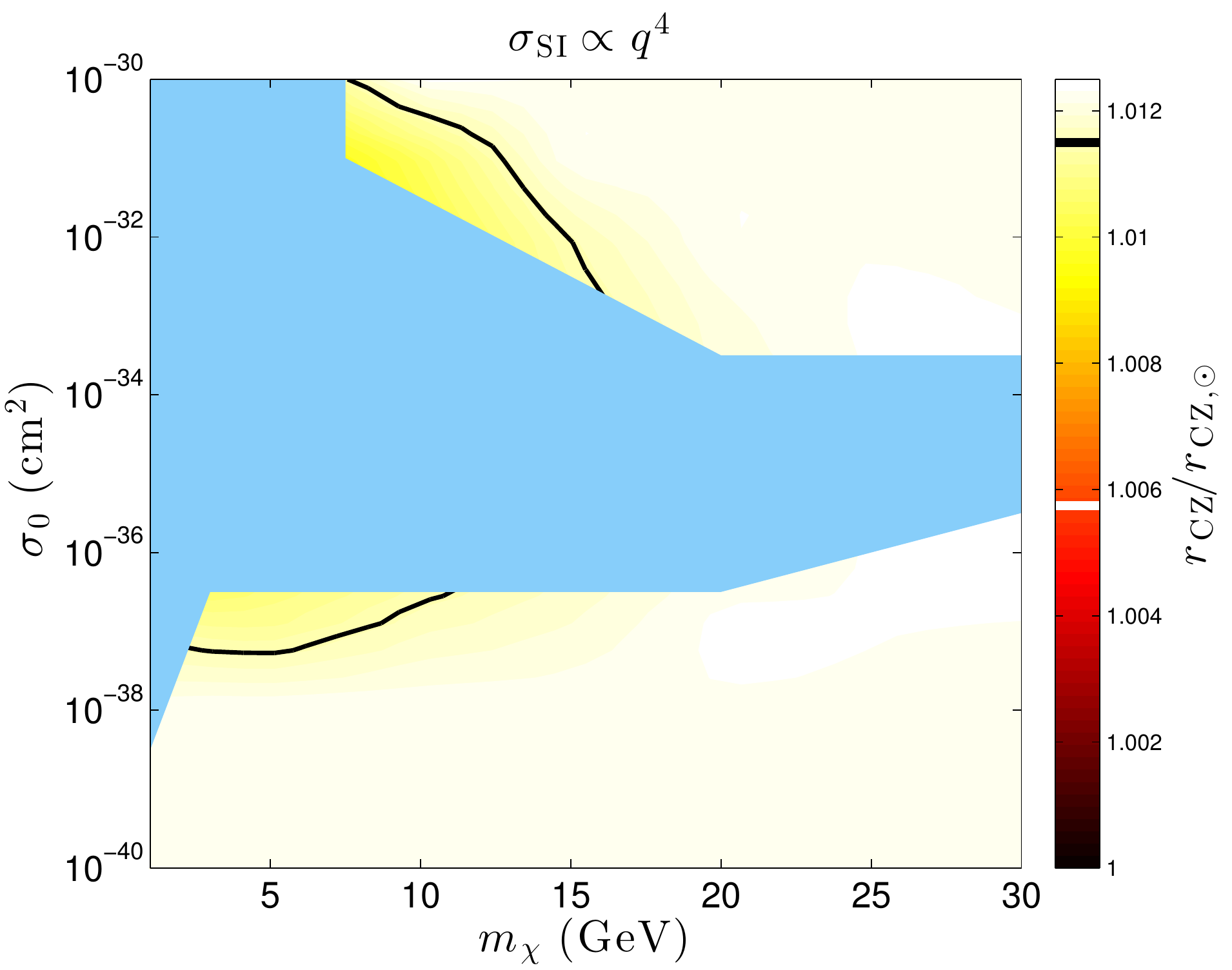} \\
\end{tabular}
\caption{Ratio between the modelled and measured location of the bottom of the convection zone $r_{\rm CZ}$, for spin-independent couplings. Darker regions represent a better fit to the observed value than the Standard Solar Model. The white and black lines represent the contours at which the predicted value falls within $1\sigma$ and $2\sigma$ of the measured value, respectively. The theoretical uncertainty on $r_{\rm CZ}$ (0.004\,$R_\odot$) is much larger than the experimental error (0.001\,$R_\odot$), so the former dominates when we add them in quadrature.}
\label{SIrc}
\end{figure}

\begin{figure}[!p]
\begin{tabular}{c@{\hspace{0.04\textwidth}}c}
\multicolumn{2}{c}{\includegraphics[height = 0.32\textwidth]{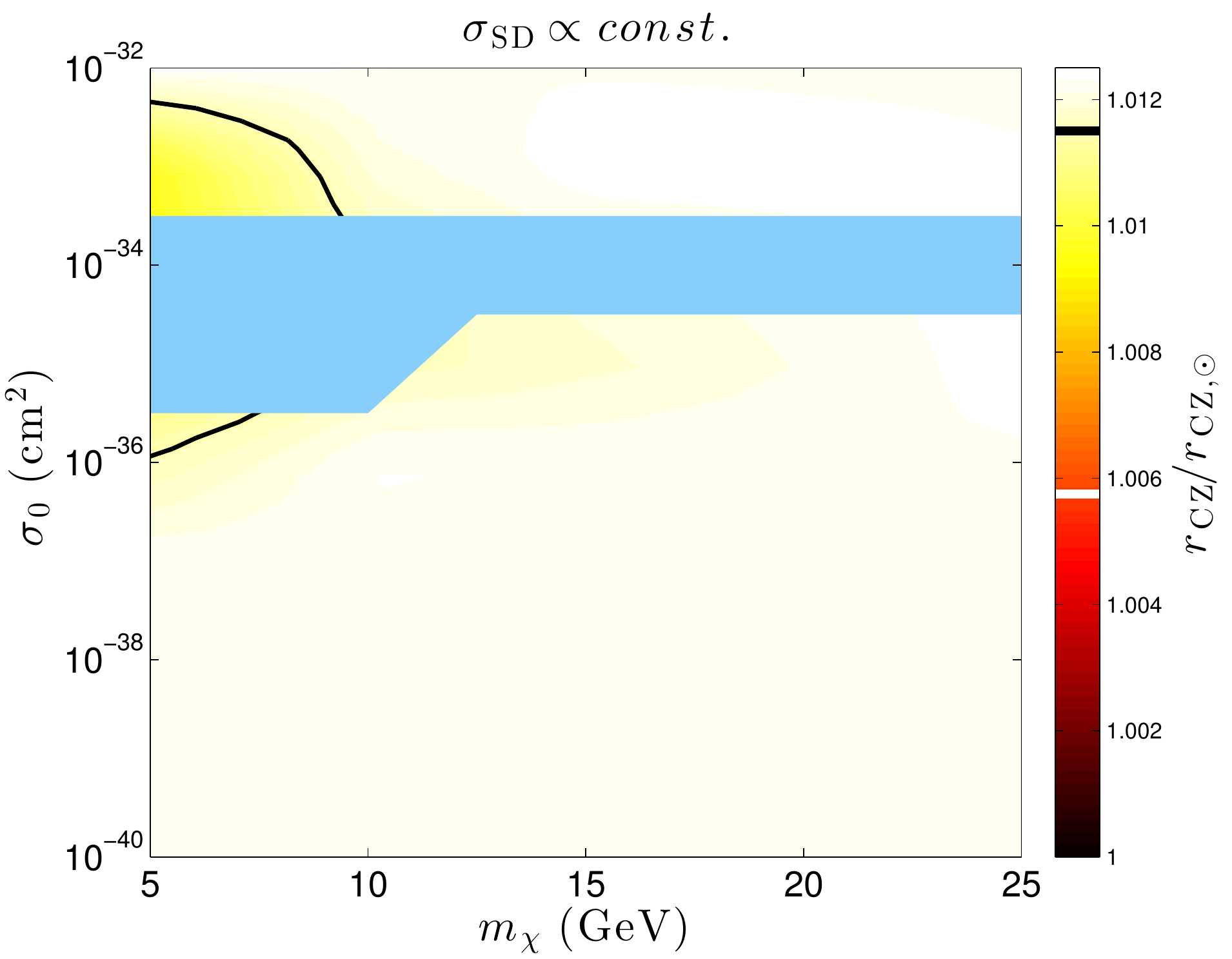}} \\
\includegraphics[height = 0.32\textwidth]{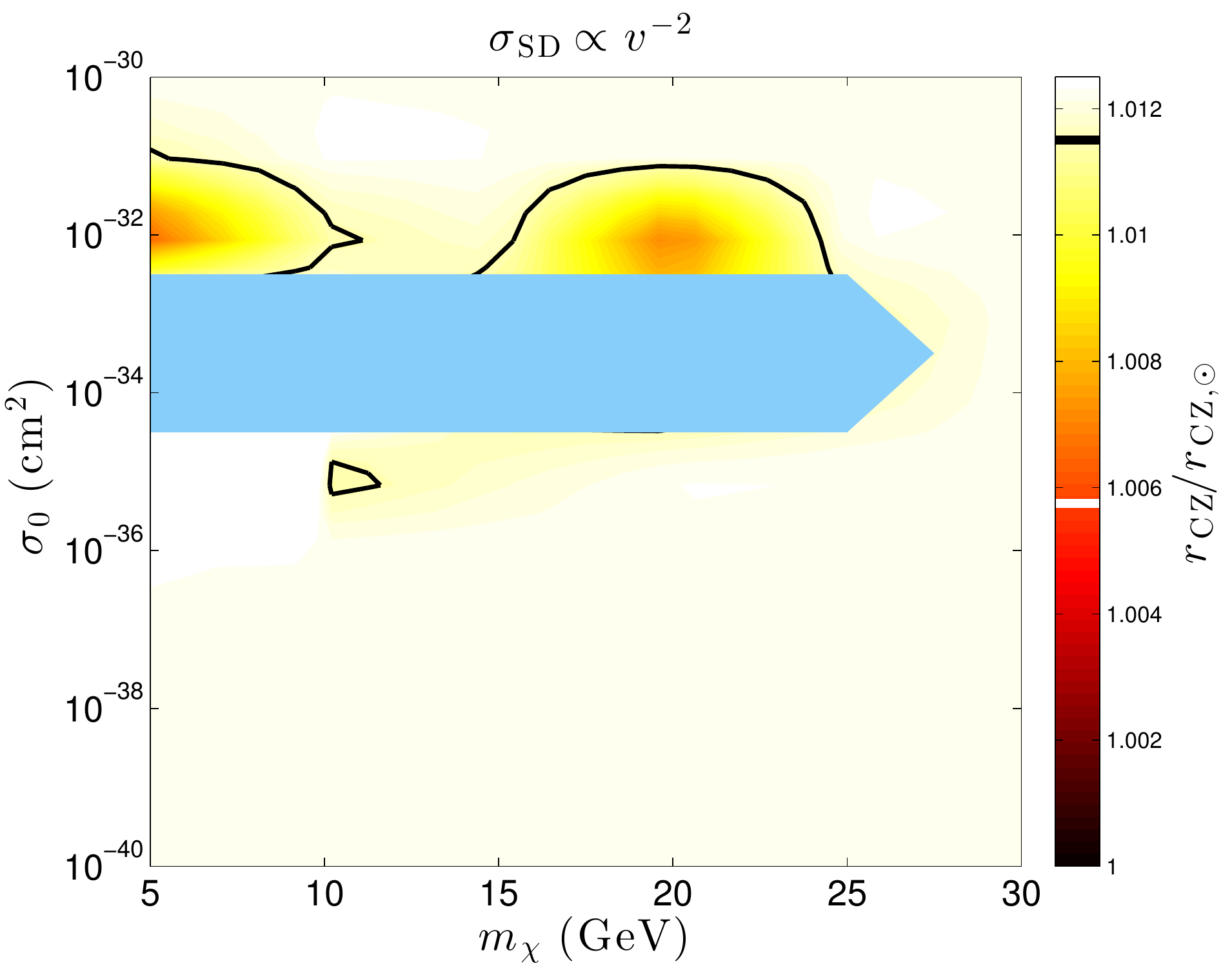} & \includegraphics[height = 0.32\textwidth]{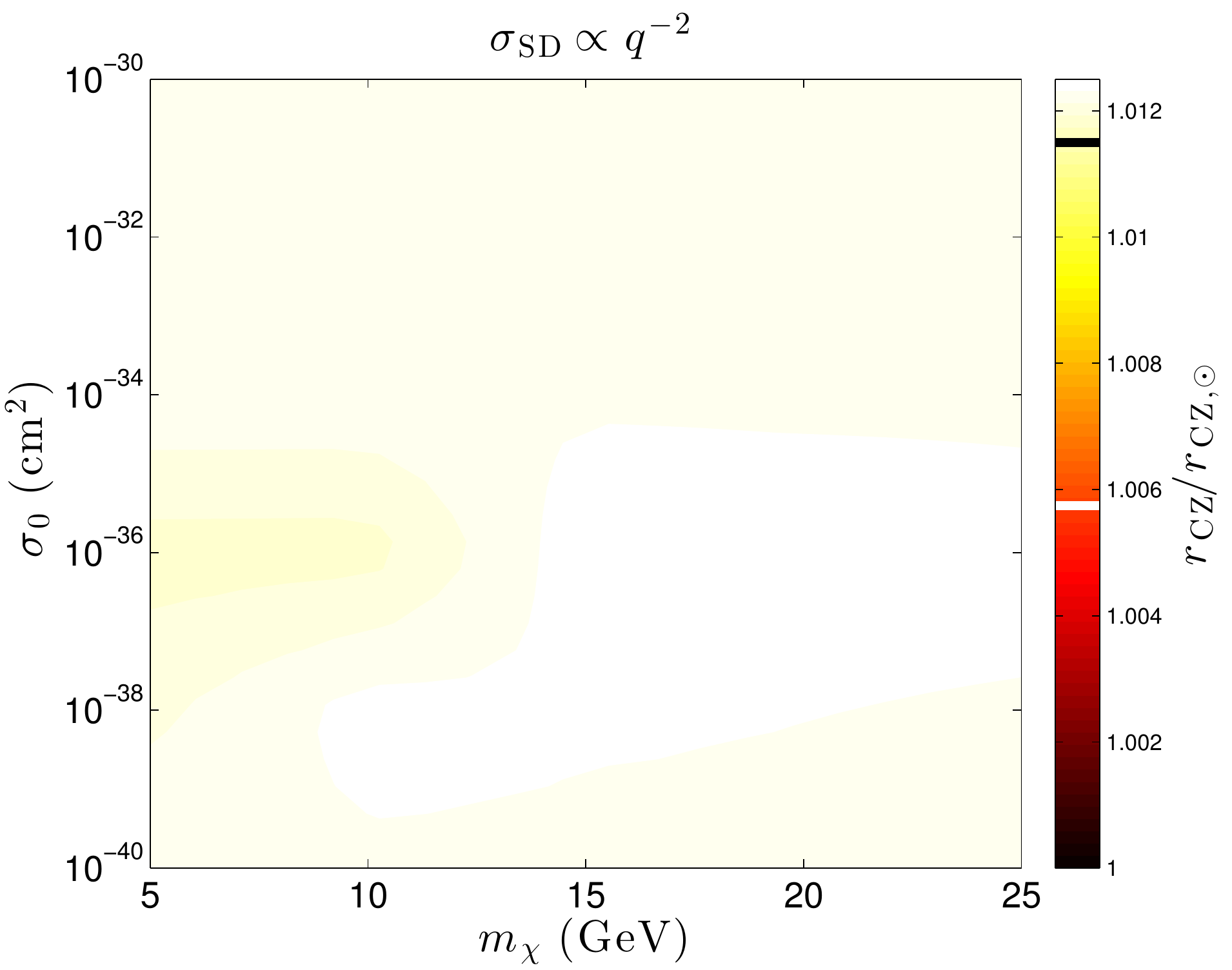} \\
\includegraphics[height = 0.32\textwidth]{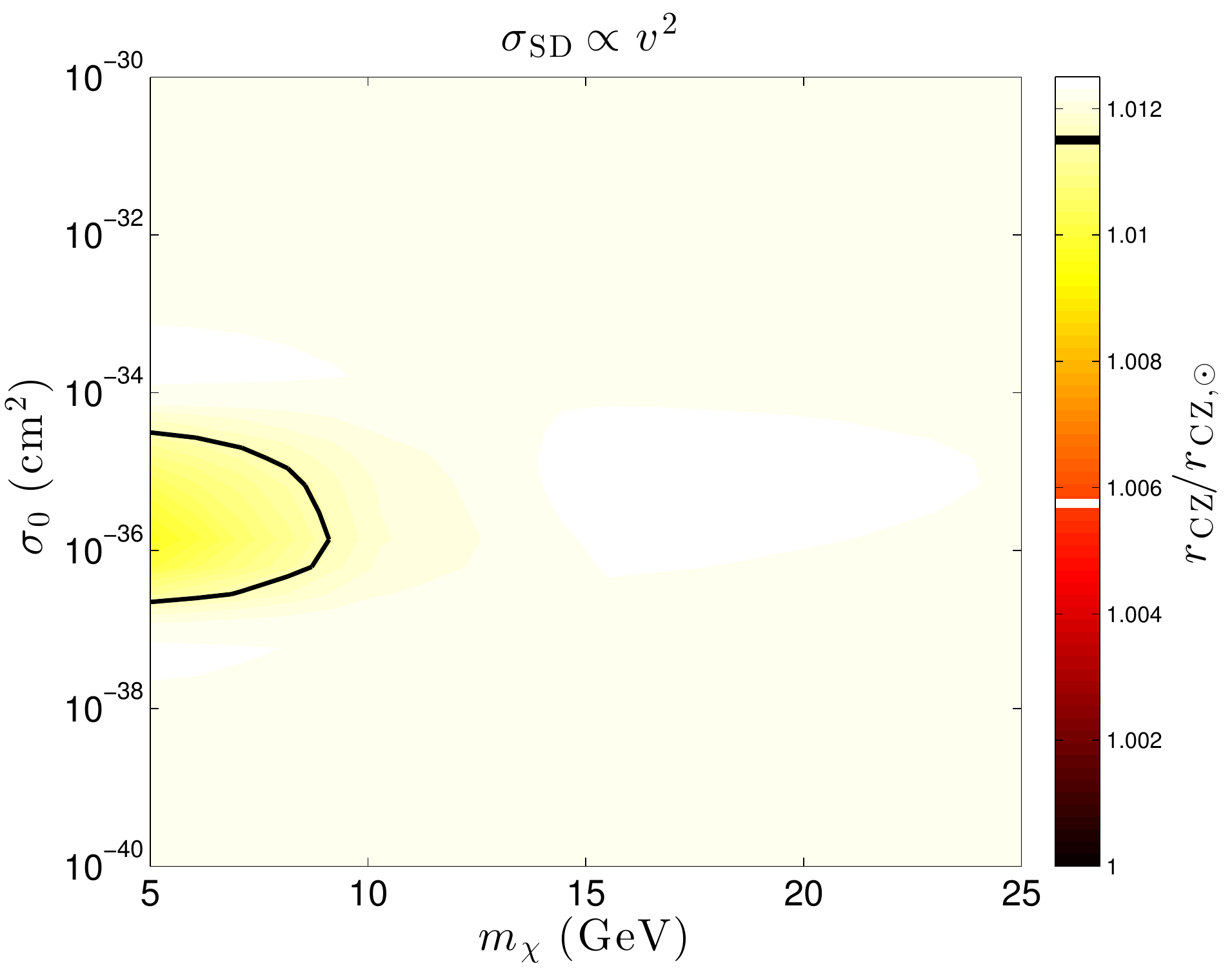} & \includegraphics[height = 0.32\textwidth]{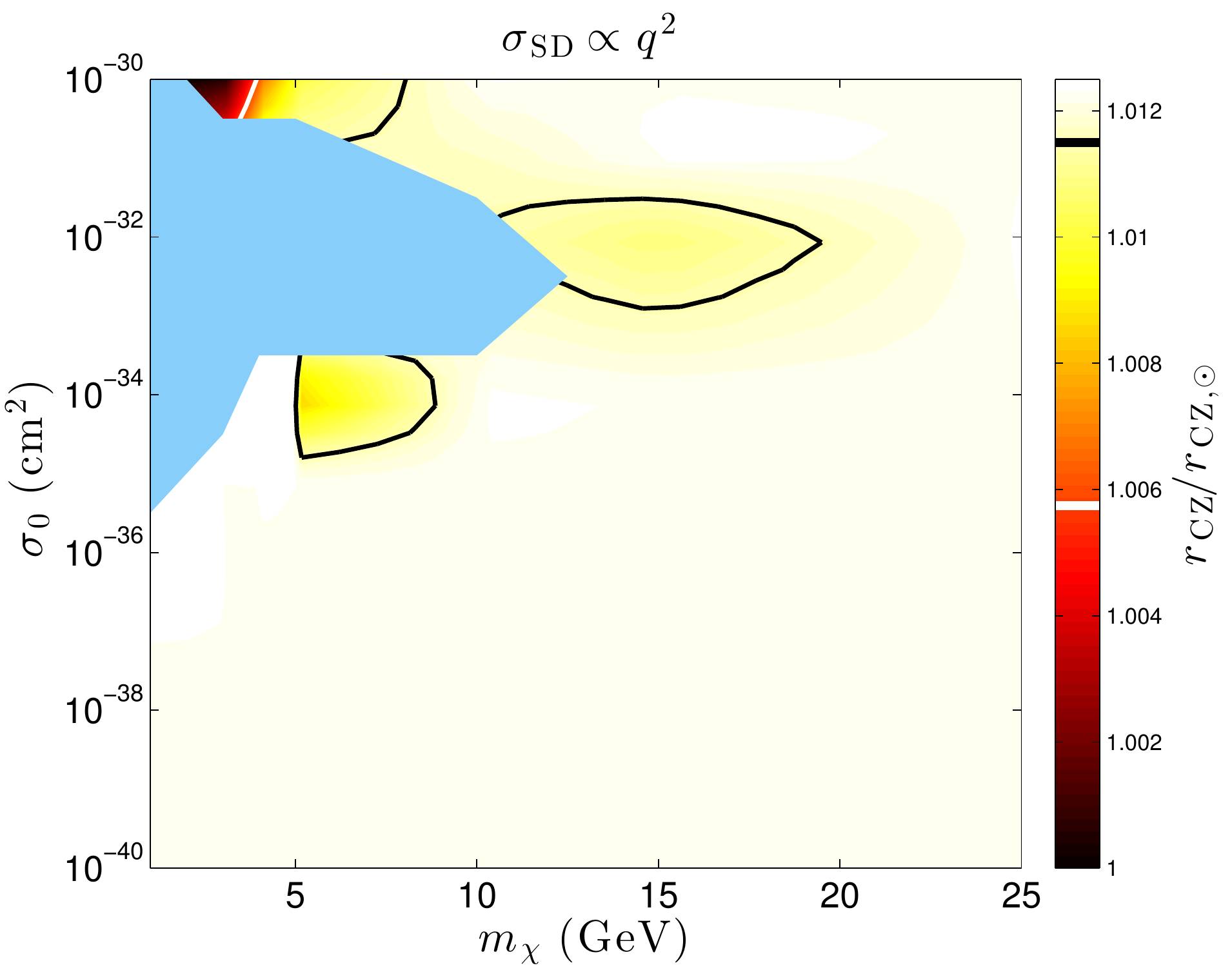} \\
\includegraphics[height = 0.32\textwidth]{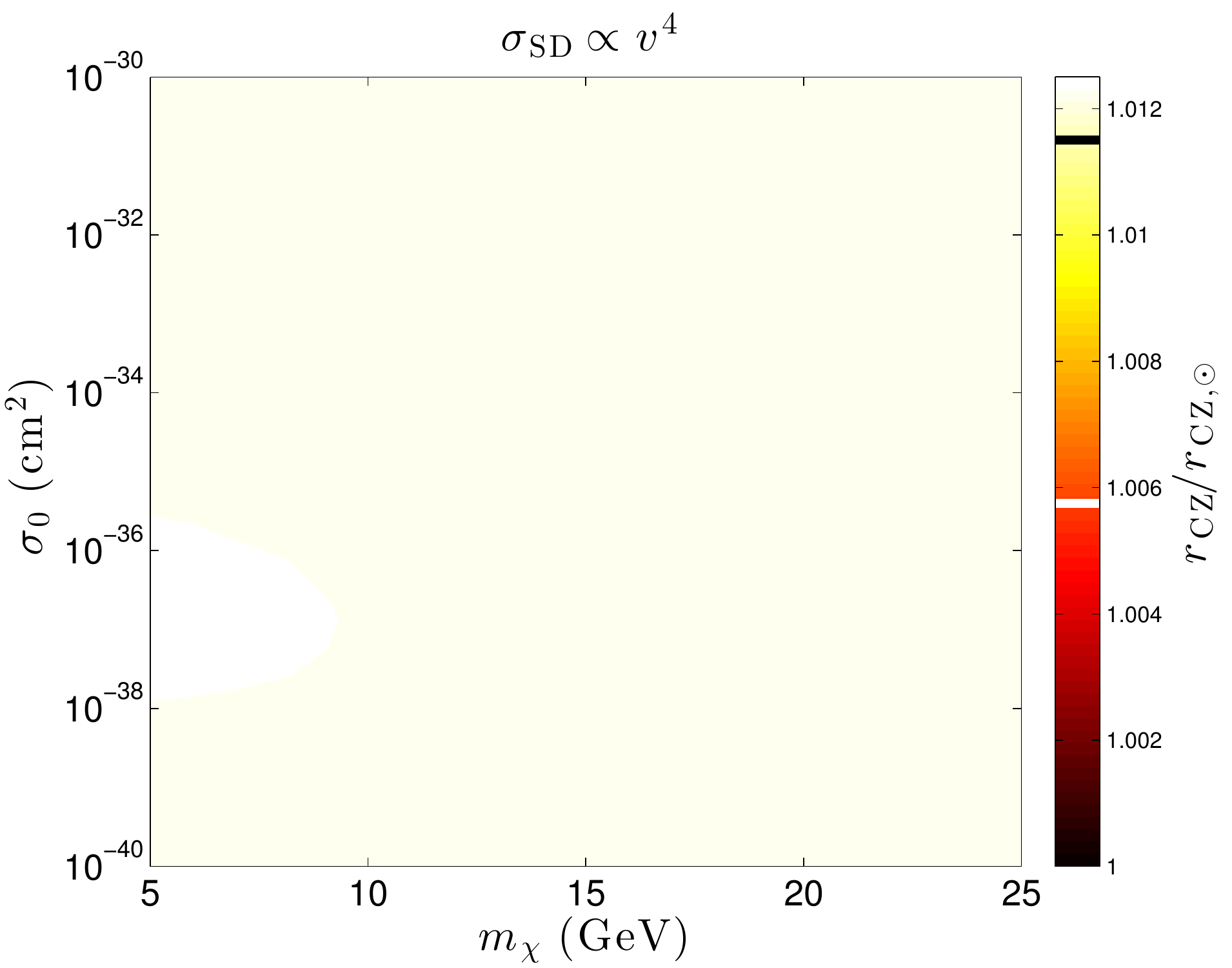} & \includegraphics[height = 0.32\textwidth]{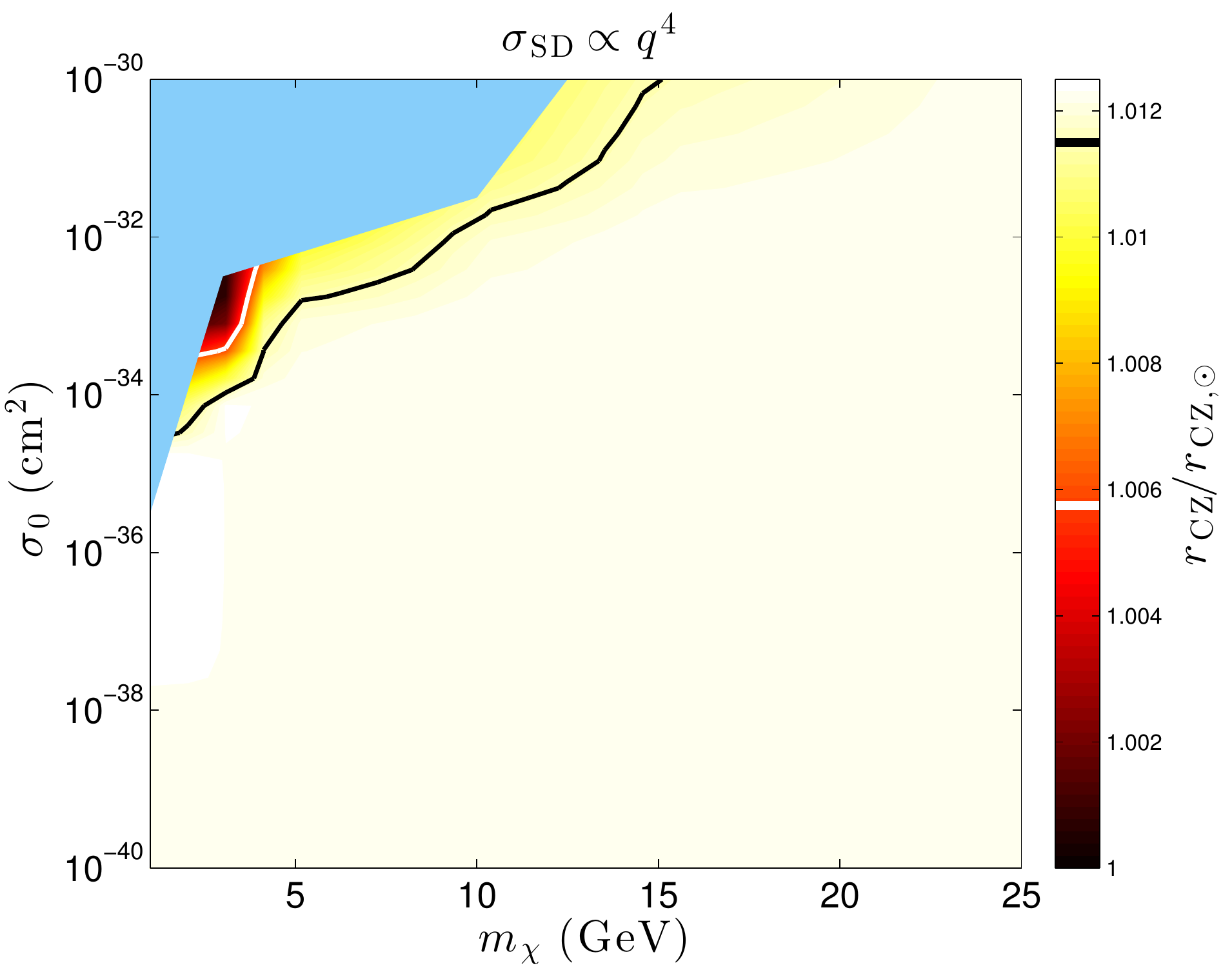} \\
\end{tabular}
\caption{As per Fig.\ \ref{SIrc}, but for spin-dependent couplings.}
\label{SDrc}
\end{figure}

In Figs.\ \ref{SIrc} and \ref{SDrc}, we show the ratio of the location of the lower boundary of the convection zone predicted by each model to the value inferred from helioseismology, $r_{\mathrm{CZ},\odot} = 0.713\,R_\odot$.

\FloatBarrier

\noindent Although the error on the helioseismological inference is just 0.001\,$R_\odot$, the theoretical error on the modelled $r_{\mathrm{CZ},\odot}$ is much larger: $\sigma_{\rm CZ,\odot,th} = 0.004\,R_\odot$. Adding these errors in quadrature, we plot in white and black the contours containing the regions where $r_{\mathrm{CZ},\odot}$ is within 1$\sigma$ and $2 \sigma$, respectively, of the inferred value.

As can be seen in Figs.\ \ref{SIrc} and \ref{SDrc}, the difference between the depth of the convection zone in the Standard Solar Model ($r_{\mathrm{CZ},\odot} = 0.722\,R_\odot$) and the depth inferred from helioseismology amounts to almost a 3$\sigma$ discrepancy. Except for a small region at high $q^2$ cross-section ($\sigma_0 = 10^{-33} $\,cm$^2$), spin-dependent DM struggles to bring $r_{\rm CZ}$ to within much better than 2$\sigma$ of the inferred value. However, energy conduction from SI interactions does substantially better: $\vrel^{\pm 2}$ and $q^2$ models can produce good agreement with the observed value of $r_{\mathrm{CZ},\odot}$, leading in some regions to agreement at better than 1$\sigma$.

\subsection{Surface helium abundance}
\label{sec:helium}

The surface helium abundance $Y_s$ is another observable that has fallen into disagreement with observations since the revision of the solar composition. Our SSM prediction is $Y_s = 0.2356$, whereas the observed value is 0.2485 $\pm 0.0034$.  With theoretical errors of $\pm 0.0035$ taken into account, this amounts to a discrepancy of $3\sigma$. The addition of dark matter does very little to change $Y_s$, and for most models the discrepancy remains at the $3\sigma$ level.  For some very few cases (not shown), it actually worsens the discrepancy by up to an additional $\sim$$2\sigma$. This only occurs for models where the fit to sound speed observables is also substantially worsened, notably for large SD $q^2$ and $q^4$ cross-sections. The reduction in $Y_s$ is caused by a corresponding increase in $X_i$, the initial hydrogen fraction.  The increase in $X_i$ is demanded by the reduction of the core temperature, which requires a greater amount of hydrogen to be present in the core in order to maintain the same nuclear burning rate as in the SSM, and to thereby match the observed solar luminosity $L_\odot$.

Because it changes little, we do not show the contour plots for the surface He abundance. For completeness though, we include it anyway in our full likelihood computation in Sec.\ \ref{sec:combined}.

\begin{figure}[t]
\begin{tabular}{c c}
\includegraphics[width=0.45\textwidth]{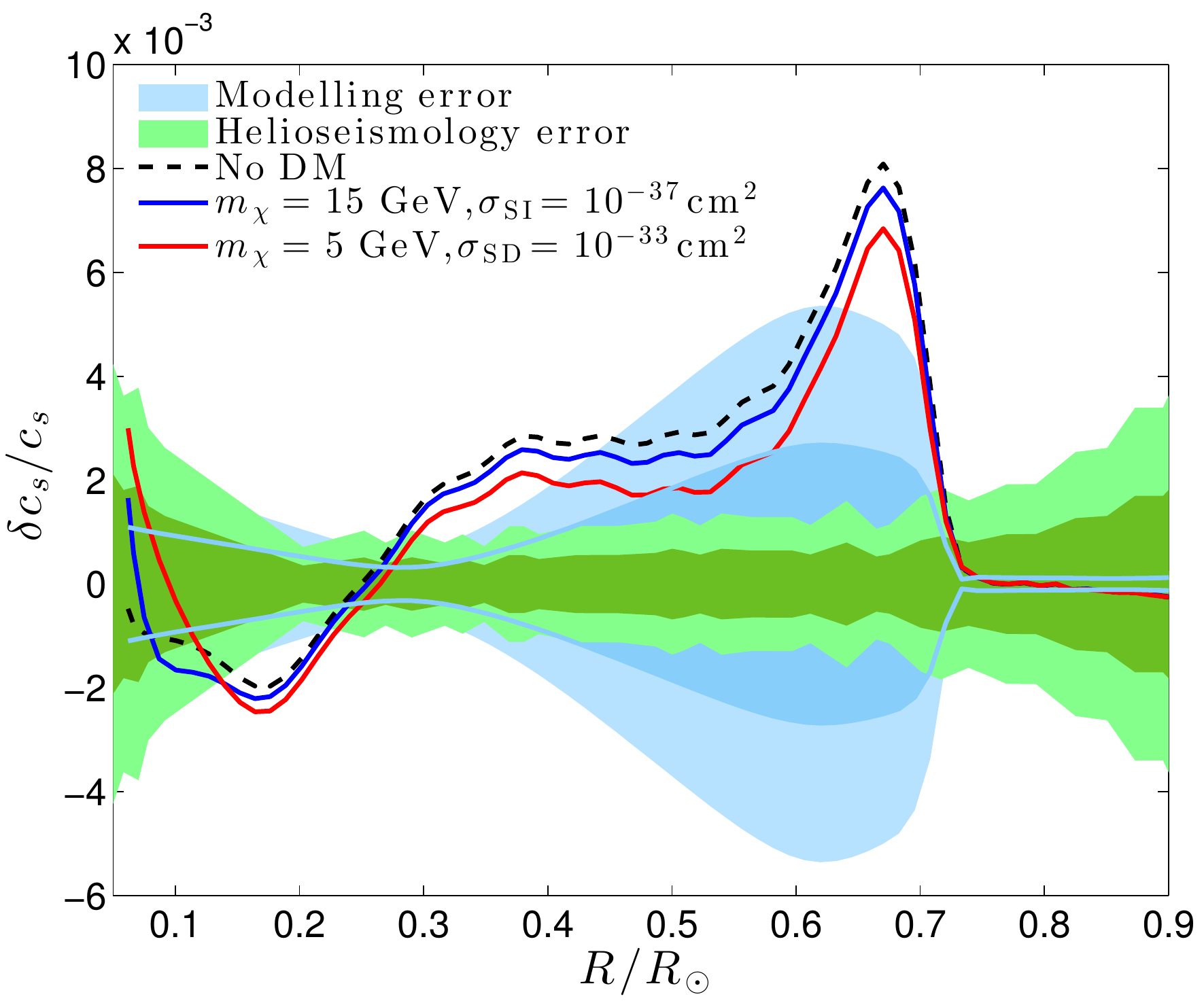}&\includegraphics[width=0.45\textwidth]{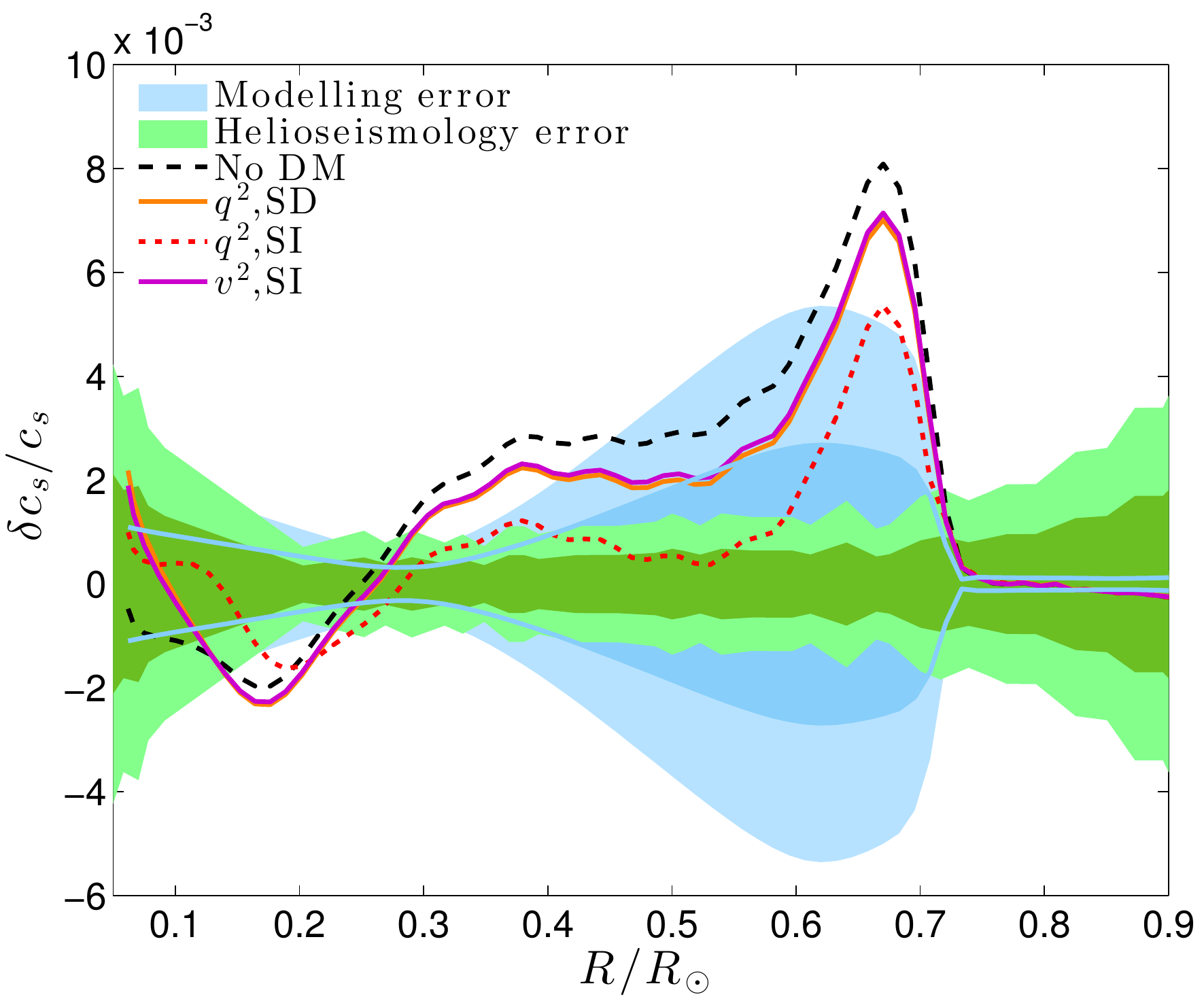}
\end{tabular}
\caption{Deviation of radial sound speed profile from values inferred from helioseismology. Left: the best overall fits with constant spin-dependent (SD) and spin-independent (SI) cross-sections. Right: best-fit models of the three couplings returning the best overall $p$-values (see Tab.~\ref{best_fit_tab}): ($q^2$, SD): $m_\chi = 5$ GeV, $\sigma_0 = 10^{-30}$ cm$^{2}$; ($q^2$, SI): $m_\chi = 3$ GeV, $\sigma_0 = 10^{-37}$ cm$^{2}$; and ($v^2$, SI): $m_\chi = 5$ GeV, $\sigma_0 = 10^{-35}$ cm$^{2}$.  }
\label{csfig}
\end{figure}

\subsection{Sound speed profile}
\label{sec:soundspeed}

\begin{figure}[p]
\begin{tabular}{c@{\hspace{0.04\textwidth}}c}
\multicolumn{2}{c}{\includegraphics[height = 0.32\textwidth]{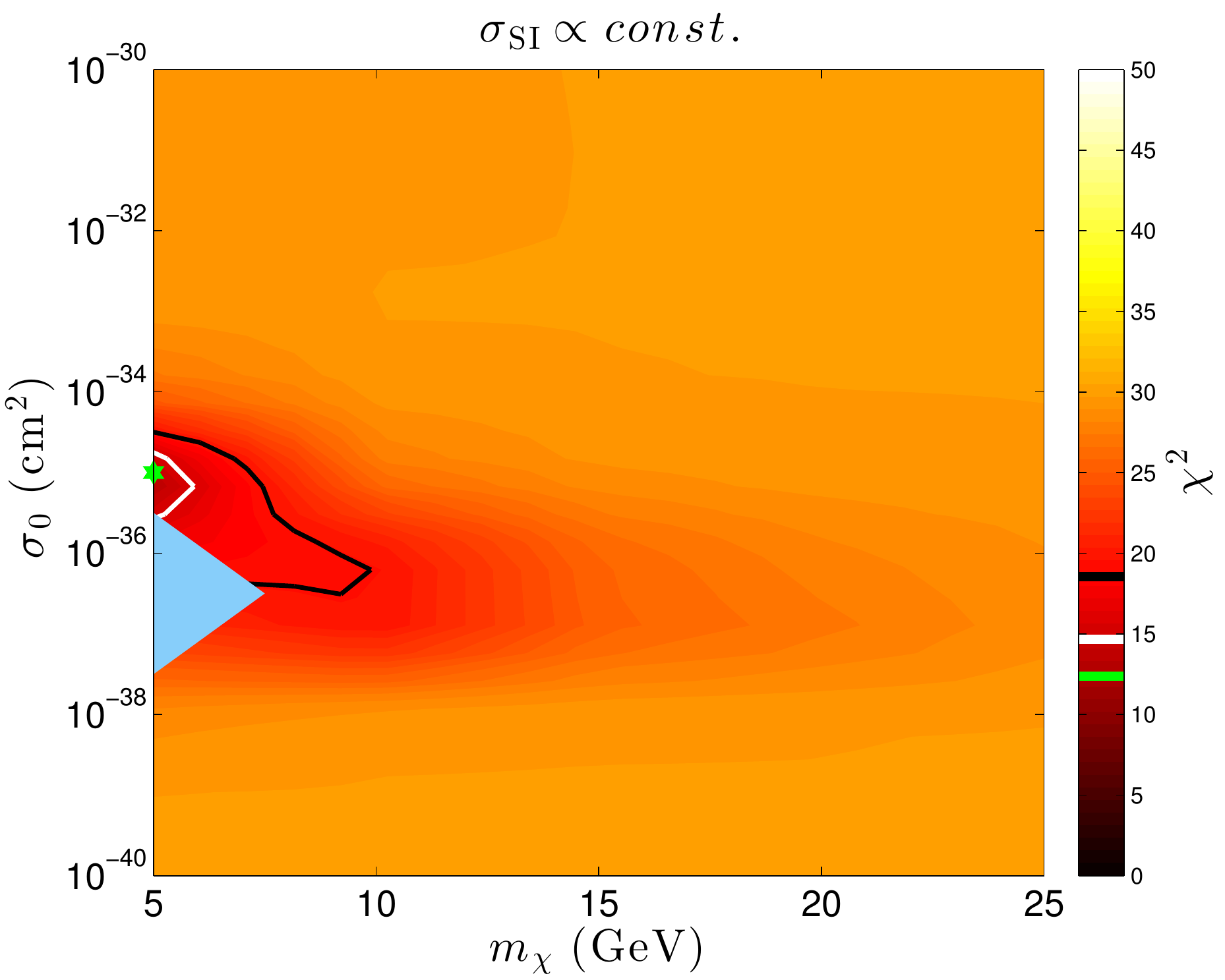}} \\
\includegraphics[height = 0.32\textwidth]{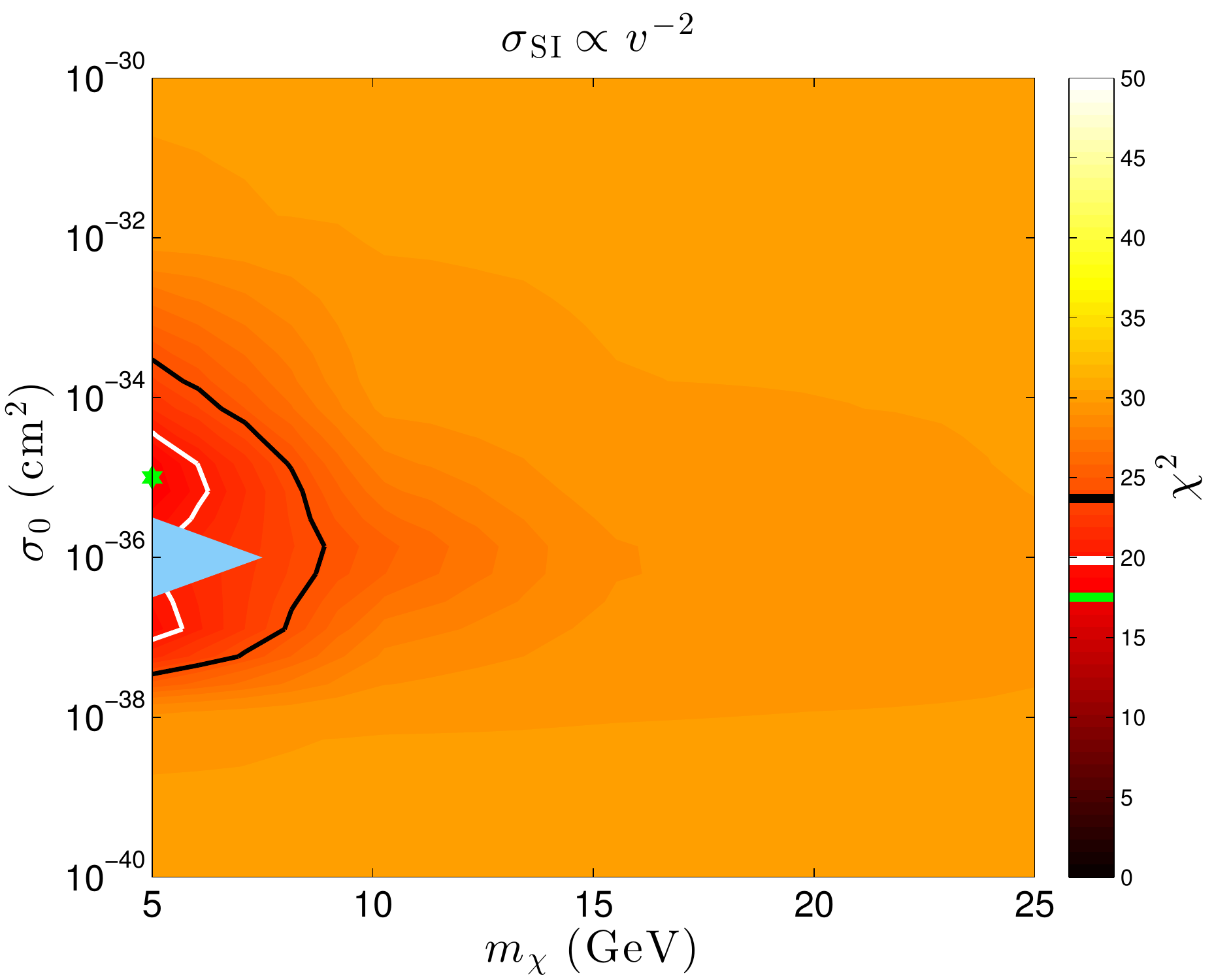} & \includegraphics[height = 0.32\textwidth]{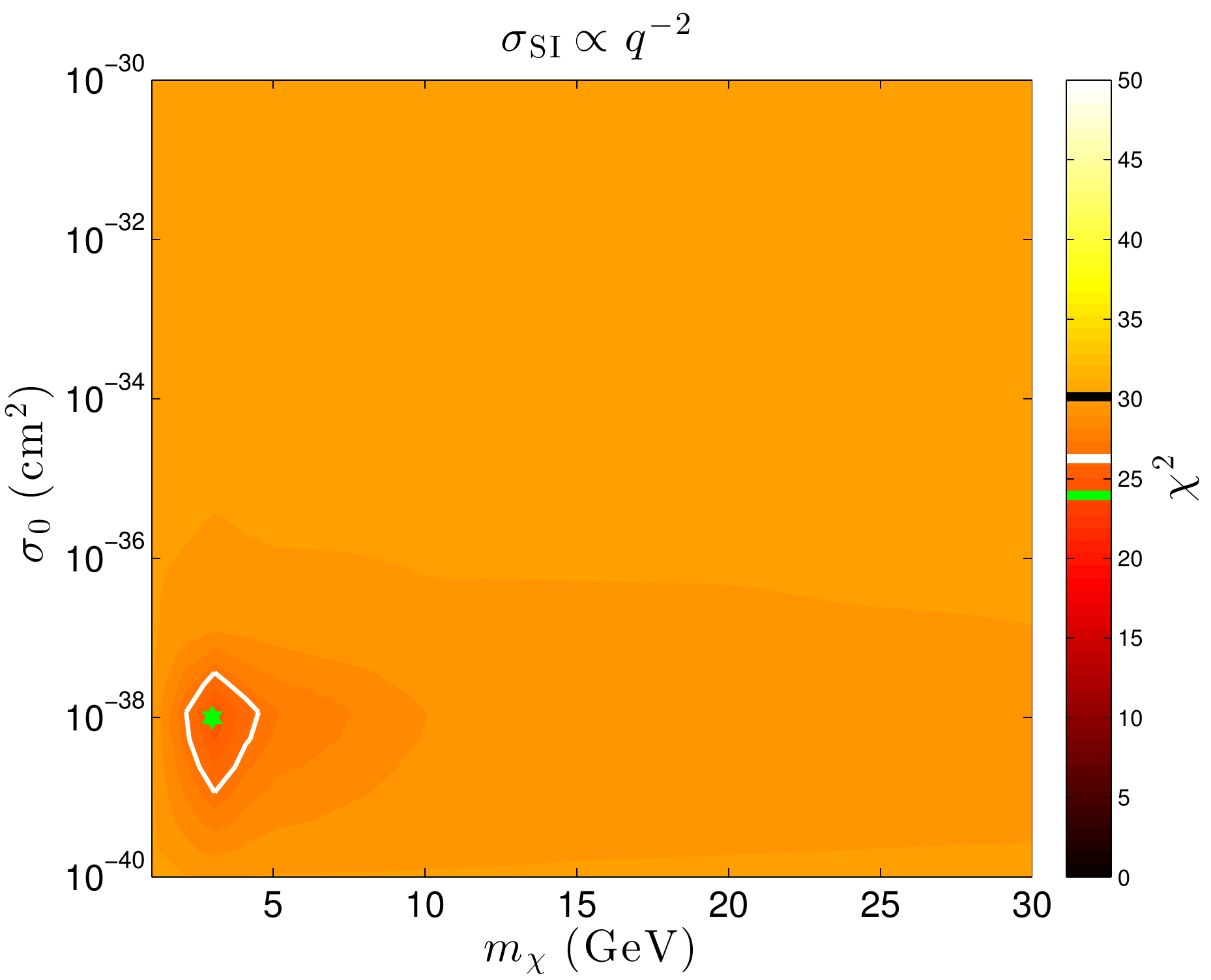} \\
\includegraphics[height = 0.32\textwidth]{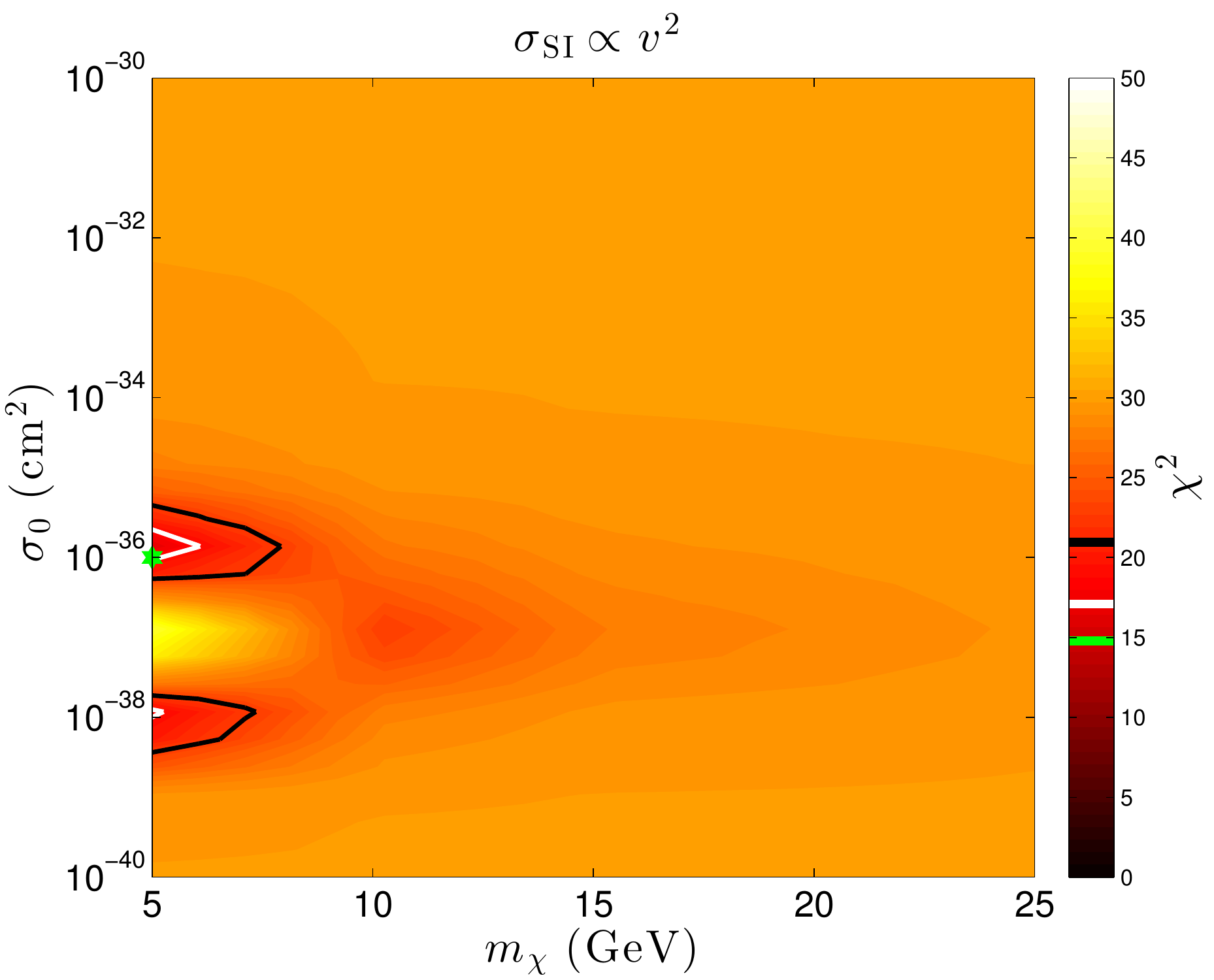} & \includegraphics[height = 0.32\textwidth]{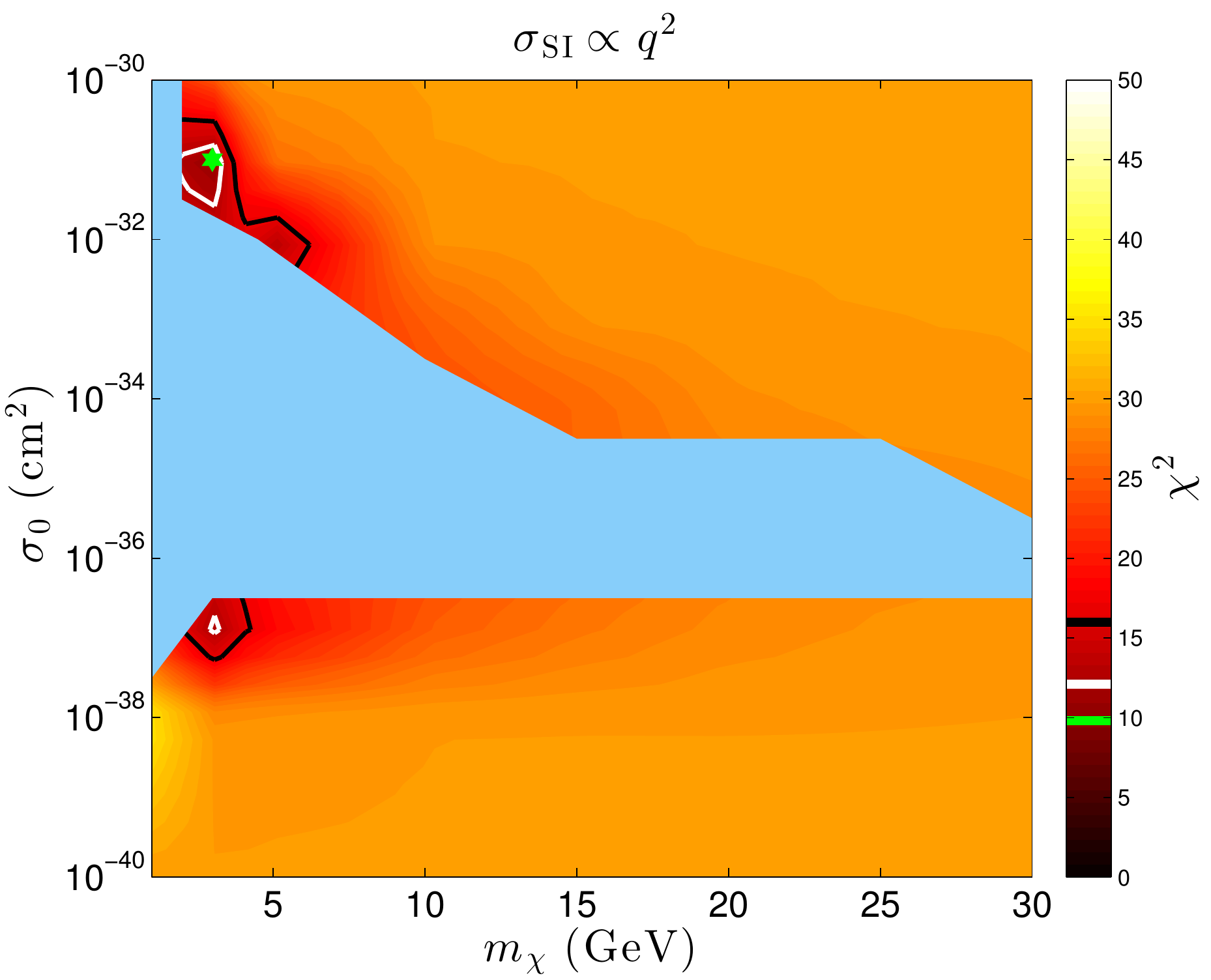} \\
\includegraphics[height = 0.32\textwidth]{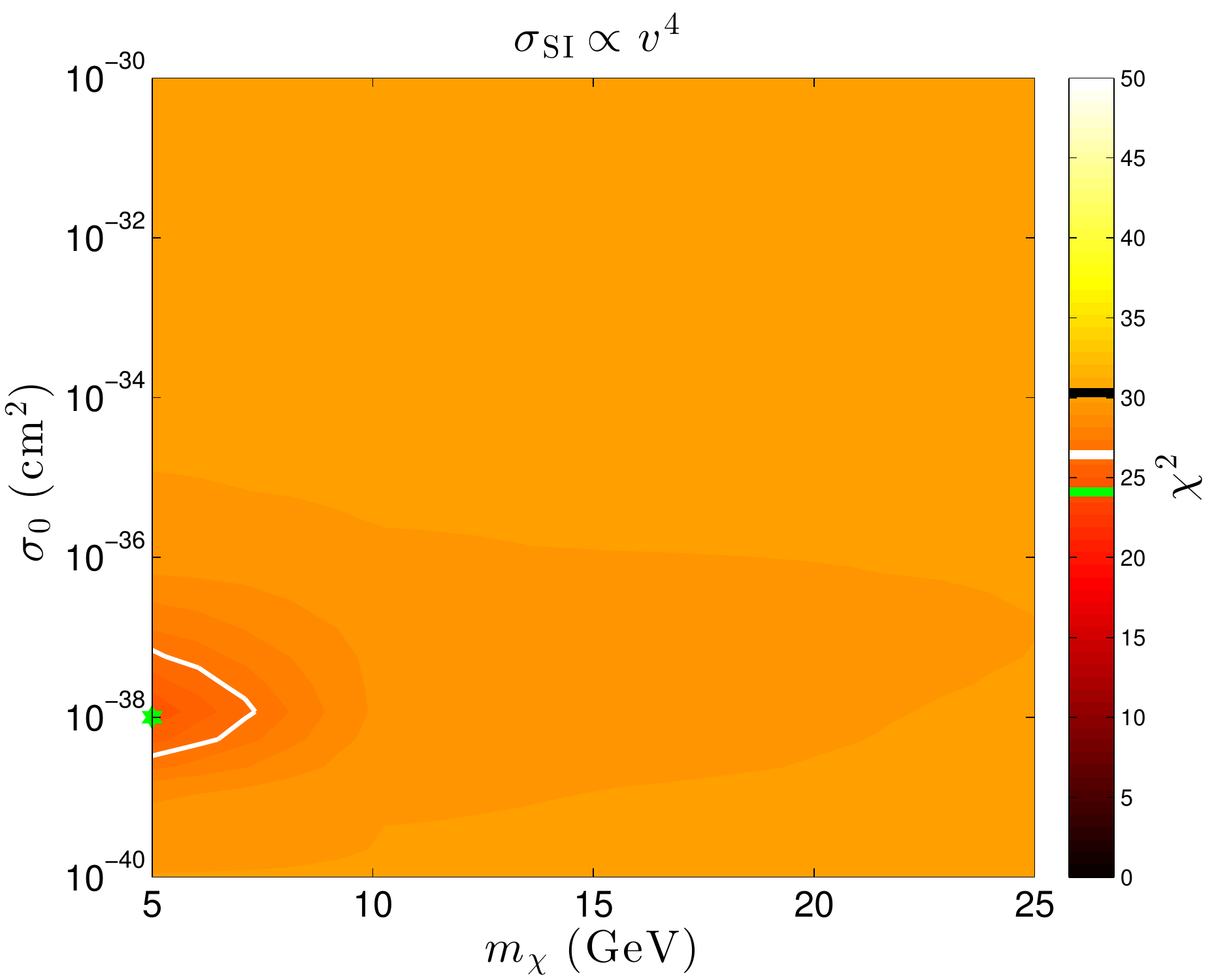} & \includegraphics[height = 0.32\textwidth]{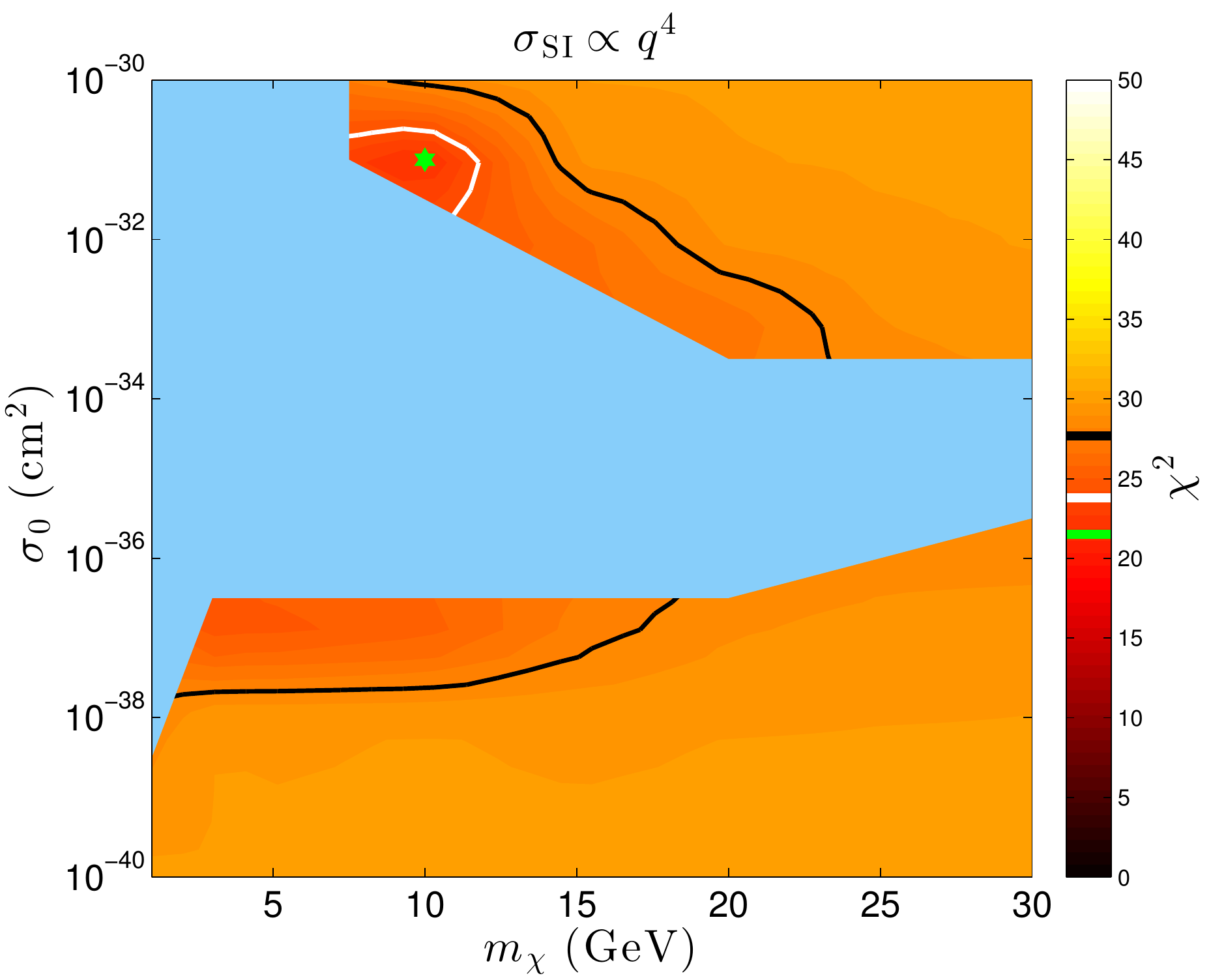} \\
\end{tabular}
\caption{Effective sound-speed $\chi_{c_s}^2$ defined in Eq.\ \ref{cschisq}, showing the improvement in goodness-of-fit of the radial sound speed profile in solar models with different types of spin-independent ADM. The $\chi^2$ value of the best fit point is indicated by a green line on the colour bar and is shown as a green star in each figure. $\Delta \chi^2 = 2.3$ and 6.18 contours, corresponding to 1$\sigma$ and 2$\sigma$ deviations from the best fit, are respectively shown in white and black.}
\label{SIchisqf_csf}
\end{figure}

\begin{figure}[p]
\begin{tabular}{c@{\hspace{0.04\textwidth}}c}
\multicolumn{2}{c}{\includegraphics[height = 0.32\textwidth]{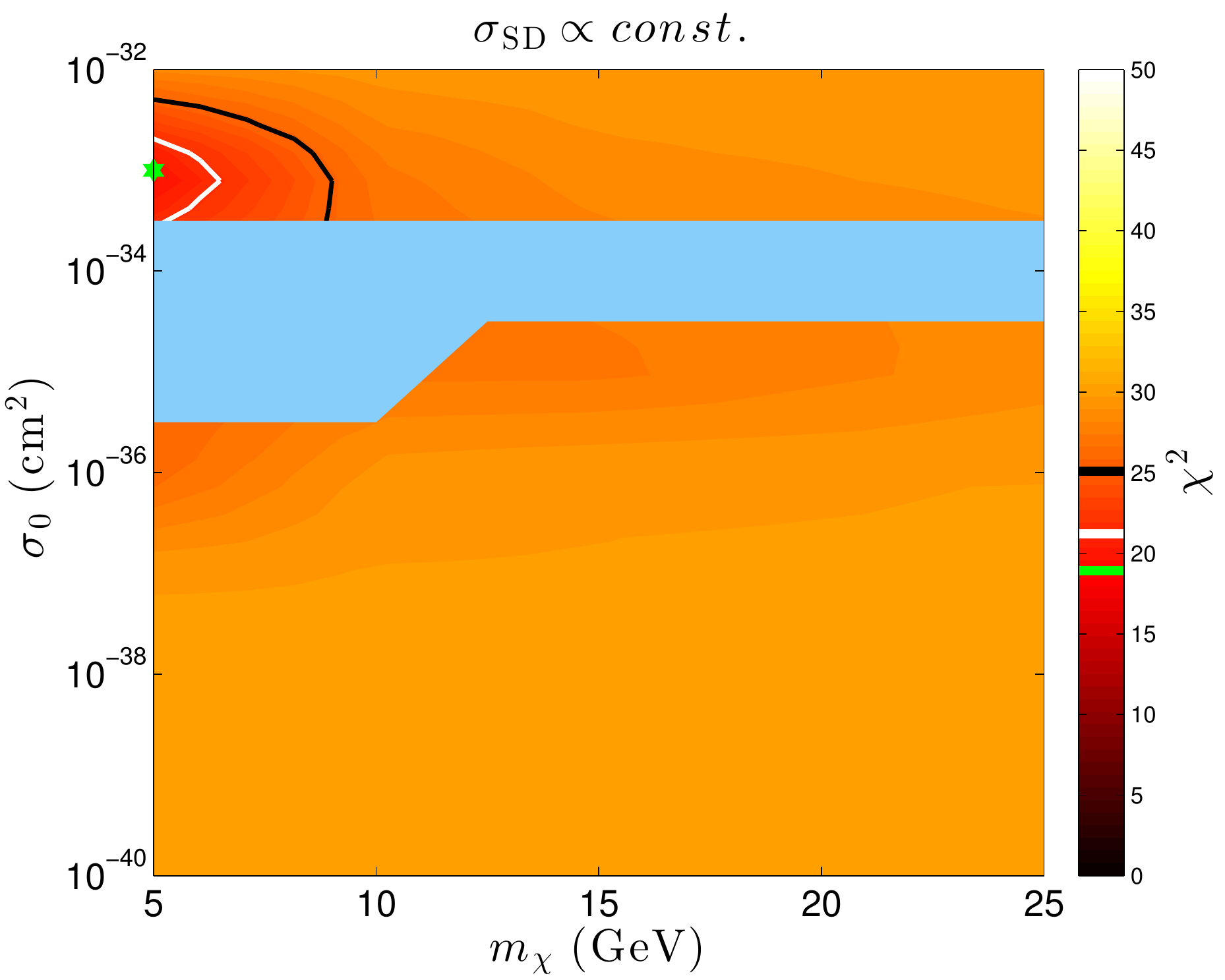}} \\
\includegraphics[height = 0.32\textwidth]{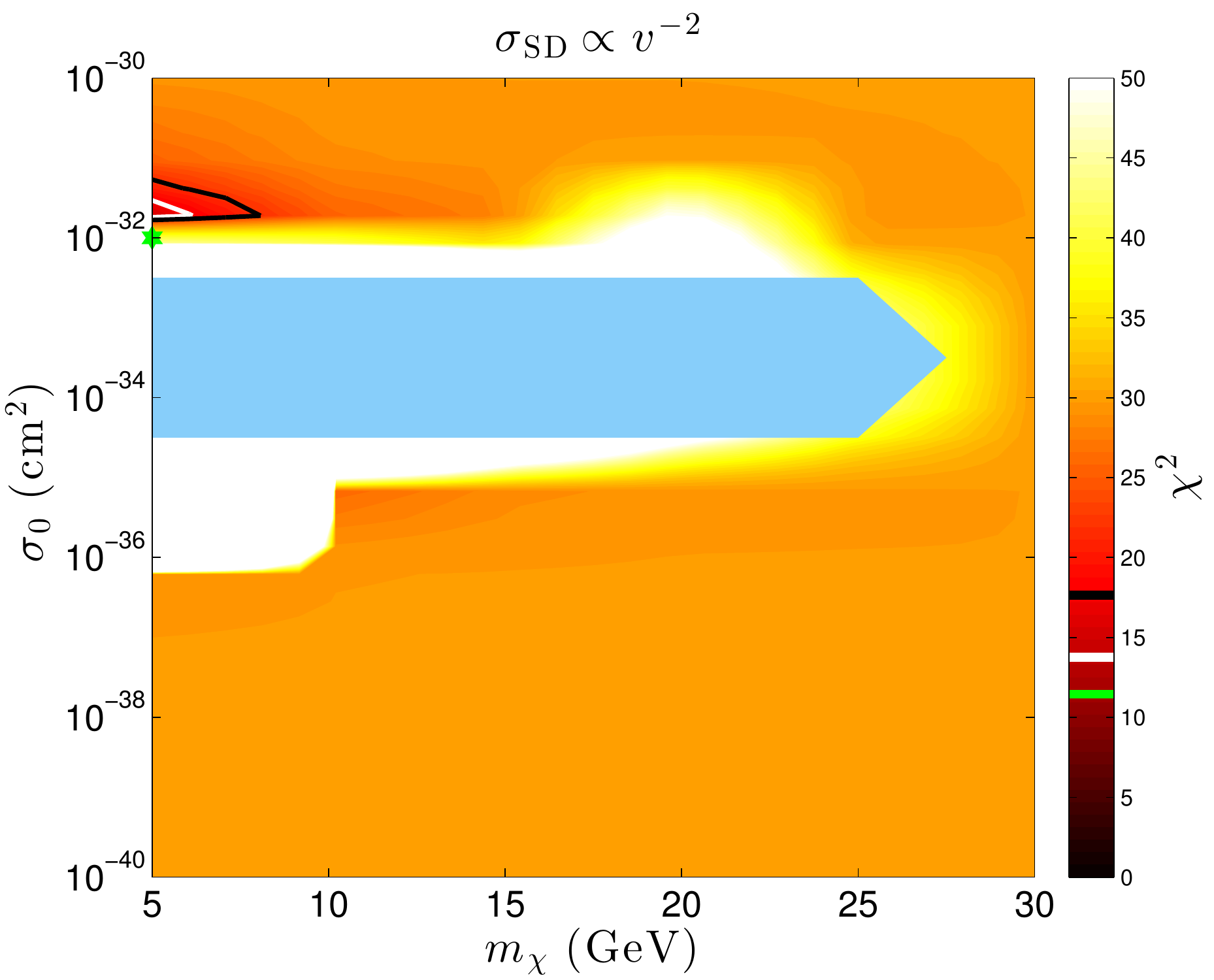} & \includegraphics[height = 0.32\textwidth]{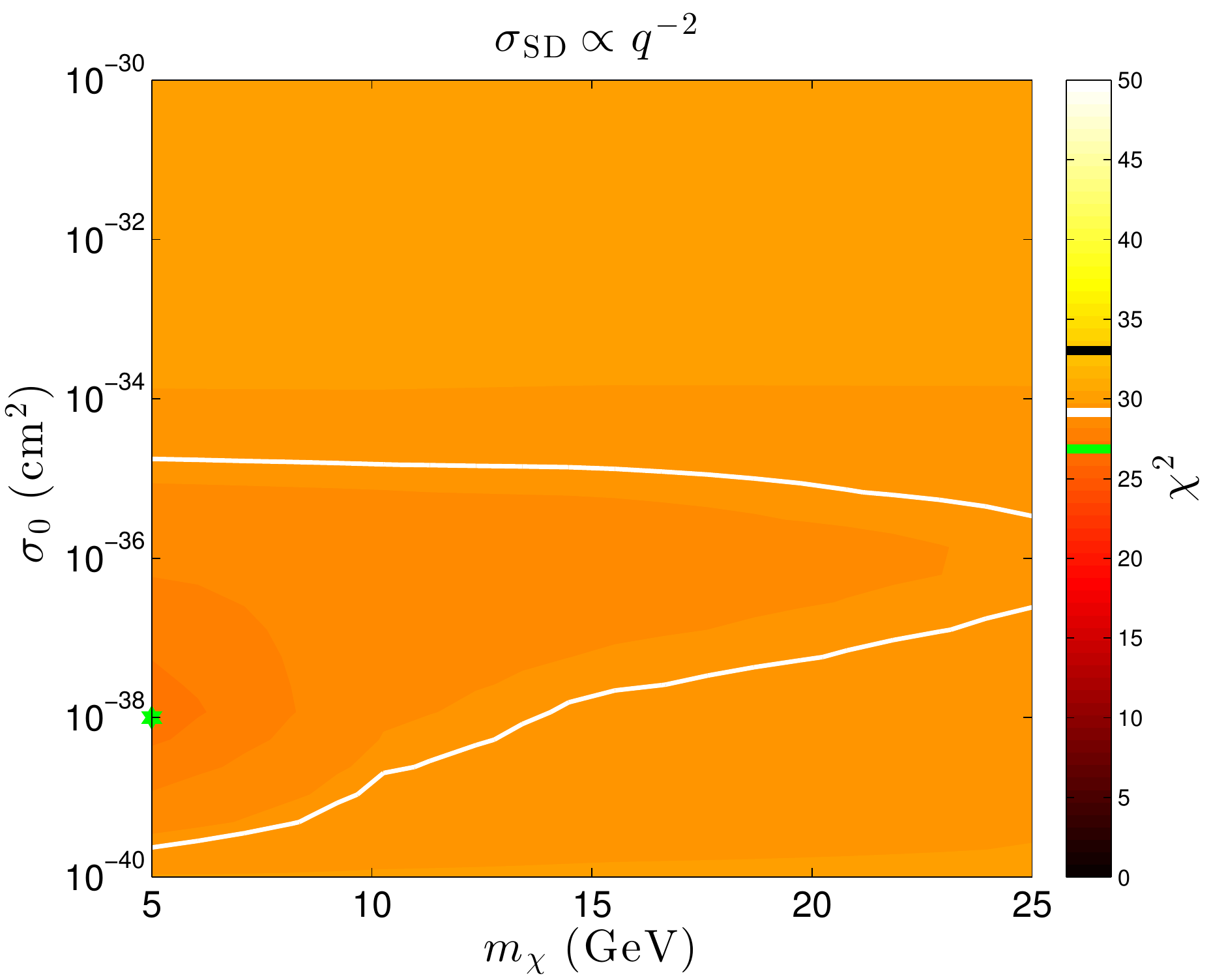} \\
\includegraphics[height = 0.32\textwidth]{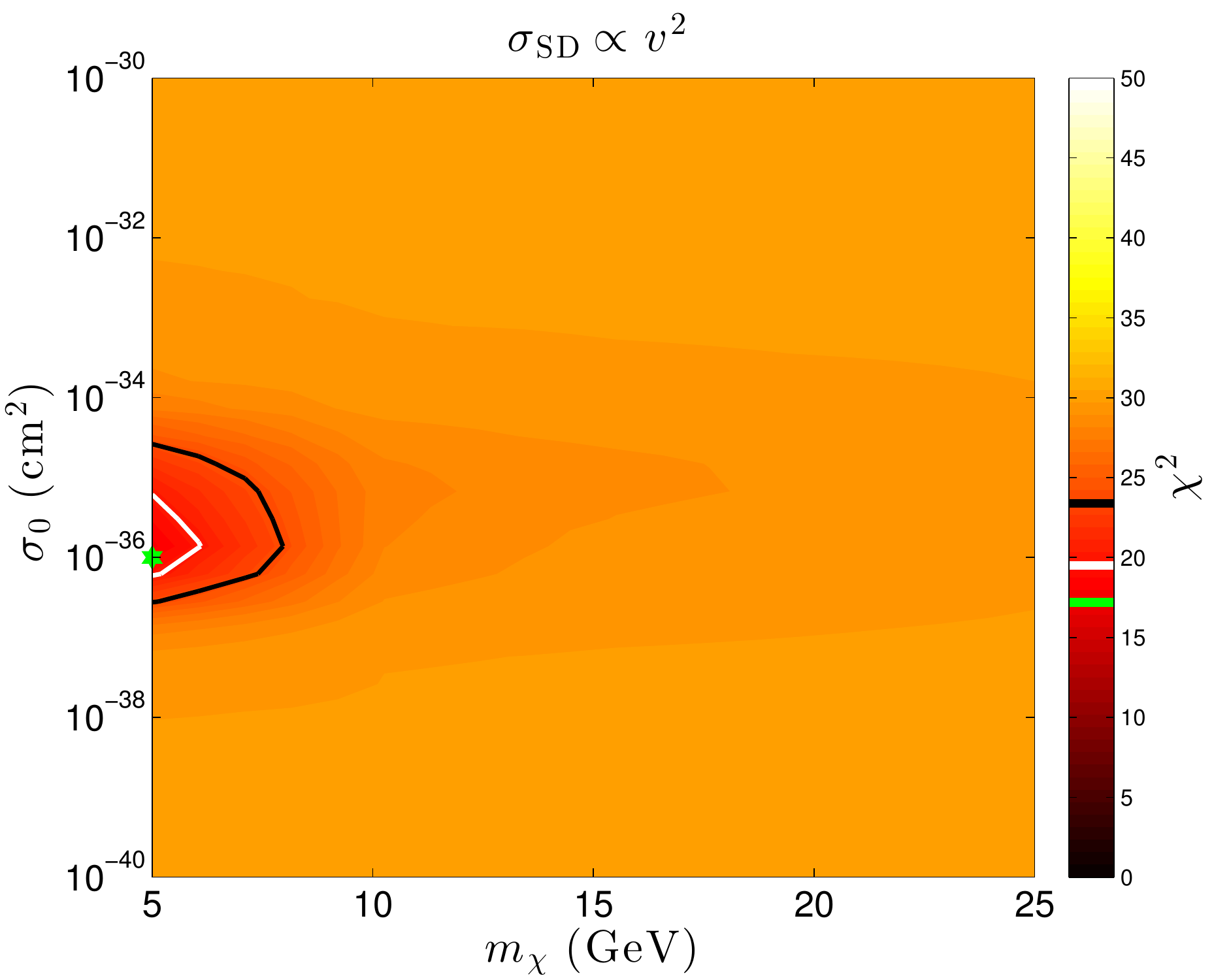} & \includegraphics[height = 0.32\textwidth]{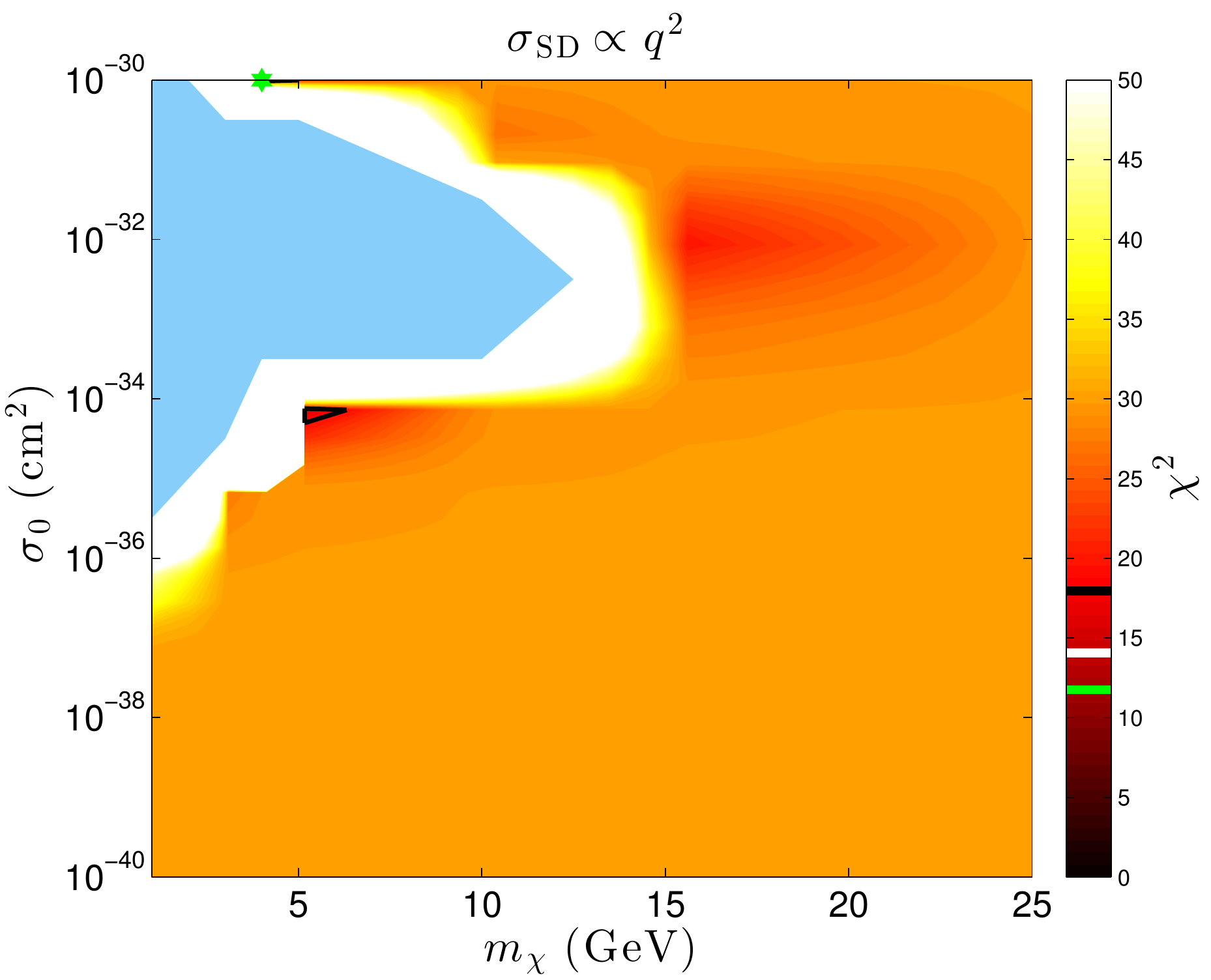} \\
\includegraphics[height = 0.32\textwidth]{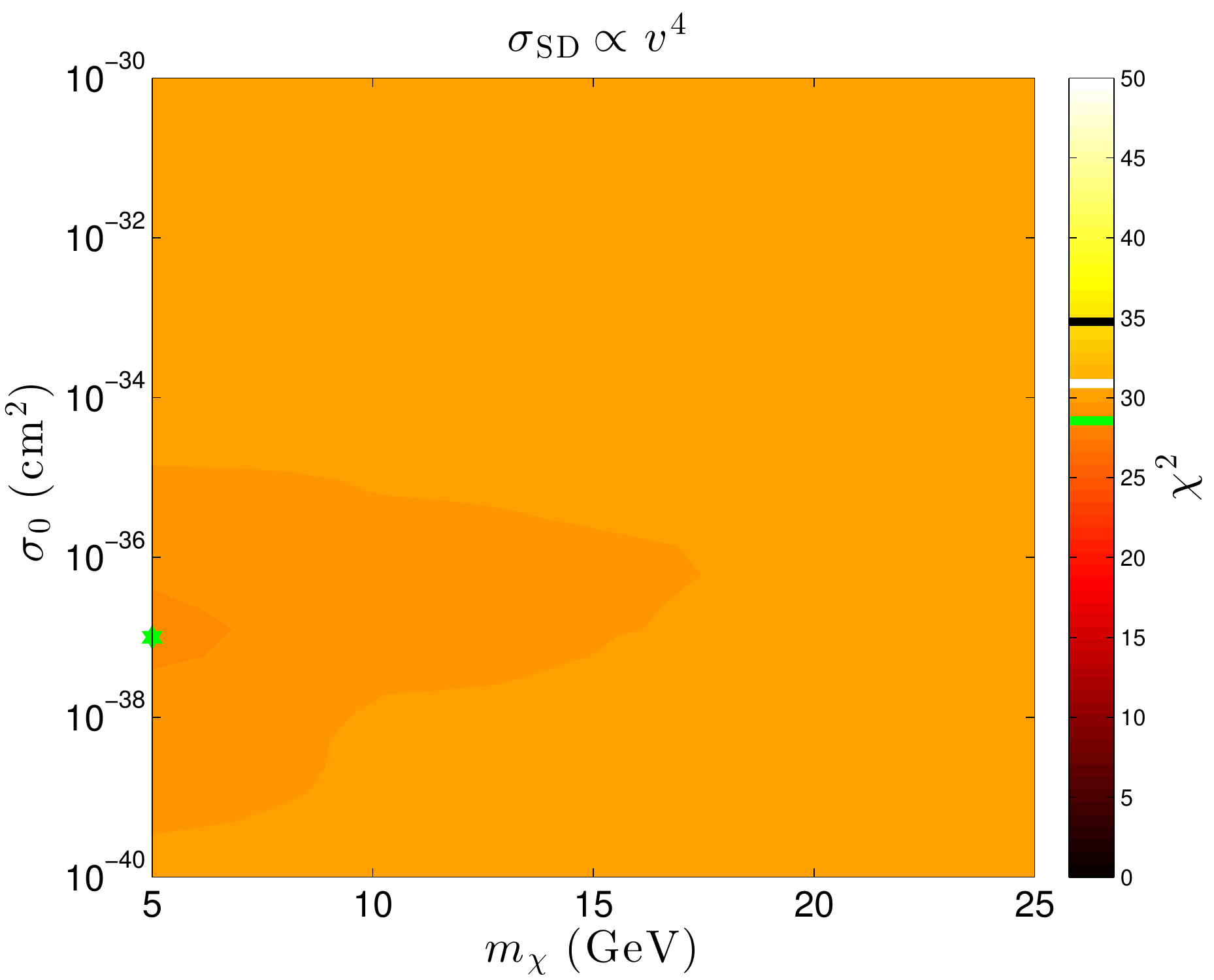} & \includegraphics[height = 0.32\textwidth]{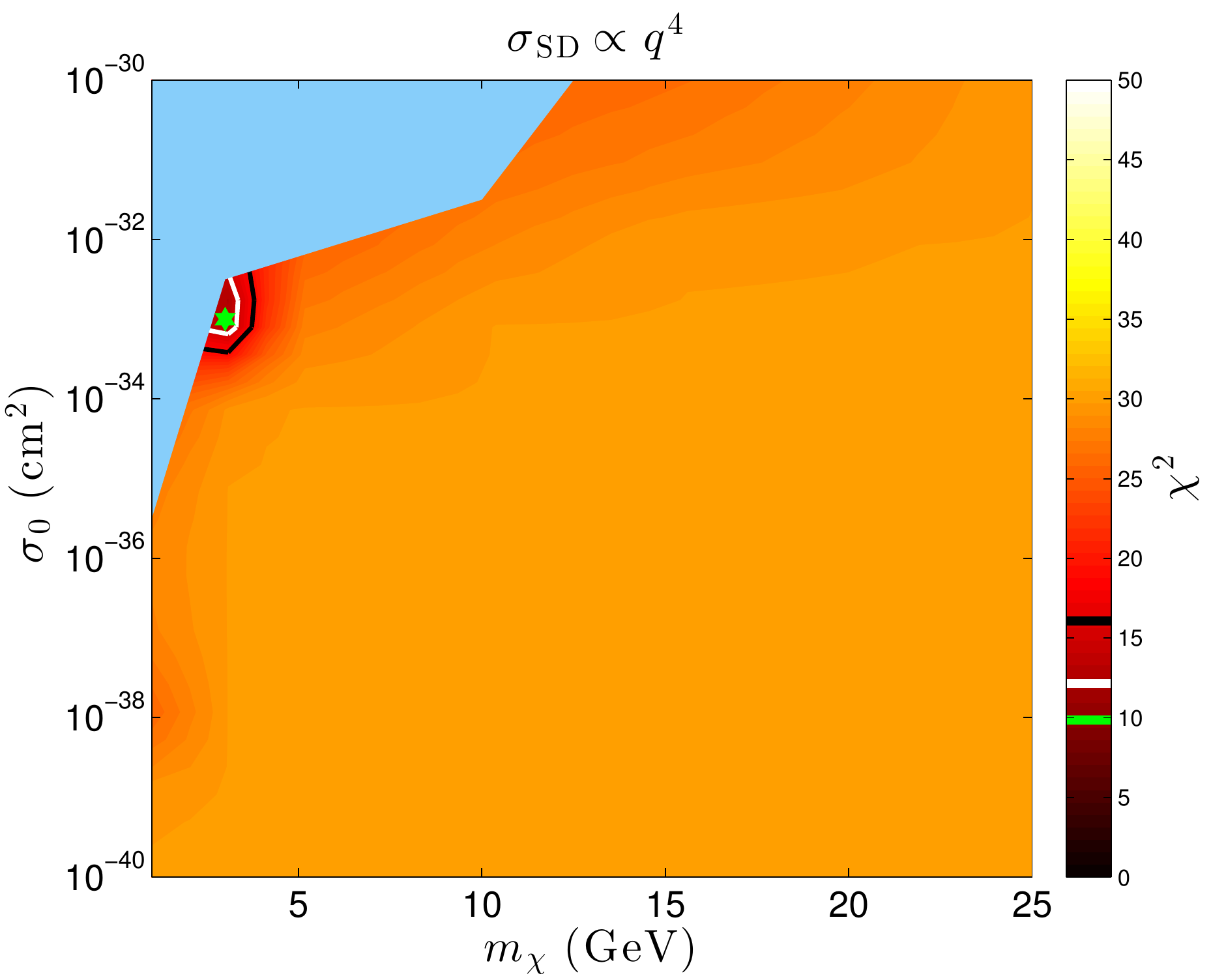} \\
\end{tabular}
\caption{Same as Fig.\ \ref{SIchisqf_csf}, but for spin-dependent couplings. }
\label{SDchisqf_csf}
\end{figure}

In Fig.~\ref{csfig} we show the deviation of the modelled radial sound speed of some models that produced the best overall fits with respect to the measured values from helioseismology. To illustrate the agreement of each model with the observed sound speed profile, we construct an effective chi-squared measure
\begin{equation}
\chi^2_{c_{\rm s} } = \sum_{r_i} \frac{\left[c_{\rm s,model}(r_i) - c_{\rm s,hel.}(r_i)\right]^2}{\sigma_{c_{\rm s,hel.}}^2(r_i)}
\label{cschisq}
\end{equation}
where the sum goes over 5 equally-spaced radial points between $r = 0.1\,R_\odot$ and $0.67\,R_\odot$.  We do not include radii below $0.1\,R_\odot$ because the reconstructed sound speed in this region is highly uncertain due to the low number of low-degree (low-$l$) p-modes reaching the innermost radii of the solar core \cite{basu:2009}.  The upper limit of the range is the location of the largest discrepancy in the sound speed of the Standard Solar Model, at the base of the convection zone; above this point essentially all models agree very well with the observed sound speed because the temperature gradient is adiabatic and therefore does not depend on the detailed composition of the Sun.  We added errors from modelling and inversions for each point in quadrature. We obtained modelling errors by using models for which one input parameter was varied at a time to obtain partial derivatives at each radial point, and then combined quadratically given the uncertainty in the AGSS09 abundances and errors in each parameter quoted in \cite{Serenelli2013}. The errors on inversions were taken from \cite{1997APh.....7...77D}. The values of this $\chi^2$, meant to show the relative improvement to the sound speed modelling, are shown in Figs.\ \ref{SIchisqf_csf} and \ref{SDchisqf_csf}.

From these figures, along with the sound speed profiles illustrated in Fig.\ \ref{csfig}, it is clear that the addition of momentum or velocity-dependent dark matter to the solar model does indeed help alleviate the discrepancy between modelling and observation in some specific cases. We will return to these cases in Sec.\ \ref{sec:bestcases}.

\subsection{Frequency separation ratios}
\label{sec:smallfreq}

The inverted sound speed profile $c_s(r)$ obtained from helioseismological measurements is not very accurate near the solar core because not enough low-$l$ p-modes are available for a precise and accurate inversion. Instead, information about the solar core can be gained by using the so-called frequency separation ratios. A large advantage of using these ratios is that, unlike individual frequencies, they are not affected by the detailed structure of the solar surface, which is poorly described by solar models \cite{roxburgh:2003}. This is because for radial order $n \gg 1$ and low angular degree $l$, the surface effects are functions of the eigenfrequency, so they cancel out when considering frequency differences between modes of similar frequencies. In addition, by taking ratios of appropriate frequency differences, the solar core structure becomes the dominant effect in the observed signal \cite{Basu:2006vh,bison2007}. In particular, two very useful quantities are the frequency separation ratios 
\begin{equation}
r_{02}(n) = \frac{d_{02}(n)}{\Delta_1(n)}, \, \, \, \, r_{13}(n) = \frac{d_{13}(n)}{\Delta_0(n+1)},
\label{eq:rdef}
\end{equation}
where
\begin{equation}
d_{l,l+2}(n) \equiv \nu_{n,l} - \nu_{n-1,l+2} \simeq -(4l + 6) \frac{\Delta_l(n)}{4 \pi^2 \nu_{n,l}}\int_0^{R_\odot} \frac{dc_s}{dr}\frac{dr}{r}.
\label{eq:ddef}
\end{equation}

\begin{figure}[p]
\begin{tabular}{c c}
\includegraphics[width=0.5\textwidth]{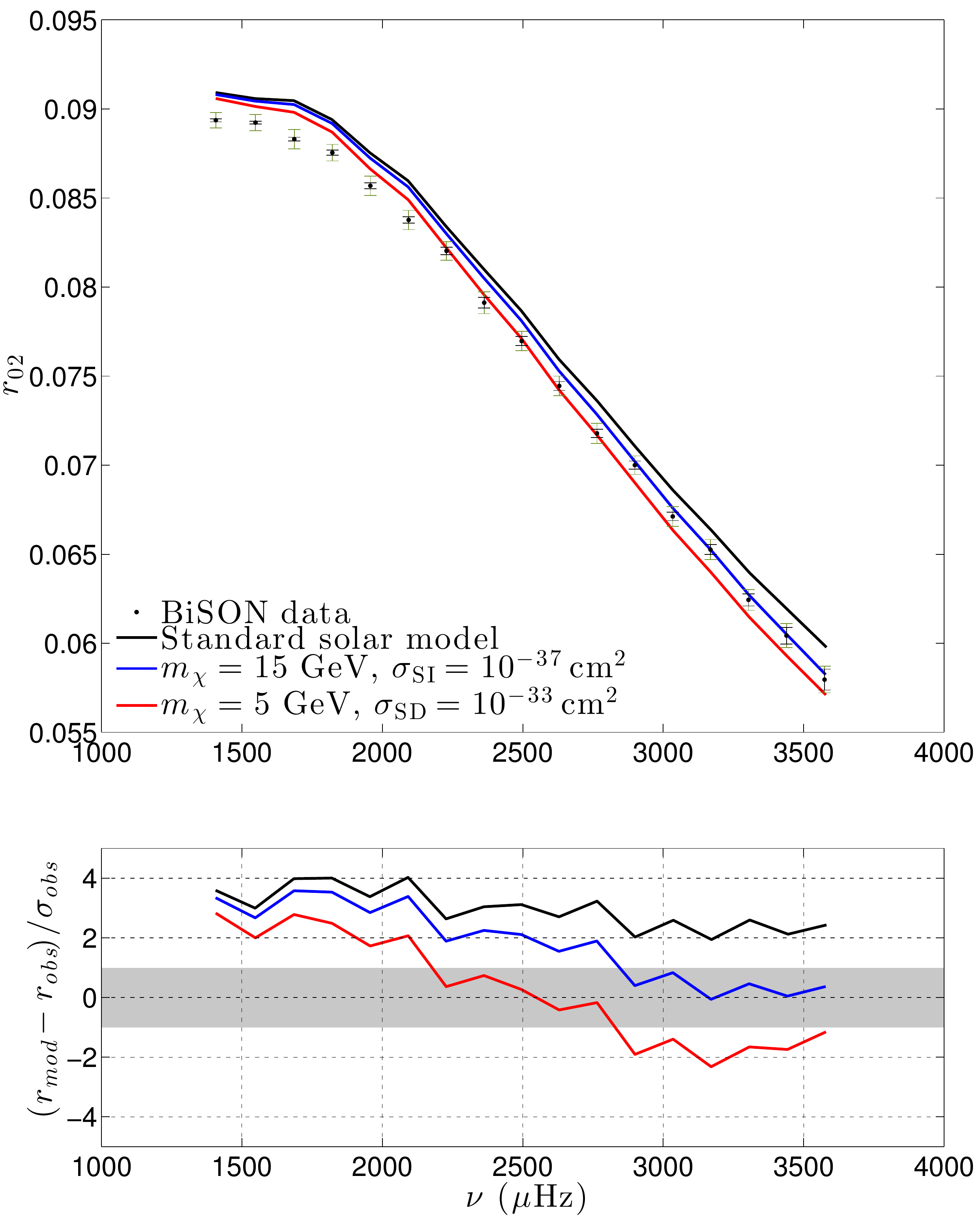} & \includegraphics[width=0.5\textwidth]{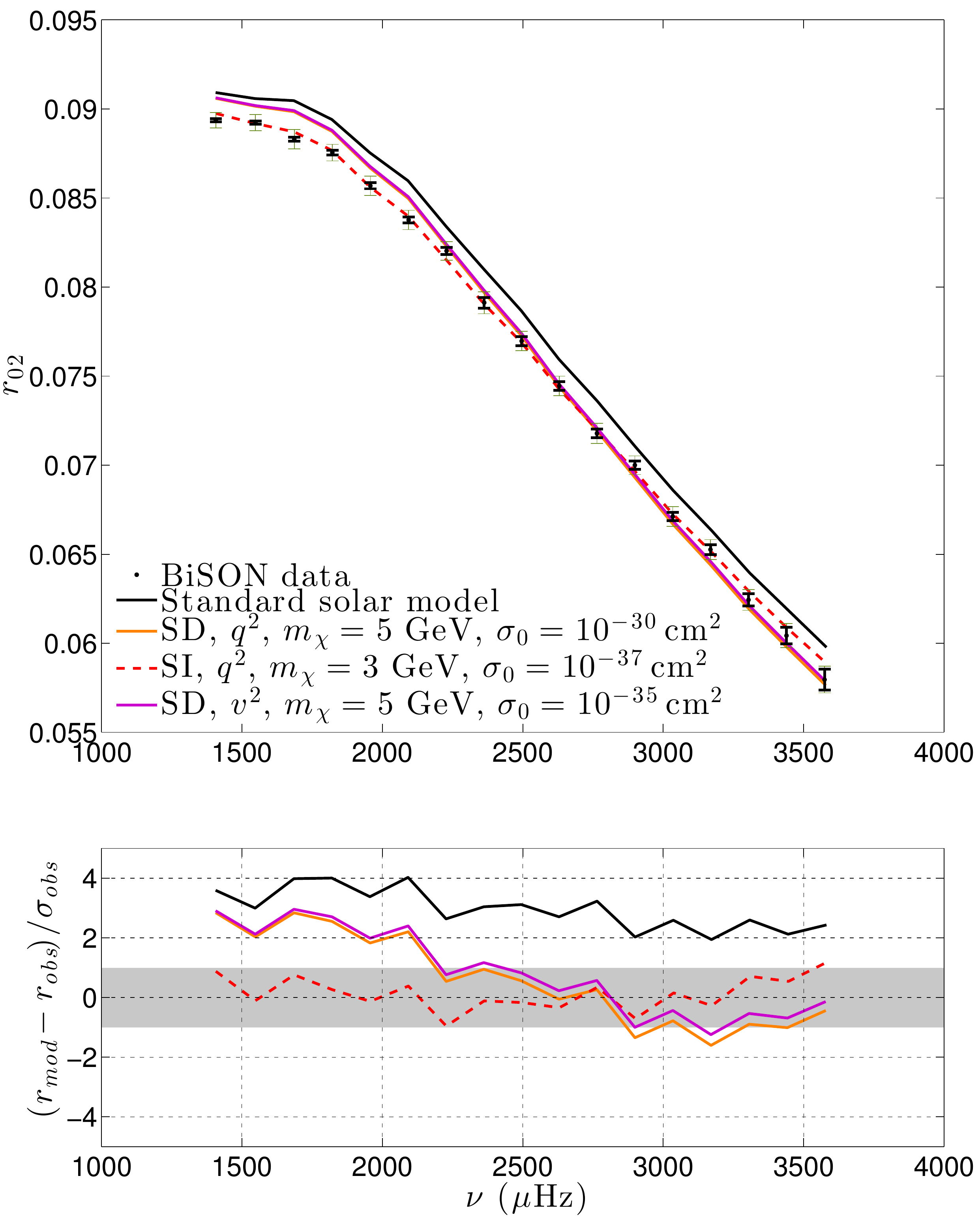} \\
\includegraphics[width=0.5\textwidth]{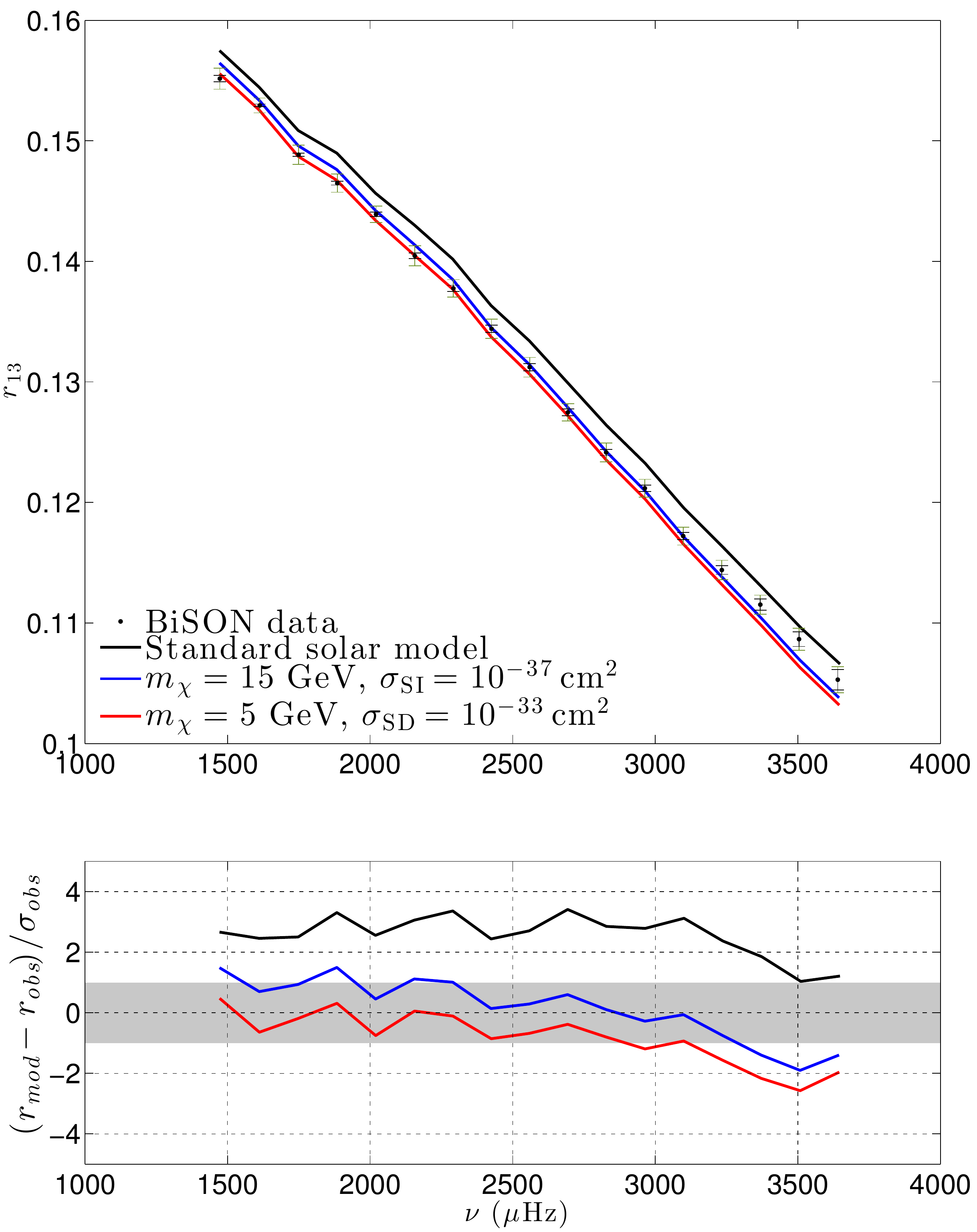} & \includegraphics[width=0.5\textwidth]{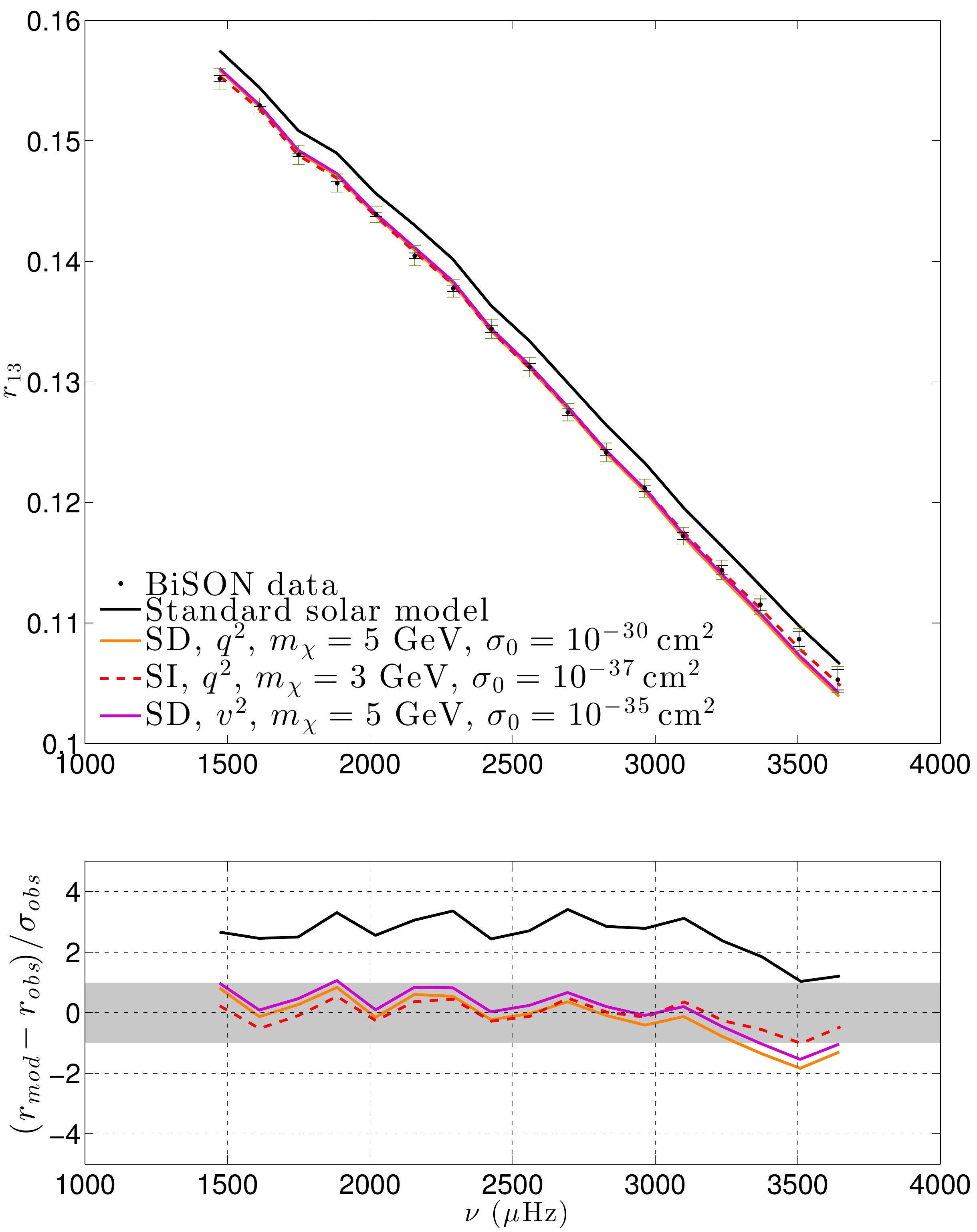} \\
\end{tabular}
\caption{Small frequency separations $r_{02}$ (top) and $r_{13}$ (bottom) as defined in Eq.\ \ref{eq:rdef}, for the best fit constant (left) and generalised form-factor dark matter models (right). The latter correspond to the best-fit models of the three couplings returning the best overall $p$-values (see Tab.~\ref{best_fit_tab}). Predictions are compared with helioseismological observations from the BiSON experiment \cite{Basu:2006vh}. Inner black error bars correspond to observational error, whereas outer (green) bars also include modelling error. Below each figure we show the residuals with respect to BiSON data, in units of the total error. The $m_\chi = 3$ GeV, $q^2$ case with $\sigma_0 = 10^{-37}$ cm$^2$ yields the best improvement, bringing the largest discrepancy from nearly 4$\sigma$ to little more than a standard deviation.}
\label{fig:smallfreq}
\end{figure}

\FloatBarrier
\noindent and $\Delta_l(n) \equiv \nu_{n,l} - \nu_{n-1,l}$. The sound speed gradient is weighted by $1/r$ in the integral above, so $r_{02}(n)$ and $r_{13}(n)$ are most sensitive to changes there.

We show a set of these ratios in Fig.\ \ref{fig:smallfreq} for the same examples shown in Fig.\ \ref{csfig}. These are compared with the values of $r_{02}(n)$ and $r_{13}(n)$ measured by the BiSON experiment \cite{bison2007}. Model uncertainties are computed following the same procedure as for the sound speed. As can be seen in the residual subplots, the SSM overestimates each of these ratios by as much as 4$\sigma$, once modelling errors are included. Thermal transport by ADM can smooth the sound speed gradient, yielding smaller differences in Eq.\ \ref{eq:ddef}. As in the case of the sound speed profile, the best improvement is provided by $q^2$ scattering with $m_\chi = 3$ GeV, $\sigma_0 = 10^{-37}$ cm$^2$.  In this model, the largest discrepancy in $r_{02}$ and $r_{13}$ falls to barely $1\sigma$.

\begin{figure}[p]
\begin{tabular}{c@{\hspace{0.04\textwidth}}c}
\multicolumn{2}{c}{\includegraphics[height = 0.32\textwidth]{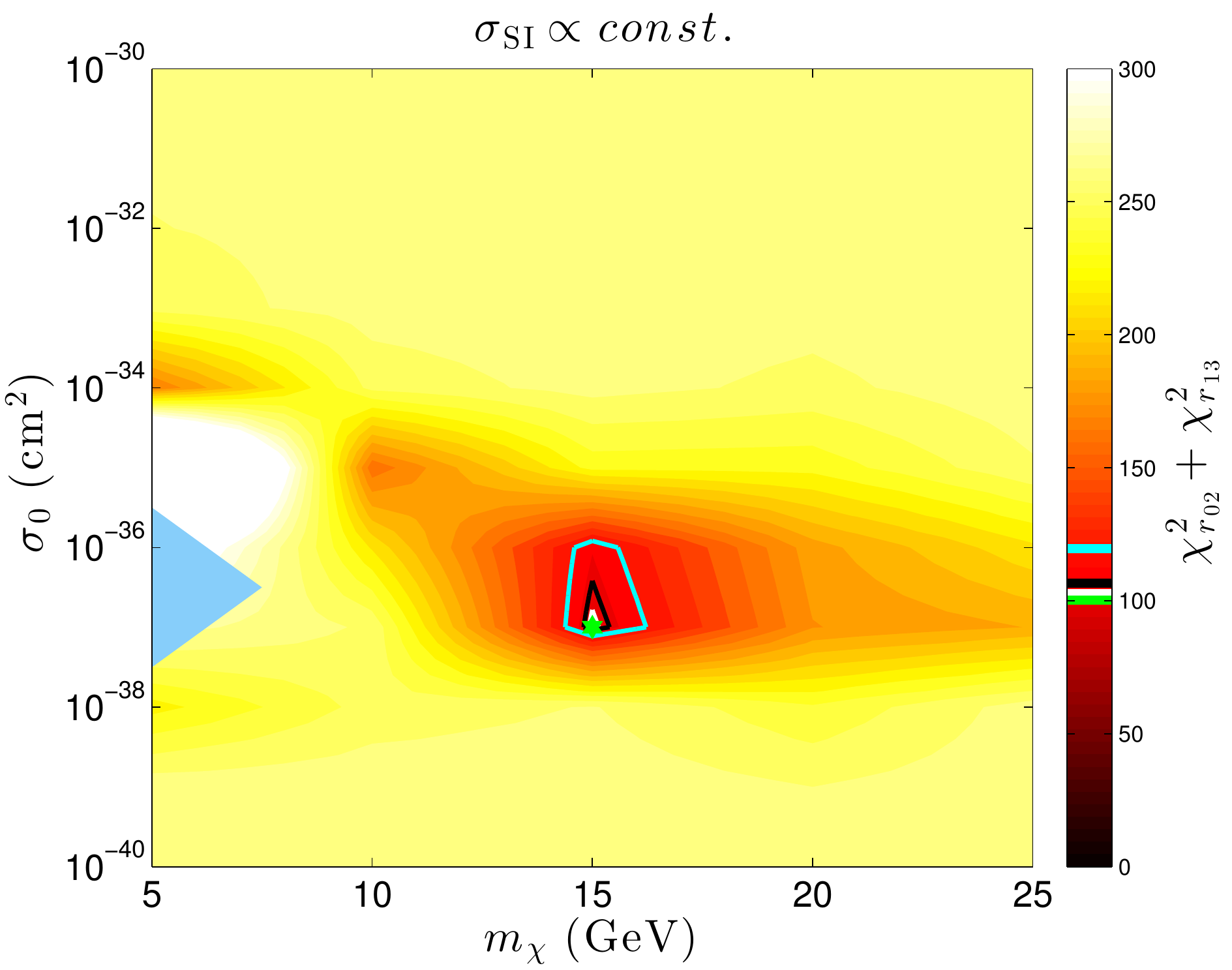}} \\
\includegraphics[height = 0.32\textwidth]{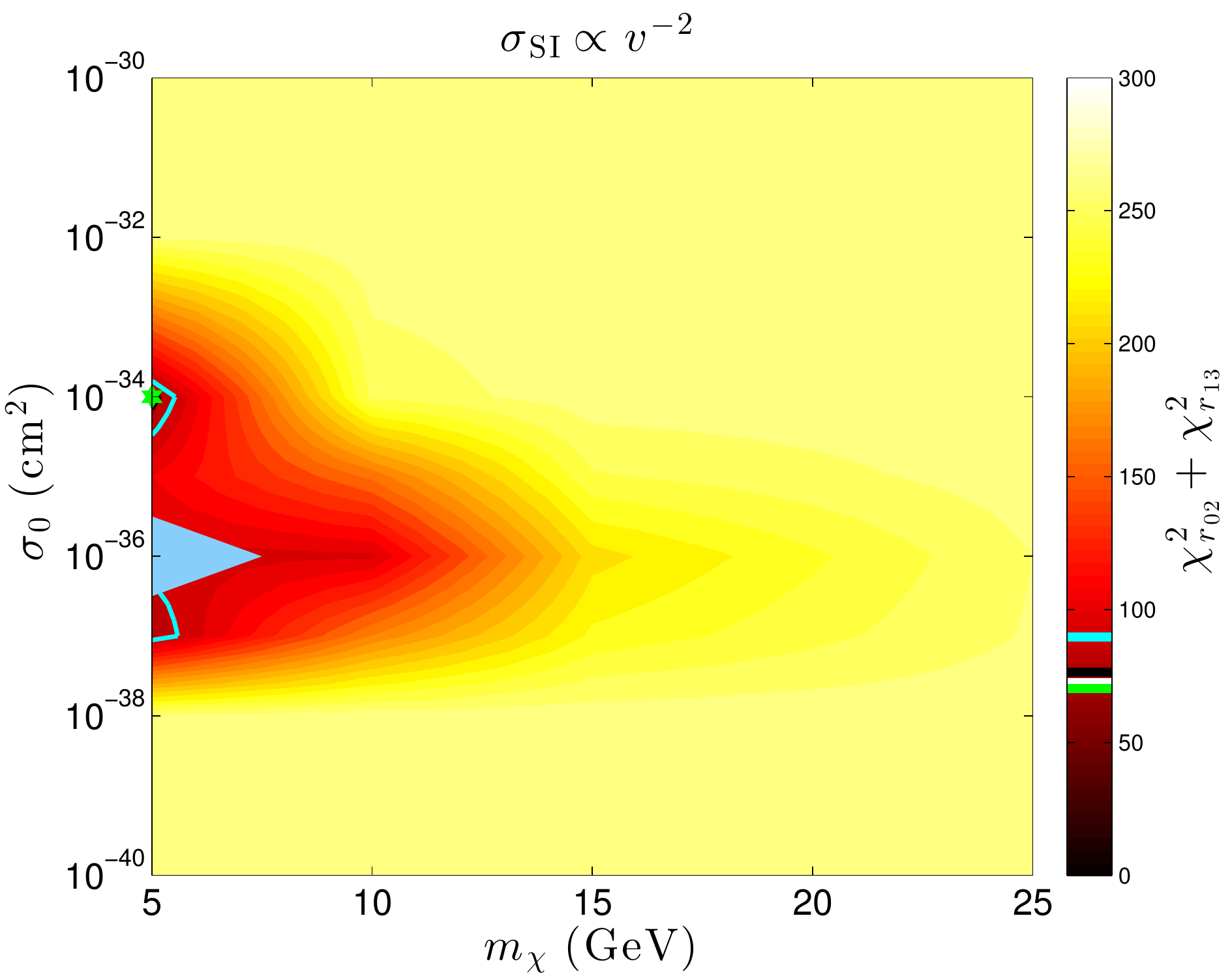} & \includegraphics[height = 0.32\textwidth]{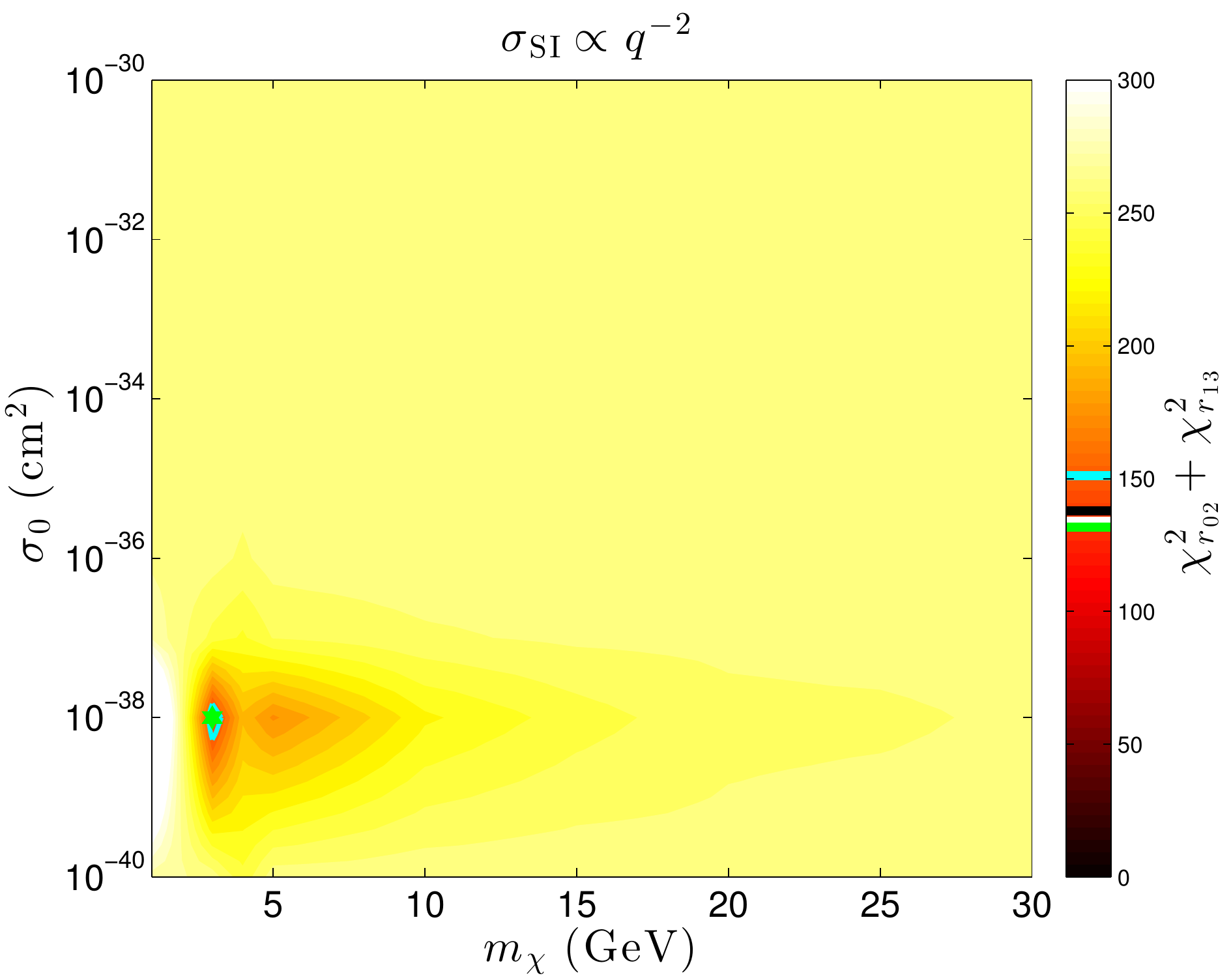} \\
\includegraphics[height = 0.32\textwidth]{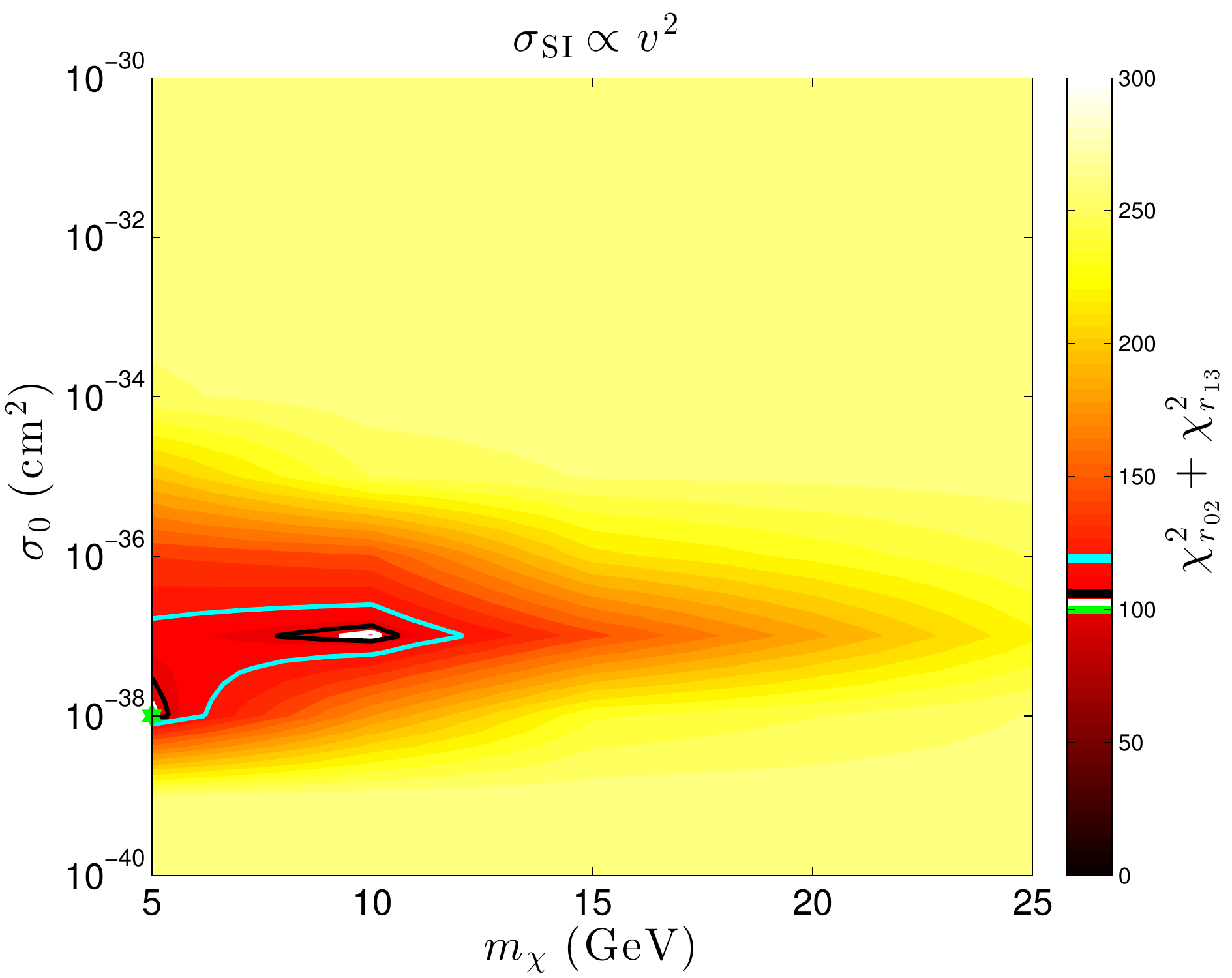} & \includegraphics[height = 0.32\textwidth]{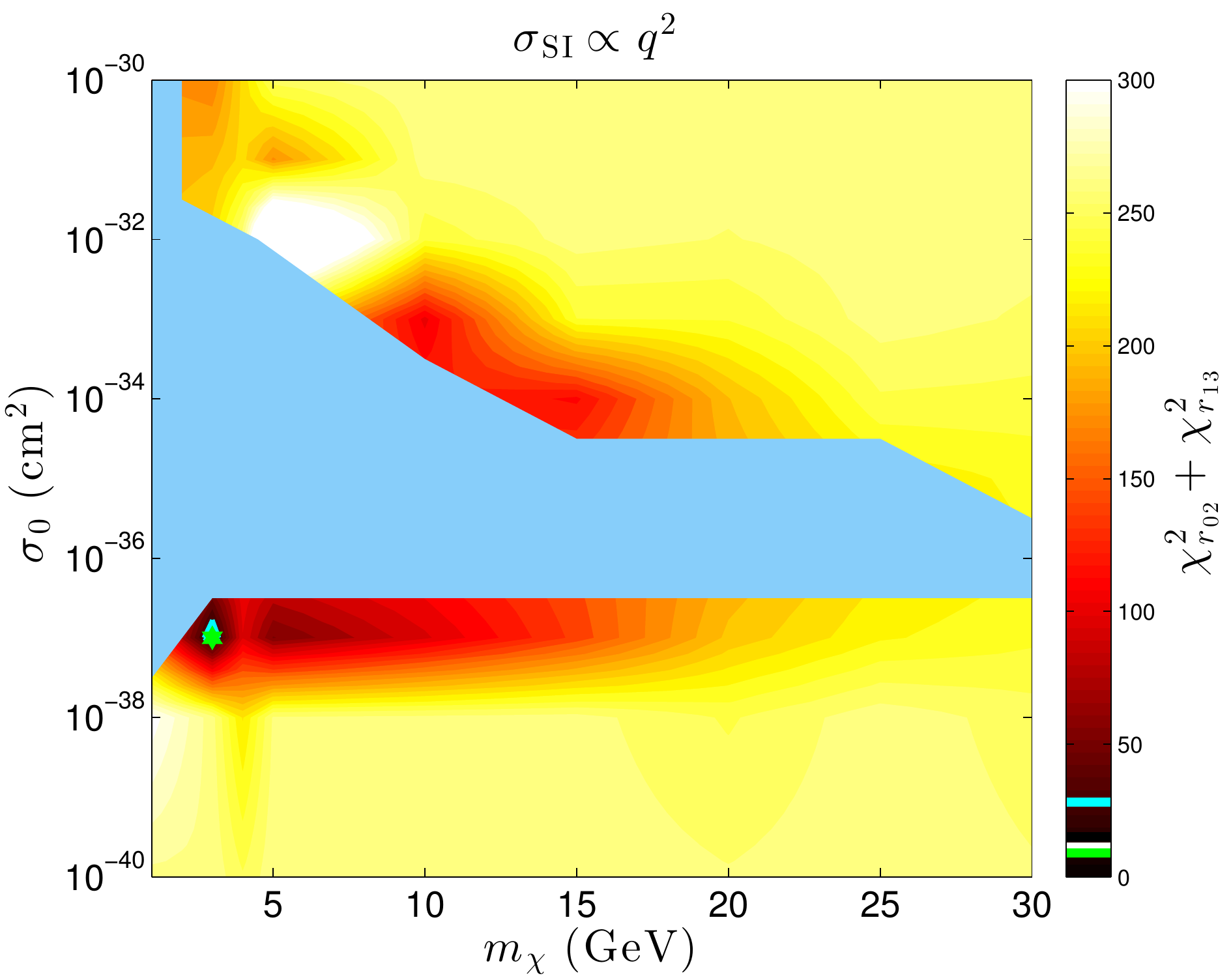} \\
\includegraphics[height = 0.32\textwidth]{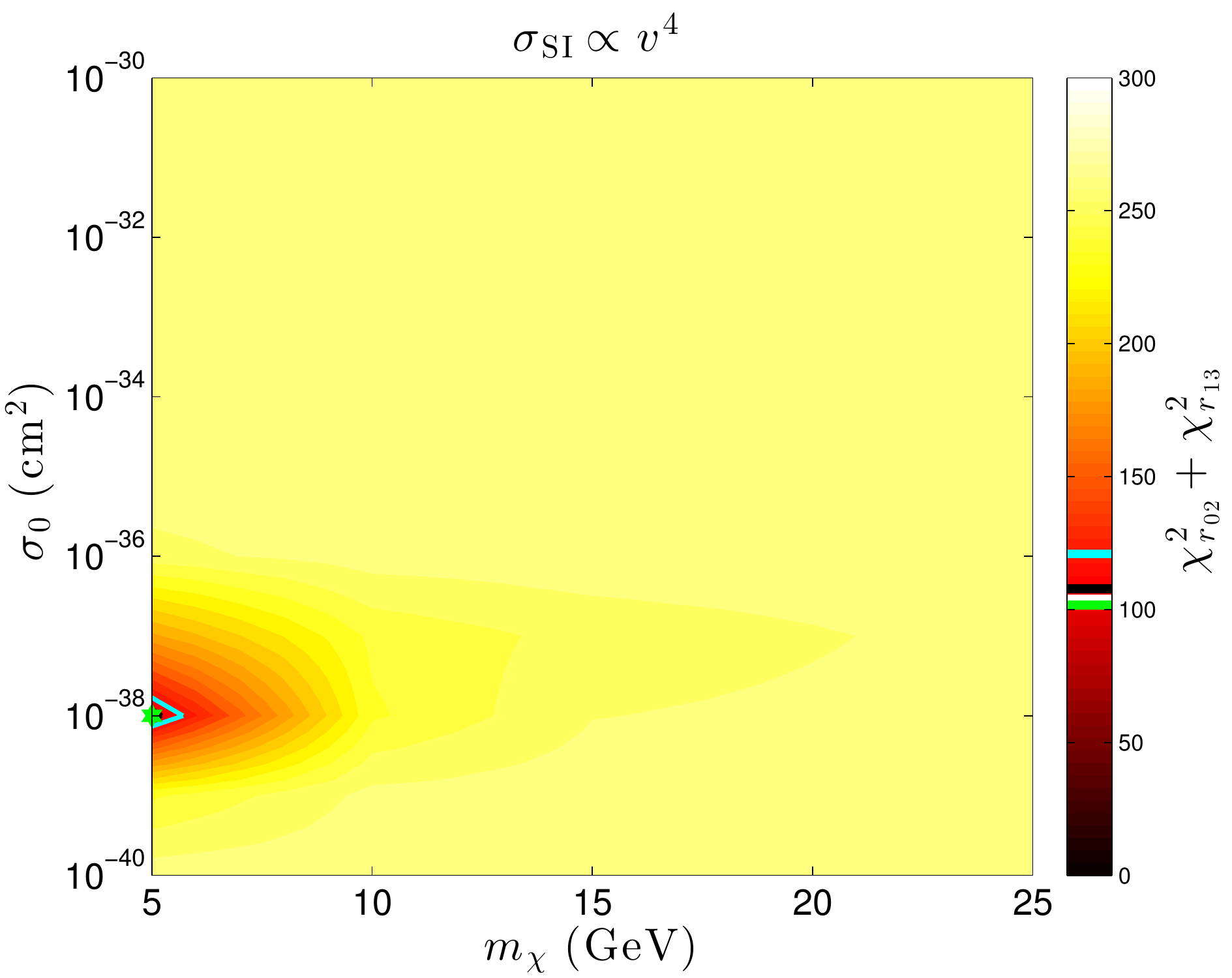} & \includegraphics[height = 0.32\textwidth]{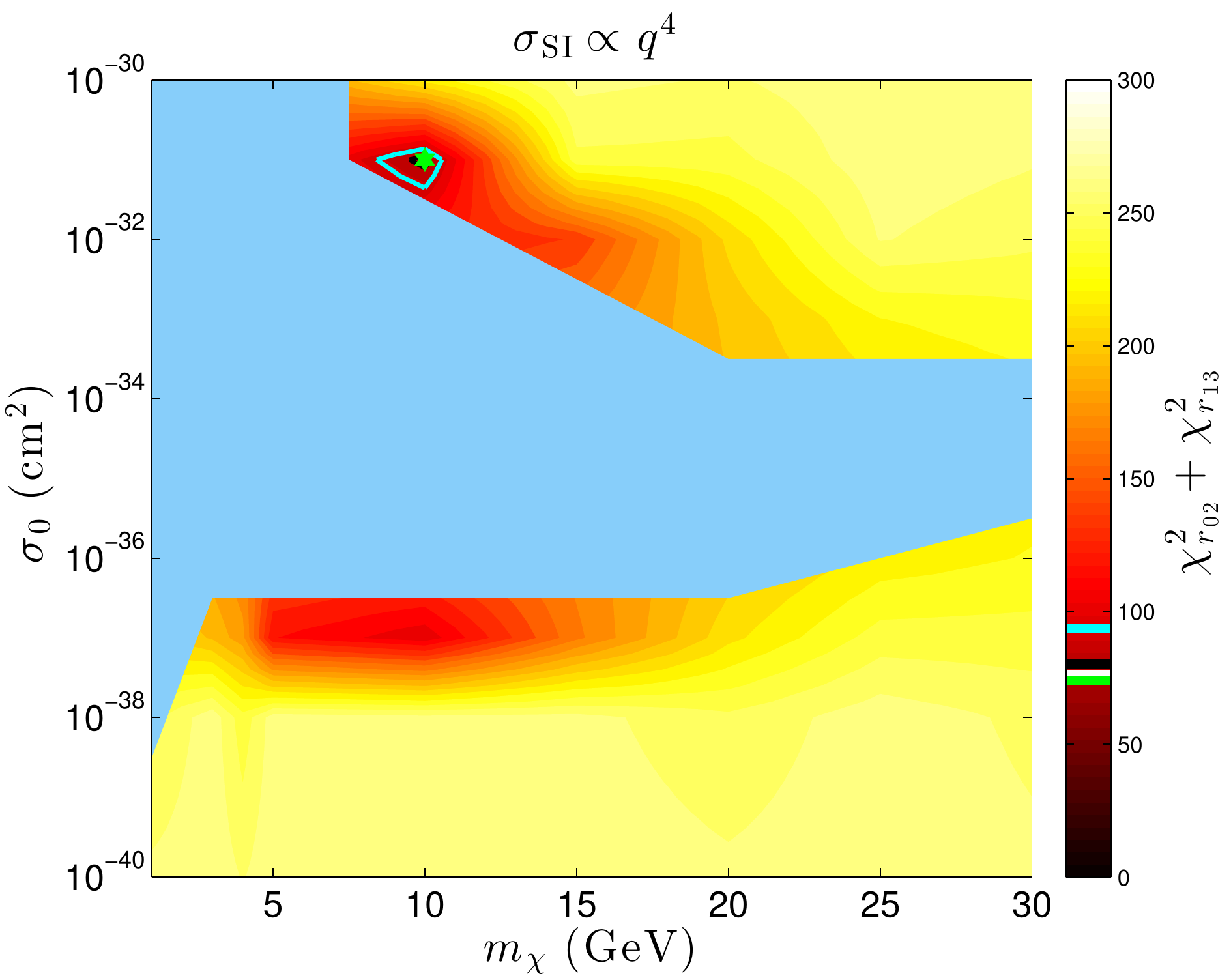} \\
\end{tabular}
\caption{Combined likelihood $\chi^2_{r_{02}} + \chi^2_{r_{13}}$ for the small frequency separations defined in Eq.\ \ref{eq:rdef}, for different models of ADM with spin-independent scattering cross sections. The green line on the colour bar shows the $\chi^2$ value of the best fit, indicated by a green star in each panel, whereas white, black and cyan lines show the preferred regions at 1, 2 and 4 $\sigma$ confidence levels, respectively. At low masses, many DM models can yield improvements in the overall chi-squared with respect to the Standard Solar Model, but the $q^2$ model provides a clear best fit.}
\label{SIr0213}
\end{figure}

\begin{figure}[p]
\begin{tabular}{c@{\hspace{0.04\textwidth}}c}
\multicolumn{2}{c}{\includegraphics[height = 0.32\textwidth]{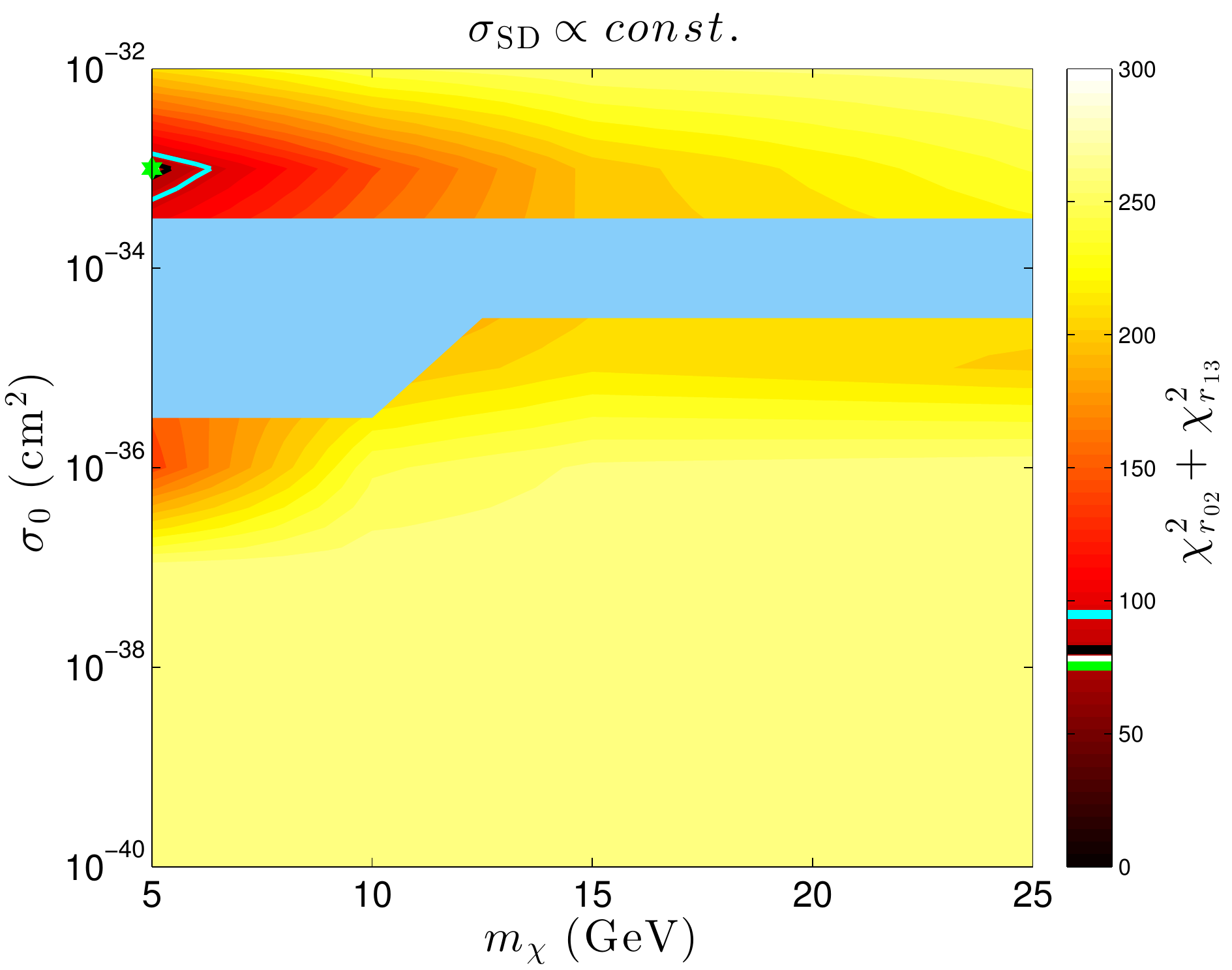}} \\
\includegraphics[height = 0.32\textwidth]{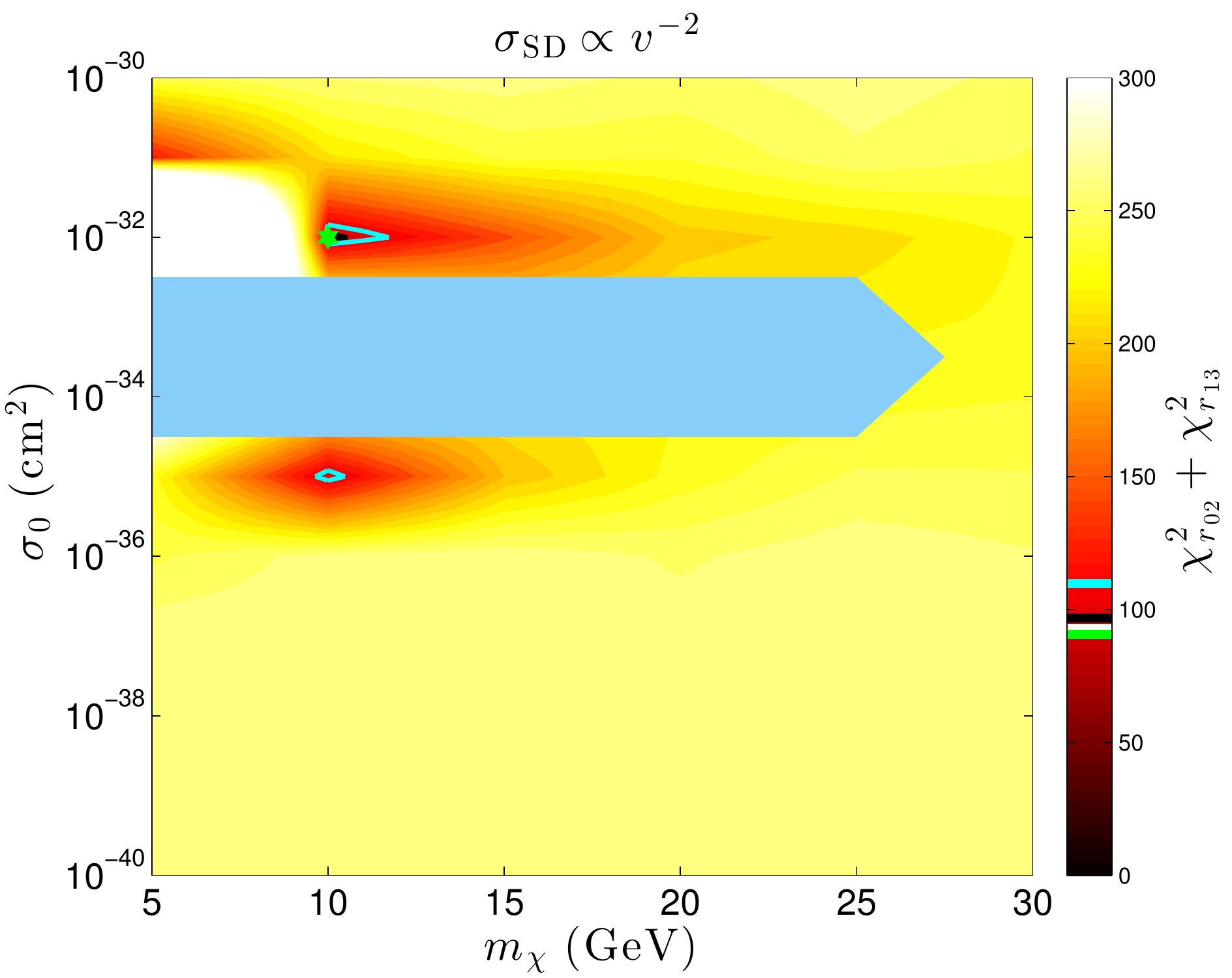} & \includegraphics[height = 0.32\textwidth]{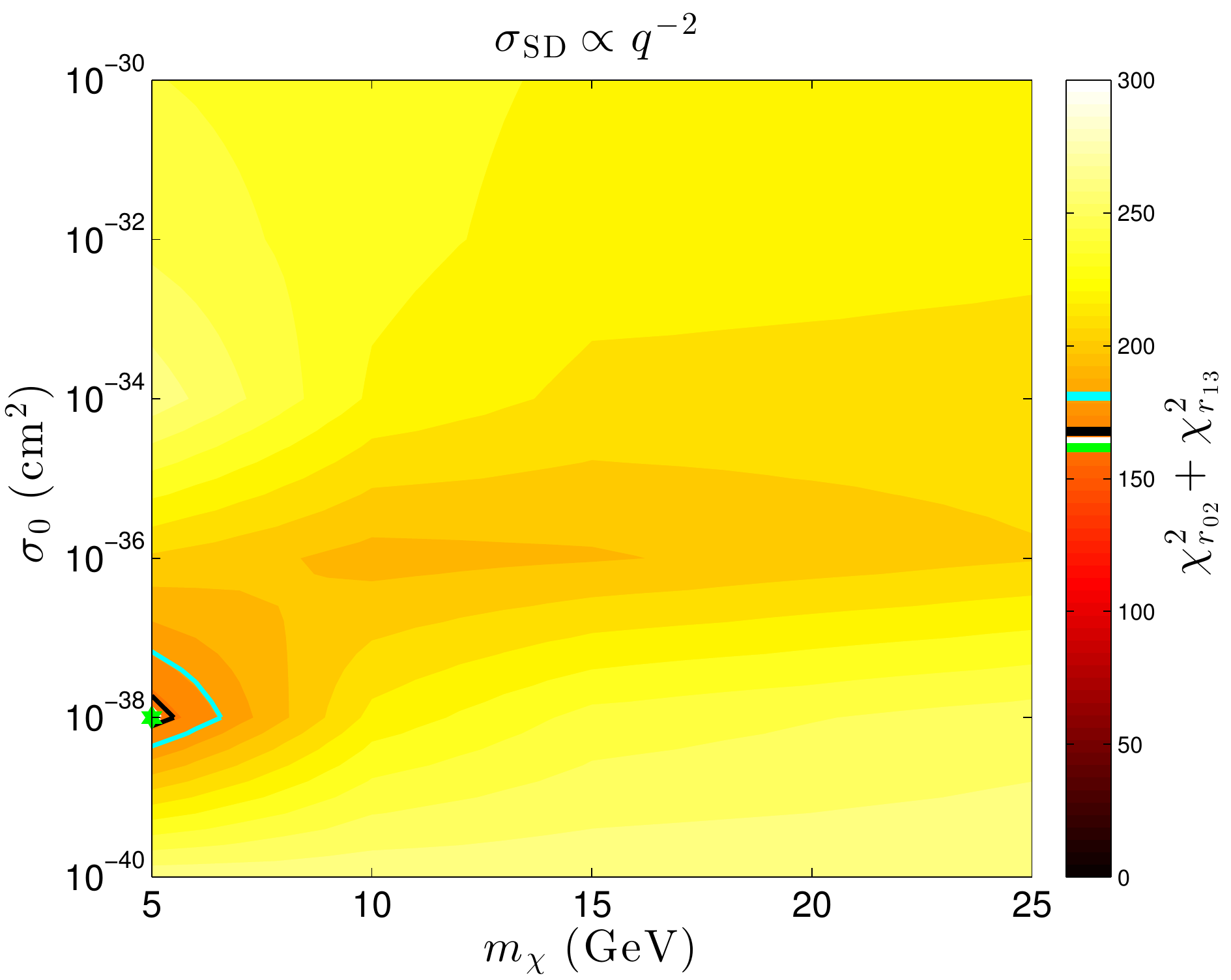} \\
\includegraphics[height = 0.32\textwidth]{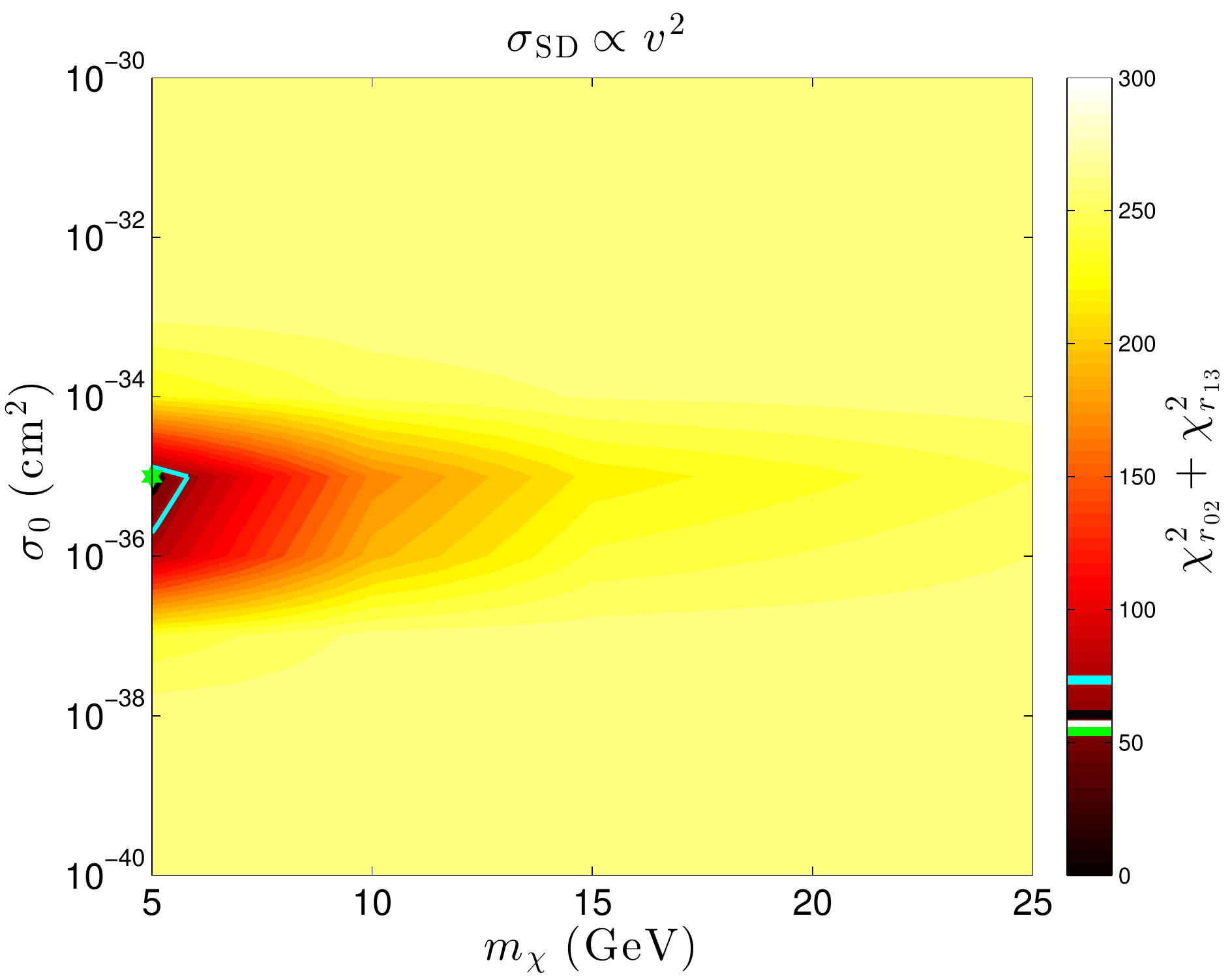} & \includegraphics[height = 0.32\textwidth]{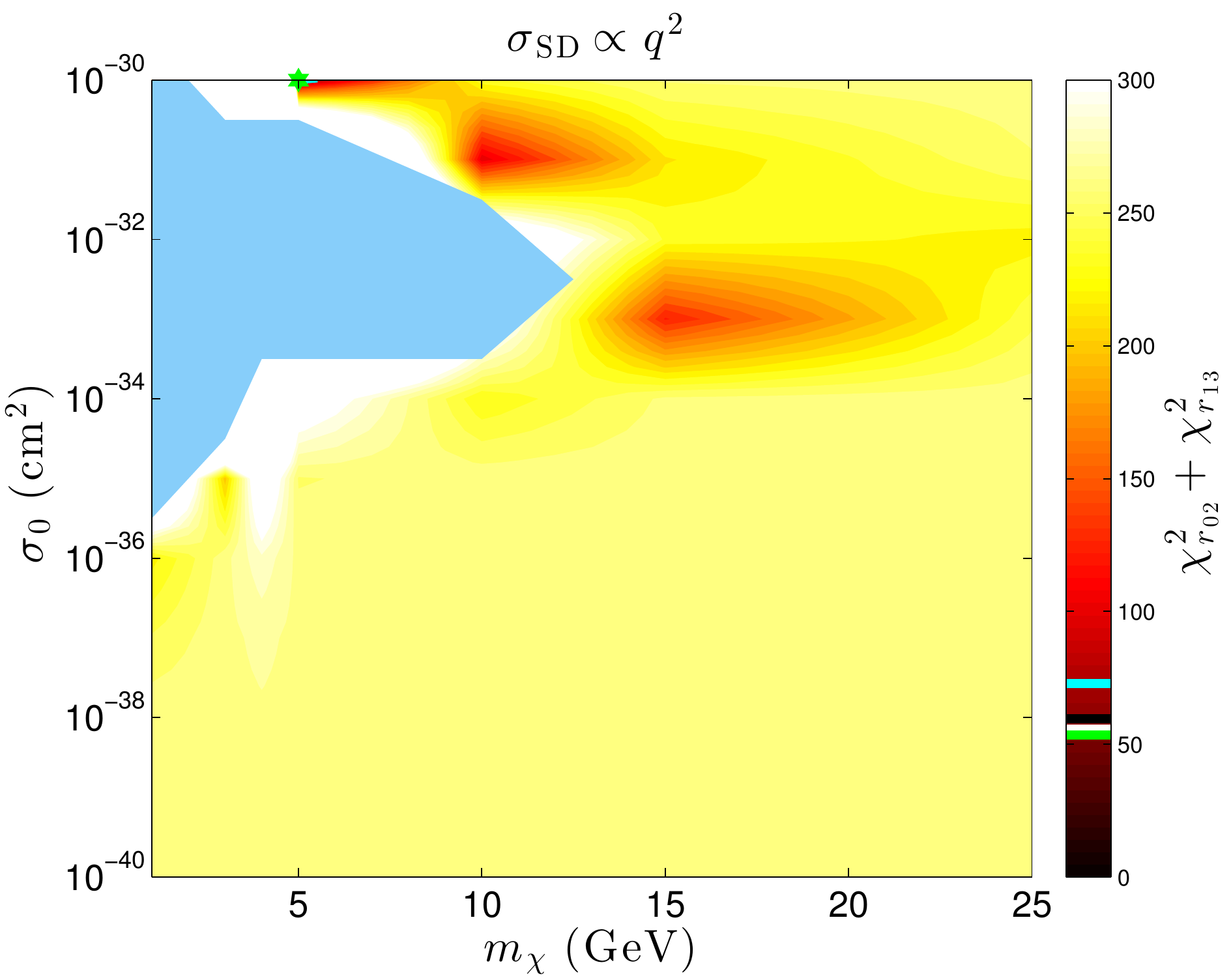} \\
\includegraphics[height = 0.32\textwidth]{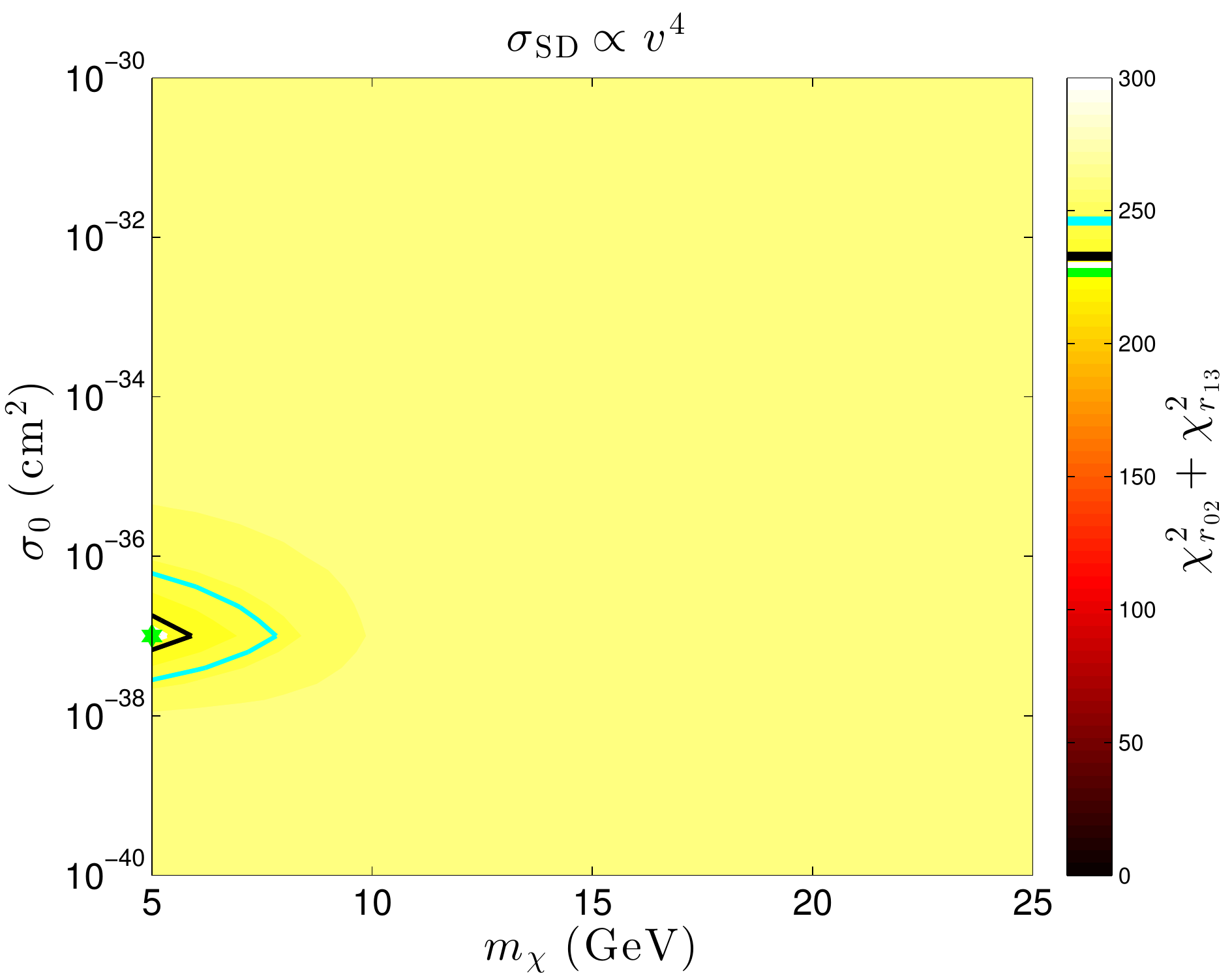} & \includegraphics[height = 0.32\textwidth]{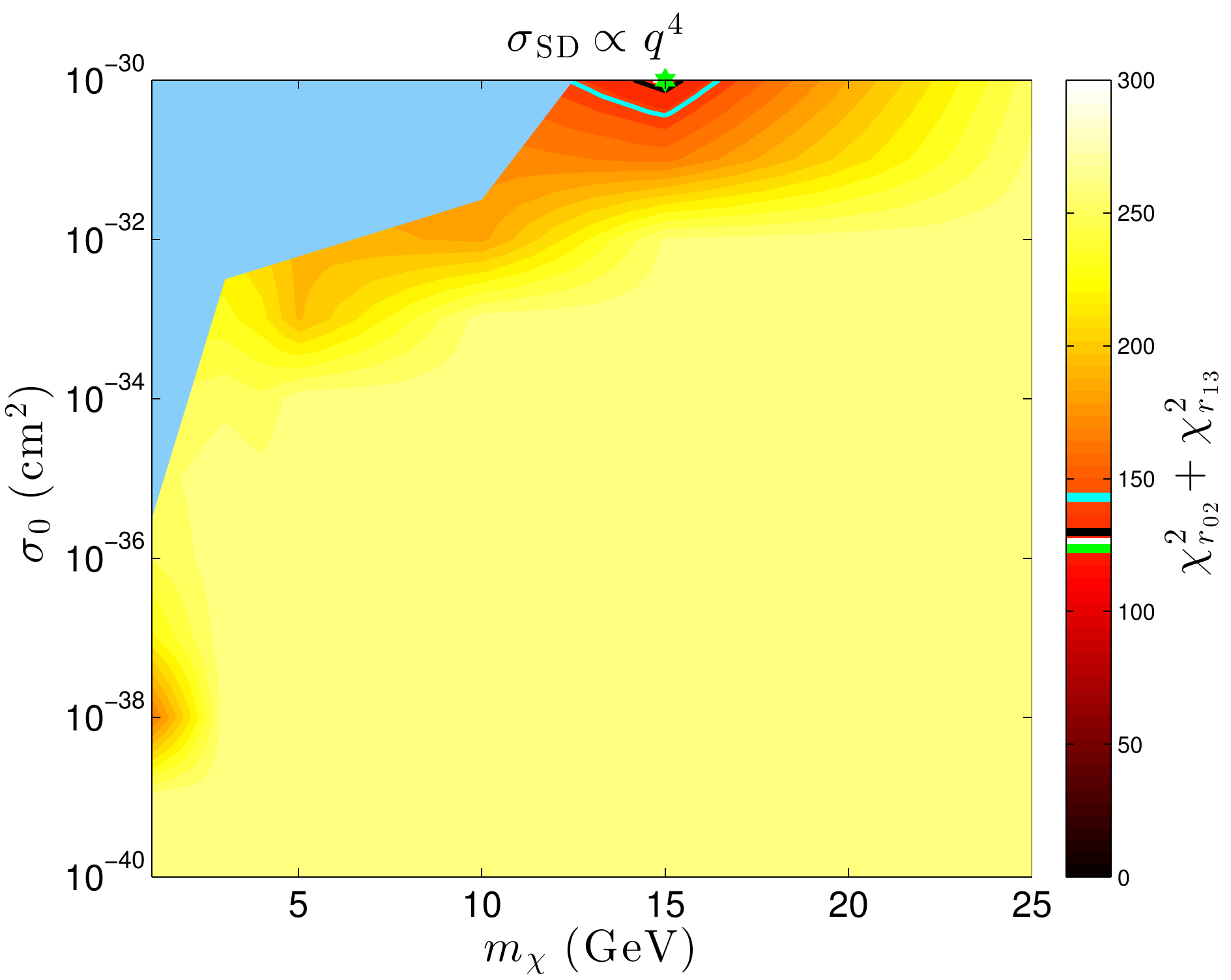} \\
\end{tabular}
\caption{Same as Fig.\ \ref{SIr0213}, but for spin-dependent couplings.  }
\label{SDr0213}
\end{figure}

In Figs. \ref{SIr0213} and \ref{SDr0213}, we show the overall ability of different models to fit the observed frequency separations.  We quantify this with the combined chi-squared $\chi^2_{r_{02}} + \chi^2_{r_{13}}$, where
\begin{equation}
\chi^2_{r_{\ell,\ell+2}} =  \sum_n \frac{[r_{\ell,\ell+2,{\rm th.}}(n) - r_{\ell,\ell+2,{\rm obs.}}(n)]^2}{\sigma^2_{\rm obs.}(n) + \sigma^2_{\rm th.}(n)}.
\end{equation}
Formally, there is a correlation between values of $r_{02}(n)$ and $r_{13}(n)$ with common eigenfrequencies. In practice, however, the correlation is minimal (the Pearson correlation coefficient is always lower than 0.1), so we assume they are independent when computing $\chi^2_{r_{02}} + \chi^2_{r_{13}}$.
Figs. \ref{SIr0213} and \ref{SDr0213} show that all SI and SD couplings do exhibit areas of parameter space where the small frequency separations can be brought into better agreement with observations than in the SSM.  In many cases however, these are not the same parameter combinations preferred by other observables.  The notable exception to this is the $q^2$ SI coupling, which also clearly produces the best agreement with the observed frequency separations.

\subsection{Combined limits}
\label{sec:combined}

It is clear from the results we have presented in Secs.\ \ref{sec:neutrinos}--\ref{sec:smallfreq} that the disagreement between model predictions and helioseismology can be ameliorated, at the cost of a slightly worse agreement with neutrino fluxes. Surface helium Y$_s $ is also affected to a lesser extent. It is therefore instructive to construct an overall likelihood function based on the fit of these data by our models. We define the combined chi-squared as:
\begin{eqnarray}
\chi^2 &=& \frac{(1 - \phi^\nu_{\rm B,obs}/\phi^\nu_{\rm B})^2}{\sigma_{\rm B}^2} + \frac{(1 - \phi^\nu_{\rm Be,obs}/\phi^\nu_{\rm Be})^2}{\sigma_{\rm Be}^2} \nonumber  \\
&&+ \frac{(r_{\rm CZ}  - r_{\rm CZ,obs} )^2}{\sigma_{\rm CZ} ^2} + \frac{(Y_{\rm S} - Y_{\rm S,obs} )^2}{\sigma_{Y_{\rm S}}^2} \nonumber \\
&&+ \chi^2_{r_{02}} + \chi^2_{r_{13}} .
 \label{eq:fullchisq}
\end{eqnarray}
where $ \phi^\nu_{\rm B,obs} =5.00\times10^6\,\hbox{cm$^{-2}$s$^{-1}$}$, $\phi^\nu_{\rm Be, obs} =4.82\times10^9 \hbox{cm$^{-2}$s$^{-1}$}$, $r_{\rm CZ,obs}  = 0.713\,R_\odot$ and $Y_{\rm S,obs}  = 0.2485$. The uncertainties include both the observational and modelling errors, added in quadrature. These are given individually in the last two lines of Table \ref{best_fit_tab}. We include the small frequency separations $r_{\ell,\ell + 2}$, but not the sound speed profile, as the latter is less precise and is correlated with the former. 

\begin{figure}[p]
\begin{tabular}{c@{\hspace{0.04\textwidth}}c}
\multicolumn{2}{c}{\includegraphics[height = 0.32\textwidth]{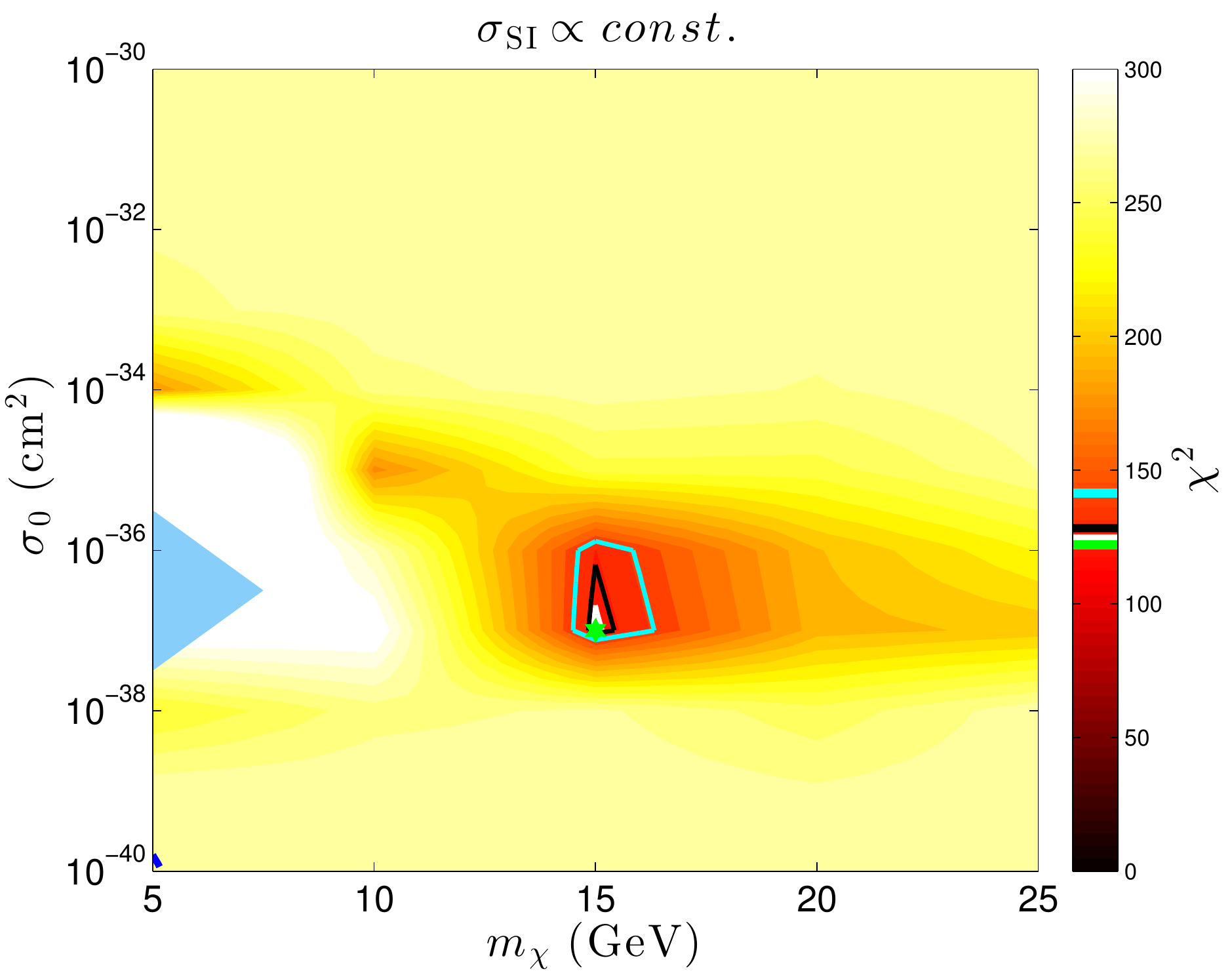}} \\
\includegraphics[height = 0.32\textwidth]{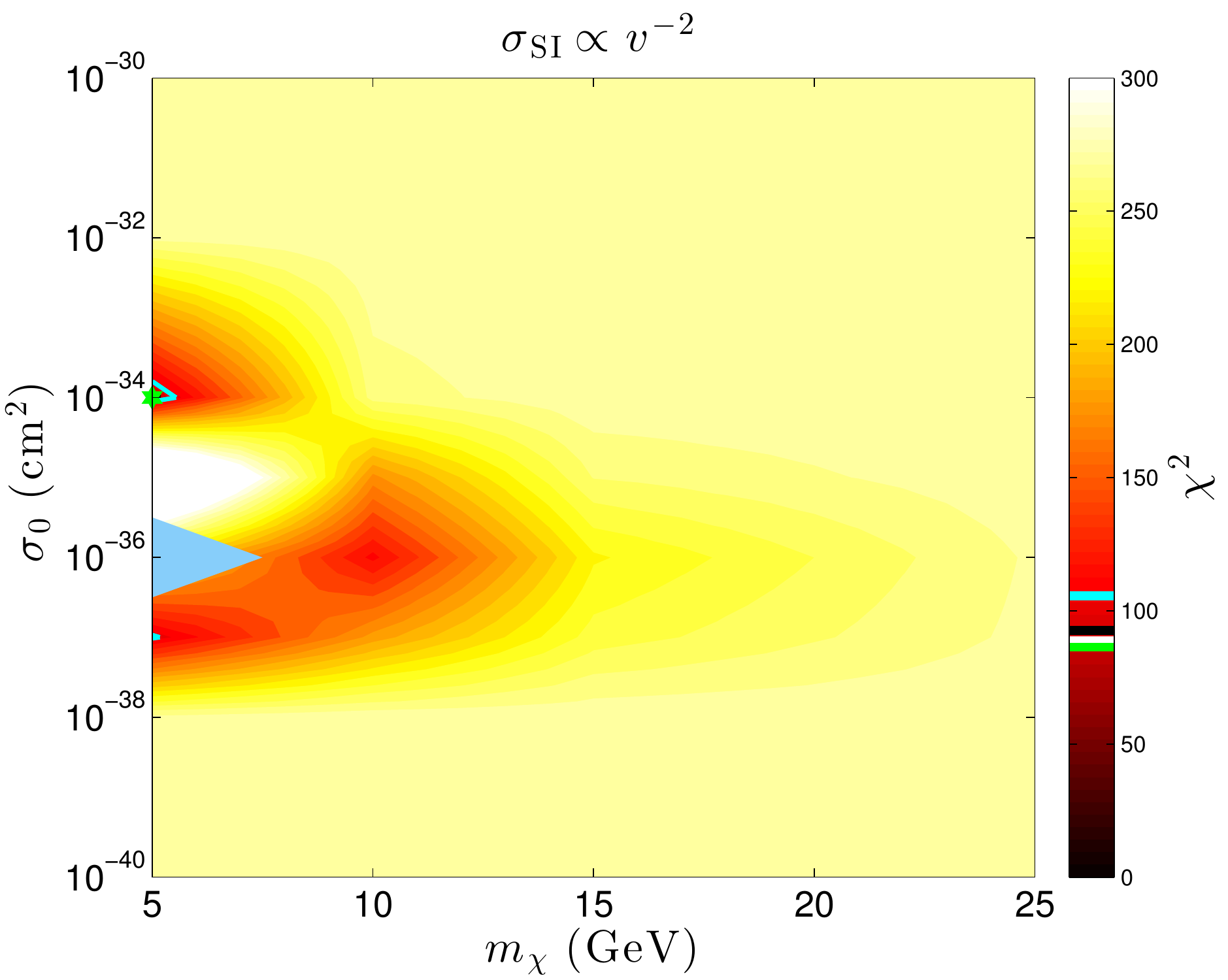} & \includegraphics[height = 0.32\textwidth]{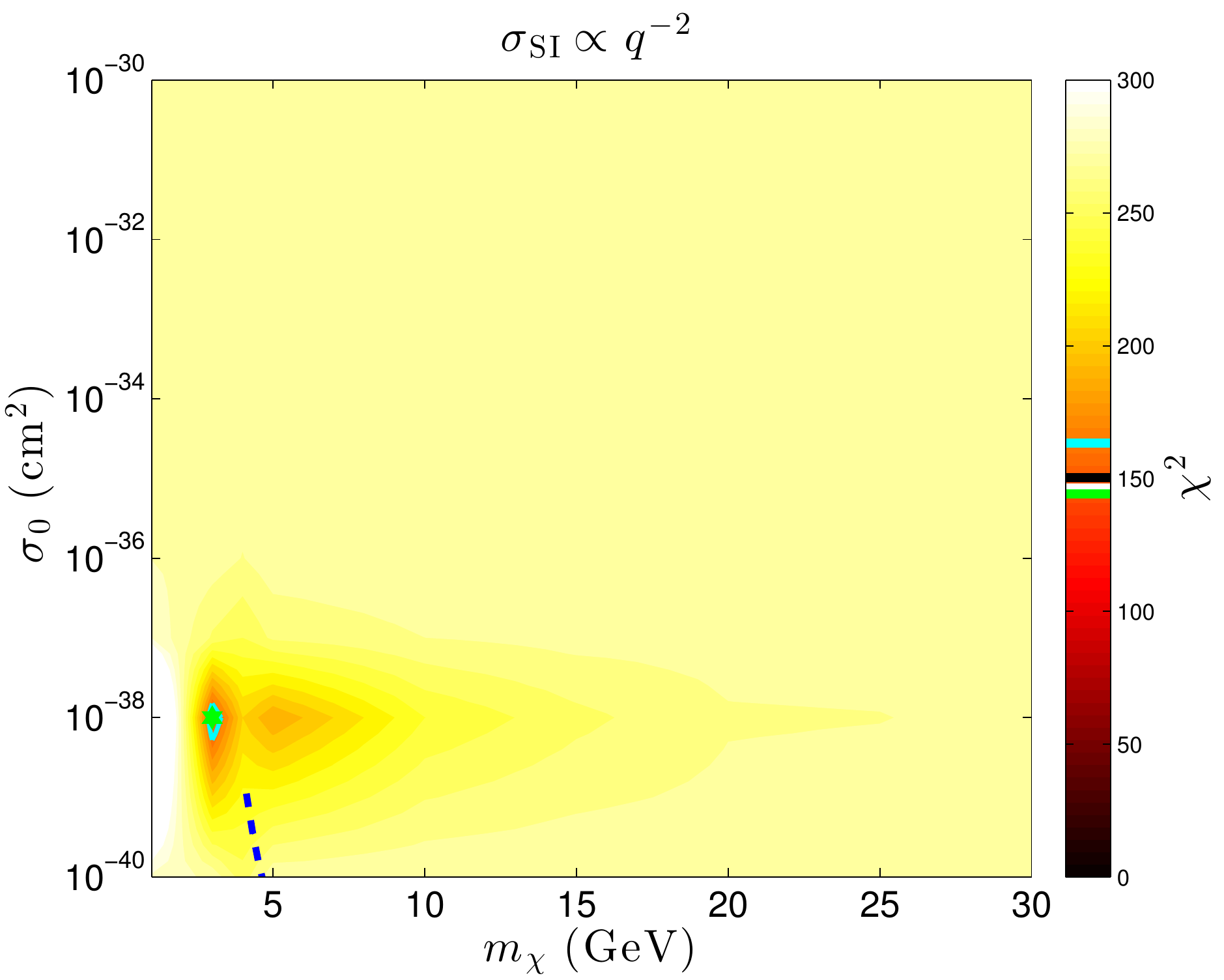} \\
\includegraphics[height = 0.32\textwidth]{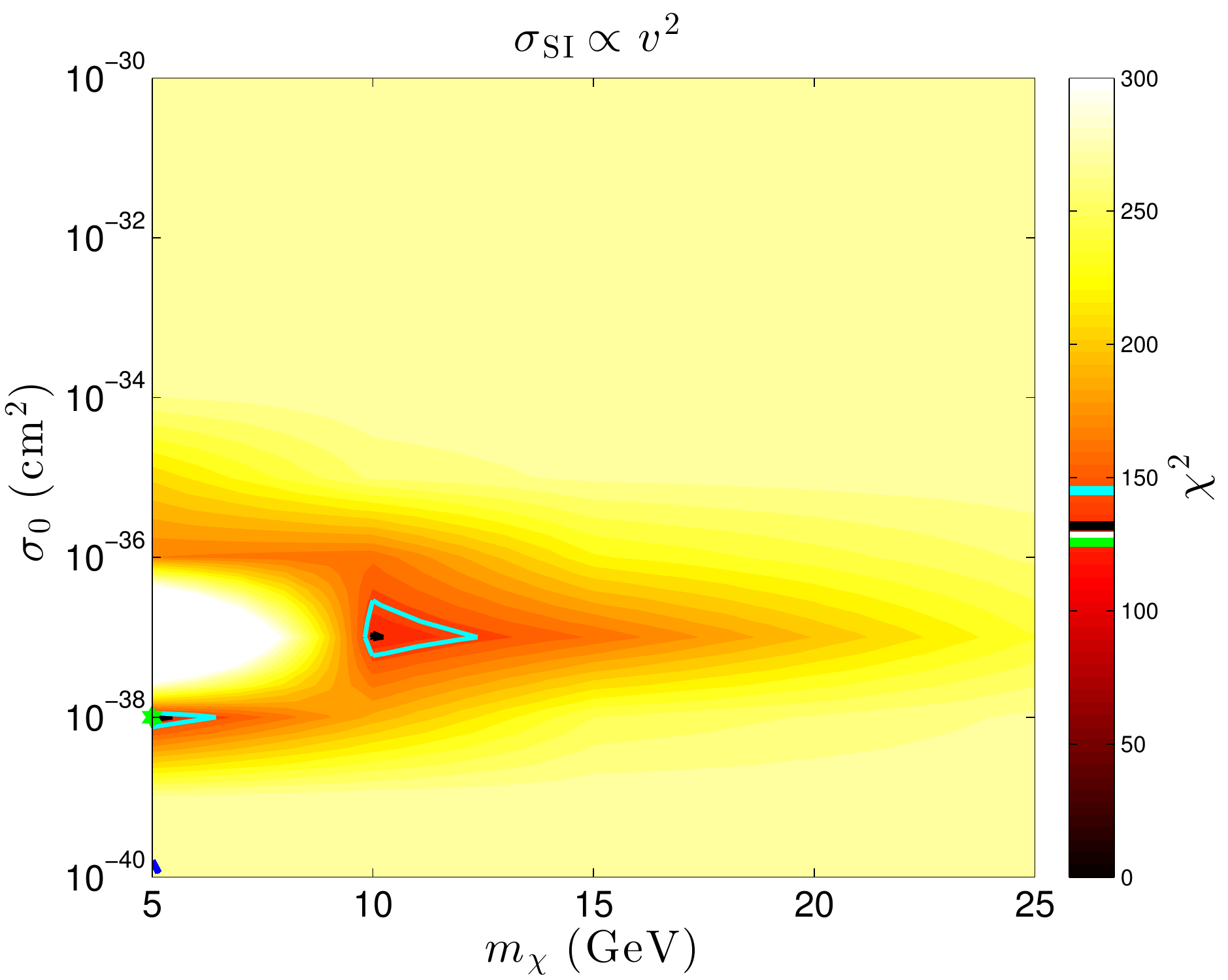} & \includegraphics[height = 0.32\textwidth]{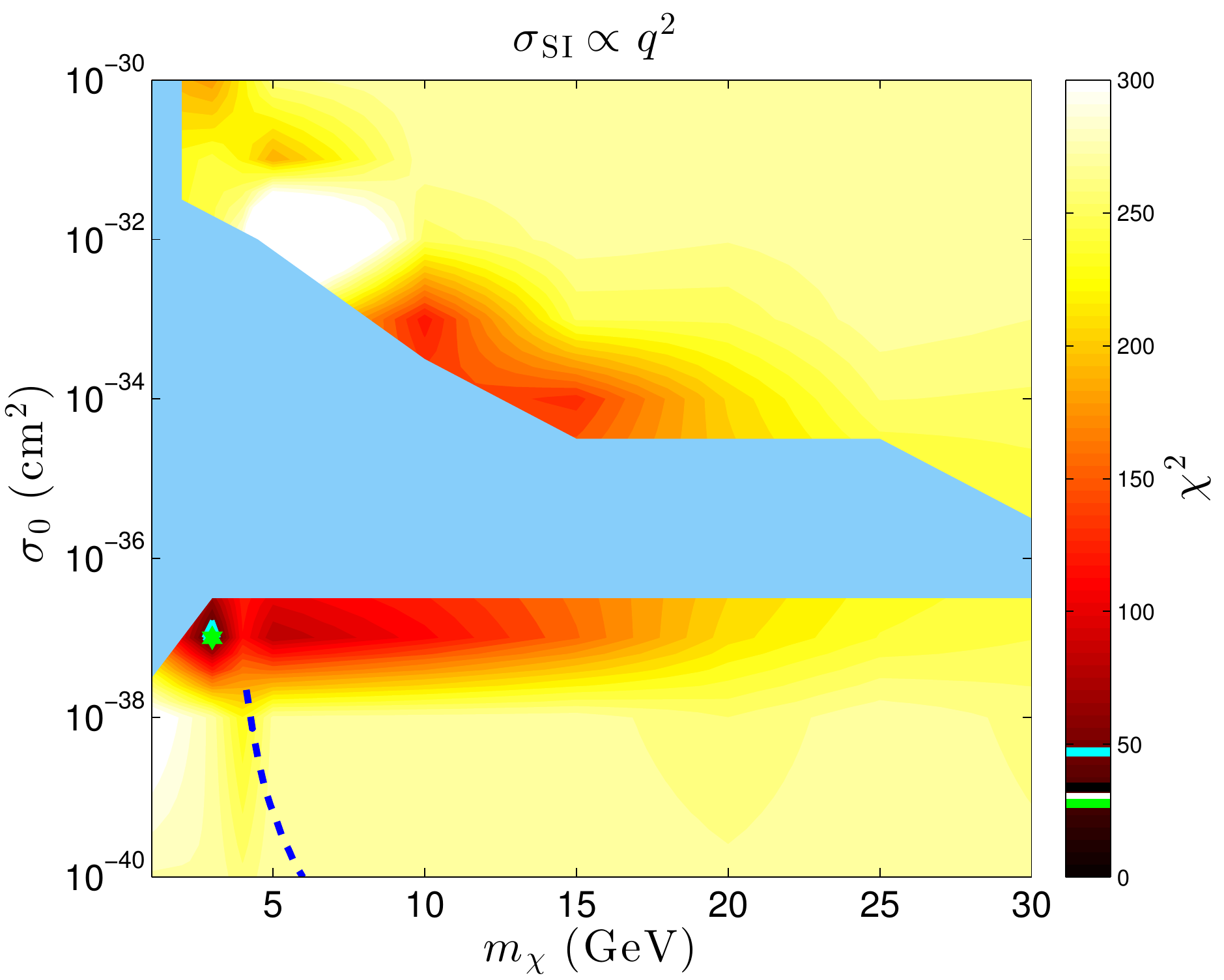} \\
\includegraphics[height = 0.32\textwidth]{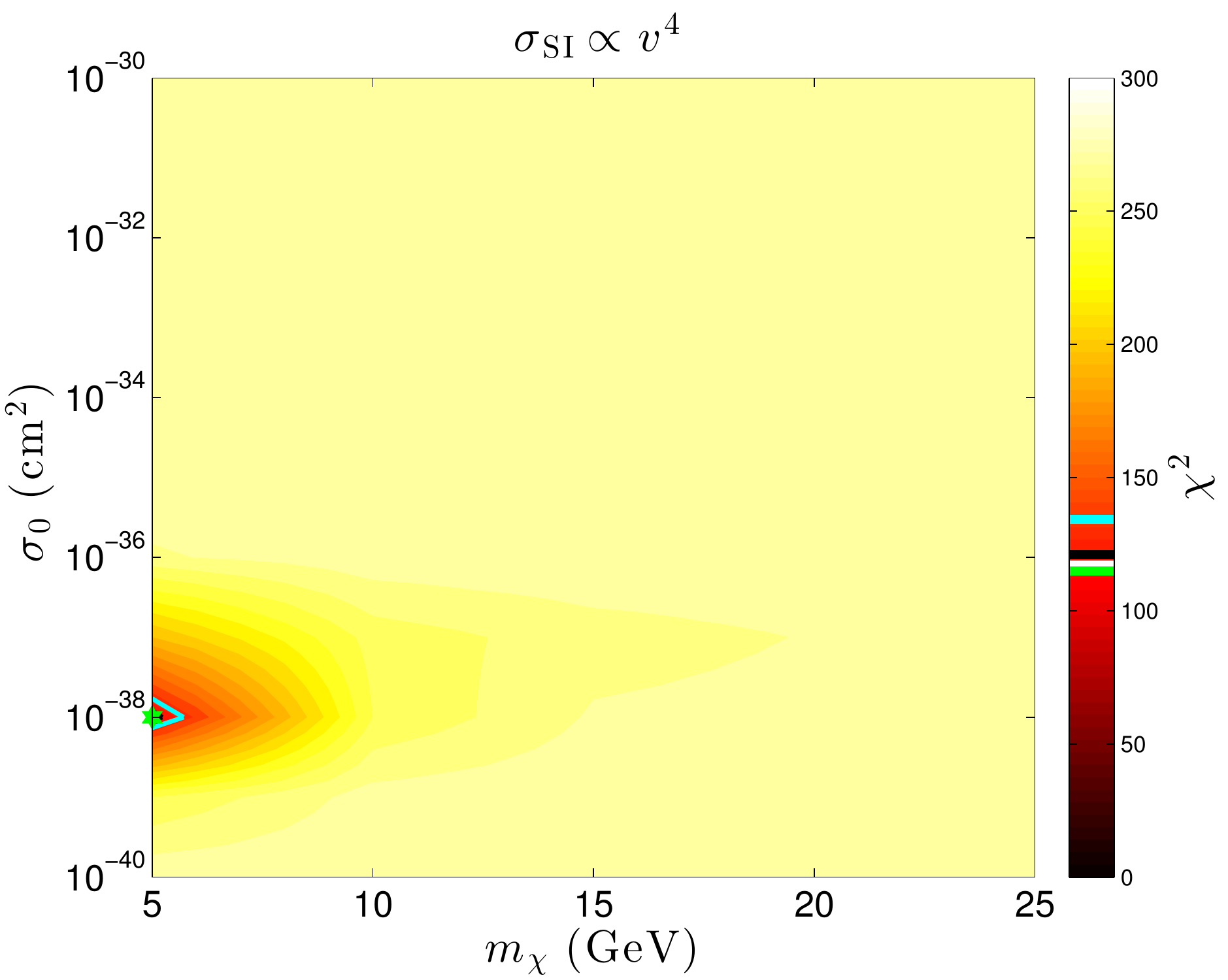} & \includegraphics[height = 0.32\textwidth]{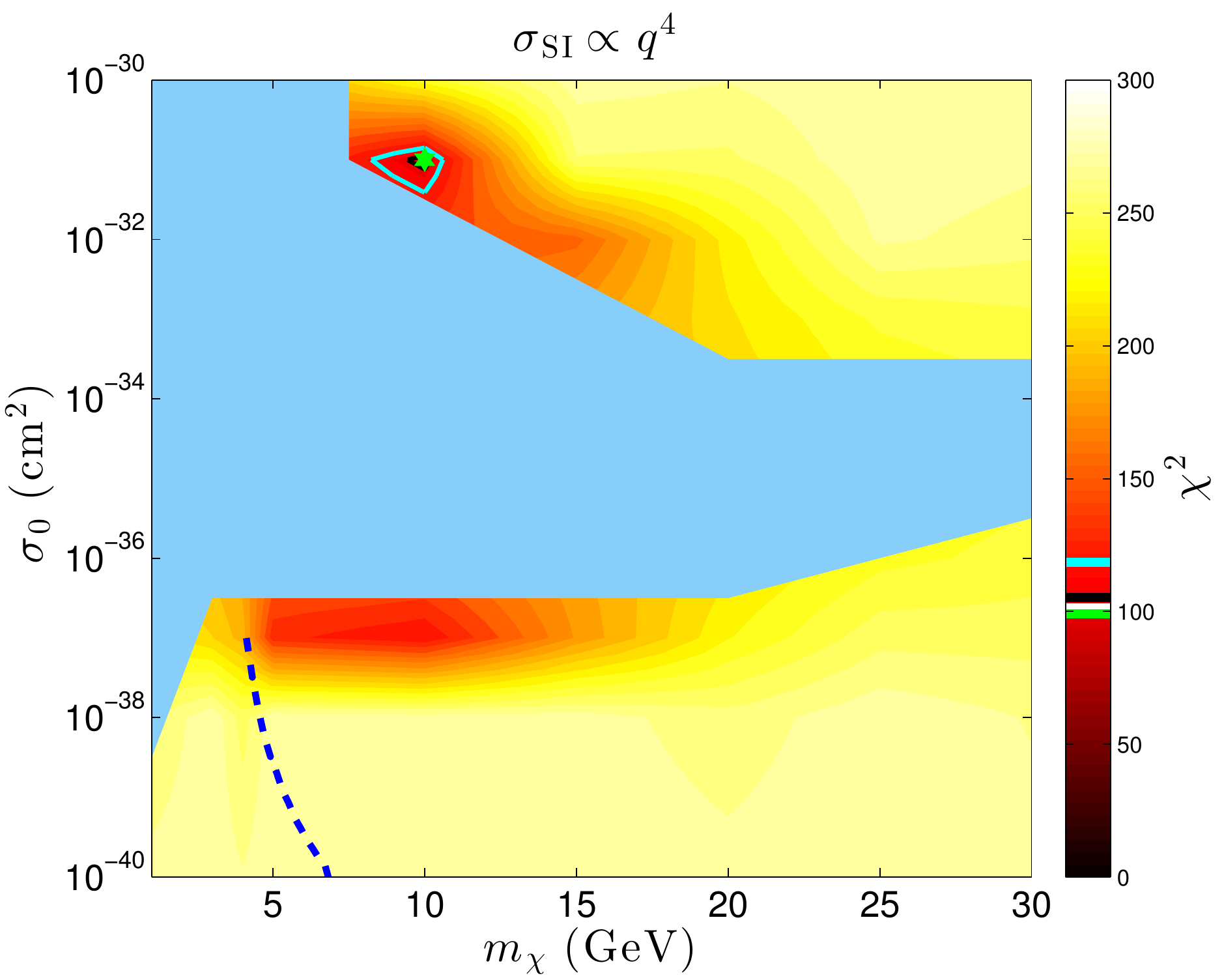} \\
\end{tabular}
\caption{Full combined likelihood (Eq.\ \ref{eq:fullchisq}) incorporating $^8$B and $^7$Be neutrino flux measurements, the surface helium abundance $Y_{\rm S}$, the radius of the convection zone and the small frequency separations. The $\chi^2$ value of the best fit point is shown as a green star, and indicated by a green line on the colour bar. $\Delta \chi^2 = 2.3$, 6.18 and 19.33 contours corresponding to 1$\sigma$,  2$\sigma$ and 4$\sigma$ deviations from the best fit are shown in white, black and cyan, respectively.}
\label{SIchisq}
\end{figure}

\begin{figure}[p]
\begin{tabular}{c@{\hspace{0.04\textwidth}}c}
\multicolumn{2}{c}{\includegraphics[height = 0.32\textwidth]{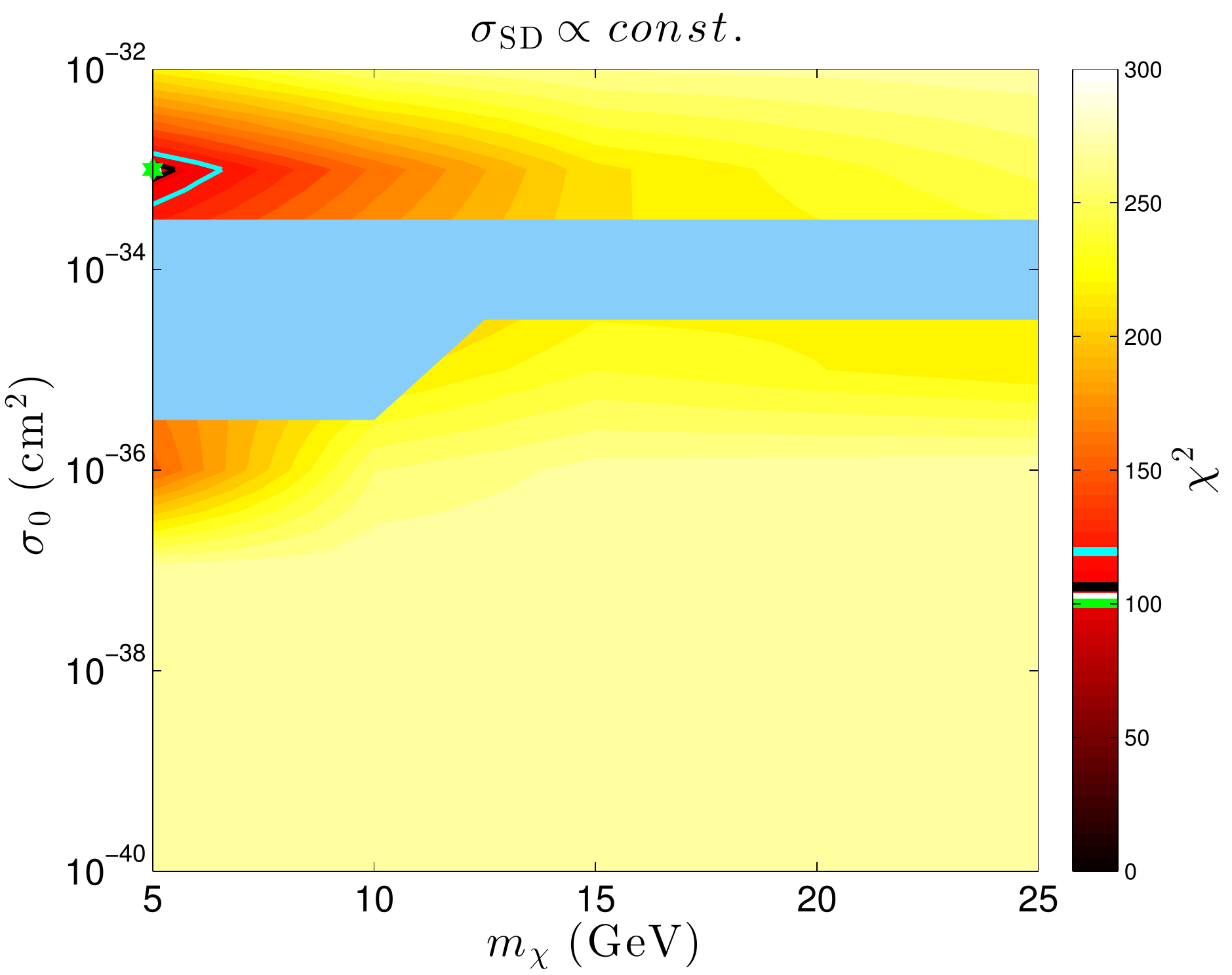}} \\
\includegraphics[height = 0.32\textwidth]{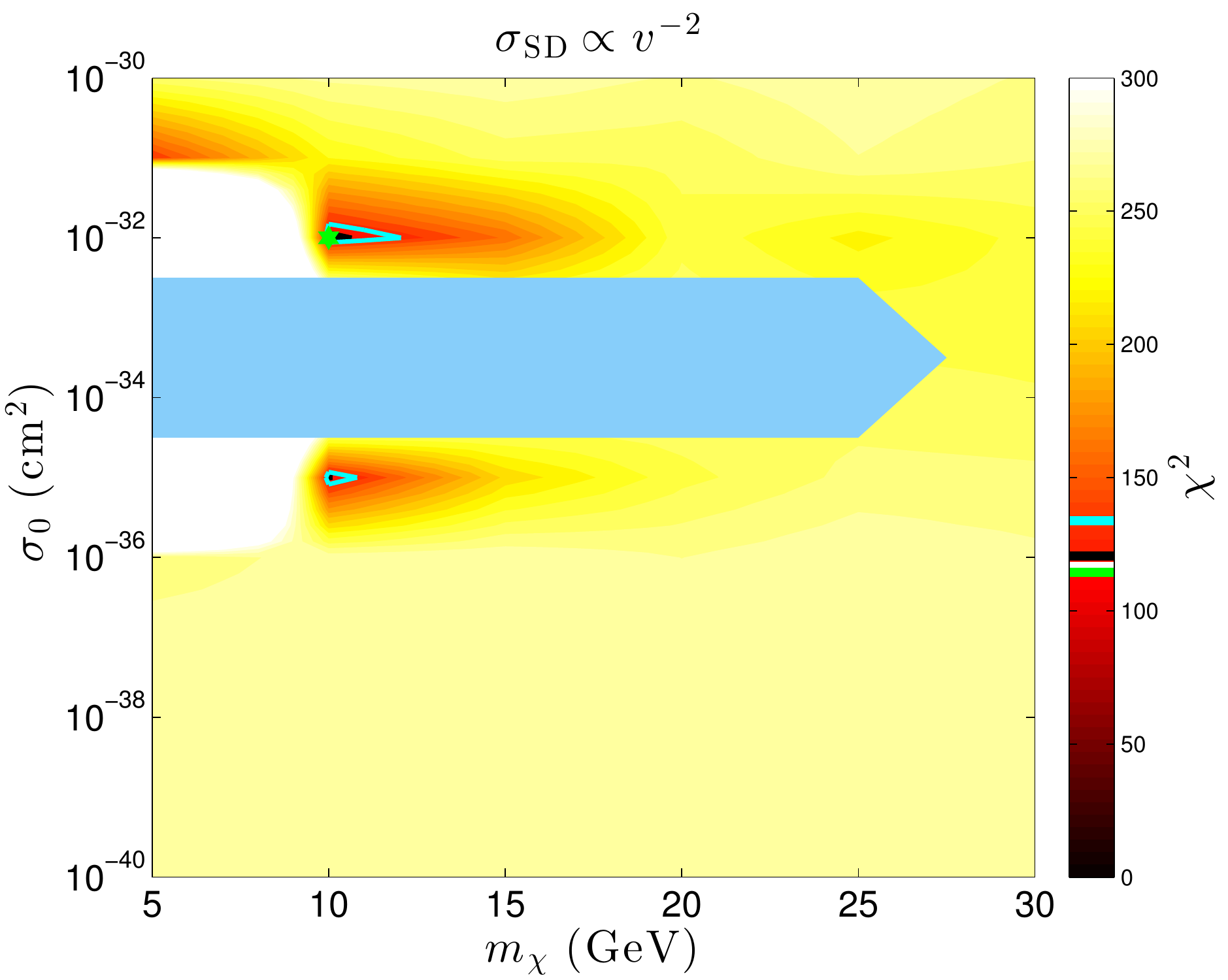} & \includegraphics[height = 0.32\textwidth]{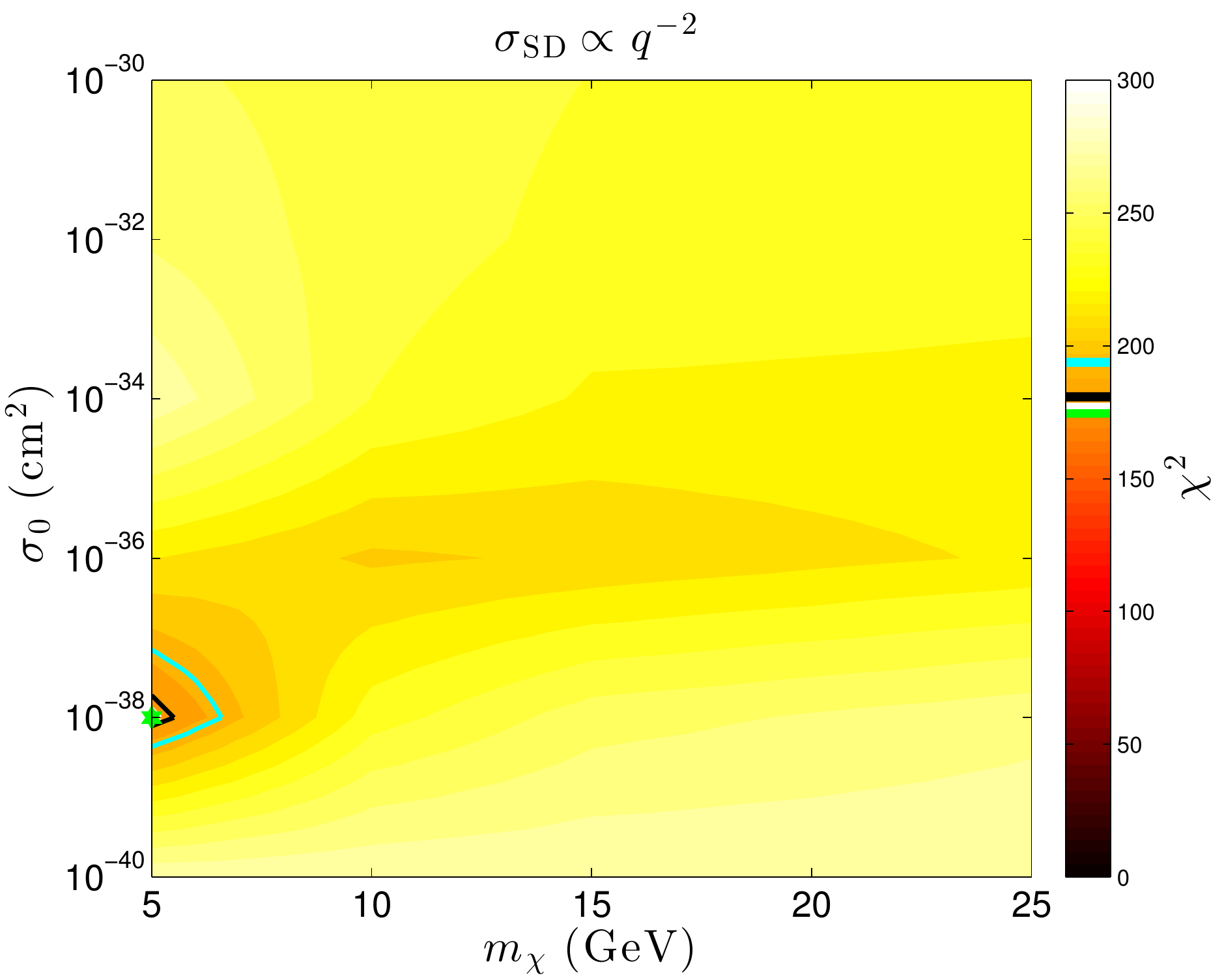} \\
\includegraphics[height = 0.32\textwidth]{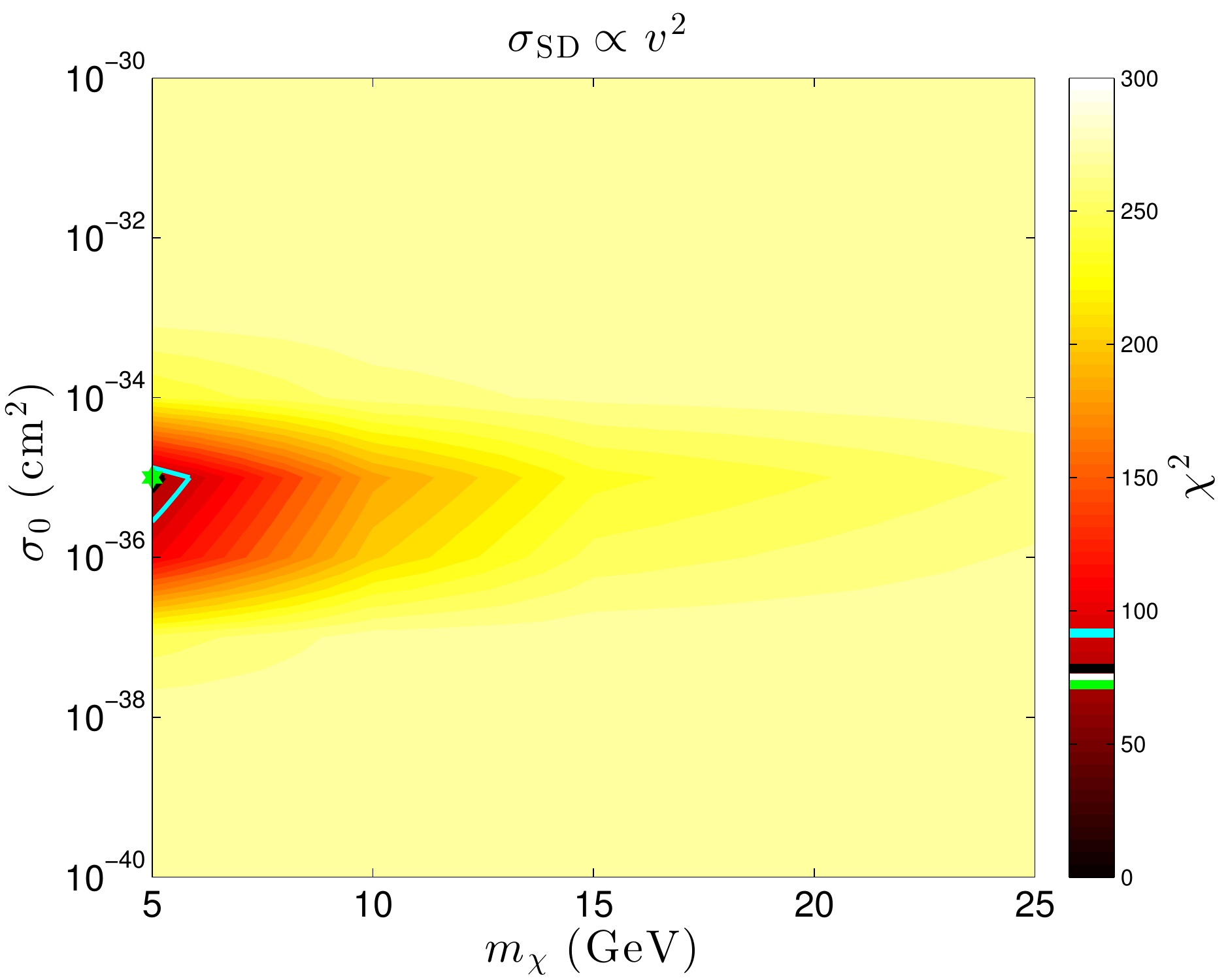} & \includegraphics[height = 0.32\textwidth]{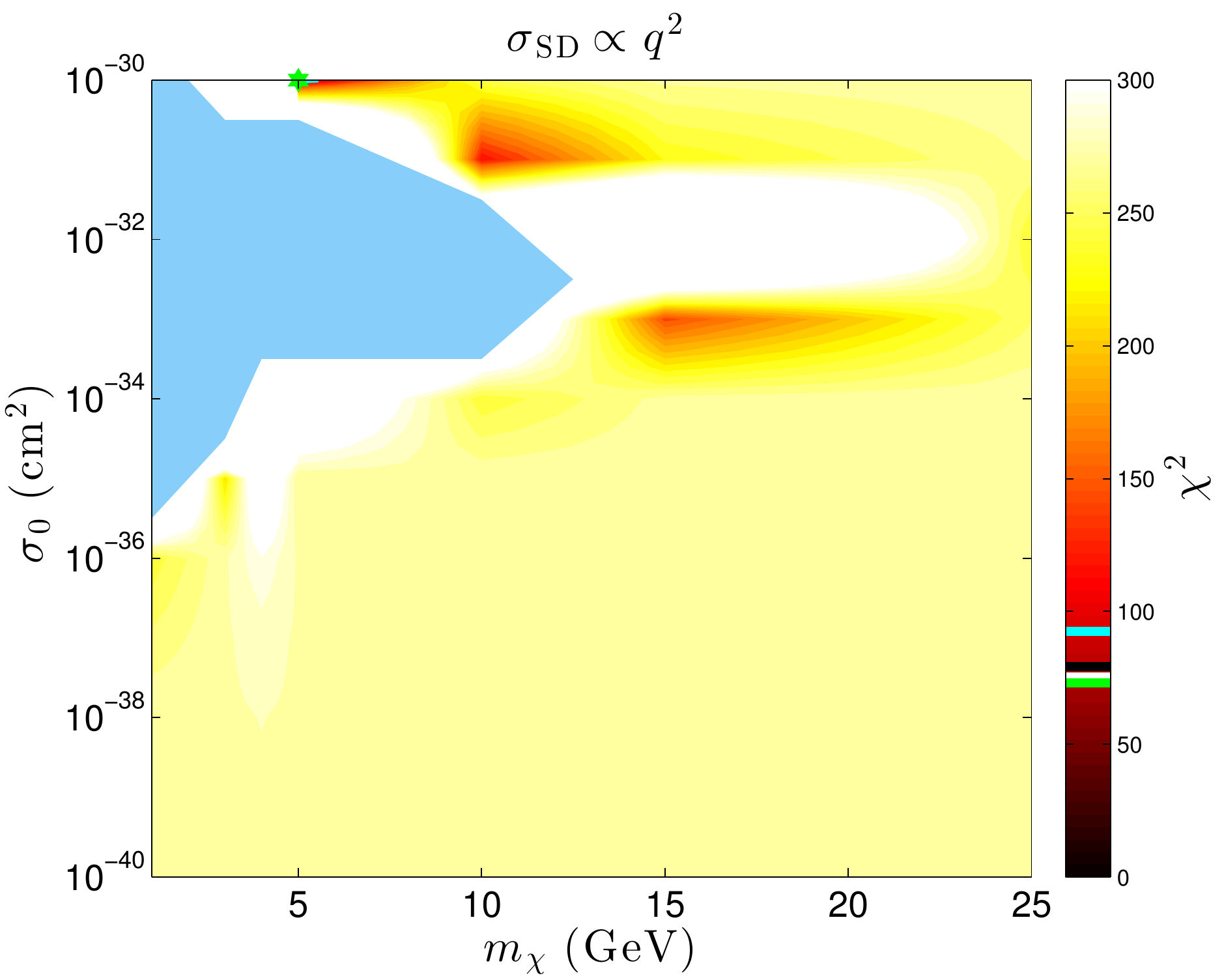} \\
\includegraphics[height = 0.32\textwidth]{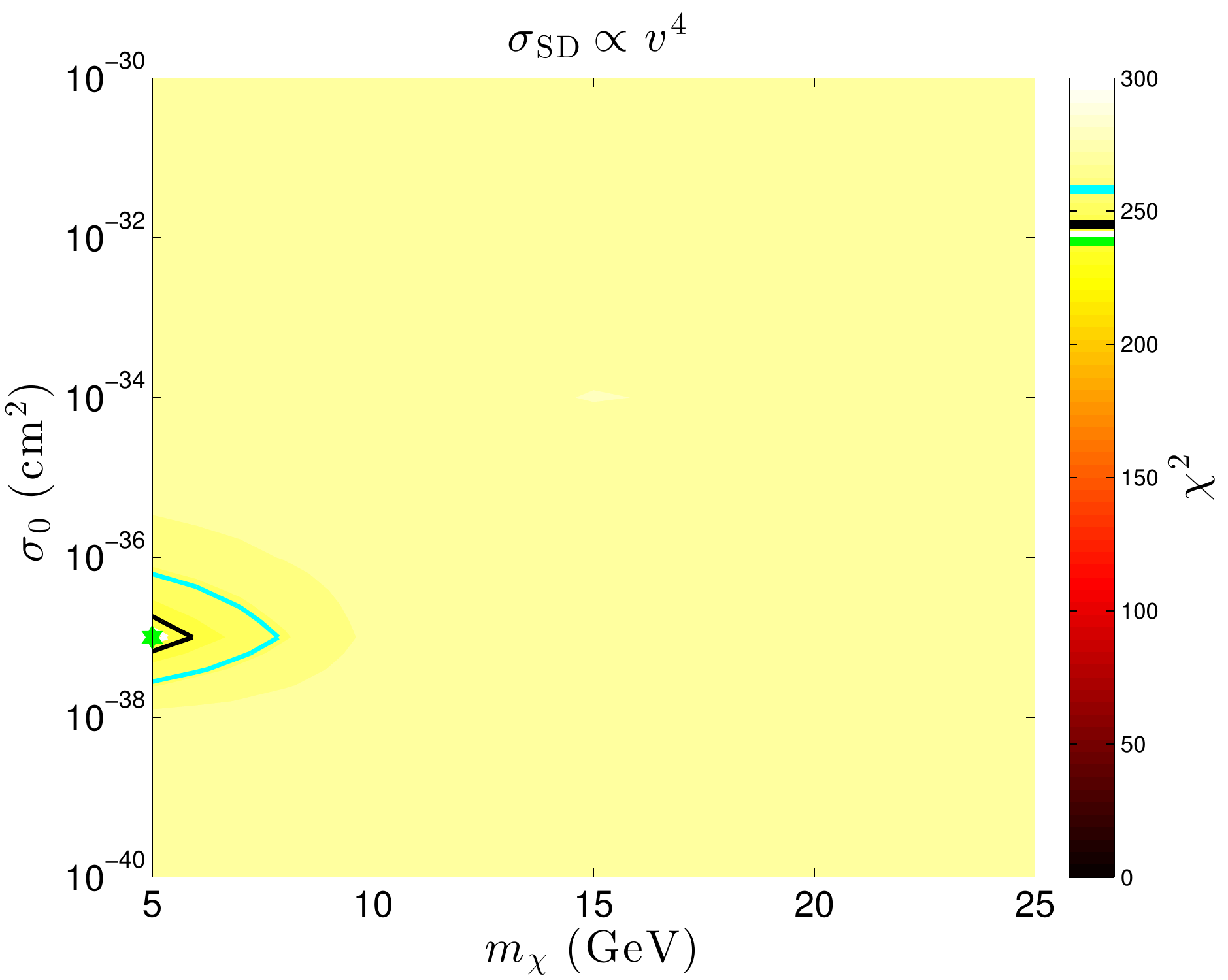} & \includegraphics[height = 0.32\textwidth]{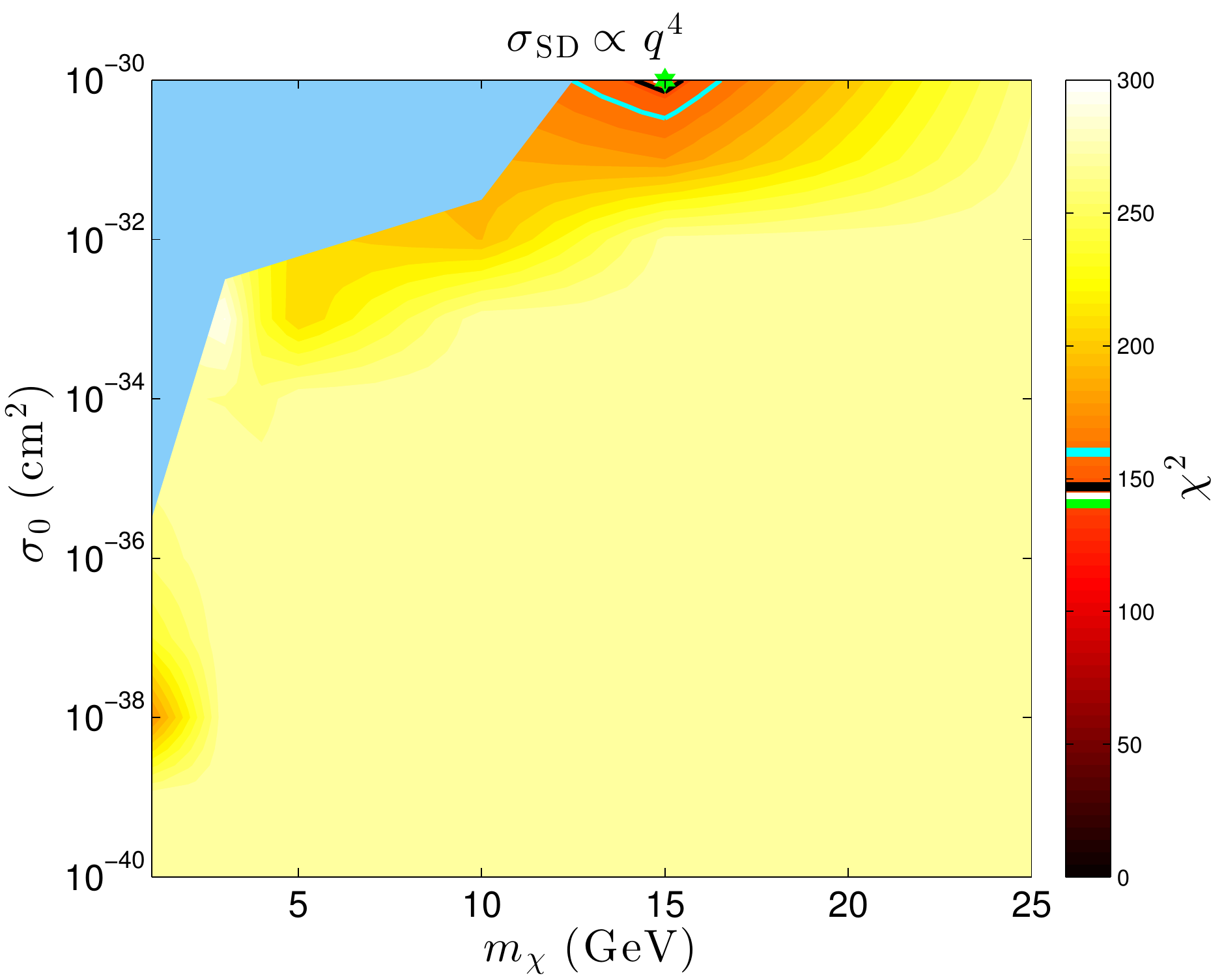} \\
\end{tabular}
\caption{Same as Fig.\ \ref{SIchisq}, but for spin-dependent couplings.  }
\label{SDchisq}
\end{figure}

We show the combined limits from Eq.\ \ref{eq:fullchisq} in Figs.\ \ref{SIchisq} and \ref{SDchisq}, comparing with limits from direct detection where such results exist (i.e.\ in the SI case; \cite{Guo2013}).  We discuss these results, and the comparison with other limits, in more detail in the following section.

\FloatBarrier

\begin{table}[ht]
\caption{Standard Solar Model (SSM) and best fit (b.f.) values for each of the models we consider, along with observable quantities. The DM mass and cross-sections are in GeV and cm$^2$, respectively. The boron-8 neutrino flux $\phi^\nu_{\mathrm{B}}$ is in units of $10^{-6}$ cm$^{-2}$ s$^{-1}$ and the beryllium-7 neutrino flux $\phi^\nu_{\mathrm{Be}}$ is expressed in $10^{-9}$ cm$^{-2}$ s$^{-1}$. The full chi-squared is defined in Eq.\ \ref{eq:fullchisq} and includes the neutrino fluxes, surface helium abundance $Y_S$, depth of the convection zone and small frequency separations.}
\label{best_fit_tab}
\vspace{1mm}
\begin{tabular*}{\textwidth}{| c |  @{\extracolsep{\fill}} l | r c c c c r r |}
\hline
\multicolumn{2}{|c|}{Model} & $(m_\chi, \sigma_0)_{b.f.}$ & $\phi^\nu_{\mathrm{B}}$ & $\phi^\nu_{\mathrm{Be}}$ & $R_{CZ}/R_\odot$ & $Y_s$ & $\chi^2$ & $p$ \phantom{100} \\ \hline \hline
\multicolumn{2}{|c|}{SSM} 		    & -- \phantom{100} & 4.95  &4.71   & 0.722  & 0.2356 & 287.8 & $ < 10^{-10}$ \\ \hline %HR
\multirow{7}{*}{$\sigma_{\rm SI}  $} &$ const.$ & (15,$10^{-37}$) & 3.48 & 4.37 & 0.721 & 0.2348 & 122.2 &$ < 10^{-10}$  \\
							&$ q^{-2}$ & (3,$10^{-38}$) & 4.51 & 4.59 & 0.721 & 0.2350 & 144.1 & $ < 10^{-10}$  \\
							&$ q^{2}$ & (3, $10^{-37}$) & 3.78 & 4.29 & 0.718 & 0.2327 & 27.5 & 0.85\phantom{100} \\ %HR
							&$ q^{4}$ & (10, $10^{-31}$) & 3.29 & 4.27 & 0.721 & 0.2348 & 98.2 & $1.1 \times 10^{-7\phantom{0}}$\\ %HR
							&$ v^{-2}$ & (5,$10^{-34}$) & 3.91 & 4.44 & 0.721 & 0.2347 & 86.4 & $5.0 \times 10^{-6\phantom{0}}$ \\
							&$ v^{2}$ & (5, $10^{-38}$) & 3.31 & 4.22 & 0.719 & 0.2337 & 135.0 & $ < 10^{-10}$ \\ %HR
							&$ v^{4}$ & (5,$10^{-38}$) & 4.30 & 4.54 & 0.721 & 0.2349 & 114.9 & $3.6\times 10^{-10}$ \\ \hline
\multirow{7}{*}{$\sigma_{\rm SD}  $} &$ const.$ & (5,$10^{-33}$) & 3.36 & 4.27 & 0.720 & 0.2341 & 100.2 & $5.8\times10^{-8\phantom{0}}$ \\
							&$ q^{-2}$ & (5,$10^{-38}$) & 4.43 & 4.59 & 0.722 & 0.2353 & 174.5 & $ < 10^{-10}$  \\
							&$ q^{2}$ & (5,$10^{-30}$) & 3.64 & 4.35 & 0.720 & 0.2343 & 73.1 & $2.5 \times 10^{-4\phantom{0}}$ \\
							&$ q^{4}$ & (15,$10^{-30}$) & 3.81 & 4.47 & 0.721 & 0.2351 & 140.7 & $ < 10^{-10}$  \\
							&$ v^{-2}$ & (10,$10^{-32}$) & 3.39 & 4.33 & 0.721 & 0.2347 & 114.5 & $4.2\times10^{-10}$ \\
							&$ v^{2}$ & (5,$10^{-35}$) & 3.75 & 4.39 & 0.721 & 0.2344 & 72.4 & $3.1\times10^{-4\phantom{0}}$ \\
							&$ v^{4}$ & (5,$10^{-37}$) & 4.80 & 4.68 & 0.722 & 0.2355 & 238.7 & $ < 10^{-10}$  \\ \hline \hline
\multicolumn{2}{|c|}{Obs.} 			& -- & 5.00  & 4.82 & 0.713  & 0.2485   & -- & -- \\ \hline
\multicolumn{2}{|c|}{Obs. error} 		& -- &  3 \% & 5\%  & 0.001 &  0.0034 & -- & -- \\ 
\multicolumn{2}{|c|}{Model error} 	& -- &  14 \% & 7\%  &  0.004 & 0.0035 & -- & -- \\ \hline
\hline
\end{tabular*}
\end{table}
In Table \ref{best_fit_tab}, we give a break-down of the values of each observable at the best-fit parameters of each model, along with the best-fit $\chi^2$ and implied $p$-value.  The best-fit regular SI and SD cases shown in Table\ \ref{best_fit_tab} differ from the best-fit cases in Ref.\ \cite{Vincent2014} because here we actually choose them on the basis of the full likelihood Eq.\ \ref{eq:fullchisq}.  In Ref.\ \cite{Vincent2014} we only computed the small-frequency separations for models that provided the best fits to all other observables.  Because the fits to different observables are inconsistent in the models with constant cross-sections, including the frequency separations shifts the best-fit masses and cross-sections.  This is to be compared to the $q^2$ SI case, where adding in the small frequencies merely makes the best-fit point even better.  

We also provide as supplementary material online a complete table of the boron and beryllium neutrino fluxes, surface helium, convective zone radius and small frequency separations for all of our models, in addition to the partial and total chi-squared values defined in Eq.\ \ref{eq:fullchisq}. This table can be used for quick lookup and interpolation for, e.g.\ global fits of DM models.

\section{Discussion}
\label{sec:discussion}

\subsection{Solving the Solar Abundance Problem}
\label{sec:bestcases}

Our results of Section \ref{sec:results} indicate that the impacts of different types of generalised form factor DM on solar observables is quite varied, and not always straightforward. What is clear, however, from Figs.\ \ref{SIchisq} and \ref{SDchisq} and Table\ \ref{best_fit_tab}, is that some models do indeed lead to a much better overall fit than the Standard Solar Model alone. Although there is always some reduction in the goodness of fit to observed neutrino fluxes, this can be accommodated in many cases by the theoretical errors. 

In Fig.\ \ref{csfig} we showed the reduction of the discrepancy between simulated and measured sound speed profiles of the best-fit models of the couplings that give the best overall improvement compared to the Standard Solar Model.  On the left, we showed the best models for constant spin-dependent and spin-independent cross-sections, and on the right we showed the best-fit models of the three best generalised form factor couplings, according to the likelihood defined in Sec.\ \ref{sec:combined}. We find that the best improvement comes from a scattering cross-section that is proportional to the square of the transferred momentum $q$. This is confirmed to very high significance by the small frequency separations, seen in Fig.\ \ref{SIr0213}.  At low masses $m_\chi \sim 3-5$ GeV, an asymmetric DM candidate with $\sigma = \sigma_0 (q/40 \mathrm{ MeV})^2$ and $\sigma_0 = 10^{-37}$ cm$^2$ leads to a substantial improvement with respect to the SSM. The best fit occurs at 3\,GeV, and leads to a convective zone depth $r_{CZ} = 0.718 R_\odot$, only 1.2$\sigma$ away from the inferred value from helioseismology. The $^8$B and $^7$Be fluxes are respectively $3.78 \times 10^6$ \,cm$^{-2}$s$^{-1}$ and $4.29 \times 10^9$ \,cm$^{-2}$s$^{-1}$, both within $2\sigma$ of the measured values. Surface helium remains low, at $Y_S = 0.233$. The large improvement to the fit to $c_s(r)$ can be seen in the right-hand panel of Fig.\ \ref{csfig}, and to the small frequency separations probing the solar core in Fig.\ \ref{fig:smallfreq}. This specific case yields an overall 6$\sigma$ improvement compared to the Standard Solar Model, as we discussed explicitly in Ref. \cite{Vincent2014}.  

In the next section, we will discuss constraints from other sectors, and show that the best-fit $q^2$ model does indeed constitute a viable candidate for reconciling solar modelling, helioseismology and spectroscopy.  It is worth noting, however, that the masked regions of non-convergence in our contour plots may hide parameter combinations that yield even better fits than this; careful work on the numerical solver is required to explore this possibility. 

Other $q$-dependent models, as well as constant and velocity-dependent models we examined, nearly all led to poor overall fits: the effect on $\phi^\nu$ and $Y_S$ is always too large, in spite of the improved fit to helioseismological measurements. Every other ADM model we examined led to some significant improvement over the SSM, but failed to provide $p$-values greater than 10$^{-3}$. As pointed out in Ref.\ \cite{Lopes14}, a $\vrel^{-2}$ SD cross-section does indeed lead to an improved fit to the sound speed profile and $r_{CZ}$ for a small region in mass and cross-section space (see Figs. \ref{SDrc} and \ref{SDchisqf_csf}). Unfortunately, this case is disfavoured by our overall likelihood, mainly due to excessive reduction in the core temperature, which is reflected in a greater-than $3\sigma$ reduction in the $^8$B neutrino fluxes, as well as significant reduction in $^7$Be neutrinos.  In fact, we find that a $\vrel^{2}$ SD cross-section yields a much more plausible overall fit to observables than $\vrel^{-2}$ SD scattering, resulting in a $p$-value of 3.1 $\times 10^{-4}$ as compared to 4.2 $\times 10^{-10}$ (vs. 0.85 for the $q^2$ SI model). It is also important to note that the region where $\vrel^{-2}$ SD improves helioseismological agreement is very close to the boundary beyond which our models failed to converge. This issue, also seen by the authors of Ref. \cite{Taoso10}, could indeed be hiding a better overall fit that is simply beyond the reach of our solver. Given the great reduction in neutrino fluxes that we observe in the bordering area however, we remain somewhat sceptical of this possibility.

\subsection{Comparison with collider and direct limits}
\label{sec:otherconstraints}

The dashed curves in Fig.\ \ref{SIchisq} show independent limits from direct detection on SI scattering, as derived in Ref.\ \cite{Guo2013}.  To the best of our knowledge, similarly model-independent limits on $q$ and $\vrel$-dependent SD couplings from direct detection are not available in the literature.  For momentum-dependent couplings, the direct limits suggest that all interesting masses (from the point of view of solar physics) above $\sim$5\,GeV are in tension with direct searches.   Curiously though, the areas that lead to the best overall improvements in solar observables ($q^n$ scattering, $m_\chi\sim3$--5\,GeV, $\sigma_{\rm SI}\sim10^{-37}$--$10^{-38}$\,cm$^2$) are some of the few that survive the direct search limits.  For velocity-dependent and constant couplings, the curves lie almost entirely below the plotted regions, indicating that in these cases, the entire section of parameter space that turns out to have an impact on solar physics is disfavoured.  

It is worth remembering, however, that limits from direct detection and solar physics probe very different parts of the dark matter halo velocity distribution: solar capture preferentially occurs from the low-velocity end of the distribution, whereas direct detection probes the high-velocity tail.  The compatibility of solar and direct searches is therefore highly sensitive to the chosen velocity distribution, and might be modified substantially by choosing something other than the standard Maxwell-Boltzmann distribution.

Limits on both momentum-dependent and velocity-dependent DM-quark couplings also exist from colliders \cite{Goodman10, Bai10, Fox12a, Cheung12}, although in this case a specific effective operator or UV-complete model must be assumed in order to calculate definite event rates.  In the classification scheme of \cite{Kumar13}(\cite{Goodman10}), operators F2(D2) and F10(D10) have leading contributions to the nuclear recoil rate from $q^2$ SI terms, operator F6(D6) from $\vrel^2$ SI terms, operators F3(D3), F6(D6) and S2(C2/R2) from $q^2$ SD terms, operator F4(D4) from $q^4$ SD terms, and operator S4(C4) from $\vrel^2$ SD terms.  For the most part, ATLAS and CMS have not focused on such operators due to the weakness of the limits on them from direct detection.  Limits from Tevatron data have been calculated by CDF \cite{CDF} on the combination of F4(D4) and the `regular' constant SI scalar coupling $\bar\chi\chi\bar{q}q$ (F1/D1), but so far only a few phenomenological groups \cite{Goodman10,Cheung12} have placed limits on the other operators using up-to-date collider data.  

From the point of view of solar physics, given the preference we see for $q^2$ SI scattering in ameliorating the Solar Abundance Problem in Fig.\ \ref{SIchisq}, operators F2(D2) and F10(D10) would seem to be the most interesting.  Existing limits on F10(D10) from ATLAS data \cite{Cheung12} are quite stringent at $m_\chi$ below a few hundred GeV, excluding mass scales of below a TeV for the physics associated with the particle mediating the interaction.  Such limits would likely make achieving our preferred reference cross-section rather difficult.  Limits on F2(D2) are much weaker though, with only mass scales below 42\,GeV excluded at $m_\chi<50$\,GeV (assuming that the contact operator approximation even remains valid in this regime).  It therefore seems feasible that a DM particle interacting with quarks by this effective coupling ($\bar\chi\gamma_5\chi\bar{q}q$) might explain the recent anomalies seen in some direct detection experiments, and help rectify the known discrepancies between theory and data in solar physics.

\subsection{Comparison with previous work}
\label{sec:comparison}
The effect of DM heat conduction in the Sun, especially by asymmetric dark matter, has been studied in several works. Most recently, the main qualitative features of such models were pointed out by \cite{Frandsen:2010yj} while accurate computations using solar models were done in Refs.\ \cite{Cumberbatch:2010hh,Lopes:2012,Taoso10}, with somewhat contradictory conclusions.  In this section we will endeavour to settle the differences. Among these studies, only \cite{Taoso10} used the proper thermal conduction mechanism of Gould and Raffelt \cite{GouldRaffelt90a}. Our energy transport and $^8$B neutrino flux computations are in good agreement with that work. However, Ref \cite{Taoso10} did not include the (small) effect of molecular settling, nor did they account for the uncertainty in the initial helium mass fraction $Y$, metallicity $Z$ or mixing length parameter $\alpha$, all of which are free parameters of solar models. Allowing $Y$, $Z$ and $\alpha$ to vary has the effect of partially compensating for the increased luminosity from DM transport, thus reducing the overall effect of DM and leading us to weaker overall constraints. 

\begin{figure}[t]
\begin{tabular}{c c}
\includegraphics[width=0.5\textwidth]{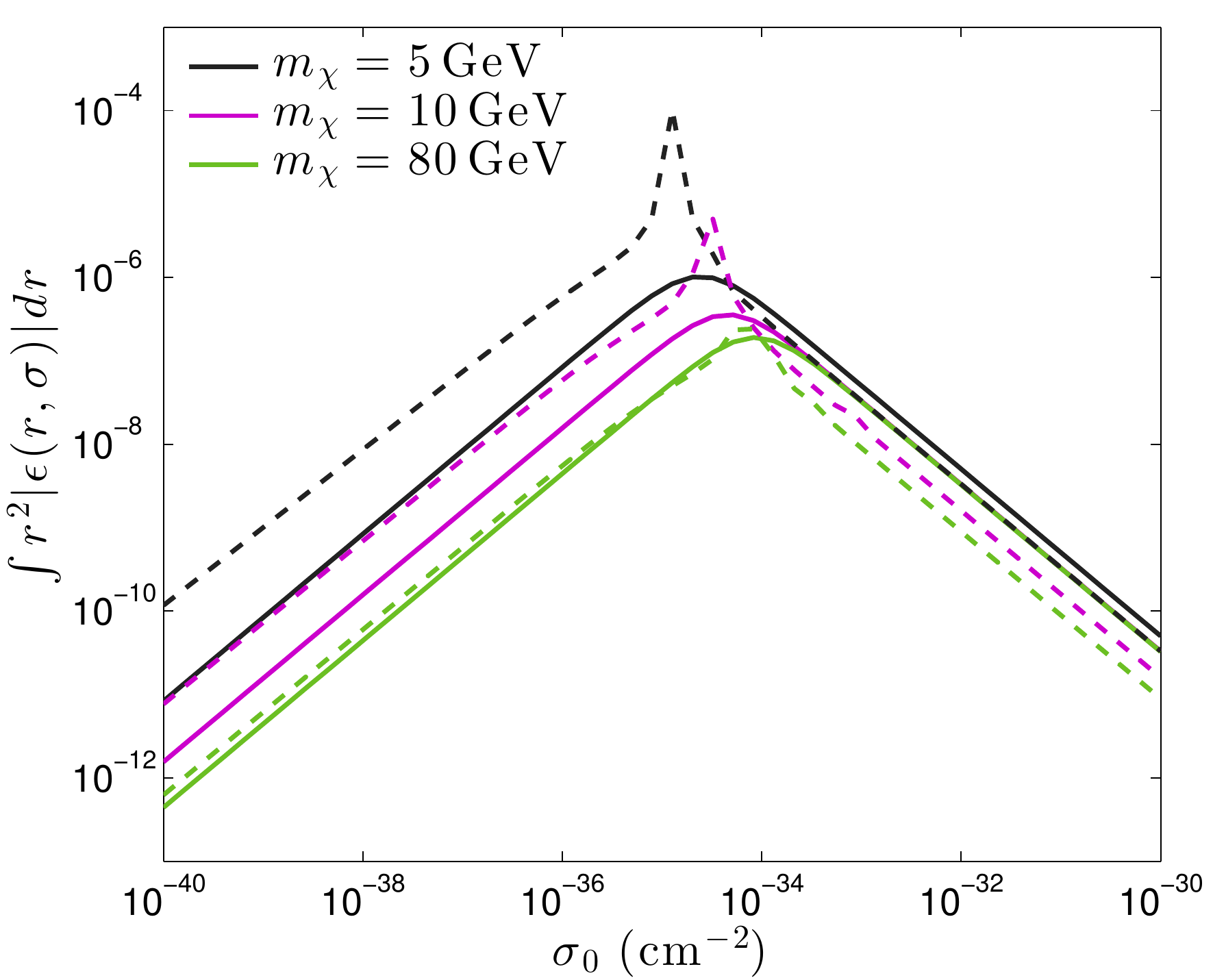}&\includegraphics[width=0.56\textwidth]{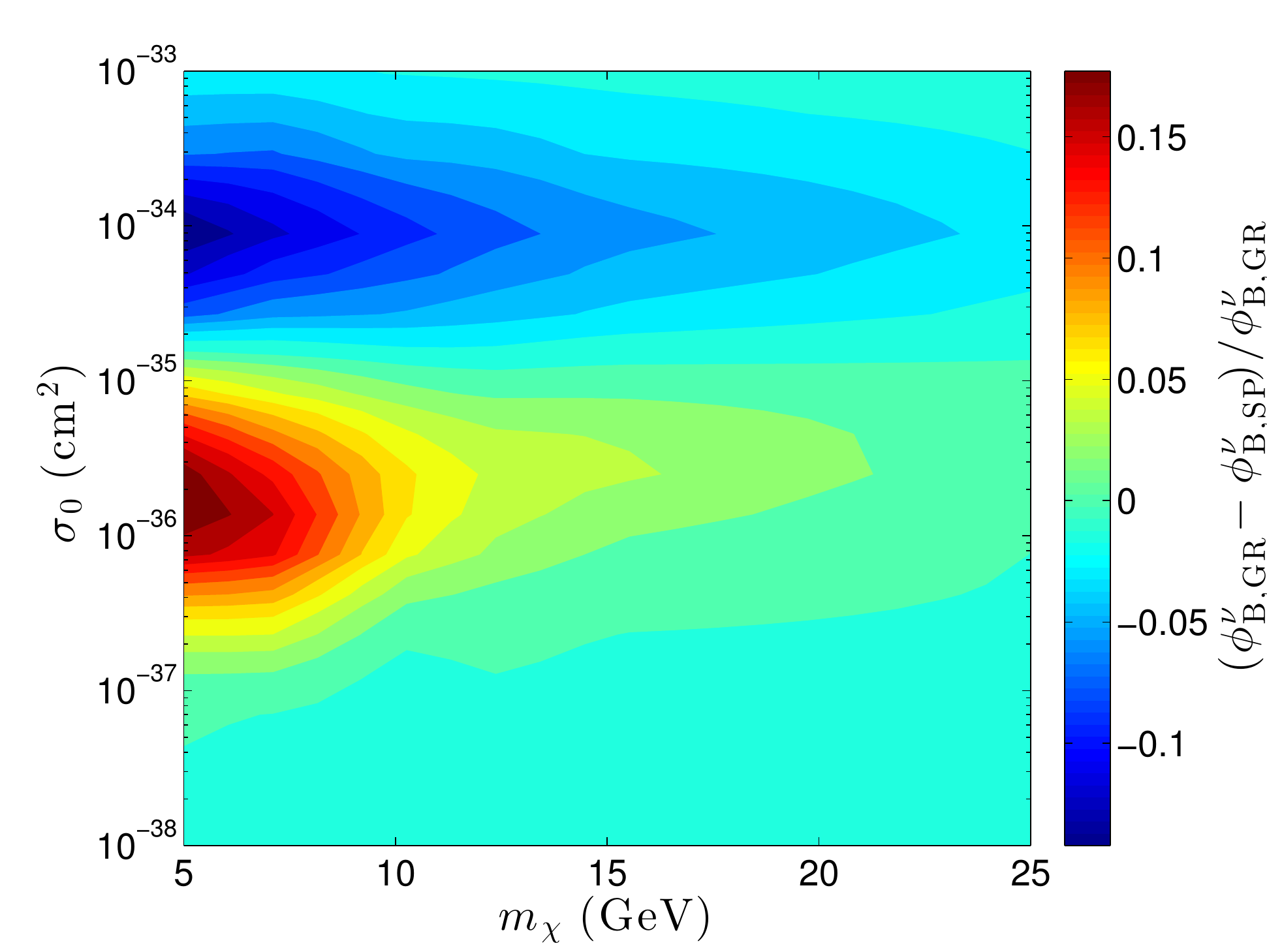}
\end{tabular}
\caption{Comparison between the Gould and Raffelt (GR) transport formalism and the incorrect Spergel and Press (SP) formalism, for a constant, spin-dependent cross-section. Left: Total energy transported by asymmetric dark matter from GR (solid lines) and SP (dashed lines) for a constant ratio of DM to baryons $n_\chi/n_{\rm b} = 10^{-15}$ in a standard solar model. The SP formalism, which was shown to be flawed by \cite{GouldRaffelt90a,GouldRaffelt90b}, overpredicts the transported energy by as much as a factor of 9 in the non-local (Knudsen) regime, while underestimating it in the LTE case. Right: comparison between predicted $^8$B fluxes once these transport mechanisms are implemented into the full \DS solar capture code. In the red region, using SP yields a larger reduction in the neutrino flux than GR. }
\label{fig:GRSP}
\end{figure}

In contrast, Refs. \cite{Cumberbatch:2010hh,Lopes:2012} used the approach developed by Spergel and Press (SP, \cite{Spergel85}). However, comparison with Monte Carlo simulations found this approach  to be flawed \cite{GouldRaffelt90b}. This is due to the violation of all three assumptions that go into the SP approach: 1) a Maxwellian velocity distribution; 2) spatial uniformity of the WIMP velocity distribution; and 3) local isotropy. To further quantify the errors introduced by the use of the SP treatment, we have implemented the SP transport mechanism into our code, following  Ref.~\cite{Lopes02a} for a constant, spin-dependent (SD) cross-section. We plot the total energy transport for a given DM population in the Sun as a function of the spin-dependent cross-section in the left-hand panel of Fig.~\ref{fig:GRSP}. Solid lines represent the accurate Gould and Raffelt (GR) approach (which we use in this paper), whereas dashed lines are the SP calculation. As in Fig.~\ref{fig:knudsen}, the peak energy transport occurs at the transition between local (LTE) and non-local (Knudsen) transport. Although both approaches roughly yield the same location for this peak, SP slightly underestimates transport in the LTE regime, whereas it significantly overpredicts transport in the non-local regime. In the large cross-section LTE regime, the SP case predicts DM luminosities that are 1.5 (for $\mx = 5$\,GeV) to 4 ($\mx = 80$\,GeV) times too small; in the Knudsen limit, this becomes an overestimate by a factor of 3 ($\mx = 80$\,GeV) to 9 ($\mx = 5$\,GeV). It is worth noting that the largest discrepancies happen very near the location of interest, where low masses and cross-sections potentially lead to effects in the Sun and simultaneously evade constraints from direct detection. In the right-hand panel of Fig.~\ref{fig:GRSP} we show the resulting discrepancy between predicted $^8$B neutrino fluxes with SP and GR transport, as computed with our solar model. The red region is where SP overpredicts an effect, whereas blue corresponds to the LTE regime where SP underestimates the energy conduction.

Two other recent papers \cite{Lopes:2014,Lopes14} have also investigated the impacts of specific dark matter models with non-constant cross-sections on solar physics.  In both these papers, the authors employed existing capture and transport schemes (SP in \cite{Lopes:2014} and GR in \cite{Lopes14}) without properly accounting for the dependence of the capture and transport processes on the kinematic structure of the cross-section.  Ref.\ \cite{Lopes:2014} examined a magnetic dipole dark matter model with a cross-section that scales as $\vrel^2q^{-2}$ relative to the standard constant case.\footnote{Here we remind readers that even the `constant' case has $\mathrm{d}\sigma/\mathrm{d}q^2 \varpropto \vrel^{-2}$, i.e.\ whenever we refer to a $q^n$ or $\vrel^n$ scaling, we mean \textit{relative} to this familiar case.}  They then applied the `momentum transfer cross-section', which amounts to multiplying the cross-section by $(1-\cos\theta)$ in order to cancel the divergence induced by forward scattering when integrating over all scattering angles to obtain a total cross section.  This process is only useful for obtaining a rough scattering rate, as it simply transmutes a momentum-dependence into an equivalent $\vrel$-dependence (see e.g.\ calculations of $\kappa$ for $q^{-2}$ couplings in \cite{VincentScott2013} for details).  The correct treatment is of course to include the full momentum- and velocity-dependent differential cross-section in the integral over the DM velocity distribution function.  In the case of Ref.\ \cite{Lopes:2014}, this treatment simply converted a $\vrel^2q^{-2}$ cross-section to a $\vrel^2\vrel^{-2}=constant$ cross-section, effectively ignoring the momentum and velocity-dependence of the model.  Similarly in Ref.\ \cite{Lopes14}, where the application of the momentum transfer cross-section treatment converted what was actually a $\vrel^2q^{-4}$ cross-section in the long-range limit into a $\vrel^2\vrel^{-4}=\vrel^{-2}$ cross-section, relative to the standard case.  The authors then took a thermal average over the velocity distribution and applied it as a constant scaling to the standard capture and transport treatments.  As we have shown previously \cite{VincentScott2013}, this is valid for the capture rate (or would be, if the cross-section were truly dependent only on $\vrel$ and not $q$), but qualitatively incorrect when dealing with conductive energy transport.  

In light of these shortcomings in the way capture and transport were modelled, and in the theoretical treatment of the scattering cross-sections involved, it is difficult to estimate how the results of Refs.\ \cite{Lopes:2014,Lopes14} would change were the kinematics of the cross-sections treated correctly.  Calculating the transfer coefficients as per Ref.\ \cite{VincentScott2013} and carrying out the full phenomenological analysis as we have here, for $\vrel^2q^{-2}$ and $\vrel^2q^{-4}$ couplings, would therefore be an interesting topic for future study.

\section{Conclusions}
\label{sec:conclusion}
In this paper, we have explored the prospects for solving the Solar Abundance Problem using generalised form factor dark matter, confirming that it can indeed provide a solution to this long-standing issue \cite{Vincent2014}.  In the process, we derived the full capture rates in the Sun for dark matter with momentum- or velocity-dependent interactions with nucleons, and used solar physics to place limits on those couplings.  This is the first extensive, physically-consistent analysis of the effects on the Sun of momentum- or velocity- dependent dark matter, taking into account the full capture kinematics, along with the self-consistent thermal conduction framework developed in Ref. \cite{VincentScott2013}. We incorporated this framework into a new state-of-the art dark solar evolution code, \DS.

Our best-fit model, asymmetric DM of $m_\chi \sim 3$ GeV and $\sigma_0 = 10^{-37}$ cm$^2$ interacting with nucleons proportionally to $q^2$, shows excellent promise for solving the Solar Abundance Problem. Our model provides a remarkable fit, and a full investigation of the effects of evaporation, which has not been done before for non-constant cross-sections, should serve as a cross-check on the consistency of such a scenario. The prospects for detecting such a particle at the next round of direct detection experiments, and in the next run of the LHC, also appear very promising.

\section*{Acknowledgements}
We thank Wan-lei Guo for providing us with numerical values of the XENON constraints for generalised form factor DM models. This work was done with partial support from NSERC (Canada), FQRNT (Qu\'ebec), STFC (UK) and European contracts FP7-PEOPLE-2011-ITN, PITN-GA-2011-289442-INVISIBLES, ESP2013-41268-R, ESP2014-56003-R (MINECO) and 2014SGR-1458 (Generalitat de Catalunya). Calculations were performed on SOM2 at IFIC (Universitat de Val\`encia -- CSIC) funded by PROMETEO/2009/116, PROMETEOII/2014/050 and FPA2011-29678.
\appendix

\bibliographystyle{JHEP_pat}
\bibliography{CandO,CObiblio,AbuGen,solarDM}

\end{document}